\documentclass[12pt]{article}

\usepackage{amsfonts}
\usepackage{amsmath}
\usepackage{amssymb}
\usepackage{array}
\usepackage{bigints}
\usepackage{booktabs}
\usepackage{chet_mod}  
\usepackage{color}
\usepackage{dsfont}
\usepackage{float}
\usepackage{framed}
\usepackage{graphicx}
\usepackage{indentfirst}

\usepackage{marginnote} 

\reversemarginpar 

\usepackage{mathrsfs}
\usepackage{multirow}
\usepackage{pdflscape}
\usepackage{setspace}
\usepackage{subdepth}
\usepackage{subfig}
\usepackage{titlesec}
\usepackage[dotinlabels]{titletoc}
\usepackage{wrapfig}
\usepackage[all]{xy}
\usepackage{young}
\usepackage[vcentermath]{youngtab}
\usepackage{relsize}
\usepackage{stackengine}
\usepackage{verbatim}
\usepackage{slashed}

\usepackage{hyperref}
\hypersetup{colorlinks=true}
\hypersetup{linkcolor=black}
\hypersetup{citecolor=black}
\hypersetup{urlcolor=black}

\def\tilde{\widetilde}
\def\t{\tilde}
\def\hat{\widehat}

\def\bar{\overline}
\def\b{\bar}

\def\epsilon{\ep}
\def\ep{\varepsilon}
\def\half {{1 \over 2}}

\def\d{\partial}

\def\grad{\nabla}
\def\mod{{\text{mod}}}

\def\tr{\mathop{\text{tr}}}

\def\SL{{\mathscr L}}
\def\SP{{\mathscr P}}

\def\CA{{\cal A}}
\def\CB{{\cal B}}
\def\CC{{\cal C}}

\def\CI{{\cal I}}
\def\CJ{{\cal J}}
\def\CK{{\cal K}}
\def\CL{{\cal L}}
\def\CM{{\cal M}}
\def\CN{{\cal N}}
\def\CO{{\cal O}}

\def\CS{{\cal S}}

\def\1{\hbox{$\mathds 1$}}
\def\R{\hbox{$\mathbb R$}}
\def\C{\hbox{$\mathbb C$}}

\def\Z{\hbox{$\mathbb Z$}}

\def\P{\hbox{$\mathbb P$}}

\title{Exploring 2-Group Global Symmetries}

\author{
Clay C\'{o}rdova,$^{1}$ Thomas T.~Dumitrescu,$^2$ and Kenneth Intriligator\hskip1pt$^3$
}

\affiliation{$^1$~{\it School of Natural Sciences, Institute for Advanced Study, Princeton, NJ 08540, USA}

$^2$~{\it Department of Physics, Harvard University, Cambridge, MA 02138, USA}

$^3$~{\it Department of Physics, University of California, San Diego, La Jolla, CA 92093, USA}
}

\abstract{
\vskip-30pt
\noindent We analyze four-dimensional quantum field theories with continuous 2-group global symmetries. At the level of their charges, such symmetries are identical to a product of continuous flavor or spacetime symmetries with a 1-form global symmetry~$U(1)^{(1)}_B$, which arises from a conserved 2-form current~$J_B^{(2)}$. Rather, 2-group symmetries are characterized by deformed current algebras, with quantized structure constants, which allow two flavor currents or stress tensors to fuse into~$J_B^{(2)}$. This leads to unconventional Ward identities, which constrain the allowed patterns of spontaneous 2-group symmetry breaking and other aspects of the renormalization group flow. If~$J_B^{(2)}$ is coupled to a 2-form background gauge field~$B^{(2)}$, the 2-group current algebra modifies the behavior of~$B^{(2)}$ under background gauge transformations. Its transformation rule takes the same form as in the Green-Schwarz mechanism, but only involves the background gauge or gravity fields that couple to the other 2-group currents. This makes it possible to partially cancel reducible 't Hooft anomalies  using Green-Schwarz counterterms for the 2-group background gauge fields. The parts that cannot be cancelled are reinterpreted as mixed, global anomalies involving~$U(1)_B^{(1)}$, which receive contributions from topological, as well as massless, degrees of freedom. Theories with 2-group symmetry are constructed by gauging an abelian flavor symmetry with suitable mixed 't Hooft anomalies, which leads to many simple and explicit examples. Some of them have dynamical string excitations that carry~$U(1)_B^{(1)}$ charge, and 2-group symmetry determines certain 't Hooft anomalies on the world sheets of these strings. Finally, we point out that holographic theories with 2-group global symmetries have a bulk description in terms of dynamical gauge fields that participate in a conventional Green-Schwarz mechanism.  
}

\date{February 2018}

\begin{document}


\maketitle 


\toc


\newsec{Introduction}[SecIntro]

In this paper we discuss four-dimensional quantum field theories (QFTs) with continuous global symmetries, such as ordinary flavor symmetries or Poincar\'e symmetry. As explained in~\cite{Gaiotto:2014kfa}, there are also generalized~$q$-form\foot{~Throughout, we use a superscript~$(q)$ in parentheses to indicate a~$q$-form.} global symmetries~$U(1)_B^{(q)}$, which arise from conserved~$(q+1)$-form currents~$J_B^{(q+1)}$. The objects that carry charge under such symmetries are~$q$-dimensional defect operators and dynamical~$q$-brane excitations. We will mostly focus on the case~$q =1$, i.e.~on~$U(1)_B^{(1)}$ global symmetries that act on line operators and strings. We will explore theories in which the associated conserved 2-form current~$J_B^{(2)}$ appears in the operator product expansion (OPE) of two 1-form flavor currents or two stress tensors. For reasons explained below, we refer to this structure, which mixes~$U(1)_B^{(1)}$ with flavor or Poincar\'e symmetries at the level of their current algebras, as a 2-group global symmetry. We will see that~2-group symmetries occur in many QFTs, including simple and familiar ones, such as massless quantum electrodynamics (QED) with multiple flavors.  

After reviewing the conserved currents and background gauge fields associated with flavor, $q$-form, and Poincar\'e symmetries (section~\ref{convgsbf}), we give a detailed introduction to continuous 2-group symmetries. We first present them from the perspective of their background gauge fields, by analogy with the Green-Schwarz (GS) mechanism (sections~\ref{gsfbgf} and~\ref{genintro}), before discussing 2-group current algebras and the associated Ward identities (section~\ref{twgpcurwi}). In section~\ref{rgphtwgp}, we summarize constraints on the allowed patterns of spontaneous 2-group symmetry breaking, as well as other aspects of renormalization group (RG) flows with 2-group symmetry, that follow from the 2-group Ward identities. 

't Hooft Anomalies for 2-group symmetries are discussed in section~\ref{gscttwgpano}. There, certain GS contact terms in two-point functions of~$J_B^{(2)}$ with abelian flavor currents, which are intimately related to GS counterterms for the 2-group background gauge fields, play a crucial role. These GS contact terms are four-dimensional analogues of the Chern-Simons contact terms that were analyzed in~\cite{Closset:2012vg,Closset:2012vp}. Some simple examples of QFTs with 2-group symmetry are summarized in section~\ref{introex}, among them multi-flavor QED, a topological quantum field theory (TQFT) with 2-group symmetry, and a theory with spontaneous 2-group breaking. In section~\ref{relwork}, we mention other work related to 2-group global symmetries in QFT, most of which focuses on discrete 2-groups. There we also summarize some results about continuous 2-group symmetries in six spacetime dimensions, which will appear in~\cite{our6dtoapp}.

The introduction gives a detailed overview of sections~\ref{SecHooft} through~\ref{SecExamples} of the paper. Some further aspects of continuous 2-group symmetries are discussed in section~\ref{secfurthasp}, including their gauging, their holographic duals, and their implications for dynamical string excitations that carry~$U(1)_B^{(1)}$ charge. This section is largely self contained and can (for the most part) be read after the introduction. Additional material appears in several appendices.

Given the length of the paper, we also mention that some familiarity with 't~Hooft anomalies (reviewed in section~\ref{SecHooft}) makes it possible to read sections~\ref{SecIntro}, \ref{SecBasics}, and~\ref{SecExamples} in sequence.

\subsec{Global Symmetries and Background Fields}[convgsbf]

We consider continuous global symmetries that arise from conserved currents.\foot{~In many situations, the existence of such currents follows from Noether's theorem.} The currents encode local features of a symmetry, such as Ward identities or 't Hooft anomalies, that are typically not visible at the level of its global charges. A useful way to access this local information is to couple the theory to non-dynamical background gauge fields, which act as classical sources for the currents. We begin by reviewing some examples of continuous global symmetries, as well as the associated currents and background gauge fields.

\begin{itemize}

\item[1.)] {\it Flavor (or 0-form) symmetries.} These are associated with a Lie group~$G$. We will also refer to them as~$0$-form global symmetries (see below), and denote them by~$G^{(0)}$. Our main example will be~$G^{(0)} = U(1)_A^{(0)}$. The associated 1-form current~$j_A^{(1)}$ (or~$j_\mu^A$, if we write out the spacetime indices), satisfies the conservation equation
\eqn{
d * j_A^{(1)} = 0~.
}[jaconsfirst]
Conserved charges~$Q_A(\Sigma_{3})$ are defined by integrating~$* j_A^{(1)}$ over~$3$-cycles~$\Sigma_{3}$, 
\eqn{
Q_A(\Sigma_{3}) = \int_{\Sigma_{3}} * j_A^{(1)}~.
}[chargedef]
The charged objects are local operators, which can be surrounded by a closed 3-cycle~$\Sigma_{3}$ in euclidean signature, and point particles, which reside on~$\Sigma_{3}$, if we think of it as a time slice in hamiltonian quantization. 

\smallskip

The appropriate classical source for~$j_A^{(1)}$ is an abelian background gauge field~$A^{(1)}$,\foot{~\label{ft:swactdef} We will use~$S = S[\CB, \psi]$ to denote the euclidean action, which can include background fields~$\CB$ and dynamical fields~$\psi$. The partition function~$Z[\CB]$, which only depends on the background fields, is given by the functional integral~$\int D\psi \, \exp(-S[\CB, \psi])$ over all dynamical fields~$\psi$. Note that~\eqref{sajcoup} means that insertions of~$j_\mu^A(x)$ are given by~$-{\delta Z \over \delta A^\mu(x)}$. By a mild abuse of language, we will refer to~$W[\CB] = - \log Z[\CB]$ as the effective action for background fields, even though it is typically non-local.}
\eqn{
S \; \supset \; \int A^{(1)} \wedge * j_A^{(1)} = \int d^4 x \, A^\mu j_\mu^A~. 
}[sajcoup]
Under a~$U(1)_A^{(0)}$ background gauge transformation, with 0-form gauge parameter~$\lambda_A^{(0)}$, the gauge field~$A^{(1)}$ shifts as follows,
\eqn{
A^{(1)} \quad \longrightarrow \quad A^{(1)} + d\lambda_A^{(0)}~.
}[atrans]
In the absence of t' Hooft anomalies (discussed below), the effective action~$W[A^{(1)}]$ (see footnote~\ref{ft:swactdef}) is invariant under~\atrans. This encodes the conservation equation~\jaconsfirst. 

\smallskip

The statement that the flavor symmetry is~$U(1)_A^{(0)}$ (rather than~$\R^{(0)}_A$) means that all charges~$Q_A$ in~\chargedef\ are integers. We can therefore take the gauge parameter~$\lambda_A^{(0)}$ in~\atrans\ to be compact, $\lambda_A^{(0)} \sim \lambda_A^{(0)} + 2 \pi$, so that~$ {1 \over 2 \pi} \int_{\Sigma_1} d \lambda_A^{(0)} \in \Z$ for any closed 1-cycle~$\Sigma_1$. Similarly, the~$U(1)_A^{(0)}$ background field strength~$F_A^{(2)} = d A^{(1)}$ can have arbitrary integer fluxes through closed 2-cycles~$\Sigma_2$, 
\eqn{
{1 \over 2\pi} \int_{\Sigma_2} F_A^{(2)}  \in \Z~.
}[fsquant]

\smallskip

More generally, the flavor symmetry~$G^{(0)}$ can be a compact Lie group. In this paper, we will limit our discussion to flavor symmetries of the form~$G^{(0)} = \prod_I G^{(0)}_I$, where every factor~$G^{(0)}_I$ is either~$U(1)^{(0)}$ or~$SU(N)^{(0)}$. The associated currents and background gauge fields are then valued in the Lie algebra of~$G^{(0)}$, and transform appropriately under the corresponding, possibly nonabelian, background gauge transformations.

\item[2.)] {\it Generalized $q$-form global symmetries} (see~\cite{Gaiotto:2014kfa} and references therein). A~$U(1)^{(q)}_B$ symmetry arises from a conserved~$(q+1)$-form current~$J_B^{(q+1)}$ (i.e.~$J^B_{\mu_1 \cdots \mu_{q+1}}$) satisfying 
\eqn{
d * J_B^{(q+1)} = 0~.
}[jqcons]
The conserved charges~$Q_B(\Sigma_{3-q})$ are now defined on~$(3-q)$-cycles~$\Sigma_{3-q}$,
\eqn{
Q_B(\Sigma_{3-q}) = \int_{\Sigma_{3-q}} * J_B^{(q+1)}~.
}[qqdef]
Charged defect operators are supported on~$q$-cycles linked by~$\Sigma_{3-q}$, and charged dynamical~$q$-brane excitations extend along~$q$ spatial dimensions transverse to~$\Sigma_{3-q}$.\foot{~These concepts are ubiquitous in the context of supersymmetry algebras and BPS branes (see~\cite{Dumitrescu:2011iu} and references therein).}

\smallskip

The appropriate classical source for~$J_B^{(q+1)}$ is an abelian~$(q+1)$-form gauge field~$B^{(q+1)}$,
\eqn{
S \; \supset \; \int B^{(q+1)} \wedge * J_B^{(q+1)} = {1 \over q!} \int d^4 x \, B^{\mu_1 \cdots \mu_{q+1}} J^B_{\mu_1 \cdots \mu_{q+1}}~.
}[bjbcoup]
A~$U(1)_B^{(q)}$ background gauge transformation, with~$q$-form gauge parameter~$\Lambda_B^{(q)}$, shifts
\eqn{
B^{(q+1)} \quad \longrightarrow \quad B^{(q+1)} +  d\Lambda_B^{(q)}~.
}[bgtshift]
The invariance of the effective action~$W[B^{(q+1)}]$ (see footnote~\ref{ft:swactdef}) under this shift is tantamount to the conservation law~\jqcons. 

\smallskip

Saying that the symmetry is~$U(1)^{(q)}_B$ (rather than~$\R_B^{(q)}$) implies that all charges~$Q_B$ in~\qqdef\ are integers, so that both~$d \Lambda^{(q)}_B$ and the gauge-invariant~$U(1)_B^{(q)}$ field strength $d B^{(q+1)}$ can have arbitrary integer fluxes through~$(q+1)$- and~$(q+2)$-cycles,
\eqn{
{1 \over 2 \pi} \int_{\Sigma_{q+1}} d \Lambda^{(q)}_B \in \Z~, \qquad {1 \over 2 \pi} \int_{\Sigma_{q+2}} d B^{(q+1)} \in \Z~.
}[LamHflux]

\smallskip

Note that the case~$q = 0$ is a standard abelian flavor symmetry. This case is special because it admits a nonabelian generalization (see point 1.) above). By contrast, $q$-form symmetries with~$q \geq 1$ are necessarily abelian~\cite{Gaiotto:2014kfa}. 

\smallskip

In this paper, we will mostly focus on the case~$q = 1$, i.e.~on 1-form symmetries~$U(1)_B^{(1)}$ that arise from conserved~$2$-form currents~$J_B^{(2)}$. The charged objects are line defects and dynamical strings. The main example will be the magnetic~$U(1)_B^{(1)}$ symmetry of a dynamical~$U(1)_c^{(0)}$ gauge theory with Maxwell field strength~$f^{(2)}_c$ and no dynamical magnetic~$U(1)_c^{(0)}$ charges. Then~$f^{(2)}_c$ satisfies the Bianchi identity, $d f^{(2)}_c = 0$, which implies the conservation equation~\jqcons\ for the magnetic 2-form current,\foot{~\label{ft:wickroti} The factor of~$i$ in the definition~$J_B^{(2)} = {i \over 2\pi} \, * f^{(2)}_c$ arises from the Wick rotation to euclidean signature. In lorentzian signature, $J_B^{(2)} = {1 \over 2\pi} *_\text{L} f^{(2)}_c$, where~$*_\text{L}$ is the lorentzian Hodge star operator, which satisfies~$*_\text{L}^2 = -1$ when acting on 2-forms.}
\eqn{
J^{(2)}_B = {i \over 2 \pi} \, * f^{(2)}_c~.
}[magcurrint]
The defects charged under~$U(1)_B^{(1)}$ are 't Hooft lines. Examples of charged dynamical excitations are Abrikosov-Nielsen-Olesen (ANO) strings.

\item[3.)] {\it Poincar\'e symmetry ($\SP$).} The associated current is the stress tensor~$T_{\mu\nu}$, which must be symmetric and conserved,
\eqn{
T_{\mu\nu} = T_{(\mu\nu)}~, \qquad \d^\mu T_{\mu\nu} = 0~.
}[tcons]
The appropriate classical sources for~$T_{\mu\nu}$ are background gravity fields, such as a riemannian background metric~$g_{\mu\nu}$ on the spacetime 4-manifold~$\CM_4$. We will describe the background gravity fields using an orthonormal frame (or vielbein)~$e^{(1)a} $, so that~$g_{\mu\nu} = \delta_{ab} e^a_\mu e^b_\nu$,  and the associated spin connection~${\omega^{(1)a }}_b$. Here~$a, b = 1, \ldots, 4$ are local frame indices.\foot{~In lorentzian signature, they are often referred to as local Lorentz indices.} An insertion of the stress tensor~$T_{\mu\nu} = e^a_\mu e^b_\nu T_{ab}$ is defined by a variational derivative of the partition function~$Z[e^{(1)a}]$ with respect to the vielbein,
\eqn{
\sqrt g \, T_{ab}(x) = - e_{b\mu} \, {\delta Z \over \delta e_\mu^a(x)}~.
}[stdef]
The gauge transformations of the background gravity fields consist of local~$SO(4)$ frame rotations (i.e.~Wick-rotated local Lorentz transformations), and diffeomorphisms (i.e.~local translations). Infinitesimally, they are parametrized by~${\delta^a}_b + {\theta^{(0)a}}_b(x) \in SO(4)$ and a vector field~$\xi^\mu(x)$. Under these transformations, the vielbein shifts as follows,
\eqn{
e^{(1)a} \quad \longrightarrow \quad e^{(1)a} - {\theta^{(0)a}}_b \, e^{(1)b} + \CL_\xi e^{(1)a}~,
}[vielbeingt]
where~$\CL_\xi$ is the Lie derivative along the vector field~$\xi^\mu$. Invariance of the effective action~$W[e^{(1)}] = - \log Z[e^{(1)}]$ (see footnote~\ref{ft:swactdef}) under~\vielbeingt\ encodes the fact that the stress tensor is symmetric and conserved, as in~\tcons.

\end{itemize}

As emphasized in~\cite{Gaiotto:2014kfa}, there are many similarities between ordinary~$0$-form flavor symmetries and higher~$q$-form symmetries. In addition to those that are apparent from the review above, we recall the following additional parallels:
\begin{itemize}
\item A continuous~$q$-form symmetry may be unbroken or spontaneously broken by the vacuum. In the latter case there are massless Nambu-Goldstone (NG) bosons. For 0-form flavor symmetries, these are the familiar abelian or nonabelian NG scalars; for~$q \geq 1$, they are suitable higher-spin particles.\foot{~As discussed in~\cite{Gaiotto:2014kfa}, only~$0$- and~$1$-form symmetries can be spontaneously broken in four dimensions.}  For instance, the NG boson for a spontaneously broken 1-form symmetry is a free photon. In the deep IR, it is described by a free~$U(1)_c^{(0)}$ gauge theory with Maxwell field strength~$f^{(2)}_c$. As in~\eqref{magcurrint}, the 2-form current~$J_B^{(2)}$ is linear in~$f^{(2)}_c$ and creates a one-photon state.

\item 't Hooft anomalies manifest as c-number shifts of the effective action~$W[A^{(1)}, B^{(q+1)}]$ under the gauge transformations~\eqref{atrans} and~\eqref{bgtshift} of the various background gauge fields. Consequently, they also modify the conservation equations~\eqref{jaconsfirst} and~\eqref{jqcons} in the presence of such background fields. This constitutes an obstruction to gauging the symmetry (see below). 't Hooft anomalies are subject to matching: they do not change along RG flows and must be reproduced in any effective description of the theory, e.g.~in terms of ultraviolet (UV) or infrared (IR) degrees of freedom. 

\item If a~$q$-form symmetry is free of 't Hooft anomalies, it can be gauged by promoting the background gauge field~$B^{(q+1)}$ to a dynamical gauge field~$b^{(q+1)}$ and doing the functional integral over gauge orbits of~$b^{(q+1)}$.\foot{~Throughout, we denote background fields by uppercase letters and dynamical fields by lowercase letters.} 

\item $q$-form global symmetries can be emergent, accidental symmetries in the IR, even if they are explicitly broken in the UV. This is standard for ordinary flavor symmetries. Another example is the emergent magnetic 1-form symmetry (with current~\eqref{magcurrint}) that arises upon higgsing a dynamical nonabelian gauge theory to a~$U(1)_c^{(0)}$ subgroup. 

\end{itemize}

All of these statements also apply to Poincar\'e symmetry~$\SP$, but we will only consider relativistic continuum QFTs, for which~$\SP$ is an exact symmetry at all energies. We will also assume the existence of a Poincar\'e-invariant vacuum, so that~$\SP$ is not spontaneously broken. We will encounter 't Hooft anomalies involving Poincar\'e symmetry, and the associated gravity background fields, but we will not contemplate making these fields dynamical.

\subsec{2-Group Symmetry:~a Green-Schwarz Mechanism for Background Fields}[gsfbgf]

The background gauge fields reviewed in section~\ref{convgsbf} do not mix under their respective background gauge transformations~\eqref{atrans}, \eqref{bgtshift}, \eqref{vielbeingt}. In this paper, we explore global symmetries that allow such mixings. The simplest example involves the mixing of a background 1-form gauge field~$A^{(1)}$ for a~$U(1)_A^{(0)}$ 0-form flavor symmetry with a background 2-form gauge field~$B^{(2)}$ for a~$U(1)_B^{(1)}$ 1-form symmetry. The transformation rule for~$A^{(1)}$ in~\eqref{atrans} is unchanged, but the transformation rule for~$B^{(2)}$ in~\eqref{bgtshift} now takes the following modified form,
\eqn{
B^{(2)} \quad \longrightarrow \quad B^{(2)} + d \Lambda_B^{(1)} + {\hat \kappa_A \over 2 \pi} \, \lambda_A^{(0)}  \, F_A^{(2)}~, \qquad F_A^{(2)} = d A^{(1)}~,
}[Btwogp]
where~$\hat \kappa_A$ is a real constant. In section~\ref{SecGlobal}, we will show that~\Btwogp\ is only consistent if~$\hat \kappa_A$ is quantized, $\hat \kappa_A \in \Z$. In addition to conventional~$U(1)_B^{(1)}$ background gauge transformations, parametrized by~$\Lambda_B^{(1)}$, the transformation rule of~$B^{(2)}$ in~\Btwogp\ involves a shift  under~$U(1)_A^{(0)}$ background gauge transformations, with gauge parameter~$\lambda_A^{(0)}$, which is proportional to the~$U(1)_A^{(0)}$ background field strength~$F^{(2)}_A = d A^{(1)}$. It is therefore typically inconsistent to activate a non-trivial profile for~$A^{(1)}$, without also turning on~$B^{(2)}$. Many arguments in this paper can be understood in terms of this basic observation. 

The shift of~$B^{(2)}$ in~\eqref{Btwogp} under~$U(1)_A^{(0)}$ background gauge transformations takes exactly the same form as in the Green-Schwarz (GS) mechanism~\cite{Green:1984sg} (see for instance~\cite{Green:1987mn, Polchinski:1998rr} for a textbook treatment). There~$B^{(2)}$ is typically dynamical and the GS shift can be used to cancel certain mixed gauge anomalies by adding suitable GS terms to the action.\foot{~\label{bdualpifn}In four dimensions, a dynamical~$2$-form gauge field is dual to a NG scalar with a continuous shift symmetry (see also section~\ref{ssTQFT}). The GS mechanism then reduces to the statement that some anomalies can be cancelled using suitable couplings to such a NG boson. The duality does not apply to the non-dynamical background field~$B^{(2)}$.} By contrast, our~$B^{(2)}$ is a non-dynamical background field that couples to the~$2$-form current~$J_B^{(2)}$ associated with a global~$U(1)_B^{(1)}$ symmetry (see~\eqref{bjbcoup}). As is familiar from the GS mechanism, the conventional 3-form field strength~$d B^{(2)}$ is not invariant under the GS shift in~\Btwogp. Instead, we can define a different field strength~$H^{(3)}$, which is fully gauge invariant but satisfies a modified form of the Bianchi identity,
\eqn{H^{(3)}=dB^{(2)}-{\hat \kappa_A \over 2\pi} \, A^{(1)}\wedge F_A^{(2)}~, \qquad dH^{(3)}=-{\hat \kappa_A \over 2\pi} \, F_A^{(2)}\wedge F_A^{(2)}~.}[Htwoabelian]
Note that the definition of~$H^{(3)}$ involves the Chern-Simons 3-form~$\text{CS}^{(3)}(A) = A^{(1)} \wedge F_A^{(2)}$. In section~\ref{SecGlobal}, we will show that the modified Bianchi identity~\Htwoabelian\ leads to topological restrictions on the possible configurations of~$A^{(1)}$ and~$B^{(2)}$, which are reminiscent of similar constraints in string compactifications with a GS mechanism~\cite{Candelas:1985en,Strominger:1986uh}.

In this paper, we will identify and analyze explicit examples of QFTs that couple to a background field~$B^{(2)}$ that is subject to GS-like shifts, as in~\eqref{Btwogp}. We will argue that this phenomenon should be viewed as an unconventional form of global symmetry. Unlike the standard GS mechanism, which is closely associated with anomalies, it is not appropriate to think of~\Btwogp\ as an anomaly. (Nevertheless, there are several important ways in which anomalies will make an appearance below.) The relationship of GS shifts such as~\Btwogp\ to symmetries was pointed out in~\cite{Kapustin:2013uxa}, where the underlying mathematical structure was identified as a 2-group~\cite{baez2004higher} (see sections~\ref{genintro} and~\ref{relwork} for additional references). A 2-group is a higher category generalization of group. For our purposes, it will be sufficient to know that the definition of a 2-group involves three pieces of data: a 0-form flavor symmetry~$G^{(0)}$, such as~$U(1)_A^{(0)}$ above; an abelian 1-form symmetry~$G^{(1)}$, which we will always take to be~$U(1)_B^{(1)}$; and a choice of group-cohomology class~$\beta \in H^3(G^{(0)},  G^{(1)}) = H^3(G^{(0)}, U(1)_B^{(1)})$.\foot{~In the notation of~\cite{Kapustin:2013uxa}, the group cohomology class~$\beta$ (which was referred to as a Postnikov class there) belongs to~$H^3(\Pi_1, \Pi_2)$, while for us~$\Pi_1 = G^{(0)}$ and~$\Pi_2 = G^{(1)} = U(1)_B^{(1)}$. The discussion in~\cite{Kapustin:2013uxa} also involved an action~$\alpha$ of~$G^{(0)}$ on~$G^{(1)}$ (via automorphisms), which will be trivial in all of our examples.}

In this paper, we will not discuss 2-groups themselves, but rather the associated 2-group background gauge fields. In analogy with conventional background gauge fields, they can be thought of as 2-connections on suitable 2-bundles~\cite{Baez:2004in, Schreiber:2008}. In the abelian example discussed above, with~$G^{(0)} = U(1)_A^{(0)}$ and~$G^{(1)} = U(1)_B^{(1)}$, the background gauge field~$B^{(2)}$ is subject to the non-trivial GS shift in~\Btwogp. We will also refer to such GS shifts for~$B^{(2)}$ as 2-group shifts. As explained in~\cite{Kapustin:2013uxa}, the form of the 2-group shift in~\Btwogp\ and the cohomology class~$\beta \in H^3(U(1)_A^{(0)}, U(1)_B^{(1)})$ are related by descent (see section~\ref{ssecAnomPol}). To see this, recall that the group cohomology~$H^3(U(1)_A^{(0)}, U(1)_B^{(1)}) = \Z$ classifies three-dimensional Chern-Simons actions for~$U(1)_A^{(0)}$ gauge fields, which are labeled by an integer level~\cite{Dijkgraaf:1989pz}. Precisely such a Chern-Simons term appears in the definition of the modified field strength~$H^{(3)}$ in~\Htwoabelian, and the integer level~$\hat \kappa_A \in \Z$ labels the choice of cohomology class~$\beta$. Under a~$U(1)_A^{(0)}$ background gauge transformation, the Chern-Simons term in~$H^{(3)}$ shifts by an amount~$\sim \hat \kappa_A \, d \big( \lambda_A^{(0)} \, F_A^{(2)}\big)$ that is exactly compensated by the 2-group shift of~$B^{(2)}$ in~\Btwogp. Similar arguments apply if the flavor symmetry~$G^{(0)}$ is a more general, possibly nonabelian, Lie group (see section~\ref{genintro}). 

We will say that a QFT has 2-group symmetry, if it can be coupled to a 2-form background gauge field~$B^{(2)}$ that is subject to a suitable 2-group shift, in addition to its own~$U(1)_B^{(1)}$ background gauge transformations. In the example above, the 2-group shift in~\Btwogp\ is associated with a~$U(1)_A^{(0)}$ flavor symmetry and its background field~$A^{(1)}$. A QFT that couples to such background fields will be said to have abelian 2-group symmetry
\eqn{
U(1)_A^{(0)} \times_{\hat \kappa_A} U(1)_B^{(1)}~, \qquad \hat \kappa_A \in \Z~.
}[abtwgpdef]
We will refer to the integer~$\hat \kappa_A$, which determines the 2-group shift in~\Btwogp, as a 2-group structure constant. It characterizes the 2-group symmetry, somewhat like the structure constants of a Lie algebra (however, see the discussion below~\eqref{qajacom}), or the level of a Kac-Moody algebra. In particular, when~$\hat \kappa_A = 0$, the 2-group shift in~\Btwogp\ disappears and~\abtwgpdef\ decomposes into a conventional product symmetry~$U(1)_A^{(0)} \times U(1)_B^{(1)}$. 

As is appropriate for a constant that determines the properties of a global symmetry, the 2-group structure constant~$\hat \kappa_A$ is a meaningful, scheme-independent property of a QFT. For instance, its value cannot be changed by rescaling (or otherwise redefining) the background fields~$A^{(1)}$ and~$B^{(2)}$, because this would modify the quantization conditions in~\eqref{fsquant} and~\eqref{LamHflux}, or equivalently, because the normalization of the associated currents~$j_A^{(1)}$ and~$J_B^{(2)}$ is meaningful. Below we will see that~$\hat \kappa_A$ controls the OPE that allows two~$j_A^{(1)}$ currents to fuse into~$J_B^{(2)}$, which implies modified Ward identities for these currents in the presence of the 2-group symmetry~\abtwgpdef. This is one way to see that~$\hat \kappa_A$ is an absolute constant, which is inert under RG flow. The same conclusion also follows from the fact that~$\hat \kappa_A \in \Z$ is quantized. Furthermore, this quantization implies that~$\hat \kappa_A$ does not depend on continuously variable coupling constants. Therefore~$\hat \kappa_A$ can only arise at tree level or at one loop (see also section~\ref{introex}). 

As was already mentioned, there is a close connection between 2-group symmetries and 't Hooft anomalies for conventional global symmetries. We will now explain this connection for the abelian 2-group in~\abtwgpdef. In the process, we identify many QFTs that possess~$U(1)_A^{(0)} \times_{\hat \kappa_A} U(1)_B^{(1)}$ 2-group symmetry. We will find that such theories arise from a parent theory with conventional~$U(1)_A^{(0)} \times U(1)_C^{(0)}$ flavor symmetry and suitable mixed 't Hooft anomalies, by gauging~$U(1)_C^{(0)}$,\foot{~A similar phenomenon for discrete symmetries was described in~\cite{Tachikawa:2017gyf}.} which involves the following substitutions, 
\eqn{
U(1)_C^{(0)} \; \rightarrow \; U(1)_c^{(0)}~, \qquad C^{(1)} \; \rightarrow \; c^{(1)}~, \qquad F_C^{(2)} = d C^{(1)} \; \rightarrow \; f_c^{(2)} = d c^{(1)}~.
}[gaugeuonec]

The possible 't Hooft anomalies for a~$U(1)_A^{(0)} \times U(1)_C^{(0)}$ flavor symmetry are conveniently summarized by an anomaly 6-form polynomial~$\CI^{(6)}$ that depends on the background field strengths~$F_{A,C}^{(2)}$ and four anomaly coefficients, $\kappa_{A^3}, \kappa_{A^2C}, \kappa_{AC^2}$, $\kappa_{C^3}$,\foot{~Here~$\kappa_{A^3}$ arises from the three-point function of~$j_A^{(1)}$, while~$\kappa_{A^2C}$ can be extracted from a three-point function involving two~$U(1)_A^{(0)}$ currents and one~$U(1)_C^{(0)}$ current, and similarly for~$\kappa_{AC^2}$, $\kappa_{C^3}$. In theories of free fermions, these correlators reduce to the standard triangle diagrams.}
\eqna{
\CI^{(6)} = {1 \over (2\pi)^3} \, \bigg(  {\kappa_{A^3} \over 3 !}  ~ F^{(2)}_A & \wedge  F^{(2)}_A  \wedge F^{(2)}_A   + {\kappa_{A^2 C} \over 2 !}  ~ F^{(2)}_A \wedge F^{(2)}_A  \wedge F^{(2)}_C  \\
&  + { \kappa_{A C^2} \over 2!} ~  F^{(2)}_A \wedge F^{(2)}_C   \wedge F^{(2)}_C  + {\kappa_{C^3} \over 3!} ~ F^{(2)}_C \wedge F^{(2)}_C \wedge F^{(2)}_C \bigg)~. 
}[anompolint]
The anomaly polynomial encodes (via the descent equations, see section~\ref{ssecAnomPol}) the anomalous c-number shift of the effective action~$W[A^{(1)}, C^{(1)}]$ under~$U(1)_A^{(0)}$ and~$U(1)_C^{(0)}$ background gauge transformations. Since we would like to gauge~$U(1)_C^{(0)}$, it should not give rise to any such shift, and this requires~$\kappa_{C^3} = 0$. We further assume that the~$U(1)_A^{(0)}$ current does not suffer from an Adler-Bell-Jackiw (ABJ) anomaly of the form~$d * j_A^{(1)} \sim \kappa_{AC^2} \, f_c^{(2)} \wedge f_c^{(2)}$ after gauging, and hence we also set~$\kappa_{AC^2} = 0$.  After adjusting various counterterms, the remaining anomalous shifts of the effective action~$W[A^{(1)}, C^{(1)}]$ arise solely from~$U(1)_A^{(0)}$ background gauge transformations,
\eqna{
W[A^{(1)} & + d \lambda_A^{(0)},  C^{(1)} + d  \lambda_C^{(0)}] = \cr
& = W[A^{(1)}, C^{(1)}] + {i \over 4 \pi^2} \int \, \lambda_A^{(0)} \left({\kappa_{A^3} \over 3!} \, F_A^{(2)} \wedge F_A^{(2)} + {\kappa_{A^2 C} \over 2 !} \, F_A^{(2)} \wedge F_C^{(2)}\right)~.
}[wacshift]
The term~$\sim \kappa_{A^3}$ is a conventional~$U(1)_A^{(0)}$ 't Hooft anomaly. It remains a c-number after gauging~$U(1)_C^{(0)}$, although its status as an anomaly must be reevaluated (see section~\ref{gscttwgpano}). However, the term~$\sim \kappa_{A^2 C}$ also involves~$F_C^{(2)}$, which becomes the dynamical field-strength operator~$f_c^{(2)}$ after gauging~$U(1)_C^{(0)}$ (see~\gaugeuonec). Such an operator-valued shift is unacceptable,\foot{~It is tempting to refer to such operator-valued shifts as anomalies. For instance, the ABJ anomaly $d * j_A^{(1)} \sim \kappa_{AC^2} \, f_c^{(2)} \wedge f_c^{(2)}$ arises from a similar operator-valued shift (see section~\ref{ssecBC}). However, we would like to avoid conflating phenomena that involve operator-valued shifts with 't Hooft anomalies, i.e.~c-number shifts of the effective action, which are physically distinct. For this reason we will not refer to operator-valued shifts as anomalies, except in the case of the ABJ anomaly, where this terminology is the unavoidable standard.} because the effective action~$W$ should only depend on background fields. 

The resolution of this apparent paradox is that the dynamical~$U(1)_c^{(0)}$ gauge theory has a new current -- the magnetic 2-form current~$J_B^{(2)} = {i \over 2 \pi} \, * f_c^{(2)}$ introduced in~\eqref{magcurrint} -- and hence also a new background gauge field~$B^{(2)}$, which couples to~$J_B^{(2)}$ as in~\eqref{bjbcoup},
\eqn{
S \; \supset \; \int B^{(2)} \wedge * J_B^{(2)} = {i \over 2 \pi} \int B^{(2)} \wedge f^{(2)}_c~.
}[Bfcoup]
If we interpret this BF coupling as a GS term, we can cancel the operator-valued term $\sim \kappa_{A^2 C} \, \lambda_A^{(0)} \, F_A^{(2)} \wedge f_c^{(2)}$ that arises from~\wacshift\ by assigning a 2-group shift to~$B^{(2)}$, as in~\Btwogp,
\eqn{
B^{(2)} \quad \longrightarrow \quad B^{(2)} + {\hat \kappa_A \over 2 \pi} \, \lambda_A^{(0)} \, F_A^{(2)}~, \qquad \hat \kappa_A = - \half \, \kappa_{A^2 C}~.  
}[btwgpbiscoef]
Therefore the theory has abelian 2-group symmetry~\abtwgpdef. The 2-group structure constant~$\hat \kappa_A$ is determined by the mixed 't Hooft anomaly coefficient~$\kappa_{A^2 C}$ of the parent theory. (Despite the factor of~$-\half$ in~\btwgpbiscoef, $\hat \kappa_A$ is always an integer, see appendix~\ref{AppQuantAnom}.) Therefore any theory with~$U(1)_A^{(0)} \times U(1)_C^{(0)}$ flavor symmetry and a~$\kappa_{A^2 C}$ 't Hooft anomaly (as well as~$\kappa_{AC^2} = \kappa_{C^3} = 0$) gives rise to a theory with abelian 2-group symmetry upon gauging~$U(1)_C^{(0)}$. As we will explain in section~\ref{SsecGauging}, this construction has an inverse: if we gauge~$U(1)_B^{(1)}$ in a theory with~$U(1)_A^{(0)} \times_{\hat \kappa_A} U(1)_B^{(1)}$ 2-group symmetry, we recover the parent theory with~$U(1)_A^{(0)} \times U(1)_C^{(0)}$ flavor symmetry and a~$\kappa_{A^2 C} = - 2 \, \hat \kappa_A$ 't Hooft anomaly.

\subsec{More General 2-Group and~$n$-Group Symmetries}[genintro]

In this paper, we will encounter several generalizations of the abelian 2-group symmetry $U(1)_A^{(0)} \times_{\hat \kappa_A} U(1)_B^{(1)}$ described above. (Some additional possibilities are mentioned in section~\ref{relwork}.) All of them are background-field versions of GS mechanisms, which lead to modifications of the 3-form field strength and its Bianchi identity, as in~\eqref{Htwoabelian}. Moreover, all of these 2-group symmetries (with 2-group structure constants denoted by~$\hat \kappa$'s) arise by gauging a~$U(1)_C^{(0)}$ flavor symmetry in a parent theory with suitable mixed 't Hooft anomalies (where the anomaly coefficients are denoted by~$\kappa$'s):
\begin{itemize}
\item {\it Abelian 2-group symmetry~$\big(\prod_I U(1)_I^{(0)}\big) \times_{\hat \kappa_{IJ}} U(1)_B^{(1)}$ of higher rank:}  If the flavor symmetry is~$G^{(0)} = \prod_I U(1)_I^{(0)}$, with background gauge fields~$A_I^{(1)}$ that transform as~$A^{(1)}_I \rightarrow A^{(1)}_I + d \lambda_I^{(0)}$, the 2-group shift of~$B^{(2)}$ in~\Btwogp\ can be modified as follows, 
\eqn{
B^{(2)} \quad \longrightarrow \quad B^{(2)} + d \Lambda_B^{(1)} + {1 \over 2\pi} \sum_{I, J} \hat \kappa_{IJ} \, \lambda_I^{(0)} \,  F_J^{(2)}~, \quad F_I^{(2)} = d A_I^{(1)}~, \quad \hat \kappa_{IJ} \in \Z~.
}[multabbshift]
The 2-group structure constants~$\hat \kappa_{IJ} = \hat \kappa_{(IJ)}$ now define a symmetric matrix. The higher-rank abelian 2-group symmetry~$\big(\prod_I U(1)_I^{(0)}\big) \times_{\hat \kappa_{IJ}} U(1)_B^{(1)}$ arises by gauging~$U(1)_C^{(0)}$ in a parent theory with mixed~$U(1)_I^{(0)}$-$U(1)_J^{(0)}$-$U(1)_C^{(0)}$ 't Hooft anomalies and anomaly coefficients $\kappa_{IJC} = - 2 \, \hat \kappa_{IJ}$.\foot{~The 't Hooft anomaly coefficient~$\kappa_{IJC}$ can be extracted from a three-point function involving one~$U(1)_I^{(0)}$ current, one~$U(1)_J^{(0)}$ current, and one~$U(1)_C^{(0)}$ current.} 

\item {\it Nonabelian~$SU(N)_A^{(0)} \times_{\hat \kappa_A} U(1)_B^{(1)}$ 2-group symmetry:} A nonabelian flavor symmetry~$G^{(0)}$ can be embedded inside a nonabelian 2-group. For simplicity, we only consider $G^{(0)} = SU(N)^{(0)}_A$. We can then assign the following 2-group shift to~$B^{(2)}$,
\eqn{
B^{(2)} \quad \longrightarrow \quad B^{(2)} + d\Lambda_B^{(1)} + {\hat \kappa_A \over 4 \pi} \, \tr\big(\lambda_A^{(0)} \, d A^{(1)}\big)~, \qquad \hat \kappa_A \in \Z~.
}[nonabbint]
The background gauge field~$A^{(1)}$ and the gauge parameter~$\lambda_A^{(0)}$ are valued in the Lie algebra of~$SU(N)_A^{(0)}$, over which~$\tr$ is a suitable trace (see section~\ref{ssecNonabHooft}). The 2-group symmetry~$SU(N)_A^{(0)} \times_{\hat \kappa_A} U(1)_B^{(1)}$ arises upon gauging~$U(1)_C^{(0)}$ in a theory with a mixed~$SU(N)_A^{(0)}$-$U(1)_C^{(0)}$ 't Hooft anomaly and anomaly coefficient~$\kappa_{A^2 C} = \hat \kappa_A$.\foot{~The anomaly coefficient~$\kappa_{A^2 C}$ is encoded in a three-point function of two~$SU(N)_A^{(0)}$ currents and one~$U(1)_C^{(0)}$ current.} 

\item {\it Poincar\'e 2-group symmetry~$\SP \times_{\hat \kappa_\SP} U(1)_B^{(1)}$:} The 2-group shift of~$B^{(2)}$ can also include background gravity fields, i.e.~the background gauge fields of Poincar\'e symmetry~$\SP$,
\eqn{
B^{(2)} \quad \longrightarrow \quad B^{(2)} + d \Lambda_B^{(1)} + {\hat \kappa_\SP \over 16 \pi} \, \tr \big(\theta^{(0)} \, d \omega^{(1)}\big)~, \qquad \hat \kappa_\SP \in \Z~.
}[ptwgpint]
Here~${\theta^{(0)a}}_b$ is a local~$SO(4)$ frame rotation (see~\eqref{vielbeingt}), ${\omega^{(1)a}}_b$ is the spin connection, and~$\tr$ is a trace over the indices~$a,b$. The 2-group symmetry~$\SP \times_{\hat \kappa_\SP} U(1)_B^{(1)}$ arises upon gauging~$U(1)_C^{(0)}$ in a theory with a mixed~$U(1)_C^{(0)}$-$\SP$ 't Hooft anomaly~$\kappa_{C \SP^2} = - 6 \, \hat \kappa_\SP$.\foot{~A mixed~$U(1)_C^{(0)}$-Poincar\'e anomaly is often referred to as a~$U(1)_C^{(0)}$-gravity anomaly. The corresponding anomaly coefficient~$\kappa_{C \SP^2}$ can be extracted from a three-point function involving the~$U(1)_C^{(0)}$ current and two stress tensors. In appendix~\ref{AppQuantAnom}, we show that~$\hat \kappa_\SP = -{1 \over 6} \,\kappa_{C\SP^2}$ is an integer whenever~$U(1)_C^{(0)}$ can be gauged.} Therefore spacetime symmetries can also be embedded inside 2-group symmetries.

\end{itemize}

We will also encounter more general $n$-group symmetries. By this we (somewhat loosely) mean a symmetry with an~$n$-form background gauge field~$B^{(n)}$ that shifts under gauge transformations associated with other~$q$-form or gravitational background fields. For instance, we will encounter a theory with 3-group symmetry in section~\ref{ssGM}. Here we would briefly like to comment on the case of~$d$-group symmetry in~$d$ spacetime dimensions (see for instance~\cite{Sharpe:2015mja, Tachikawa:2017gyf}).\foot{~We thank N. Seiberg and Y. Tachikawa for a related discussion.} This involves a background gauge field~$B^{(d)}$ associated with a~$U(1)_B^{(d-1)}$ symmetry. Since a conserved~$d$-form current is necessarily a constant multiple of the volume form,\foot{~Such constants give rise to space-filling charges, which can have non-trivial effects. For instance, they can deform certain supersymmetry algebras~\cite{Dumitrescu:2011iu}. } $B^{(d)}$ only couples to the identity operator. In a sense, it is therefore extraneous to the theory. For instance, any 't Hooft anomaly (reducible or not) can be cancelled via a suitable GS (or $d$-group) shift for~$B^{(d)}$.\foot{~Note that this does not spoil 't Hooft anomaly matching, since the contribution of~$B^{(d)}$ to the anomalies, which is encoded by its~$d$-group (or GS) shift, does not change under RG flow. Thus~$B^{(d)}$ plays a role analogous to that of the spectators in 't Hooft's original argument for anomaly matching~\cite{tHooft:1979rat}.} While this might seem slightly artificial for an intrinsically~$d$-dimensional theory, it can happen naturally for theories that arise from a higher-dimensional parent theory. We will see examples of this in sections~\ref{ssecDimRed} and~\ref{SecDefects}.

\subsec{2-Group Current Algebras and Ward Identities}[twgpcurwi]

Continuous global symmetries imply Ward identities for correlation functions that involve the associated conserved currents. Although Ward identities imply the selection rules enforced by the global charges, they also encode the local implications of the symmetry. After reviewing Ward identities for ordinary symmetries, we explain how these Ward identities are modified in the presence of 2-group symmetry. For simplicity, we focus on the abelian 2-group~$U(1)_A^{(0)} \times_{\hat \kappa_A} U(1)_B^{(1)}$ in~\abtwgpdef. 

In the absence of background fields, the~$U(1)_A^{(0)}$ flavor current~$j_A^{(1)}$ is a conserved operator, which satisfies~\eqref{jaconsfirst}. This operator equation is valid inside correlation functions at separated points, i.e.~as long as the current does not collide with other operators. However, at coincident points the conservation equation may be modified by~$\delta$-function contact terms. These contact terms can be c-numbers associated with 't Hooft anomalies (see section~\ref{gscttwgpano}), or they can involve non-trivial operators. The latter case implies a Ward identity. To see this, consider the contact term that arises when~$\d^\mu j_\mu^A(x)$ collides with a local operator~$\CO(y)$ (e.g.~a Lorentz scalar) that carries charge~$q_A$ under the~$U(1)_A^{(0)}$ flavor symmetry, 
\eqn{
\d^\mu j_\mu^A(x) \CO(y) = i q_A \delta^{(4)}(x-y) \CO(y)~. 
}[joope]
This equation should be understood as an OPE, which applies whenever~$\d^\mu j_\mu^A(x)$ collides with a charged operator inside a correlation function. For instance, \joope\ implies the standard~$U(1)_A^{(0)}$ Ward identity (here the conjugate~$\CO^\dagger$ has charge~$-q_A$),
\eqn{
\d^\mu \langle j_\mu^A(x) \CO^\dagger(y)  \CO(z) \rangle = \left(- i q_A \delta^{(4)}(x-y) + i q_A \delta^{(4)}(x-z)\right)\langle \CO^\dagger(y) \CO(z)\rangle~.
}[jodowi]
Integrating over~$x$ implies a selection rule for the~$U(1)_A^{(0)}$ charges: they must sum to zero. Note that~\joope\ and~\jodowi\ imply that~$\CO$ appears in the OPE of~$j_A^{(1)}$ with~$\CO$. 

The OPE in~\joope, and hence all Ward identities that follow from it, such as~\jodowi, is encoded in the transformation rules of background fields under~$U(1)_A^{(0)}$ background gauge transformations. In addition to the~$U(1)_A^{(0)}$ background gauge field~$A^{(1)}$, which couples to~$j_A^{(1)}$ as in~\eqref{sajcoup}, we must also include a complex source~$\CS_\CO$ that couples to~$\CO$ and~$\CO^\dagger$,
\eqn{
S \; \supset \; \int d^4 x \left(A^\mu j_\mu^A + \CS_\CO^\dagger \CO + \CS_\CO \CO^\dagger\right)~.
}[jajo]
Here both~$\CO$ and~$\CS_\CO$ carry~$U(1)_A^{(0)}$ charge~$q_A$, so that the effective action~$W[A^{(1)}, \CS_\CO]$ is invariant under~$U(1)_A^{(0)}$ background gauge transformations of the form~$A^{(1)} \rightarrow A^{(1)} + d \lambda_A^{(0)}$ and~$\CS_\CO \rightarrow e^{i q_A \lambda_A^{(0)}} \CS_\CO$. Substituting into~\jajo\ gives the (non-) conservation equation
\eqn{
\d^\mu j_\mu^A = i q_A \,\CS_\CO \, \CO^\dagger - i q_A \, \CS_\CO^\dagger \,\CO~.
}[jancJ]
Note that the right-hand side vanishes when~$\CS_\CO = 0$, consistent with the fact that~$j_A^{(1)}$ is a conserved operator. Taking a variational derivative~$-{\delta \over \delta \CS^\dagger_\CO(y)}$ of~\jancJ\ inserts~$\CO(y)$ on the left-hand side (see~\jajo) and reproduces the operator-valued contact term in~\joope.

We will now repeat the preceding discussion for the abelian 2-group background gauge fields~$A^{(1)}$ and~$B^{(2)}$, which couple to the currents~$j_A^{(1)}$ and~$J_B^{(2)}$ as in~\eqref{sajcoup} and~\eqref{bjbcoup},
\eqn{
S \; \supset \; \int A^{(1)} \wedge * j_A^{(1)} + B^{(2)} \wedge * J_B^{(2)} = \int d^4 x \, \left(A^\mu j^A_\mu + \half B^{\mu\nu} J^B_{\mu\nu}\right)~.
}[jajbcoup]
In the presence of~$U(1)_A^{(0)} \times_{\hat \kappa_A} U(1)_B^{(1)}$ 2-group symmetry, $A^{(1)}$ transforms like a conventional 1-form gauge field, but~$B^{(2)}$ is subject to a non-trivial 2-group (or GS) shift~\Btwogp,
\eqn{
A^{(1)} \; \rightarrow \; A^{(1)} + d \lambda_A^{(0)}~, \qquad B^{(2)} \; \rightarrow \; B^{(2)} + d \Lambda_B^{(1)} + {\hat \kappa_A \over 2 \pi} \, \lambda_A^{(0)} \, F_A^{(2)}~.
}[abtwgpintro]
Together with~\jajbcoup, the invariance of the effective action~$W[A^{(1)}, B^{(2)}]$ under these gauge transformations (which holds in the absence of 't Hooft anomalies, see section~\ref{gscttwgpano}) implies the following 2-group (non-) conservation equations for the currents,\foot{~\label{ft:lornoncons} When the Lorentz indices are written out, the non-conservation equation for~$j_A^{(1)}$ takes the form
\eqn{
\d^\mu j^A_\mu = {\hat \kappa_A \over 4 \pi} \, F_A^{\mu\nu}\, J^B_{\mu\nu}~.
}[jancind]
}
\eqn{
d * J_B^{(2)} = 0~, \qquad d * j_A^{(1)} = {\hat \kappa_A \over 2 \pi} \, F_A^{(2)} \wedge * J_B^{(2)}~.
}[modconseq]
Thus~$J_B^{(2)}$ remains exactly conserved, but conservation of~$j_A^{(1)}$ is broken by the operator~$J_B^{(2)}$ in the presence of a~$U(1)_A^{(0)}$ background field strength~$F_A^{(2)}$. Just as in~\jancJ, this effect disappears when~$F_A^{(2)} = 0$, so that~$j_A^{(1)}$ remains a conserved 1-form current in the absence of background fields, and at separated points.

If we follow the discussion after~\jancJ\ and take a variational derivative~$- {\delta \over \delta A^\nu(y)}$ of~\modconseq\ (or equivalently, of~\eqref{jancind} in footnote~\ref{ft:lornoncons}), we find an operator-valued contact term proportional to~$J_B^{(2)}$ in the OPE of~$\d^\mu j_\mu^A(x) $ with another~$U(1)_A^{(0)}$ current~$j_\nu^A(y)$,
\eqn{
\d^\mu j^A_\mu(x) j^A_\nu(y) = { \hat \kappa_A \over 2 \pi} \, \d^\lambda \delta^{(4)}(x-y) \, J^B_{\nu\lambda}(y)~.
}[djjope]
Unlike in the conventional Ward identity~\joope, no operator in~\djjope\ is charged under either~$U(1)_A^{(0)}$ or~$U(1)_B^{(1)}$. Integrating~\djjope\ with respect to~$x$, and recalling~\eqref{chargedef}, gives
\eqn{
[Q_A, j^A_\nu(y)] = -{\hat \kappa_A\over 2 \pi} \, \d^\lambda J^B_{\nu\lambda}(y) = 0~.
}[qajacom]
The right-hand side has the form of an improvement term,\foot{~Improvement terms for conserved currents are automatically conserved and do not contribute to the associated charges. For a 1-form current~$j_\mu^A$, the most general improvement term takes the form~$\d^\nu U_{\mu\nu}$, where~$U_{\mu\nu} = U_{[\mu\nu]}$ is a 2-form. In differential form notation, $j_A^{(1)} \supset * d U^{(2)}$, see section~\ref{sssnonprimcurr}.} but vanishes because~$J_B^{(2)}$ is a conserved 2-form current. It follows that the 2-group symmetry~$U(1)_A^{(0)} \times_{\hat \kappa_A} U(1)_B^{(1)}$ does not modify the global charges. In order to distinguish it from a conventional product symmetry~$U(1)_A^{(0)} \times U(1)_B^{(1)}$, we can either study the response to the background gauge fields~$A^{(1)}$ and~$B^{(2)}$, as we have done previously, or examine the OPE in~\djjope. The 2-group (non-) conservation equations in~\modconseq\ and the OPE in~\djjope\  have analogues for the other continuous 2-group symmetries summarized in section~\genintro. 

The OPE~\djjope\ leads to 2-group Ward identities for the currents~$j_A^{(1)}$ and~$J_B^{(2)}$. The simplest example involves the~$\langle j_A^{(1)} j_A^{(1)} J_B^{(2)}\rangle$ three-point function, which characterizes the fusion of two~$j_A^{(1)}$ currents into~$J_B^{(2)}$. For this reason we refer to it as the characteristic three-point function of the abelian 2-group symmetry~$U(1)_A^{(0)} \times_{\hat \kappa_A} U(1)_B^{(1)}$. It follows from~\modconseq\ that~$J_B^{(2)}$ is exactly conserved inside the characteristic three-point function. By contrast, the non-conservation of~$j_A^{(1)}$ at coincident points, which is captured by the OPE~\djjope, implies
\eqn{
{\d \over \d x_\mu} \big\langle j_\mu^A(x) j_\nu^A(y) J_{\rho\sigma}^B (z)\big\rangle = {\hat \kappa_A \over 2\pi} \, \d^\lambda \delta^{(4)}(x-y) \, \big\langle J^B_{\nu\lambda}(y) J^B_{\rho\sigma} (z) \big \rangle~.
}[twgpwiint]
This Ward identity is central to our analysis of QFTs with continuous 2-group symmetry. 

In section~\ref{secCharThrPt}, we solve~\twgpwiint\ for the characteristic three-point function in momentum space, $\langle j_\mu^A(p) j_\nu^A(q) J^B_{\rho\sigma}(-p-q)\rangle$. The analysis is simplified by choosing the momenta so that~$p^2 = q^2 = (p+q)^2 = Q^2$, where~$Q$ is a Lorentz-scalar with dimensions of energy. We parametrize all momentum-space correlators in terms of dimensionless structure functions of~$Q^2 \over M^2$ (here~$M$ is some mass scale), which multiply independent Lorentz structures. The structure that is responsible for the right-hand side of the Ward identity~\twgpwiint\ is given by 
\eqn{
\langle j_\mu^A(p) j_\nu^A(q) J^B_{\rho\sigma}(-p-q)\rangle \; \supset \; {\hat \kappa_A \over 2 \pi Q^2} \, {\mathsf J} \left({Q^2 \over M^2}\right) \Big(\delta_{\mu\rho} (p_\nu + q_\nu)(p_\rho-q_\sigma) + \text{permutations}\Big)~.
}[jjJmomintro]
Here~$\mathsf J\left({p^2 \over M^2}\right)$ is the structure function that controls the momentum-space two-point function~$\langle J^B_{\mu\nu}(p) J^B_{\rho\sigma}(-p)\rangle$. Note the pole~$\sim {\hat \kappa_A \over Q^2}$ that multiplies~$\mathsf J\left({Q^2 \over M^2}\right)$ in~\jjJmomintro. This non-analytic behavior in momentum space contributes to the position-space~$\langle j_A^{(1)} j_A^{(1)} J_B^{(2)}\rangle$ three-point function at separated points. Generically, this implies that the 2-group structure constant~$\hat \kappa_A$ can be extracted from the characteristic three-point function at separated points. Equivalently, $\hat \kappa_A$ controls the fusion of two~$j_A^{(1)}$ currents into~$J_B^{(2)}$. An important exception occurs when~$J_B^{(2)}$ is a redundant operator, which satisfies~$J_B^{(2)} = 0$ at separated points, but may have non-trivial contact terms. Then the structure function~$\mathsf J \left({p^2 \over M^2}\right)$, and hence the right-hand side of~\jjJmomintro, vanishes identically.\foot{~More precisely, if~$J_B^{(2)}$ is a redundant operator, then its momentum-space two-point function~$\langle J_{\mu\nu}^B(p) J^B_{\rho\sigma}(-p)\rangle$ is a polynomial in~$p$, which can be set to zero using local counterterms.} Nevertheless, the theory may possess 2-group symmetry. A simple example is a deformed version of~$\Z_p$ gauge theory, introduced in section~\ref{introex}, which is a TQFT with~$U(1)_A^{(0)} \times_{\hat \kappa_A} U(1)_B^{(1)}$ 2-group symmetry, even though all of its local operators are redundant.

\subsec{RG Flows and Phases of Theories with 2-Group Symmetry}[rgphtwgp]

In section~\ref{SecCurrents}, we use the 2-group Ward identity~\twgpwiint\ for the characteristic~$\langle j_A^{(1)} j_A^{(1)} J_B^{(2)}\rangle$ three-point function to analyze general aspects of RG flows with 2-group symmetry, such as~$U(1)_A^{(0)} \times_{\hat \kappa_A} U(1)_B^{(1)}$. (Additional constraints on possible RG flows arise from 't Hooft anomalies for 2-group symmetries and their matching, see section~\ref{gscttwgpano}.) This is facilitated by the well-motivated assumption that the UV and IR endpoints of such RG flows are conformal field theories (CFTs). In particular, the structure of unitary conformal representations implies that a conformal primary 2-form current~$J_B^{(2)}$ must in fact be proportional to a free Maxwell field strength~$f^{(2)}$, which satisfies~$d f^{(2)} = d* f^{(2)} = 0$ (see e.g.~\cite{Mack:1975je,Minwalla:1997ka,Cordova:2016emh}). A general theme, which will emerge from various points of view, is that~$U(1)_A^{(0)}$ does not behave like a good subgroup symmetry of the full 2-group~$U(1)_A^{(0)} \times_{\hat \kappa_A} U(1)_B^{(1)}$, while~$U(1)_B^{(1)}$ does.\foot{~\label{ft:localsgp}This is not precise, because 2-group symmetry does not modify the charge algebra, as explained below~\qajacom. Rather, the~$U(1)_A^{(0)}$ current algebra is not a good subalgebra of the~$U(1)_A^{(0)} \times_{\hat \kappa_A} U(1)_B^{(1)}$ 2-group current algebra. Since this qualification is tedious, we will usually omit it and simply say that~$U(1)_A^{(0)}$ is not a good subgroup of~$U(1)_A^{(0)} \times_{\hat \kappa_A} U(1)_B^{(1)}$.} This is reflected in various properties of the RG flow, as well as the allowed realizations of~$U(1)_A^{(0)} \times_{\hat \kappa_A} U(1)_B^{(1)}$ 2-group symmetry (with~$\hat \kappa_A \neq 0$) in the IR: 
\begin{itemize}
\item Just as any other symmetry, 2-group symmetry can be unbroken or spontaneously broken by the vacuum. In the former case, we find that~$J_B^{(2)}$ must flow to a redundant operator (i.e.~$J_B^{(2)} = 0$ at separated points) in the deep IR. In the latter case, we show that the only allowed patterns of spontaneous 2-group breaking are
\eqn{
U(1)_A^{(0)} \times_{\hat \kappa_A} U(1)_B^{(1)} \; \rightarrow \; U(1)_B^{(1)} \qquad \text{or} \qquad U(1)_A^{(0)} \times_{\hat \kappa_A} U(1)_B^{(1)} \; \rightarrow \; \text{nothing}~.
}[ssbpat]
The fact that the symmetry cannot break to~$U(1)_A^{(0)}$ is a manifestation of the general theme according to which~$U(1)_A^{(0)}$ does not behave like a good subgroup of the 2-group~$U(1)_A^{(0)} \times_{\hat \kappa_A} U(1)_B^{(1)}$ (see footnote~\ref{ft:localsgp}). As in the unbroken case, the first scenario in~\ssbpat\ requires~$J_B^{(2)}$ to be redundant in the deep IR. In the second scenario, both~$j_A^{(1)}$ and~$J_B^{(2)}$ are non-trivial operators in the IR. They create the~$U(1)_A^{(0)}$ NG scalar and the~$U(1)_B^{(1)}$ NG photon from the vacuum (see the discussion at the end of section~\ref{convgsbf}). The resulting model of spontaneous 2-group breaking is further discussed in section~\ref{introex}. 

\item We will find that exact 2-group symmetry (with non-redundant~$J_B^{(2)}$)  is not compatible with conventional UV completions that have CFT fixed points at short distances. However, such UV completions can exist if the 2-group symmetry is an emergent, accidental symmetry of the IR theory, which is explicitly broken at short distances. In the context of this scenario, we argue in favor of an approximate inequality between the energy scales~$E^\text{UV} (j_A^{(1)})$ and~$E^\text{UV}(J_B^{(2)})$ at which the operators~$j_A^{(1)}$ and~$J_B^{(2)}$ emerge as approximately conserved currents, 
\eqn{
E^\text{UV} (J_B^{(2)}) \; \gtrsim\;  E^\text{UV}(j_A^{(1)})~.
}[emergeineqint]
This inequality states that the~$U(1)_B^{(1)}$ symmetry must emerge before~$U(1)_A^{(0)}$ can emerge, in line with the general theme that the former is a good subgroup of the 2-group~$U(1)_A^{(0)} \times_{\hat \kappa_A} U(1)_B^{(1)}$, while the latter is not. The reason~\emergeineqint\ is not a sharp inequality is that the emergence scales~$E^\text{UV}(J^{(2)}_B)$ and~$E^\text{UV}(j^{(1)}_A)$ are themselves not sharply defined.

\item An inequality similar to~\eqref{emergeineqint} exists for Poincar\'e 2-group symmetry~$\SP \times_{\hat \kappa_\SP} U(1)_B^{(1)}$, which was introduced around~\eqref{ptwgpint}. Now the stress tensor~$T_{\mu\nu}$ plays the role of~$j_A^{(1)}$ in~\eqref{emergeineqint}. In continuum theories that are relativistic and local at all energy scales, the stress tensor~$T_{\mu\nu}$ should be exactly conserved, rather than emergent. Such theories can therefore only realize exact Poincar\'e 2-group symmetry. As was already mentioned, this is incompatible with standard UV completions that involve a CFT fixed point.

\end{itemize}

\subsec{Green-Schwarz Contact Terms and 2-Group 't Hooft Anomalies}[gscttwgpano]

In section~\ref{SecGSAnom}, we examine 't Hooft anomalies in the presence of 2-group symmetries, such as~$U(1)_A^{(0)} \times_{\hat \kappa_A} U(1)_B^{(1)}$. At first glance, these anomalies appear to descend from the conventional~$\kappa_{A^3}$ anomaly for the~$U(1)_A^{(0)}$ flavor symmetry (see~\anompolint\ and~\wacshift), which is reducible, but as we will see, reducible 2-group 't Hooft anomalies are qualitatively very different from conventional 't Hooft anomalies. (This is natural given the interpretation of 2-groups as background-field analogues of the GS mechanism, see sections~\gsfbgf\ and~\genintro.) We will find that the~$\kappa_{A^3}$ anomaly splits into two parts, one of which will turn out to be removable using local counterterms (and should therefore not be viewed as a genuine anomaly), while the other part remains a genuine 't Hooft anomaly, but only due to global considerations. Conventional 't Hooft anomalies for continuous symmetries only receive contributions from massless, local degrees of freedom. By contrast, the sensitivity of reducible 2-group 't Hooft anomalies to global issues enables them to also receive contributions from non-trivial TQFTs.  This fact is essential to ensuring that these anomalies satisfy 't Hooft anomaly matching (see section~\ref{introex}).

A central role is played by GS counterterms for the 2-group background fields~$A^{(1)}$, $B^{(2)}$, 
\eqn{
S_\text{GS} = {i n \over 2 \pi} \int B^{(2)} \wedge F_A^{(2)}~.
}[gsctint]
This counterterm, and the associated GS contact term (see below), can be viewed as four-dimensional analogues of the three-dimensional Chern-Simons counterterms and contact terms analyzed in~\cite{Closset:2012vg,Closset:2012vp}.  The GS counterterm in~\gsctint\ has two important properties:
\begin{itemize}
\item[1.)] $S_\text{GS}$ gives rise to an anomalous c-number shift under~$U(1)_A^{(0)}$ background gauge transformations, due to the 2-group (or GS) shift of~$B^{(2)}$ in~\Btwogp. Adding~$S_\text{GS}$ to the action therefore shifts the~$\kappa_{A^3}$ 't Hooft anomaly coefficient as follows,
\eqn{
\kappa_{A^3} \quad \longrightarrow \quad \kappa_{A^3} + 6 n \, \hat \kappa_A~.
}[kaaashiftint]
\item[2.)] $S_\text{GS}$ is only invariant under large~$U(1)_B^{(1)}$ background gauge transformations, if the coefficient~$n$ is quantized, $n \in \Z$. 
\end{itemize}

Point~$1.)$ above suggests that the~$\kappa_{A^3}$ 't Hooft anomaly can be cancelled by a GS counterterm~\eqref{gsctint} with coefficient~$n = - {\kappa_{A^3} \over 6 \hat \kappa_A}$. However, point~$2.)$ shows that this is only correct if~${\kappa_{A^3} \over 6 \hat \kappa_A}$ is an integer. Its fractional part~${\kappa_{A^3} \over 6 \hat \kappa_A}~(\mod~1)$ can only be absorbed by a GS counterterm with fractional coefficient~$n$, which in turn gives rise to an 't Hooft anomaly under large~$U(1)_B^{(1)}$ background gauge transformations. The upshot is that the~$\kappa_{A^3}$ 't Hooft anomaly for a conventional~$U(1)_A^{(0)}$ flavor symmetry is truncated -- but not completely obliterated -- in the presence of 2-group symmetry: its fractional part~${\kappa_{A^3} \over 6 \hat \kappa_A}~(\mod~1)$ survives, but it is reinterpreted as a mixed anomaly that arises from a clash between~$U(1)_A^{(0)}$ and large~$U(1)_B^{(0)}$ background gauge transformations; by contrast, the integer part of~${\kappa_{A^3} \over 6 \hat \kappa_A}$ is scheme dependent and can be adjusted, or set to zero, using a GS counterterm~\gsctint\ with a properly quantized coefficient~$n \in \Z$. The GS counterterm~\gsctint\ can similarly truncate other reducible 't Hooft anomalies, if~$B^{(2)}$ undergoes additional GS shifts arising from other 2-group symmetries (see section~\ref{genintro}). By contrast, 2-group symmetry does not truncate irreducible 't Hooft anomalies, as is familiar from the GS mechanism. 

The GS counterterm~\gsctint\ is closely related to the the two-point function of the currents~$J_B^{(2)}$ and~$j_A^{(1)}$. In momentum space (see appendix~\ref{AppMomSpace}),
\eqn{
\langle J_{\mu\nu}^B(p) j_\rho^A(-p)\rangle = -{1 \over 2 \pi} \, {\mathsf K} \left({p^2 \over M^2}\right) \, \ep_{\mu\nu\rho\lambda} p^\lambda~.
}[Jjmomint]
Here~${\mathsf K} \left({p^2 \over M^2}\right)$ is a real, dimensionless structure function, and~$M$ is some mass scale. The non-trivial momentum dependence of~${\mathsf K} \left({p^2 \over M^2}\right)$ is scheme independent and contributes to the~$\langle J_B^{(2)} j_A^{(1)}\rangle$ two-point function in position space at separated points. In a scale-invariant theory~$M$ should not appear, so that~${\mathsf K} \left({p^2 \over M^2}\right) = \mathsf K$ reduces to a constant. Fourier-transforming~\Jjmomint\ back to position space then gives rise to a contact term,\foot{~The constant~$\mathsf K$ can be thought of as a four-dimensional analogue of the Hall conductivity.}
\eqn{
\langle J^B_{\mu\nu}(x) j^A_\rho(0)\rangle =  {i \mathsf K \over 2 \pi} \, \ep_{\mu\nu\rho\lambda} \d^\lambda \delta^{(4)}(x)~.
}[gscottintro]
We refer to~\eqref{gscottintro} as a GS contact term (in analogy to Chern-Simons contact terms~\cite{Closset:2012vg,Closset:2012vp}). Such a term can arise in CFTs, and even in TQFTs, where the currents are redundant. 

It is occasionally useful (though imprecise, see section~\ref{secGScts}) to think of the GS contact term~\gscottintro\ as a (potentially fractional) GS term in the background-field effective action,
\eqn{
W[A^{(1)}, B^{(2)}] \; \supset \; {i \mathsf K \over 2 \pi} \int B^{(2)} \wedge F_A^{(2)}~. 
}[wgsctintro]
Varying such a term with respect to the background gauge fields reproduces~\gscottintro. 
Comparing with~\gsctint, we see that a properly quantized GS counterterm shifts~$\mathsf K$ by an integer,
\eqn{
\mathsf K \quad \longrightarrow \quad \mathsf K + n~, \qquad n \in \Z~.
}[kgsctshift]
Therefore only the fractional part~$\mathsf K~(\mod~1)$ of the GS contact term is scheme-independent. Such a fractional part can only arise from non-trivial massless or topological degrees of freedom. For instance, a topological~$\Z_p$ gauge theory can give~$\mathsf K \in {1 \over p} \, \Z$~(see section~\ref{introex}). By contrast, in fully gapped theories, without any dynamical degrees of freedom at long distances, the effective action~$W$ for the background fields only consists of genuine local counterterms, such as properly quantized GS counterterms~\gsctint. In such theories the scheme-independent fractional part~$\mathsf K~(\mod~1)$ necessarily vanishes. 

In a theory with~$U(1)_A^{(0)} \times_{\hat \kappa_A} U(1)_B^{(1)}$ 2-group symmetry, the GS contact term in~\wgsctintro\ contributes (via the 2-group shift of~$B^{(2)}$ in~\Btwogp) an amount~$6 \, \mathsf K \, \hat \kappa_A$ to the~$\kappa_{A^3}$ 't Hooft anomaly. The scheme-dependent contribution from the integer part of~$\mathsf K$ was already discussed after~\eqref{kaaashiftint} above. We now see that the scheme-independent fractional part~$\mathsf K~(\mod~1)$ of the GS contact term also induces a scheme-independent contribution to the~$\kappa_{A^3}$ 't Hooft anomaly. As we have already mentioned, $\mathsf K~(\mod~1)$ can receive contributions from non-trivial TQFTs. In the presence of 2-group symmetry, such TQFTs can therefore also contribute to reducible 't Hooft anomalies (see section~\ref{introex}). As was already emphasized above, this is in stark contrast to conventional 't Hooft anomalies, which are only activated by massless, local degrees of freedom. 

In section~\ref{sssHoftAbTwGp}, we elaborate on the preceding discussion, and compare it to a detailed analysis of the anomalous Ward identity satisfied by the~$\langle j_A^{(1)} j_A^{(1)} j_A^{(1)}\rangle$ three-point function in the presence of 2-group symmetry. (A review of the conventional case~\cite{Frishman:1980dq, Coleman:1982yg}, without 2-group symmetry, can be found in section~\ref{secModHooft}.) Recall from section~\twgpcurwi\ that the 2-group OPE~\djjope\ leads to the 2-group Ward identity~\twgpwiint\ for the characteristic~$\langle j_A^{(1)} j_A^{(1)} J_B^{(2)}\rangle$ three-point function. Similarly, applying the OPE~\djjope\ to the~$\langle j_A^{(1)} j_A^{(1)} j_A^{(1)}\rangle$ correlator leads to an anomalous 2-group Ward identity of the schematic from  $\langle (d * j_A^{(1)}) \, j_A^{(1)} \, j_A^{(1)} \rangle \sim \hat \kappa_A \, \langle J_B^{(2)} j_A^{(1)}\rangle + \kappa_{A^3}$. (Here we have omitted all~$\delta$-functions.) As above, the~$\langle J_B^{(2)} j_A^{(1)}\rangle$ correlator in~\Jjmomint\ naturally makes an appearance. This will play an important role in section~\ref{sssHoftAbTwGp}.

\subsec{Summary of Examples}[introex]

In section~\ref{SecExamples}, we analyze a variety of simple, explicit QFTs with 2-group symmetry. As we saw in sections~\gsfbgf\ and~\genintro, theories with 2-group symmetry can be constructed from suitable parent theories with global symmetries and mixed 't Hooft anomalies, by gauging a~$U(1)_C^{(0)}$ flavor symmetry. (In section~\ref{SsecGauging}, we explain how this construction can be inverted.)  In the examples we consider, the 't Hooft anomalies of the parent theories arise either at one loop, from massless fermions, or a tree level, from NG bosons for spontaneously broken flavor symmetries. In the latter case, the 2-group symmetries that emerge after gauging~$U(1)_C^{(0)}$ are visible classically. By contrast, in the former case the 2-group deformation of the global symmetry arises as a one-loop quantum effect.

The first set of examples (discussed in section~\ref{SubsecQEDexamples}) involves massless multi-flavor QED, i.e. $U(1)_c^{(0)}$ gauge theory with~$N_f$ massless Dirac flavors of charge~$q$. The model has flavor symmetry~$G^{(0)} = SU(N_f)_L^{(0)} \times SU(N_f)^{(0)}_R$, where the left~$(L)$ and right~$(R)$ symmetries only act on Weyl fermions of~$U(1)_c^{(0)}$ charge~$q$ and~$-q$, respectively. As explained around~\eqref{nonabbint}, the mixed~$SU(N)_{L \,,\, R}^{(0)}\,$-$U(1)_C^{(0)}$ 't Hooft anomalies~$\kappa_{L^2 C} = - \kappa_{R^2 C} = q$ of the ungauged parent theory imply that multi-flavor QED possesses the following 2-group symmetry,\foot{~The~$\kappa_{L^2 C}$ and~$\kappa_{R^2 C}$ 't Hooft anomalies are due to standard fermion triangle diagrams, with two~$SU(N)_{L, R}^{(0)}$ currents and one~$U(1)_C^{(0)}$ current at the vertices.}
\eqn{
\left(SU(N_f)_L^{(0)} \times SU(N_f)_R^{(0)}\right) \times_{\hat \kappa_L\, , \, \hat \kappa_R} U(1)_B^{(1)}~, \qquad \hat \kappa_L = - \hat \kappa_R = q~.
}[qedtwgpint]
The~$U(1)_B^{(1)}$ symmetry arises from the magnetic 2-form current $J_B^{(2)} = {i \over 2 \pi} \, * f_c^{(2)}$, where~$f_c^{(2)}$ is the field strength of the dynamical~$U(1)_c^{(0)}$ gauge field (see~\eqref{magcurrint} and~\eqref{Bfcoup}). We can also consider various~$U(1)_A^{(0)} \subset G^{(0)}$ flavor subgroups, some of which belong to abelian 2-group symmetries~$U(1)_A^{(0)} \times_{\hat \kappa_A} U(1)_B^{(1)}$ that are embedded inside the nonabelian 2-group~\qedtwgpint.  

As was mentioned in section~\ref{rgphtwgp}, exact 2-group symmetry is not compatible with conventional UV completions. In multi-flavor QED, and other examples below, this is closely related to the fact that the~$U(1)_c^{(0)}$ gauge coupling has a Landau pole at high energies. Conventional UV completions are possible if the 2-group symmetry is emergent. In section~\ref{sssQEDemerge}, we recall some simple, asymptotically-free nonabelian gauge theories that flow to multi-flavor QED after their gauge symmetry is higgsed to~$U(1)_c^{(0)}$. In these models, the 2-group symmetry~\qedtwgpint\ emerges at low energies, below the scale of higgsing. 

The QED-like~$U(1)_c^{(0)}$ gauge theories discussed above are vector-like. A qualitatively different set of examples is furnished by chiral~$U(1)_c^{(0)}$ gauge theories.\foot{~An important chiral gauge theory that arises in nature is the standard model of particle physics. It has a dynamical~$U(1)_Y^{(0)}$ hypercharge gauge symmetry, as well as an abelian flavor symmetry~$U(1)_{B-L}^{(0)}$, which is free of ABJ anomalies (although it is likely broken by irrelevant operators). The left-handed Weyl fermions of the standard model, their~$U(1)_Y^{(0)} \times U(1)_{B-L}^{(0)}$ charges~$(Q_Y, Q_{B-L})$, and their multiplicities (which are due to their quantum numbers under the~$SU(3)^{(0)}_\text{color} \times SU(2)_\text{weak}^{(0)}$ gauge symmetry, as well as the fact that there are three generations) take the following form, 
\eqn{
q_\alpha = 18 \cdot \left({1 \over 6}, {1 \over 3}\right)~,~u_\alpha = 9 \cdot \left(-{2 \over 3}, -{1 \over 3}\right)~,~d_\alpha = 9 \cdot \left({1 \over 3}, -{1 \over 3}\right)~,~\ell_\alpha = 6 \cdot \left(-\half, -1\right)~,~e_\alpha = 3 \cdot (1, 1)~.
}
This leads to the well-known fact that~$\kappa_{(B-L)^2 Y} = \kappa_{Y \SP^2} = 0$, and hence the standard model does not have an abelian 2-group symmetry involving~$U(1)_{B-L}^{(0)}$, or Poincar\'e 2-group symmetry.} One of the simplest examples (discussed in section~\ref{SubsecFermats}) has four Weyl fermions~$\psi^i_\alpha~(i = 1, \ldots, 4)$ with the following~$U(1)_c^{(0)}$ gauge charges,
\eqn{
q_C^1 = 3~, \qquad q_C^2 = 4~, \qquad q_C^3 = 5~, \qquad q_C^4 = -6~,
}[fermachint]
which satisfy the gauge-anomaly cancellation condition~$\kappa_{C^3} = \sum_{i = 1}^4 (q_C^i)^3 = 0$ in a non-trivial way. Since this cubic constraint on the integers~$q_C^i$ is an example of a Fermat equation, we refer to the~$U(1)_c^{(0)}$ gauge theory based on~\fermachint\ as a Fermat model. This model has a mixed~$U(1)_C^{(0)}$-Poincar\'e ($\SP$) 't Hooft anomaly~$\kappa_{C \SP^2} = \sum_{i = 1}^4 q_C^i = 6$.\foot{~The~$\kappa_{C\SP^2}$ 't Hooft anomaly arises from a fermion triangle diagram with one~$U(1)_C^{(0)}$ current and two stress tensors at the vertices.} As discussed around~\eqref{ptwgpint}, this mixed anomaly gives rise to Poincar\'e 2-group symmetry upon gauging~$U(1)_C^{(0)}$, with properly quantized 2-group structure constant~$\hat \kappa_\SP = -{1 \over 6} \kappa_{C \SP^2} = -1$ (see appendix~\ref{AppQuantAnom}),
\eqn{
\SP \times_{\hat \kappa_\SP} U(1)_B^{(1)}~, \qquad \hat \kappa_\SP = - 1~.
}[fmptgint]
The Fermat model thus suffers from the obstruction to UV completion mentioned in the last bullet point of section~\ref{rgphtwgp}. The model also has an abelian flavor symmetry of rank two, $G^{(0)} = U(1)_X^{(0)} \times U(1)_Y^{(0)}$. Prior to gauging~$U(1)_C^{(0)}$ there are mixed 't Hooft anomalies~$\kappa_{IJC}~(I, J \in \{X, Y\})$.\foot{~These 't Hooft anomalies come from fermion triangles with a~$U(1)_I^{(0)}$ current, a~$U(1)_J^{(0)}$ current, and a~$U(1)_C^{(0)}$ current at each vertex. Here~$I, J \in \{X, Y\}$.} Once we gauge~$U(1)_C^{(0)}$, $G^{(0)}$ remains free of ABJ anomalies and participates in a higher-rank abelian 2-group~$G^{(0)} \times_{\hat \kappa_{IJ}} U(1)_B^{(1)}$, with a symmetric matrix of 2-group structure constants~$\hat \kappa_{IJ} = - \half \kappa_{IJ C}$. Together with~\fmptgint, this furnishes the full 2-group symmetry of the Fermat model.

We also consider deformations of the QED-like and Fermat models discussed above and study the resulting RG flows. In section~\ref{SecExamples}, we focus on deformations that involve an additional complex scalar field~$\phi$ and various Yukawa couplings. (A different kind of deformation, which involves gauging the two-group background gauge fields, is discussed in section~\ref{SsecGauging}.) These deformations allow us to exhibit explicit examples of the different possible IR phases for theories with 2-group symmetry that were mentioned in section~\rgphtwgp. For instance, we find RG flows that preserve abelian~$U(1)_A^{(0)} \times_{\hat \kappa_A} U(1)_B^{(1)}$ or Poincar\'e $\SP \times_{\hat\kappa_\SP} U(1)_B^{(1)}$ 2-group symmetry and lead to a gapped theory in the IR.\foot{~Some of these theories admit dynamical string excitations that are charged under the~$U(1)_B^{(1)}$ global symmetry. These are discussed in section~\ref{SecDefects}.} In some cases, 't Hooft anomaly matching for reducible 2-group anomalies (see section~\gscttwgpano) requires contributions from a non-trivial TQFT in the IR. We now briefly explain how this works for gapped RG flows with abelian 2-group symmetry~$U(1)_A^{(0)} \times_{\hat \kappa_A} U(1)_B^{(1)}$ (see section~\ref{ssTQFT} for details and generalizations). 

All our gapped examples arise by higgsing the~$U(1)_c^{(0)}$ gauge symmetry with a complex scalar Higgs field~$\phi$ of~$U(1)_c^{(0)}$ charge~$q_C \neq 0$ and~$U(1)_A^{(0)}$ flavor charge~$q_A$. The low-energy TQFT is a dynamical~$\Z_{|q_C|}$ gauge theory. As explained in~\cite{Maldacena:2001ss,Banks:2010zn,Kapustin:2014gua}, this theory has a convenient presentation as a dynamical BF theory, which facilitates the coupling to~$U(1)_A^{(0)}$ and~$U(1)_B^{(0)}$ background gauge fields. This leads to the following quadratic action, 
\eqn{
S_\text{BF}[A^{(1)}, B^{(2)}, b^{(2)}, c^{(1)}] = {i q_C \over 2 \pi} \int b^{(2)} \wedge f_c^{(2)} + {i q_A \over 2 \pi} \int b^{(2)} \wedge F_A^{(2)} + {i \over 2 \pi} \int B^{(2)} \wedge f_c^{(2)}~.
}[bfdynint]
Here~$b^{(2)}$ is a dynamical~$U(1)_b^{(1)}$ 2-form gauge field, and~$c^{(1)}$ is the dynamical~$U(1)_c^{(0)}$ gauge field. The 1-form gauge fields~$A^{(1)}$ and~$c^{(1)}$ transform in a standard fashion under~$U(1)_A^{(0)}$ and~$U(1)_c^{(0)}$. Similarly, the 2-form gauge fields~$B^{(2)}$ and~$b^{(2)}$ are subject to~$U(1)_B^{(1)}$ and~$U(1)_b^{(1)}$ 1-form gauge transformations.  Note that the background gauge fields~$A^{(1)}$ and~$B^{(2)}$ couple to the conserved currents~$j_A^{(1)} =  {i q_A \over 2\pi}  \, * db^{(2)}$ and~$J_B^{(2)} = {i \over 2\pi} \, * f_c^{(2)}$, which vanish if we use the equations of motion~$db^{(2)} = f_c^{(2)} = 0$ that hold in the absence of background fields.\foot{~More generally, the on-shell currents are given by c-number terms in the background fields, and hence their correlation functions can have non-trivial contact terms, see below.} The currents are therefore redundant operators, and hence the continuous~$U(1)_A^{(0)}$ and~$U(1)_B^{(1)}$ symmetries do not act on the non-trivial line or surface operators of the theory.\foot{~As was explained in~\cite{Banks:2010zn, Kapustin:2014gua, Gaiotto:2014kfa}, the~$\Z_{|q_C|}$ gauge theory described by the BF theory in~\bfdynint\ has a discrete~$\Z_{|q_C|}^{(1)}$ 1-form symmetry that acts on the~$|q_C|$ distinct Wilson lines of~$a^{(1)}$, and a~$\Z_{|q_C|}^{(2)}$ 2-form symmetry that acts on the~$|q_C|$ distinct Wilson surfaces of~$b^{(2)}$.}

We can deform the model~\bfdynint\ by declaring that~$b^{(2)}$ and~$B^{(2)}$ are also subject to the following GS shifts (parametrized by~$\alpha$) under~$U(1)_A^{(0)}$ background gauge transformations,
\eqn{
b^{(2)}  \; \rightarrow \; b^{(2)}  + {\alpha \over 2 \pi} \, \lambda_A^{(0)} \, F_A^{(2)}~, \qquad B^{(2)} \; \rightarrow \; B^{(2)} + {\hat \kappa_A \over 2 \pi} \, \lambda_A^{(0)} \, F_A^{(2)}~, \qquad \hat \kappa_A = - \alpha q_C~.
}[bBshiftint]
The action~\bfdynint\ is invariant under these shifts, up to a c-number 't Hooft anomaly. The GS shift of the background field~$B^{(2)}$ shows that the deformed model has~$U(1)_A^{(0)} \times_{\hat \kappa_A} U(1)_B^{(1)}$ 2-group symmetry. The anomalous c-number shift contributes an amount~$6 \left(-{q_A \over q_C}\right) \hat \kappa_A$ to the 't Hooft anomaly coefficient~$\kappa_{A^3}$ (see~\anompolint\ and~\wacshift). In section~\ref{gscttwgpano}, we saw that TQFTs with 2-group symmetry contribute precisely~$6 \, \mathsf K \, \hat \kappa_A$ to the~$\kappa_{A^3}$ 't Hooft anomaly. Here~$\mathsf K$ is the GS contact term in~\gscottintro\ and~\wgsctintro. It can be checked that~\bfdynint\ gives rise to just such a contact term, with the correct value~$\mathsf K = -{q_A \over q_C}$, for instance by integrating out the dynamical fields (this requires some care, see sections~\ref{secGScts} and~\ref{ssTQFT}).  

Finally, we would like to mention a simple model which arises in the deep IR of RG flows that spontaneously break the entire 2-group~$U(1)_A^{(0)} \times_{\hat \kappa_A} U(1)_B^{(1)}$ (see~\eqref{ssbpat}). In the absence of background fields, the model consists of a free~$U(1)_A^{(0)}$ NG scalar~$\chi$, and a free Maxwell field~$f_c^{(2)}$, which furnishes the~$U(1)_B^{(1)}$ NG boson. For this reason, we refer to it as the Goldstone-Maxwell (GM) model.  The coupling of the dynamical fields to the background fields~$A^{(1)}$ and~$B^{(2)}$ proceeds via the following quadratic action,
\eqna{
S_\text{GM}[A^{(1)}, B^{(1)}, \chi, c^{(1)}] =~& v^2 \int \left(d \chi - A^{(1)} \right) \wedge * \left(d \chi - A^{(1)}\right) + {1 \over 2 e^2} \int f^{(2)}_c \wedge * f^{(2)}_c  \cr
&  + {i \over 2\pi} \int \left(B^{(2)} - {\hat \kappa_A \over 2 \pi} \, \chi\,  F_A^{(2)} \right) \wedge f^{(2)}_c~.
}[gmint]
The NG scalar shifts as~$\chi \rightarrow \chi + \lambda_A^{(0)}$ under~$U(1)_A^{(0)}$ background gauge transformations. The second line of~\gmint\ is only invariant if we also declare that~$B^{(2)}$ undergoes a 2-group shift, $B^{(2)} \rightarrow B^{(2)} + {\hat \kappa_A \over 2 \pi} \, \lambda_A^{(0)} \, F_A^{(2)}$. This shows that the model has~$U(1)_A^{(0)} \times_{\hat \kappa_A} U(1)_B^{(2)}$ 2-group symmetry. The GM model is further discussed in section~\ref{ssGM}. As we will see there, its 2-group symmetry is in fact embedded in an even large 3-group symmetry (see section~\genintro).\foot{~The 3-group symmetry of the GM model is essential to making its 't Hooft anomalies compatible with anomaly inflow from a five-dimensional bulk.} 

The deformed BF theory described by~\bfdynint, \bBshiftint\ and the GM model~\gmint\ illustrate the general point emphasized below~\qajacom: the presence or absence of 2-group symmetry can only be detected if we know how the dynamical fields couple to the background gauge fields~$A^{(1)}$ and~$B^{(2)}$. (Equivalently, if we know the associated currents.) Without this additional data, the models cannot be distinguished from conventional~$\Z_{|q_C|}$ gauge theory, or from the theory of a free NG boson and Maxwell field, which do not possess 2-group symmetry.

\subsec{Related Work}[relwork]

The continuous 2-group symmetries analyzed in this paper have much in common with their discrete counterparts. Most discussions of 2-groups in the literature have focused on the discrete case (an exception is~\cite{Sharpe:2015mja}). In this context, the authors of~\cite{Kapustin:2013uxa} pointed out the relation between GS shifts for background fields and 2-group symmetries (see section~\ref{gsfbgf}), following earlier related work~\cite{Gukov:2013zka, Kapustin:2013qsa}. Possible 't Hooft anomalies for such symmetries were analyzed in~\cite{Kapustin:2013uxa,Kapustin:2014zva,Thorngren:2015gtw}. Other recent discussions of discrete 2-group (and higher~$n$-group) symmetries in QFT appear in\cite{Bhardwaj:2016clt,Tachikawa:2017gyf}. Many phenomena that occur for continuous 2-group symmetries also happen in the discrete case. For instance, the fact that 2-group symmetries arise by gauging a~$U(1)_C^{(0)}$ flavor symmetry with suitable mixed 't Hooft anomalies (see sections~\gsfbgf\ and~\genintro) has a discrete analogue~\cite{Tachikawa:2017gyf}. Other phenomena that arise in both cases are the truncation of certain 2-group 't Hooft anomalies, and the fact that TQFTs can contribute to such anomalies (see section~\gscttwgpano and~\cite{Kapustin:2013uxa,Kapustin:2014zva,Thorngren:2015gtw}). A detailed analysis of these, and other, aspects of discrete 2-groups will appear in~\cite{BeniniToApp}. Finally, we would like to point out that some of the phenomena described in~\cite{etingof2009fusion, Barkeshli:2014cna,Barkeshli:2017rzd}, which also involve a group-cohomology class in~$H^{3}(G^{(0)}, G^{(1)})$ (see section~\ref{gsfbgf}), and which are sometimes referred to as anomalies, can be understood in terms of discrete 2-group symmetries. Some comments appear in~\cite{Thorngren:2015gtw, Tachikawa:2017gyf}, see~\cite{BeniniToApp} for a detailed discussion. As we have emphasized in the continuous case, it is more appropriate to think of 2-group symmetries as unconventional global symmetries, rather than as anomalies. This distinction is especially important if the 2-group symmetries have 't Hooft anomalies of their own (see section~\ref{gscttwgpano}).  

A powerful handle on theories with continuous 2-group symmetries is furnished by the associated 2-group currents and their Ward identities (see sections~\twgpcurwi, \rgphtwgp, and~\gscttwgpano). In the present paper, we focus on theories with continuous 2-group symmetries in four dimensions. The six-dimensional case will be analyzed in~\cite{our6dtoapp}. Here we briefly summarize some of the results. 

In six dimensions, 't Hooft anomalies for continuous flavor and spacetime symmetries first appear in four-point functions of the associated  currents. This allows for a richer structure of mixed anomalies than exists in four dimensions. For instance, there are mixed anomalies that involve two different nonabelian flavor symmetries, with two nonabelian currents of each kind appearing in the anomalous four-point function. (It is also possible to replace some of the flavor currents by stress tensors.) By generalizing the arguments in sections~\gsfbgf\ and~\genintro, we find that gauging one of these nonabelian flavor symmetries leads to a 2-group whose~$U(1)_B^{(1)}$ subgroup arises from a 2-form current constructed out of the dynamical nonabelian field strength~$f^{(2)}$,
\eqn{
J_B^{(2)} ~ \sim ~* \tr\left(f^{(2)} \wedge f^{(2)}\right)~.
}[jbsixdint]
This shows that nonabelian gauge theories in six dimensions can have 2-group symmetry. 

Even though six-dimensional gauge theories are IR-free effective field theories, some supersymmetric examples have known UV completions as little string theories, or as superconformal theories (SCFTs). The former can possess 2-group symmetries. For instance, the little string theory~\cite{Seiberg:1997zk} that arises from~$N$ small~$SO(32)$ instantons~\cite{Witten:1995gx} in the heterotic string provides a six-dimensional UV completion for a particular~$Sp(N)^{(0)}$ gauge theory with $(1,0)$ supersymmetry and suitable matter. In this theory, the 2-form current~\jbsixdint, which involves the~$Sp(N)^{(0)}$ field strength~$f^{(2)}$, is associated with the string charge of the little string theory. The known anomaly structure of this theory (see for instance~\cite{Schwarz:1995zw}) implies that this 2-form current participates in a 2-group, together with the~$SO(32)^{(0)}$  flavor symmetry of the theory, as well as with six-dimensional Poincar\'e symmetry~$\SP$. 

By contrast, six-dimensional SCFTs do not admit 2-group symmetries~\cite{our6dtoapp}, because they cannot posses a conformal primary 2-form current, such as~\jbsixdint. This follows from the fact that conserved 2-form currents, which reside in short representations of conformal symmetry, cannot be embedded into any unitary representation of the six-dimensional superconformal algebras~\cite{Cordova:2016emh}. In~\cite{our6dtoapp}, we use these observations to justify the prescription of~\cite{Ohmori:2014kda} for extracting the 't Hooft anomalies of an SCFT from the low-energy theory on its tensor branch (if such a branch exists). Together with the fact that some of these 't Hooft anomalies determine the~$a$-type Weyl anomaly~\cite{Cordova:2015vwa, Cordova:2015fha}, this can be used to prove that the~$a$-anomaly of any six-dimensional SCFT with a tensor branch is positive, $a > 0$~\cite{our6dtoapp}.

\newsec{Review of 't Hooft Anomalies for Conventional Symmetries}[SecHooft]

In this section we review 't Hooft anomalies for the continuous global symmetries summarized in section~\ref{convgsbf}. Anomalies for 2-group symmetries will be discussed in section~\ref{SecGSAnom}. 

\subsec{Generalities}[ssecGenHooft]

QFTs with global symmetries can have 't Hooft anomalies.\foot{~Various aspects of anomalies are nicely reviewed in~\cite{AlvarezGaume:1985ex, Weinberg:1996kr,Harvey:2005it}.} One way to exhibit such anomalies involves coupling the theory to background gauge fields~$\CB$ for the global symmetries.  By adjusting local counterterms for these background fields, it is sometimes possible to make the partition function~$Z[\CB]$ of the theory invariant under background gauge transformations~$\CB \rightarrow \CB + \delta \CB$. 't Hooft anomalies arise when this is not possible, in which case the effective action~$W[\CB] = - \log Z[\CB]$ for background fields (see footnote~\ref{ft:swactdef}) is not gauge invariant,
\eqn{
W[\CB + \delta \CB] = W[\CB] + \CA[\CB]~.
}[anomfundef]
The anomaly~$\CA[\CB]$ is a local c-number functional of the background fields (roughly, because there is a sense in which it arises from physics at very short distances), which satisfies the Wess-Zumino consistency conditions~\cite{Wess:1971yu}.\foot{~\label{conscovanom} For this reason, the functional~$\CA$ in~\anomfundef\ is sometimes called the consistent anomaly. It should be distinguished from the covariant form of the anomaly~\cite{Bardeen:1984pm}. Throughout, we only discuss the consistent anomaly.} Moreover, $\CA[\CB]$ vanishes when the background fields~$\CB$ are turned off. In particular, the symmetry is unbroken in this case. This should be contrasted with a distinct (but related) phenomenon --  the Adler-Bell-Jackiw (ABJ) anomaly~\cite{Adler:1969gk, Bell:1969ts} -- which does not vanish in the absence of background fields. For instance, an ABJ anomaly is responsible for the non-conservation of the axial current~$j^\mu_\text{axial}$ in massless QED, which  satisfies the operator equation~$\d_\mu j^\mu_\text{axial} \sim * ( f^{(2)} \wedge f^{(2)})$ (here~$f^{(2)}$ is the Maxwell field strength operator of the dynamical electromagnetic field). 

The anomaly functional~$\CA[\CB]$ in~\anomfundef\ can be modified by adding local counterterms in the background fields to the action~$S$, and hence to~$W[\CB]$. This can change the presentation of the anomaly (as we will see in examples below), but genuine anomalies cannot be removed using local counterterms. Two reasons for the enduring interest in 't Hooft anomalies are that they obey strong non-renormalization theorems (as in~\cite{Adler:1969er}) and are subject to anomaly matching~\cite{tHooft:1979rat, Frishman:1980dq, Coleman:1982yg}. By anomaly matching we mean that 't Hooft anomalies must be reproduced in all effective descriptions of a given theory.

\subsec{Anomaly Polynomials, Descent, Inflow, and Counterterms}[ssecAnomPol]

We are interested in 't Hooft anomalies for continuous symmetries. In $d$ spacetime dimensions these are conveniently summarized by a~$d+2$-form anomaly polynomial~$\CI^{(d+2)}[\CB]$. Here we imagine extending the background gauge fields~$\CB$, and their gauge transformations~$\delta \CB$, to $d+2$ dimensions. Then~$\CI^{(d+2)}[\CB]$ is a gauge-invariant polynomial in background field strengths and curvatures constructed out of various characteristic classes (see below), which determines the anomalous shift~$\CA[\CB]$ of the $d$-dimensional effective action~$W[\CB]$ in~\anomfundef\ via the descent equations (see for instance~\cite{AlvarezGaume:1985ex, Weinberg:1996kr,Harvey:2005it} and references therein),
\eqn{
\CA[\CB] = 2 \pi i \int_{\CM_d}  \CI^{(d)}[\CB, \delta \CB]~, \quad d \CI^{(d)}[\CB, \delta \CB] = \delta \CI^{(d+1)}[\CB]~, \quad d\CI^{(d+1)}[\CB] = \CI^{(d+2)}[\CB]~.
}[descenteq]
Here~$\CM_d$ is the $d$-dimensional spacetime manifold and~$\CI^{(d)}[\CB, \delta \CB]$, $\CI^{(d+1)}[\CB]$ are local expressions in the background fields (and, in the case of~$\CI^{(d)}$, also the gauge parameters~$\delta \CB$). On a closed $(d+1)$-manifold~$\CM_{d+1}$, the euclidean action~$S_{d+1}[\CB] =  2 \pi i \int_{\CM_{d+1}} \CI^{(d+1)}[\CB]$ is gauge invariant modulo~$2 \pi i \Z$, so that~$e^{- S_{d+1}}$ is gauge invariant. However, if~$\CM_{d+1}$ is a manifold with boundary~$\d \CM_{d+1} = \CM_d$, the action~$S_{d+1}[\CB]$ induces the anomaly~$\CA[\CB]$ on the boundary~$\CM_d$ by anomaly inflow from the~$(d+1)$-dimensional bulk.\foot{~\label{zeroinflow} Alternatively, we can take the action in the $(d+1)$-dimensional bulk to be~$-S_{d+1}[\CB]$, which contributes~$-\CA$ to the boundary anomaly. Since anomaly of the~$d$-dimensional boundary theory is~$+\CA$, the combined bulk-boundary system is then invariant under background gauge transformations.} It is believed that all 't Hooft anomalies in local QFTs should admit a description in terms of anomaly inflow, once the symmetries and background fields have been correctly identified (see section~\ref{ssGM}).  

An anomaly polynomial~$\CI^{(d+2)}_\text{red.}$ is called reducible if it factorizes into a product of closed, gauge-invariant anomaly polynomials~$\CJ^{(p)}$ and~$\CK^{(d+2-p)}$ of lower degree,
\eqn{
\CI^{(d+2)}_\text{red.} = \CJ^{(p)} \wedge \CK^{(d+2 -p)}~, \qquad d\CJ^{(p)}=d\CK^{(d+2-p)} = 0~.
}[redanom]
The first step of the descent procedure described around~\descenteq\ involves removing an exterior derivative~$d$ from the anomaly polynomial. However, this is ambiguous for the reducible anomaly~$\CI^{(d+2)}_\text{red.}$ in~\redanom, because we can remove an exterior derivative from either factor. 

To see this explicitly, consider the the first descendants~$\CI^{(d+1)}_\text{red.}$, $\CJ^{(p-1)}$, $\CK^{(d+1-p)}$ of the anomaly polynomials in~\redanom,
\eqn{
\CI^{(d+2)}_\text{red.} = d \CI^{(d+1)}_\text{red.}~, \qquad \CJ^{(p)} = d \CJ^{(p-1)}~, \qquad \CK^{(d+2-p)} = d \CK^{(d+1-p)}~.
}[gendesc]
The ambiguity described above leads to an expression for~$\CI^{(d+1)}_\text{red.}$ that depends on an undetermined real parameter~$s$, 
\eqna{
 \CI^{(d+1)}_\text{red.} &= \left(1+ s (-1)^{p-1} \right) \, \CJ^{(p-1)} \wedge \CK^{(d+2-p)} + s \, \CJ^{(p)} \wedge \CK^{(d+1-p)} \cr
 & = \CJ^{(p-1)} \wedge \CK^{(d+2-p)} + s \, d \left(\CJ^{(p-1)} \wedge \CK^{(d+1-p)} \right)~, \qquad s \in \R~.
}[gendescii]
The free parameter~$s$ multiplies an exact term in the~$(d+1)$-dimensional anomaly-inflow action~$2 \pi i \int_{\CM_{d+1}} \CI^{(d+1)}_\text{red.}$, and hence it corresponds to a local counterterm in~$d$ dimensions, 
\eqn{
S_\text{C.T.}[\CB] =  2 \pi i \, s \int_{\CM_d}  \CJ^{(p-1)} \wedge \CK^{(d+1-p)}~.
}[gendct]
Adjusting such counterterms modifies the presentation of reducible anomalies. This will play an important role below. 

We now briefly sketch the basic ingredients that make up the anomaly 6-form polynomial~$\CI^{(6)}$ in~$d = 4$ spacetime dimensions. (A detailed discussion appears in the subsections below.) Since the anomaly polynomial must be gauge invariant, it naturally involves background field strengths and curvatures, which assemble into various characteristic classes: 
\begin{itemize}

\item An ordinary~$U(1)_I^{(0)}$ flavor symmetry contributes to~$\CI^{(6)}$ via the first Chern class $c_1(F_I^{(2)}) = {1 \over 2 \pi} \, F^{(2)}_I$, where~$F_I^{(2)} = d A_I^{(1)}$ is the associated background field strength.

\item An~$SU(N)_A^{(0)}$ flavor symmetry contributes to~$\CI^{(6)}$ through the Chern classes $c_k(F_A^{(2)}) = {1 \over (2 \pi)^k} \, \tr \left( (F_A^{(2)})^k\right)$, with~$k \geq 2$, which are~$2k$-forms constructed from the~$SU(N)_A^{(0)}$ background field strength~$F_A^{(2)}$. Here~$c_k(F_A^{(2)})$ is independent if there is an~$SU(N)$ Casimir of order~$k$. This is always true for~$k = 2$, but for~$k = 3$ it requires~$N \geq 3$. 
\item Poincar\'e symmetry contributes to~$\CI^{(6)}$ via Pontryagin classes~$p_k \sim \tr \left((R^{(2)})^{2k}\right)$, which are~$4k$-forms constructed from the Riemann curvature 2-form~${R^{(2)a}}_b$. Here~$a, b$ are~$SO(4)$ frame indices, and~$\tr$ denotes a trace over such indices.

\item $U(1)_B^{(q)}$ symmetries, with~$q \geq 1$, contribute to~$\CI^{(6)}$ via the field strength $d B^{(q+1)}$, which is invariant under~$q$-form background gauge transformations~\eqref{bgtshift} of~$B^{(q+1)}$. 

\end{itemize}
\noindent Schematically (in particular, omitting all prefactors, which are explained in detail below) these ingredients can be used to construct the following candidate anomalies:
\begin{itemize}
\item[a)] Abelian flavor symmetries can contribute mixed~$U(1)_I^{(0)}$-$U(1)_J^{(0)}$-$U(1)_K^{(0)}$ anomalies,
\eqn{
\sum_{I, J, K} \, \kappa_{IJK} \, c_1\big(F_I^{(2)}\big) \wedge c_1\big(F_J^{(2)}\big) \wedge c_1\big(F_K^{(2)}\big) \; \subset \; \CI^{(6)}~.
}[ijkabanom]
Here the indices~$I, J , K$ may coincide, e.g.~$\kappa_{III} = \kappa_{I^3}$ denotes a cubic~$U(1)_I^{(0)}$ anomaly. Note that~\ijkabanom\ is always a reducible anomaly (see~\redanom).
\item[b)]  There can be reducible, mixed~$SU(N)^{(0)}$-$U(1)_I^{(0)}$ anomalies of the form
\eqn{
\sum_I \, \kappa_{A^2 I} \, c_2 \big(F_A^{(2)}\big) \wedge c_1\big(F_I^{(2)}\big) \; \subset \; \CI^{(6)}~.
}[sunsunuone]
\item[c)] If~$N \geq 3$, there can be an irreducible cubic~$SU(N)^{(0)}$ anomaly,
\eqn{
\kappa_{A^3} \, c_3\big(F_A^{(2)}\big) \; \subset \; \CI^{(6)}~.
}[sunirred]
\item[d)]  There can be reducible, mixed~$U(1)_I^{(0)}$-Poincar\'e ($\SP$) anomalies of the form
\eqn{
\sum_I \, \kappa_{I \, \SP^2} \, c_1\big(F_I^{(2)}\big) \wedge p_1 \; \subset \; \CI^{(6)}~.
}[ipoincanom]
\item[e)] There can be reducible, mixed~$U(1)_B^{(1)}$-$U(1)_{B'}^{(1)}$ anomalies, 
\eqn{
\kappa_{B B'} \, d B^{(2)} \wedge dB'^{(2)} \; \subset \; \CI^{(6)}~. 
}[bbpanom]
Note that the left-hand side vanishes by antisymmetry if~$B = B'$. The anomaly~\bbpanom\ therefore requires two distinct 1-form global symmetries. We will only encounter this situation in the context of free Maxwell theory (see appendix~\ref{appHooftFreeMax} and section~\ref{ssGM}), which has both an electric and a magnetic 1-form symmetry~\cite{Gaiotto:2014kfa}. In section~\ref{ssGM}, we also discuss a mixed anomaly of the form~$c_1(F_A^{(2)}\big) \wedge d \Theta^{(3)} \; \subset \; \CI^{(6)}$, where~$U(1)_A^{(0)}$ is an ordinary flavor symmetry, while~$\Theta^{(3)}$ is the 3-form background gauge field for a~$U(1)_\Theta^{(2)}$ 2-form global symmetry. 
\end{itemize}
\noindent Note that there is no candidate anomaly that mixes a~$U(1)_B^{(1)}$ symmetry with ordinary flavor symmetries, or with Poincar\'e symmetry. This will be important in section~\ref{sssHoftAbTwGp}.

\subsec{Abelian Flavor Symmetries and Background Gauge Fields}[SecDescabelian]

We first consider four-dimensional theories with abelian~$0$-form flavor symmetry 
\eqn{
G^{(0)} = U(1)^{(0)}_A \times U(1)^{(0)}_C~.
}[uonac]
The corresponding~$1$-form background gauge fields are~$A^{(1)}$ and~$C^{(1)}$; their field strengths are~$F^{(2)}_A = dA^{(1)}$ and~$F_C^{(2)} = dC^{(1)}$. We will eventually gauge~$U(1)^{(0)}_C$, but throughout this section it will be a global symmetry. The most general anomaly~$6$-form~$\CI^{(6)}$ that can be constructed using~$F^{(2)}_A$ and $F^{(2)}_C$ takes the form~\eqref{ijkabanom}, with~$I, J, K \in \{A, C\}$. Explicitly, 
\eqna{
\CI^{(6)} = {1 \over (2\pi)^3} \, \bigg(  {\kappa_{A^3} \over 3 !}  ~ F^{(2)}_A & \wedge  F^{(2)}_A  \wedge F^{(2)}_A   + {\kappa_{A^2 C} \over 2 !}  ~ F^{(2)}_A \wedge F^{(2)}_A  \wedge F^{(2)}_C  \\
&  + { \kappa_{A C^2} \over 2!} ~  F^{(2)}_A \wedge F^{(2)}_C   \wedge F^{(2)}_C  + {\kappa_{C^3} \over 3!} ~ F^{(2)}_C \wedge F^{(2)}_C \wedge F^{(2)}_C \bigg)~. 
}[acanompol]
Here the different~$\kappa$'s are real constants --  the anomaly coefficients -- that can be extracted from the various three-point functions of the~$U(1)^{(0)}_A$ and~$U(1)^{(0)}_C$ currents. A set of Weyl fermions~$\psi_\alpha^i$ with~$U(1)^{(0)}_A$ and~$U(1)^{(0)}_C$ charges~$q^i_A$ and~$q^i_C$ contribute 
\eqn{
\CI^{(6)} = \sum_i \exp\left({q^i_A \over 2 \pi}\, F_A^{(2)} + {q^i_C \over 2 \pi} \, F_C^{(2)}\right) \bigg|_\text{6-form}~.
}[]
Expanding the exponential and comparing the~$6$-form terms with~\acanompol\ leads to
\eqna{
& \kappa_{A^3} = \sum_i \left(q^i_A\right)^3~, \hskip44pt \kappa_{A^2 C} =  \sum_i \left(q^i_A\right)^2 q_C^i~,\\
& \kappa_{A C^2} =  \sum_i q^i_A \left(q_C^i\right)^2~, \qquad \kappa_{C^3} =  \sum_i \left(q^i_C\right)^3~.
}[freefermks]
This result also follows from a direct evaluation of the various current three-point functions, which here reduce to anomalous fermion triangles. Note that the anomaly coefficients in~\freefermks\ are sums and products of~$U(1)$ charges, and hence integers. As is reviewed in appendix~\ref{AppQuantAnom}, this quantization is a general feature of 't Hooft anomaly coefficients,  which can be argued without appealing to free fermions. The fact that the anomaly coefficients are quantized explains their rigidity under RG flows, as well as the non-renormalization theorem of~\cite{Adler:1969er}.\foot{~Since they are quantized, the 't Hooft anomaly coefficients cannot depend on any continuous coupling constants (which can be promoted to background fields), and hence they are one-loop exact. This is similar to the argument of~\cite{Closset:2012vp} for the non-renormalization of Chern-Simons terms~\cite{Coleman:1985zi}. 
}

All anomalies in~\eqref{acanompol} are reducible, and hence the discussion around~\redanom\ applies. In particular, every term in~$\CI^{(6)}$ that involves both~$F^{(2)}_A$ and~$F^{(2)}_C$ leads to a one-parameter ambiguity in~$\CI^{(5)}$, as in~\eqref{gendescii}. For instance, applying descent to the term proportional to~$F^{(2)}_A \wedge F^{(2)}_A \wedge F^{(2)}_C$ in~$\CI^{(6)}$ leads to the following terms in~$\CI^{(5)}$,
\eqna{
\CI^{(5)} \; \supset \;  {\kappa_{A^2  C} \over 2 (2\pi)^3}  \,  A^{(1)} \wedge F^{(2)}_A \wedge F^{(2)}_C + s \, d \big(A^{(1)} \wedge F^{(2)}_A \wedge C^{(1)} \big) ~, \qquad s \in \R~.
}[sparam]
The ambiguity parametrized by~$s$ is an exact~$5$-form, and hence it corresponds to a local counterterms in four dimensions, 
\eqn{
S_\text{C.T.}[A^{(1)}, C^{(1)}] = 2 \pi i \, s  \int_{\CM_4} \text{CS}^{(3)}(A) \wedge C^{(1)}~, \qquad \text{CS}^{(3)}(A) = A^{(1)} \wedge F_A^{(2)}~.
}[csacct]
Here~$\text{CS}^{(3)}(A)$ denotes the Chern-Simons~$3$-form.  Adjusting the counterterm~\csacct\ amounts to dialing the parameter~$s$ in~\sparam.   In terms of the general expressions~\eqref{gendescii} and~\eqref{gendct}, this example has~$\CJ^{(p-1)} = \CJ^{(3)} \sim \text{CS}^{(3)}(A)$ and~$\CK^{(d+1-p)} = \CK^{(1)} \sim C^{(1)}$. A similar ambiguity, parametrized by~$t \in \R$, arises when we apply the descent procedure to the term proportional to~$F^{(2)}_A \wedge F^{(2)}_C \wedge F^{(2)}_C$ in~$\CI^{(6)}$.

In summary, the descent 5-form~$\CI^{(5)}$ that arises from the anomaly polynomial~$\CI^{(6)}$ in~\acanompol\ is given by
\eqna{
 \CI^{(5)}  =~&  {1 \over (2 \pi)^3}  \bigg({\kappa_{A^3} \over 3!} \, A^{(1)} \wedge F^{(2)}_A \wedge F^{(2)}_A + {\kappa_{A^2 C} \over 2!} \, A^{(1)} \wedge F^{(2)}_A \wedge F^{(2)}_C + {\kappa_{A C^2} \over 2!} \, A^{(1)} \wedge F^{(2)}_C \wedge F^{(2)}_C \cr
& + {\kappa_{C^3} \over 3!} \, C^{(1)} \wedge F^{(2)}_C \wedge F^{(2)}_C\bigg)   + s \, d \big( A^{(1)} \wedge F^{(2)}_A \wedge C^{(1)}\big) + t \, d \big(A^{(1)} \wedge C^{(1)} \wedge F^{(2)}_C\big)~. 
}[fullifive]
As explained above, the coefficients~$s, t \in \R$ of the exact terms in~$\CI^{(5)}$ can be adjusted using local counterterms in four dimensions. We now use~\descenteq\ to compute the anomalies~$\CA_A$ and~$\CA_C$ under~$U(1)_A^{(0)}$ and~$U(1)_C^{(0)}$ background gauge transformations, parametrized by~$\lambda_A^{(0)}$ and~$\lambda^{(0)}_C$, that result from~\fullifive,
\eqna{
& \CA_A = {i \over 4 \pi^2} \int_{\CM_4} \lambda^{(0)}_A \, \bigg({\kappa_{A^3} \over 3!} \, F^{(2)}_A \wedge F^{(2)}_A + \left({\kappa_{A^2 C} \over 2!} - s\right) F^{(2)}_A \wedge F^{(2)}_C + \left({\kappa_{AC^2} \over 2!} - t\right) F^{(2)}_C \wedge F^{(2)}_C \bigg)~,\cr
& \CA_C = {i \over 4 \pi^2} \int_{\CM_4} \lambda^{(0)}_C \, \bigg({\kappa_{C^3} \over 3!} \, F^{(2)}_C \wedge F^{(2)}_C + s\, F^{(2)}_A \wedge F^{(2)}_A + t\, F^{(2)}_A \wedge F^{(2)}_C\bigg)~.
}[anomalies]
The couplings of the currents~$j_{A}^{(1)}$ and~$j_C^{(1)}$ to the background gauge fields~$A^{(1)}$ and~$C^{(1)}$ are normalized as in~\eqref{sajcoup}. Therefore, the anomalies in~\anomalies\ imply the following non-conservation equations,
\eqna{
& d * j^{(1)}_A = - {i \over 4 \pi^2} ~\bigg({\kappa_{A^3} \over 3!} \, F^{(2)}_A \wedge F^{(2)}_A + \left({\kappa_{A^2 C} \over 2!} - s\right) F^{(2)}_A \wedge F^{(2)}_C + \left({\kappa_{AC^2} \over 2!} - t\right) F^{(2)}_C \wedge F^{(2)}_C  \bigg)~,\cr
& d * j^{(1)}_C = - {i \over 4 \pi^2} ~\left({\kappa_{C^3} \over 3!} \, F^{(2)}_C \wedge F^{(2)}_C + s \, F^{(2)}_A \wedge F^{(2)}_A  + t \, F^{(2)}_A \wedge F^{(2)}_C \right)~.
}[anomconseq]
Note that~\anomalies\ and~\anomconseq\ do not to treat~$U(1)_A^{(0)}$ and~$U(1)_C^{(0)}$ symmetrically for generic~$s,t$. The symmetry can be restored by choosing~$s = {1 \over 4} \kappa_{A^2 C}$ and~$t = {1 \over 4} \kappa_{AC^2}$. 

In section~\ref{SecBasics} we would like to gauge~$U(1)_C^{(0)}$. We must then ensure that~$U(1)_C^{(0)}$ gauge transformations are completely anomaly free, i.e.~that the anomalous shift~$\CA_C = 0$ in~\eqref{anomalies} vanishes, and hence that~$ d * j^{(1)}_C =0$ in~\eqref{anomconseq}. This is only possible if the cubic~$U(1)_C^{(0)}$ anomaly vanishes,
\eqn{
\kappa_{C^3} = 0~,
}[cccvanish]
but it also requires adjusting the counterterms so that
\eqn{
s = t = 0~.
}[ctfix]
Once this is done, the form of the~$U(1)_A^{(0)}$ anomaly~$\CA_A$ in~\anomalies\ is completely fixed,
\eqn{
\CA_A = {i \over 4 \pi^2} \int_{\CM_4} \lambda^{(0)}_A \, \bigg({\kappa_{A^3} \over 3!} \, F^{(2)}_A \wedge F^{(2)}_A + {\kappa_{A^2 C} \over 2!}\, F^{(2)}_A \wedge F^{(2)}_C + {\kappa_{AC^2} \over 2!} \, F^{(2)}_C \wedge F^{(2)}_C \bigg)~,
}[aanomctfix]
and the corresponding non-conservation equation in~\anomconseq\ is
\eqn{
d * j^{(1)}_A = - {i \over 4 \pi^2} ~\bigg({\kappa_{A^3} \over 3!} \, F^{(2)}_A \wedge F^{(2)}_A + {\kappa_{A^2 C} \over 2!} \, F^{(2)}_A \wedge F^{(2)}_C + {\kappa_{AC^2} \over 2!} \, F^{(2)}_C \wedge F^{(2)}_C  \bigg)~.
}[anonconseqctfix]
These equations will important in section~\ref{SecBasics}. Having fixed the counterterms to render~$U(1)_C^{(0)}$ anomaly free (and hence gaugeable), the non-vanishing of either mixed anomaly coefficient, $\kappa_{A^2 C}$ or~$\kappa_{A C^2}$, obstructs the further gauging of~$U(1)_A^{(0)}$, even if~$\kappa_{A^3} = 0$. 

For future reference, we present the analogue of the~$\kappa_{A^2 C}$ anomaly in~\aanomctfix\  for higher-rank abelian flavor symmetries of the form
\eqn{
G^{(0)} = \prod_I U(1)_I^{(0)} \times U(1)_C^{(0)}~.
}[uiuc]
The relevant terms in the anomaly polynomial are (see~\eqref{ijkabanom})
\eqn{
\CI^{(6)} \;\supset \; {1 \over 2! (2 \pi)^3} \sum_{I, J} \kappa_{IJ C} \, F_I^{(2)} \wedge F_J^{(2)} \wedge F_C^{(2)}~, \qquad \kappa_{IJ C} = \kappa_{(IJ) C}~.
}[ijcanompol]
By suitably adjusting the counterterms that arise in the context of these reducible anomalies, we can choose a symmetric presentation for the descent 5-form, 
\eqn{
\CI^{(5)} \; \supset \; {1 \over 2! (2 \pi)^3} \, \sum_{I, J}  \, \kappa_{IJC}  \, A_{(I}^{(1)} \wedge F_{J)} ^{(2)} \wedge F_C^{(2)}~. 
}[ijcifive]
The resulting anomalous shift~$\CA_I$ of the effective action under a~$U(1)_I^{(0)}$ background gauge transformation, parametrized by~$\lambda_I^{(0)}$, is then given by
\eqn{
\CA_I = {i \over 4 \pi^2} \sum_J {\kappa_{IJ C} \over 2 !} \int_{\CM_4} \lambda_I^{(0)} \, F_J^{(2)} \wedge F_C^{(2)}~.
}[aivar]
If we set~$I = J = A$ and write~$\kappa_{A A C} = \kappa_{A^2 C}$, we reproduce the corresponding term in~\aanomctfix.

\subsec{Nonabelian Flavor Symmetries and Background Gauge Fields}[ssecNonabHooft]

We now generalize the discussion of the previous subsection to include nonabelian~$0$-form symmetries. For simplicity, we focus on flavor symmetries of the from
\eqn{
G^{(0)} = SU(N)^{(0)}_A \times U(1)_C^{(0)}~.
}[sunuone] 
For the~$SU(N)_A^{(0)}$ background gauge fields, we follow the conventions of~\cite{AlvarezGaume:1984dr}, and write
\eqn{
A^{(1)} = A^{(1)a} \, t_a~, \qquad a = 1, \, \ldots, \, N^2 -1~,
}[nonabconv]
where the~$t_a$ are antihermitian~$SU(N)_A^{(0)}$ generators in the fundamental representation (i.e.~they are~$N \times N$ matrices), which are normalized so that~$\tr(t_a t_b) = - \half \delta_{ab}$. The field strength~$2$-form is then given by
\eqn{
F^{(2)}_A = dA^{(1)} + A^{(1)} \wedge A^{(1)}~, 
}[nonabfdef]
with a commutator implicit in the second term on the right-hand side. An infinitesimal~$SU(N)_A^{(0)}$ group element is parametrized by~$\1 + \lambda^{(0)}_A$, with
\eqn{
\lambda^{(0)}_A = \lambda^{(0)a}_A t_a~, \qquad  \lambda^{(0)a}_A  \in \R~.
}[nonablambdef]
An infinitesimal background gauge transformation then acts via the following shifts,
\eqn{
A^{(1)} \; \rightarrow \; A^{(1)} + d \lambda^{(0)}_A + [A^{(1)}, \lambda^{(0)}_A]~, \qquad F_A^{(2)} \; \rightarrow \; F^{(2)}_A +  [F^{(2)}_A, \lambda^{(0)}_A]~.
}[nonabgtdef]

The most general anomaly~$6$-form that can be constructed out of~$SU(N)_A^{(0)}$ and~$U(1)^{(0)}_C$ background fields is (see~\eqref{sunsunuone} and~\eqref{sunirred})
\eqna{
\CI^{(6)} = {1 \over (2 \pi)^3} \bigg(-{i \kappa_{A^3} \over 3!} \, \tr
\left(F^{(2)}_A  \wedge F^{(2)}_A \wedge F^{(2)}_A \right) & - {\kappa_{A^2 C} \over 2!} \, \tr\left(F^{(2)}_A  \wedge F^{(2)}_A \right) \wedge F^{(2)}_C \cr
& + {\kappa_{C^3} \over 3!} \, F^{(2)}_C \wedge F^{(2)}_C \wedge F^{(2)}_C  \bigg)~.
}[nonabsixfm]
A Weyl fermion~$\psi_\alpha$ in the fundamental representation of~$SU(N)_A^{(0)}$, with~$U(1)_C^{(0)}$ charge~$q_C$, contributes
\eqn{
\CI^{(6)} = \tr \exp\left({i\over 2 \pi}\, F^{(2)}_A\right) \exp \left({1\over 2\pi}\, F^{(2)}_C\right)\bigg|_\text{6-form}~.
}[weylsuncont]
Expanding the exponential and comparing with~\nonabsixfm\ leads to 
\eqn{
\kappa_{A^3} = 1~, \qquad \kappa_{A^2 C} = q_C~, \qquad \kappa_{C^3} = N q_C^3~.
}[sunfermionanom]
As discussed around~\eqref{sunirred}, the irreducible cubic anomaly~$\kappa_{A^3}$ is only possible if~$N\geq 3$. The abelian anomaly proportional to~$\kappa_{C^3}$ was already discussed in section~\ref{SecDescabelian} above. Since we would eventually like to gauge~$U(1)_C^{(0)}$, we will assume that it vanishes, $\kappa_{C^3} = 0$. Our primary interest is in the mixed, reducible anomaly proportional to~$\kappa_{A^2 C}$. In the remainder of this section we will therefore simplify the formulas by dropping  terms proportional to~$\kappa_{A^3}$. 

If we use
\eqn{
\tr\left(F^{(2)}_A \wedge F^{(2)}_A\right) = d \, \text{CS}^{(3)}(A)~, \qquad \text{CS}^{(3)}(A) = \tr \left(A \wedge dA + {2 \over 3} A \wedge A \wedge A\right)~,
}[nonabcsdef]
we can apply descent to the mixed term in~\nonabsixfm. This leads to the following descent~$5$-form,
\eqn{
\CI^{(5)} = - {\kappa_{A^2 C} \over 2! (2\pi)^3 } \, \text{CS}^{(3)}(A) \wedge F^{(2)}_C + s \, d \, \left(\text{CS}^{(3)}(A) \wedge C^{(1)} \right)~, \qquad s \in \R~.
}[nonabdescff]
As in the abelian case (see the discussion around~\ctfix) we set the parameter~$s$ in~\nonabdescff\ to zero using a local counterterm. The shift of the nonabelian Chern-Simons term in~\nonabcsdef\ under a background gauge transformation~\nonabgtdef\ is given by
\eqn{
\text{CS}^{(3)}(A) \quad \longrightarrow \quad \text{CS}^{(3)}(A) +  d \tr\left(\lambda_A^{(0)} d A^{(1)}\right)~.
}[nonabcsvar]
Note that this shift is linear in~$A^{(1)}$ and cannot be written in terms of the field strength~$F^{(2)}_A$ defined in~\nonabfdef. This leads to the following anomaly under~$SU(N)_A^{(0)}$ background gauge transformations, 
\eqn{
\CA_A = -{i \kappa_{A^2 C} \over 8 \pi^2} \int_{\CM_4} \tr \left(\lambda_A^{(0)} \, dA^{(1)}\right) \wedge F^{(2)}_C~.
}[nonabanom]

We normalize the coupling of the~$SU(N)_A^{(0)}$ current~$j_A^{(1)}$ to the associated background gauge field~$A^{(1)}$ as follows, 
\eqn{
S \; \supset \; \int  \, d^4 x \, A_\mu^a j^{a \mu}_A = -2 \int \tr\left(A^{(1)} \wedge * j^{(1)}_A\right)~.
}[nonabcurracoup]
The anomaly in~\nonabanom\ then leads to the following non-conservation equation,
\eqn{
d * j_A^{(1)} = {i \kappa_{A^2 C} \over 16 \pi^2} \, dA^{(1)} \wedge F_C^{(2)}~.
}[nonabnoncons]
This is nearly identical to the abelian~$\kappa_{A^2 C}$ term in~\anonconseqctfix, up to a relative factor of~$-\half$ which is due to~$\tr (t_a t_b)=-\half \delta_{ab}$.

\subsec{Poincar\'e Symmetry and Background Gravity Fields}[ssecGRHooft]

As for nonabelian gauge fields, we follow the conventions of~\cite{AlvarezGaume:1984dr} for background gravity fields. It is convenient (and, in theories with spinor fields, unavoidable) to describe gravity using an orthonormal frame~$e^a_\mu$, so that the riemannian metric is~$g_{\mu\nu} = \delta_{ab} e^a_\mu e^b_\nu$.  Here~$a,b$ and~$\mu, \nu$ are, respectively, frame indices (which are raised and lowered with~$\delta^{ab}, \delta_{ab}$) and spacetime indices (which are raised and lowered with~$g^{\mu\nu}, g_{\mu\nu}$). The indices~$a,b$ are acted on by local~$SO(4)$ frame rotations, and the indices~$\mu, \nu$ by diffeomorphisms. Together, these are the gauge transformations of gravity. An infinitesimal local frame rotation is an~$SO(4)$ group element~${\delta^a}_b + {(\theta^{(0)})^a}_b(x)$, with~$(\theta^{(0)})_{ab} = (\theta^{(0)})_{[ab]}$, and an infinitesimal diffeomorphism is parametrized by a vector field~$\xi^\mu(x)$. Under these transformations, the~1-form frame field~$e^{(1)a} = e^a_\mu dx^\mu$ shifts as follows,
\eqn{
e^{(1)a} \quad \longrightarrow \quad e^{(1)a}  - {(\theta^{(0)})^a}_b e^{(1)b} + \CL_\xi e^{(1)a}~,
}[deltagrave]
where~$\CL_\xi$ is the Lie derivative along the vector field~$\xi^\mu$. We will also need the spin connection~$1$-form~${\omega^{(1)a}}_b$, which is defined by the relations
\eqn{
de^{(1)a} + {\omega^{(1)a}}_b \wedge e^{(1)b} = 0~, \qquad \omega^{(1)}_{ab} = \omega^{(1)}_{[ab]}~,
}[omegadef]
as well as the Riemann curvature~$2$-form,
\eqn{
{R^{(2)a}}_b = d {\omega^{(1)a}}_b + {\omega^{(1)a}}_c \wedge {\omega^{(1)c}}_b~, \qquad R^{(2)}_{ab} = R^{(2)}_{[ab]}~.
}[riemdef]
Both~${\omega^{(1)a}}_b$ and~${R^{(2)a}}_b$ are valued in the~$SO(4)$ Lie algebra. Under a local frame rotation~\deltagrave, parametrized by~${(\theta^{(0)})^a}_b$, the spin connection and the Riemann curvature shift as follows,
\eqna{
& {\omega^{(1)a}}_b \quad \longrightarrow \quad {\omega^{(1)a}}_b + d {(\theta^{(0)})^a}_b + {\omega^{(1)a}}_c \, {(\theta^{(0)})^c}_b - {(\theta^{(0)})^a}_c \, {\omega^{(1)c}}_b~,\cr
& {R^{(2)a}}_b \quad \longrightarrow \quad {R^{(2)a}}_b +   {R^{(2)a}}_c \, {(\theta^{(0)})^c}_b - {(\theta^{(0)})^a}_c \, {R^{(2)c}}_b~.
}[lortomegar]
Note the similarity between~\riemdef, \lortomegar\ and the corresponding formulas~\nonabfdef, \nonabgtdef\ for nonabelian gauge fields. The former can be obtained from the latter by interpreting frame indices as fundamental~$SO(4)$ gauge indices and replacing~$A^{(1)} \rightarrow \omega^{(1)}$, $F^{(2)}_A \rightarrow R^{(2)}$, and~$\lambda_A^{(0)} \rightarrow \theta^{(0)}$.

In a gravitational background, an insertion of the stress tensor~$T_{\mu\nu} = e^a_\mu e^b_\nu T_{ab}$ is defined as the response to a variation in the frame field, i.e.~it is a functional derivative of the partition function, 
\eqn{
\sqrt{g} \, T_{ab}(x) = - e_{b \mu} \, {\delta Z \over \delta e^a_\mu(x)}~.
}[tdef]
Note that this definition of~$T_{ab}$ is not obviously symmetric in~$a,b$. Together with the transformation rule of the vielbein in~\deltagrave, it implies the following shift of the effective action~$W$ under local frame rotations and diffeomorphisms,
\eqn{
W \quad \longrightarrow \quad W - \int \sqrt g \, d^4 x\,  \theta^{(0) ab} \, T_{[ab]} - \int \sqrt g \, d^4 x\,  \xi_\nu \grad_\mu T^{\mu\nu}~.
}[inflordiff]
In the absence of anomalies, this shift vanishes and the stress tensor is symmetric and (covariantly) conserved. 

It is a non-trivial fact that one can always regulate a QFT in such a way as to preserve invariance under either local frame rotations or diffeomorphisms. It is therefore always possible to set one of the terms in~\eqref{inflordiff} to zero by adjusting certain local counterterms in the background gravity fields~\cite{Bardeen:1984pm, AlvarezGaume:1984dr}.  For instance, as in~\cite{AlvarezGaume:1983ig}, one can choose to preserve invariance under local frame rotations. It then follows from~\inflordiff\ that~$T_{ab} = T_{(ab)}$ is symmetric, but potentially not conserved due to a diffeomorphism 't Hooft anomaly. For our purposes, it is more convenient to assume that the counterterms have been chosen to preserve diffeomorphisms. There may then be an 't Hooft anomaly associated with local frame rotations. In this case~\inflordiff\ implies that the stress tensor is conserved, but may develop an antisymmetric part~$T_{[ab]}$ in the presence of suitable background fields. 

The most general anomaly~$6$-form that can be constructed out of background~$U(1)_C^{(0)}$ and gravity fields (see~\eqref{ipoincanom}) is given by
\eqn{
\CI^{(6)} = {1 \over (2 \pi)^3}  \bigg( {\kappa_{C\SP^2} \over 48}  \, \tr\left(R^{(2)} \wedge R^{(2)}\right) \wedge F^{(2)}_C + {\kappa_{C^3} \over 3!} \, F^{(2)}_C \wedge F^{(2)}_C \wedge F_C^{(2)} \bigg)~.
}[cgravanompol]
Here we use~$\tr$ to denote a trace over~$SO(4)$ frame indices, so that 
\eqn{
\tr \left(R^{(2)} \wedge R^{(2)}\right) = {R^{(2)a}}_b \wedge {R^{(2)b}}_a~.
}[rwedger]
In our conventions, a collection of Weyl fermions~$\psi^i_\alpha$ with~$U(1)_C^{(0)}$ charges~$q_C^i$ contributes
\eqn{
\CI^{(6)} = \sum_i \hat A \, \exp\left({q_C^i \over 2 \pi} F_C^{(2)}\right) \bigg|_\text{6-form}~, \qquad \hat A = 1 + {1 \over 192 \pi^2} \tr\left(R^{(2)} \wedge R^{(2)}\right) + \cdots~.
}[frepsianomgr]
Here~$\hat A$ is the Dirac genus (see appendix~\ref{AppQuantAnom}).  Comparing with~\rwedger\ then implies that 
\eqn{
\kappa_{C\SP^2} = \sum_i q_C^i~, \qquad \kappa_{C^3} =  \sum_i \left(q_C^i\right)^3~.
}[weylfgrava]
Below, we would like to gauge~$U(1)_C^{(0)}$, so we assume that~$\kappa_{C^3} = 0$. 

Note that (up to an overall sign) the mixed~$U(1)_C^{(0)}$-$\SP$ anomaly in~\cgravanompol\ takes the same form as the mixed~$U(1)_C^{(0)}$-$SU(N)_A^{(0)}$ anomaly in~\nonabsixfm, after substituting
\eqn{
\kappa_{A^2 C} \; \rightarrow \; {\kappa_{C\SP^2} \over 24}~, \qquad F^{(2)}_A \; \rightarrow \; R^{(2)}~.
}[nonabtog]
We can therefore follow the same steps that were described there (including adjusting a 
certain counterterm proportional to~$\text{CS}^{(3)}(\omega) \wedge C^{(1)}$, where~$\text{CS}^{(3)}(\omega)$ is the gravitational Chern-Simons term defined in~\eqref{gravcs} below, to ensure that~$U(1)_C^{(0)}$ is free of anomalies) to obtain the following descent~$5$-form from~\cgravanompol,
\eqn{
\CI^{(5)} = {\kappa_{C\SP^2} \over 48 (2\pi )^3} \, \text{CS}^{(3)}(\omega) \wedge F^{(2)}_C~.
}[gravdescff]
Here the gravitational Chern-Simons term~$\text{CS}^{(3)}(\omega)$ is given by
\eqn{
\text{CS}^{(3)}(\omega) =\tr \left(\omega^{(1)} \wedge d \omega^{(1)} + {2 \over 3} \, \omega^{(1)} \wedge \omega^{(1)} \wedge \omega^{(1)}\right)~,
}[gravcs]
and it satisfies
\eqn{
d \, \text{CS}^{(3)}(\omega) = \tr\left(R^{(2)} \wedge R^{(2)}\right)~.
}[dgravcsisrr]
Using~\lortomegar, the variation of the Chern-Simons term~\gravcs\ under a local frame rotation parametrized by~$\theta^{(0)}$ is given by the gravitational analogue of~\nonabcsvar,
\eqn{
\text{CS}^{(3)}(\omega) \quad \longrightarrow \quad \text{CS}^{(3)}(\omega) +   d \tr\left(\theta^{(0)} \, d \omega^{(1)}\right)~.
}[csvar]
This allows us to determine the anomaly in Poincar\'e symmetry from~\gravdescff,
\eqn{
\CA_\SP = {i \kappa_{C\SP^2} \over 192 \pi^2} \int_{\CM_4} \tr\left(\theta^{(0)} \, d \omega^{(1)} \right) \wedge F_C^{(2)}~. 
}[gravanom]
This anomaly follows (up to an overall sign) from the nonabelian formula~\nonabanom, if we substitute~$\kappa_{A^2 C} \rightarrow {\kappa_{C \SP^2} \over 24}$, as in~\nonabtog, as well as~$\lambda_A^{(0)} \rightarrow \theta^{(0)}$, and~$A^{(1)} \rightarrow \omega^{(1)}$. Comparing~\gravanom\ with the definition of the stress tensor in~\tdef\ leads to
\eqn{
T_{[ab]} = { i \kappa_{C\SP^2} \over 192 \pi^2} \, * \left(d \omega^{(1)}_{ab} \wedge F^{(2)}_C\right)~. 
}[tasympart]
The fact that the stress tensor develops and antisymmetric part in the presence of background fields is the analogue of the anomalous non-conservation equations~\anonconseqctfix\ and~\nonabnoncons.

\newsec{2-Group Symmetries from Mixed 't Hooft Anomalies}[SecBasics]

Here we elaborate on sections~\ref{gsfbgf} and~\ref{genintro}, where it was pointed out that theories with continuous 2-group symmetries arise from parent theories with a~$U(1)_C^{(0)}$ flavor symmetry and suitable mixed 't Hooft anomalies, by gauging~$U(1)_C^{(0)}$.\foot{~See~\cite{Tachikawa:2017gyf} for a discrete analogue.} We review and expand on the simplest abelian case discussed in section~\ref{gsfbgf}, before explaining the origin of the more general abelian, nonabelian, and Poincar\'e 2-group symmetries summarized in section~\ref{genintro}.

\subsec{Constructing the Simplest Abelian 2-Groups}[ssecBC] 

As in sections~\ref{gsfbgf} and~\ref{SecDescabelian}, we first consider parent theories with the following abelian~$0$-form flavor symmetry,
\eqn{
G^{(0)} = U(1)_A^{(0)} \times U(1)_C^{(0)}~.
}[uoneacii]
The corresponding background fields are~$A^{(1)}$ and~$C^{(1)}$. For now we ignore all other background fields, including gauge fields for possible nonabelian flavor symmetries, or gravity. (They are discussed in section~\ref{ssecBCgen} below.) As in~\eqref{gaugeuonec}, we would like to gauge~$U(1)_C^{(0)}$, by promoting the background gauge field~$C^{(1)}$ and its field strength~$F^{(2)}_C$ to dynamical fields,
\eqn{
U(1)_C^{(0)} \; \rightarrow \; U(1)_c^{(0)}~, \qquad C^{(1)} \; \rightarrow \;  c^{(1)}~, \qquad F^{(2)}_C \; \rightarrow \; f_c^{(2)}~.
}[gaugec]
We then perform the functional integral over gauge orbits of~$c^{(1)}$. This typically requires adding a suitably positive-definite quadratic action,
\eqn{
S \; \supset \; {1 \over 2 e^2} \int \,  \, f^{(2)}_c \wedge * f^{(2)}_c + {i \theta \over 8 \pi^2} \int \, f^{(2)}_c \wedge f^{(2)}_c~.
}[maxkin]
Here~$e$ is the gauge coupling, and we have also included a~$\theta$-term. Since the theories we are interested in generally contain fermions, we will always take the spacetime manifold~$\CM_4$ to be spin. Therefore, ${1 \over 8 \pi^2} \int_{\CM_4} f^{(2)}_c \wedge f^{(2)}_c \in \Z$, so that~$\theta \sim \theta + 2 \pi$ has standard periodicity.

As was explained in section~\ref{SecDescabelian}, it is only possible to gauge~$U(1)_C^{(0)}$ if~$\kappa_{C^3} = 0$ (see~\cccvanish) and if the counterterms are adjusted as in~\ctfix. The anomalous c-number shift~$\CA_A$ under~$U(1)_A^{(0)}$ background gauge transformations (parametrized by~$\lambda_A^{(0)}$) and the non-conservation equation for~$j_A^{(1)}$ are then given by~\aanomctfix, \anonconseqctfix, which we repeat here, 
\eqna{
& \CA_A = {i \over 4 \pi^2} \int \lambda^{(0)}_A \, \bigg({\kappa_{A^3} \over 3!} \, F^{(2)}_A \wedge F^{(2)}_A + {\kappa_{A^2 C} \over 2!}\, F^{(2)}_A \wedge F^{(2)}_C + {\kappa_{AC^2} \over 2!} \, F^{(2)}_C \wedge F^{(2)}_C \bigg)~,\cr
& d * j^{(1)}_A = - {i \over 4 \pi^2} ~\bigg({\kappa_{A^3} \over 3!} \, F^{(2)}_A \wedge F^{(2)}_A + {\kappa_{A^2 C} \over 2!} \, F^{(2)}_A \wedge F^{(2)}_C + {\kappa_{AC^2} \over 2!} \, F^{(2)}_C \wedge F^{(2)}_C  \bigg)~.
}[aanomnonconsrep]
Upon gauging, the background field strength~$F_C^{(2)}$ turns into the operator~$f_c^{(2)}$ (see~\gaugec). This converts the anomalous shifts proportional to~$\kappa_{A^2C}$ and~$\kappa_{AC^2}$ in~\aanomnonconsrep\ from c-numbers into operators. (The term proportional to~$\kappa_{A^3}$ remains a c-number, but its status as an 't Hooft anomaly changes, see section~\ref{sssHoftAbTwGp}.) Unlike 't Hooft anomalies, such operator-valued shifts cannot be thought of as variations of the~c-number effective action~$W[\CB]$ for background fields~$\CB$. In the remainder of this section we explain how to correctly account for such operator-valued shifts. As in section~\ref{gsfbgf}, some of them give rise to 2-group symmetries.  

We first examine the mixed~$\kappa_{AC^2}$ anomaly in~\aanomnonconsrep. Upon gauging~$U(1)_C^{(0)}$, it gives rise to an ABJ anomaly for the~$U(1)_A^{(0)}$ current (see the comments below~\anompolint\ and~\anomfundef),
\eqn{
d * j_A^{(1)} \; \supset \; -{i\kappa_{AC^2} \over 8 \pi^2} \, f_c^{(2)} \wedge f_c^{(2)}~.
}[abjnoncons]
Since~$f^{(2)}_c \wedge f^{(2)}_c$ is a nontrivial operator, the ABJ anomaly violates current conservation, even in the absence of background fields and at separated points inside correlation functions. The ABJ non-conservation equation~\abjnoncons\ is associated with the following operator-valued shift, which arises upon substituting~$F^{(2)}_C \rightarrow f_c^{(2)}$ into~$\CA_A$ (see~\aanomnonconsrep),
\eqn{
\CA_A\big(F^{(2)}_C \rightarrow f_c^{(2)}\big) \; \supset \; {i \kappa_{AC^2} \over 8 \pi^2} \int \lambda_A^{(0)} \; f_c^{(2)} \wedge f_c^{(2)}~.
}[abjanom]
As was already mentioned above, such operator-valued shifts cannot be interpreted as a non-invariance of the effective action~$W[\CB]$, which is a c-number that only depends on background fields~$\CB$. Instead, they are accounted for by modifying the transformation rules of some background fields in such a way that all operator-valued shifts ultimately cancel. (There may of course still be 't Hooft anomalies that shift~$W[\CB]$ by a c-number.) Note that this does not change the dynamics of the theory. As such, it is distinct from what is typically referred to as anomaly cancellation, which involves coupling the theory to additional propagating fields. 

As is well known, the ABJ anomaly~\abjanom\ can be described by promoting the~$\theta$-angle in~\maxkin\ to a background field~$\theta(x)$ that acts as a source for the operator~$f^{(2)}_c \wedge f^{(2)}_c$. Under~$U(1)_A^{(0)}$ background gauge transformations, $\theta(x)$ shifts jointly with the gauge field~$A^{(1)}$, 
\eqn{
A^{(1)} \quad \longrightarrow \quad A^{(1)} + d \lambda_A^{(0)}~, \qquad \theta \quad \longrightarrow \quad \theta - \kappa_{A C^2} \, \lambda_A^{(0)}~.
}[athetashift]
If~$\theta$ were a dynamical scalar field, this transformation rule would mean that~$U(1)_A^{(0)}$ is spontaneously broken, and~$\theta$ would be the corresponding NG boson.\foot{~In this case, $\theta$ would be an axion and~$U(1)_A^{(0)}$ the corresponding Peccei-Quinn~\cite{Peccei:1977ur} symmetry.} Freezing~$\theta$ into a fixed background field configuration converts spontaneous into explicit breaking, because no fixed configuration~$\theta(x)$ is invariant under the shift in~\athetashift. In the remainder of this paper, we will focus on~$U(1)_A^{(0)}$ flavor symmetries that are not explicitly broken by ABJ anomalies.  We thus require
\eqn{
\kappa_{AC^2} = 0~.
}[abjzero]

We now repeat the preceding analysis for the~$\kappa_{A^2 C}$ anomaly in~\aanomnonconsrep. After we gauge $U(1)_C^{(0)}$, it leads to the following non-conservation equation for the~$U(1)_A^{(0)}$ current,
\eqn{
d * j_A^{(1)} \; \supset \; -{i \kappa_{A^2 C}  \over 8 \pi^2} \, F^{(2)}_A \wedge f_c^{(2)}~.
}[hybnoncons]
The right-hand side contains both the background field~$F^{(2)}_A$ and the operator~$f_c^{(2)}$. The current~$j_A^{(1)}$ is broken by the operator if the~$U(1)_A^{(0)}$ background field strength is non-trivial. However, if~$F^{(2)}_A = 0$ the right-hand side of~\hybnoncons\ vanishes. Thus~$j_A^{(1)}$ is a conserved current operator, i.e. it satisfies~$d* j_A^{(1)} = 0$ at separated points inside correlation functions. This is the first of many ways in which~\hybnoncons\ is fundamentally different from the ABJ anomaly reviewed above, which breaks current conservation even at separated points. It is also distinct from 't Hooft anomalies such as~\aanomnonconsrep\ (prior to gauging~$U(1)_C^{(0)}$), which only break current conservation by c-number terms in the background fields. As before, the non-conservation equation~\hybnoncons\ is associated with an operator-valued shift (see~\aanomnonconsrep),
\eqn{
\CA_A\big(F^{(2)}_C \rightarrow f_c^{(2)}\big) \; \supset \; {i \kappa_{A^2 C} \over 8 \pi^2} \int  \lambda_A^{(0)} \, F^{(2)}_A \wedge f_c^{(2)}~. 
}[ahybanom]
We must now understand which background fields can be used to cancel~\ahybanom\ at the level of the effective action~$W[\CB]$.

As explained in~\cite{Gaiotto:2014kfa}, and reviewed in section~\ref{convgsbf}, gauging~$U(1)_C^{(0)}$ gives rise to a new~1-form global symmetry:~the magnetic~$U(1)^{(1)}_B$ symmetry associated with the dynamical~$U(1)_c^{(0)}$ gauge field strength~$f^{(2)}_c$, with~$2$-form current~$J_B^{(2)}$ given by~\eqref{magcurrint} (see also footnote~\ref{ft:wickroti}),
\eqn{
J^{(2)}_B = {i \over 2 \pi} * f^{(2)}_c~.
}[magcurr]
It is conserved because~$f^{(2)}_c$ satisfies the Bianchi identity, so that~$d * J_B^{(2)} \sim d f^{(2)}_c = 0$.\foot{~Note that the electric~$1$-form symmetry of free Maxwell theory (see appendix~\ref{appHooftFreeMax} and~\cite{Gaiotto:2014kfa}) is explicitly broken in the presence of electrically charged matter, because~${1 \over e^2} \, d *f_c^{(2)} \sim  * j_C^{(1)} \neq 0$.} The magnetic 1-form charges, evaluated by integrating~${1 \over 2 \pi} \, f_c^{(2)}$ over closed~$2$-cycles~$\Sigma_2$ (see~\eqref{qqdef}), are integers because~$f_c^{(2)}$ is a~$U(1)_c^{(0)}$ field strength, so that~${1 \over 2 \pi} \int_{\Sigma_2} f^{(2)}_c \in \Z$. (This was explained around~\eqref{fsquant} for~$U(1)^{(0)}$ background gauge fields, but it also applies in the dynamical case.) As explained around~\eqref{bjbcoup} and~\eqref{Bfcoup}, the appropriate classical source for~$J^{(2)}_B$ is a~$2$-form background gauge field~$B^{(2)}$, 
\eqn{
S \; \supset \; \int B^{(2)} \wedge * J_B^{(2)} = {i \over 2 \pi} \int B^{(2)} \wedge f^{(2)}_c~.
}[bfterm]
 This is a BF-term for the background field~$B^{(2)}$ and the dynamical field~$f^{(2)}_c$. As in~\eqref{bgtshift} and~\eqref{LamHflux}, the~$2$-form~$B^{(2)}$ is subject to~$U(1)_B^{(1)}$ background gauge transformations, which are parametrized by a (locally-defined) 1-form~$\Lambda_B^{(1)}$ with suitably quantized periods,
\eqn{
B^{(2)} \quad \longrightarrow \quad B^{(2)} + d \Lambda_B^{(1)}~, \qquad {1 \over 2 \pi} \int_{\Sigma_2} d \Lambda_B^{(1)} \in \Z~.
}[blamgt]
Invariance under small~$U(1)_B^{(1)}$ background gauge transformations (for which~$\Lambda_B^{(1)}$ has trivial fluxes) captures the Bianchi identity~$d f^{(2)}_c = 0$. The possibility of large~$U(1)_B^{(1)}$ gauge transformations, under which the BF term in~\blamgt\ is also invariant, arises because the magnetic 1-form charges (measured by integrals of~${1 \over 2 \pi} \, f_c^{(2)}$) are quantized. In general, invariance under large~$U(1)_B^{(1)}$ gauge transformations requites BF terms to have quantized coefficients,~${i n \over 2 \pi} \int B^{(2)} \wedge f^{(2)}_c$ with~$n \in \Z$. This fact will play an important role in section~\ref{SecGSAnom}. 

Given that the background~$2$-form gauge field~$B^{(2)}$ in~\bfterm\ is the appropriate source for the operator~$f^{(2)}_c$, we can cancel the operator-valued shift in~\ahybanom\ by declaring that~$B^{(2)}$ undergoes a GS, or 2-group, shift under~$U(1)_A^{(0)}$ background gauge transformations. As in~\eqref{btwgpbiscoef}, this shift takes the following form,
\eqn{
B^{(2)} \quad \longrightarrow \quad B^{(2)} + {\hat \kappa_A \over 2 \pi} \, \lambda_A^{(0)} \, F_A^{(2)}~, \qquad \hat \kappa_A = - \half \, \kappa_{A^2 C}~,  
}[coeffmatch]
with the 2-group structure constant~$\hat \kappa_A$ determined by the mixed~$\kappa_{A^2 C}$ anomaly coefficient. Recall from section~\ref{gsfbgf} that~$\hat \kappa_A \in \Z$, which requires~$\kappa_{A^2 C} \in 2 \Z$. These quantization conditions are explained in section~\eqref{SecGlobal} and appendix~\ref{AppQuantAnom}. If~$B^{(2)}$ were a dynamical 2-form gauge field, the transformation rule~\coeffmatch\ would implement the conventional GS mechanism, with the BF-term~\bfterm\ playing the role of the associated GS term. As explained in section~\ref{gsfbgf}, freezing~$B^{(2)}$ into a background field instead leads to the abelian 2-group global symmetry~$U(1)_A^{(0)} \times_{\hat \kappa_A} U(1)_B^{(1)}$ in~\eqref{abtwgpdef}.

Note the similarity between~\coeffmatch\ and the shift of the~$\theta$-angle in~\athetashift, which accounts for the ABJ anomaly. However, an important difference is that the~$\theta$-angle in~\athetashift\ shifts under~$U(1)_A^{(0)}$, which indicates that the symmetry is explicitly broken. By contrast, the 2-form gauge field~$B^{(2)}$ only shifts under~$U(1)_A^{(0)}$ if the background field strength~$F_A^{(2)}$ is nonzero. This mirrors the fact that the right-hand side of the 2-group non-conservation equation~\hybnoncons\ vanishes if~$F_A^{(2)} = 0$, which ensures that~$j_A^{(1)}$ is a conserved current.

\subsec{More General Abelian, Nonabelian, and Poincar\'e 2-Groups}[ssecBCgen]

We now explain how to obtain the more general 2-group symmetries summarized in section~\ref{genintro} by gauging a~$U(1)_C^{(0)}$ flavor symmetry with suitable mixed 't Hooft anomalies. We start with the higher-rank abelian 2-group symmetries~$\big(\prod_I U(1)_I^{(0)}\big) \times_{\hat \kappa_{IJ}} U(1)_B^{(1)}$ introduced around~\eqref{multabbshift}. These arise from parent theories with flavor symmetry
\eqn{
G^{(0)} = \prod_I U(1)_I^{(0)} \times U(1)_C^{(0)}~.
}[hrabflsym]
Since we would like to gauge~$U(1)_C^{(0)}$ while preserving all~$U(1)_I^{(0)}$ symmetries, we demand that the~$\kappa_{C^3}$ gauge anomaly and all~$\kappa_{I C^2}$ ABJ anomalies vanish. As explained around~\eqref{aivar}, it is possible to choose counterterms so that operator-valued shift under~$U(1)_I^{(0)}$ background gauge transformations (parametrized by~$\lambda_I^{(0)}$) that arises after gauging~$U(1)_C^{(0)}$ is given by
\eqn{
\CA_I(F_C^{(2)} \rightarrow f_c^{(2)}) \; \supset \; {i \over 4 \pi^2} \sum_J \, {\kappa_{IJC} \over 2!} \int \lambda_I^{(0)} \, F_J^{(2)} \wedge f_c^{(2)}~.
}[aiopshift]
Here~$\kappa_{IJC} = \kappa_{(IJ)C}$ are the mixed~$U(1)_I^{(0)}$-$U(1)_J^{(0)}$-$U(1)_C^{(0)}$ 't Hooft anomaly coefficients that appear in the anomaly polynomial~\ijcanompol. In order to cancel the operator-valued shift~\aiopshift\ for all~$U(1)_I^{(0)}$ background gauge transformations, we again use the BF term in~\bfterm\ as a GS term and assign the following 2-group shift to~$B^{(2)}$ (see~\eqref{multabbshift}),
\eqn{
B^{(2)} \quad \longrightarrow \quad B^{(2)} + {1 \over 2\pi} \sum_{I, J} \hat \kappa_{IJ} \, \lambda_I^{(0)} \,  F_J^{(2)}~,  \qquad \hat \kappa_{IJ} = \hat \kappa_{(IJ)} =  - \half \kappa_{IJC}~.
}[bijshiftsect]
Now the 2-group structure constants~$\hat \kappa_{IJ}$ determine a symmetric matrix with integer entries.\foot{~It is straightforward to extend the arguments in appendix~\ref{AppQuantAnom} to show that the 't Hooft anomaly coefficients~$\kappa_{IJC} \in 2 \Z$. This always holds for the off-diagonal entries with~$I \neq J$. For the diagonal entries~$\kappa_{IIC} =  \kappa_{I^2 C}$ it follows from the assumption that the ABJ anomaly~$\kappa_{IC^2}$ vanishes.} 

We now show how a theory with nonabelian and Poincar\'e 2-group symmetry (see~\eqref{nonabbint} and~\eqref{ptwgpint}) can be constructed by gauging~$U(1)_C^{(0)}$ in a parent theory with flavor symmetry 
\eqn{
G^{(0)} = SU(N)_A^{(0)} \times U(1)_C^{(0)}~.
}[sunoneii]
The possible 't Hooft anomalies for such a theory were reviewed in sections~\ref{ssecNonabHooft} and~\ref{ssecGRHooft}. Here we focus on the following mixed terms in the anomaly~$6$-forms~\nonabsixfm\ and~\cgravanompol, which involve the~$SU(N)_A^{(0)}$ and~$U(1)_C^{(0)}$ background gauge fields~$A^{(1)}$ and~$C^{(1)}$, as well as background gravity fields,
\eqn{
\CI^{(6)} \; \supset \; {1 \over (2 \pi)^3} \bigg(- {\kappa_{A^2 C} \over 2!} \, \tr\left(F^{(2)}_A \wedge F^{(2)}_A\right) \wedge F^{(2)}_C + {\kappa_{C\SP^2} \over 48} \, \tr\left(R^{(2)} \wedge R^{(2)} \bigg) \wedge F^{(2)}_C\right)~.
}[mixedcnag]
As above, gauging~$U(1)_C^{(0)}$ is only possible if~$\kappa_{C^3} = 0$. Moreover, we must adjust the counterterms so that the operator-valued shifts under background~$SU(N)^{(0)}_A$ gauge transformations and local frame rotations that arise after gauging~$U(1)_C^{(0)}$ are given by~\nonabanom\ and~\gravanom, 
\eqna{
& \CA_A(F_C^{(2)} \rightarrow f_c^{(2)}) \; \supset \;  - {i \kappa_{A^2 C} \over 8 \pi^2} \int \tr \left(\lambda_A^{(0)} \, dA^{(1)}\right) \wedge f_c^{(2)}~, \cr
&  \CA_\SP(F_C^{(2)} \rightarrow f_c^{(2)}) \; \supset \; {i \kappa_{C\SP^2} \over 192 \pi^2} \int \tr\left(\theta^{(0)} \, d \omega^{(1)}\right) \wedge f_c^{(2)}~.
}[anomsungravii]
We also recall the corresponding non-conservation equations~\nonabnoncons\ and~\tasympart,
\eqn{
d * j_A^{(1)}  = {i \kappa_{A^2 C} \over 16 \pi^2} \, dA^{(1)} \wedge f_c^{(2)}~, \qquad T_{[ab]} = {i \kappa_{C\SP^2} \over 192 \pi^2} * \left( d \omega_{ab}^{(1)} \wedge f_c^{(2)}\right)~.
}[nonconssungravii]
Just as~\hybnoncons, these non-conservation equations have the property that their right-hand sides involve both a background field (either~$dA^{(1)}$ or~$d\omega^{(1)}$) and the operator~$f_c^{(2)}$. This ensures that~$d *j_A^{(1)} = T_{[ab]} = 0$ in the absence of background fields, or inside correlation functions at separated points. As before, these conservation equations are broken by the operator~$f_c^{(2)}$ -- either in sufficiently non-trivial backgrounds, or by~$\delta$-function contact terms inside correlation functions. 

In order to cancel the operator-valued shifts in~\anomsungravii, we utilize the BF term~\bfterm\ and assign the following 2-group shift to~$B^{(2)}$ under~$SU(N)_A^{(0)}$ background gauge transformations and local frame rotations, 
\eqn{
B^{(2)} \; \rightarrow \; B^{(2)} + {\hat \kappa_A \over 4\pi} \, \tr\left(\lambda_A^{(0)} \, dA^{(1)}\right) + {\hat \kappa_\SP \over 16 \pi} \, \tr \left(\theta^{(0)} \, d \omega^{(1)}\right)~,~ \hat \kappa_A =  \kappa_{A^2 C}~, ~ \hat \kappa_\SP = -{\kappa_{C\SP^2} \over 6}~.
}[nonabgravkhat]
As in~\eqref{nonabbint}, \eqref{ptwgpint}, this amounts to a 2-group symmetry~$\left(SU(N)_A^{(0)} \times \SP\right) \times_{\hat \kappa_A \, , \, \hat \kappa_\SP} U(1)_B^{(1)}$. In section~\ref{SecGlobal} we show that both 2-group structure constants in~\nonabgravkhat\ are quantized, $\hat \kappa_A, \hat \kappa_\SP \in \Z$, which requires~$\kappa_{A^2 C} \in \Z$ and~$\kappa_{C\SP^2} \in 6\Z$. As is explained in appendix~\ref{AppQuantAnom}, the factor of~$6$ in the quantization of~$\kappa_{C\SP^2}$ is present whenever whenever~$\kappa_{C^3} = 0$, which we had to assume in order to gauge~$U(1)_C^{(0)}$. 

The presentation of the 2-group symmetries discussed above can be modified by redefining the background fields. This is particularly natural for Poincar\'e 2-group symmetry.  As was mentioned below~\eqref{inflordiff}, 't Hooft anomalies involving background gravity fields can manifest as anomalies in local frame rotations, or in diffeomorphisms. The two presentations are related by suitable local counterterms~\cite{Bardeen:1984pm, AlvarezGaume:1983ig, AlvarezGaume:1984dr}. Before we gauge~$U(1)_C^{(0)}$, the~$\kappa_{C \SP^2}$ anomaly can therefore be viewed as involving~$U(1)_C^{(0)}$ and either (i) local frame rotations or (ii) diffeomorphisms. Above, we have chosen option (i) by assuming that diffeomorphisms are preserved. It differs from option (ii) by a counterterm that involves the~$U(1)_C^{(0)}$ background field strength, as well as background gravity fields. Once we gauge~$U(1)_C^{(0)}$, it follows from~\bfterm\ that the counterterms relating the two presentations (i) and (ii), which now involve the dynamical field strength~$f_c^{(2)}$, can be absorbed by a field redefinition that shifts~$B^{(2)}$ by background gravity fields. The Poincar\'e 2-groups that result from (i) and (ii) are therefore physically equivalent. In description (ii), the 2-form background field~$B^{(2)}$ is invariant under local frame rotations, but it undergoes a 2-group shift under diffeomorphisms.

\newsec{2-Group Currents in Conformal and Non-Conformal Theories}[SecCurrents]
  
In this section, we continue our discussion of 2-group Ward identities from sections~\ref{twgpcurwi} and~\ref{rgphtwgp} of the introduction. We use these Ward identities to analyze the possible patterns of spontaneous 2-group breaking, and other aspects of RG flows with 2-group symmetries.

\subsec{2-Group Ward Identities and Characteristic Three-Point Functions}[secCharThrPt]

In section~\ref{twgpcurwi}, we considered the abelian 2-group~$U(1)_A^{(0)} \times_{\hat \kappa_A} U(1)_B^{(1)}$ introduced in~\abtwgpdef\ and showed that the 2-group OPE~\djjope\ leads to the Ward identity in~\twgpwiint\ for the characetristic~$\langle j_A^{(1)} j_A^{(1)} J_B^{(2)}\rangle$ three-point function, 
\eqn{
{\d \over \d x_\mu} \big\langle j_\mu^A(x) j_\nu^A(y) J_{\rho\sigma}^B (z)\big\rangle = {\hat \kappa_A \over 2\pi} \, \d^\lambda \delta^{(4)}(x-y) \, \big\langle J^B_{\nu\lambda}(y) J^B_{\rho\sigma} (z) \big \rangle~.
}[jjJwardid]
Additionally, $J^B_{\rho\sigma}$ is conserved inside the correlation function, 
\eqn{
{\d \over \d z_\rho}  \big\langle j_\mu^A(x) j_\nu^A(y) J_{\rho\sigma}^B (z)\big\rangle = 0~
}[jbconswi]
We will now explain in detail how the Ward identity~\jjJwardid\ encodes the 2-group symmetry, including the structure constant~$\hat \kappa_A$, in the characteristic~$\langle j_A^{(1)} j_A^{(1)} J_B^{(2)}\rangle$ three-point function at separated points. As we will see, this is true as long as~$J_B^{(2)}$ is a non-trivial operator. An important exception occurs when~$J_B^{(2)}$ is redundant. The characteristic three-point function then vanishes at separated points, but the theory may still possess 2-group symmetry. For instance, this can happen in TQFTs, where both~$j_A^{(1)}$ and~$J_B^{(2)}$ are redundant operators (see sections~\ref{sssHoftAbTwGp} and~\ref{ssTQFT} for more details and examples).  

As in the discussion around~\jjJmomintro, we pass from position space to momentum space, where scheme-independent information is encoded in non-analytic terms. By contrast, terms that are polynomials in the momenta are typically scheme-dependent and can be modified by adjusting local counterterms. (Some exceptions are discussed in section~\ref{secGScts} below.) In momentum space,\foot{\label{momcorrdef}~Given local operators~$\CA(x)$, $\CB(y)$, $\CC(z)$, we define the momentum space two-point function~$\langle \CA(p) \CB(-p)\rangle$ and the momentum space three-point function~$\langle \CA(p)\CB(q) \CC(-p-q)\rangle$ as follows (see also appendix~\ref{AppMomSpace}),
\eqn{
\langle \CA(p) \CB(-p)\rangle = \int d^4 x \, e^{-i  p \cdot x} \, \langle \CA(x) \CB(0)\rangle~,\quad \langle \CA(p)\CB(q) \CC(-p-q)\rangle = \int d^4x \, d^4 y \, e^{-i(p \cdot x + q \cdot y)} \, \langle \CA(x) \CB(y) \CC(0)\rangle~.
}
} 
the characteristic three-point function takes the form~$\langle j^A_\mu(p) j^A_\nu(q) J^B_{\rho\sigma}(-p-q)\rangle$ (here~$p, q$ are independent euclidean momenta), and Bose symmetry implies that it is symmetric under the simultaneous exchange~$\mu\,,\, p \leftrightarrow \nu\,,\, q$. The~$J_B^{(2)}$ conservation equation~\jbconswi\ implies
\eqn{
(p+q)^\rho \, \langle j^A_\mu(p) j^A_\nu(q) J^B_{\rho\sigma}(-p-q)\rangle = 0~,
}[jbconwimom]
while the 2-group Ward identity~\jjJwardid\ takes the form
\eqn{
p^\mu \langle j^A_\mu(p) j^A_\nu(q) J^B_{\rho\sigma}(-p-q)\rangle = {\hat \kappa_A \over 2 \pi} \, p^\lambda \, \langle J^B_{\nu\lambda}(p+q) J^B_{\rho\sigma} (-p-q)\rangle~.
}[momspward]
The momentum-space two-point function~$\langle J^B_{\mu\nu}(p) J^B_{\rho\sigma}(-p)\rangle$ that appears on the right-hand side is invariant under the simultaneous Bose exchange~$\mu\nu\,,\, p \leftrightarrow \rho\sigma\,,\, -p$ and satisfies the following conservation equation,
\eqn{
p^\mu \, \langle J^B_{\mu\nu}(p) J^B_{\rho\sigma}(-p)\rangle = 0~.
}[jbmomcons]

In order to analyze these equations, it is helpful to decompose the momentum-space correlators into independent Lorentz structures, multiplied by dimensionless, Lorentz-invariant structure functions. This task is carried out in appendix~\ref{AppMomSpace}. Here we summarize the results and highlight their implications, starting with the~$\langle J_B^{(2)} J_B^{(2)} \rangle$ two-point function. As is shown in appendix~\ref{ssjbjbtwopt}, current conservation~\jbmomcons\ and Bose symmetry imply that it is determined by a single real, dimensionless structure function~$\mathsf J\left({p^2 \over M^2}\right)$, 
\eqna{
\langle J^B_{\mu\nu}(p) J^B_{\rho\sigma}(-p)\rangle = {1 \over p^2} \, {\mathsf J} \left({p^2 \over M^2}\right) \,  \Big(p_\mu p_\rho \delta_{\nu\sigma} & - p_\nu p_\rho \delta_{\mu\sigma} - p_\mu p_\sigma \delta_{\nu\rho} \cr
& + p_\nu p_\sigma \delta_{\mu\rho} - p^2 \delta_{\mu\rho} \delta_{\nu\sigma} + p^2 \delta_{\nu\rho} \delta_{\mu\sigma}\Big)~,
}[jbjbmomsp]
where~$M$ is some mass scale. Note that the overall normalization of the structure function~$\mathsf J\left({p^2 \over M^2}\right)$ is meaningful, because~$J_B^{(2)}$ is a conserved current. In a CFT, scale invariance implies that~${\mathsf J}\left({p^2 \over M^2}\right) = {\mathsf J}$ is a constant, while reflection positivity requires~${\mathsf J} \geq 0$. If this inequality is saturated, $\mathsf J = 0$, the~$\langle J_B^{(2)} J_B^{(2)}\rangle$ correlator vanishes at separated points, which happens if and only if~$J_B^{(2)}$ is a redundant operator.\foot{~More generally, $J_B^{(2)}$ is redundant whenever~$\mathsf J\left({p^2 \over M^2}\right)$ is a polynomial without a term of degree~$0$.}

The two-point function in~\jbjbmomsp\ only contains parity-even Lorentz structures, i.e.~structures without an explicit Levi-Civita~$\ep$-symbol. Since we would like to understand which terms in the characteristic three-point function~$\langle j^A_\mu(p) j^A_\nu(q) J^B_{\rho\sigma}(-p-q)\rangle$ give rise to the nontrivial right-hand side of the Ward identity~\momspward, it suffices to focus on the parity-even part of that three-point function. (The parity-odd part is necessarily annihilated by~$p^\mu$.) In appendix~\ref{ssjajaJBthreept}, we decompose the parity-even part of~$\langle j^A_\mu(p) j^A_\nu(q) J^B_{\rho\sigma}(-p-q)\rangle$ into independent Lorentz structures, multiplied by dimensionless, Lorentz-invariant structure functions.  This task is simplified by restricting the momenta~$p,q$ to special configurations, 
\eqn{
p^2 = q^2 = (p+q)^2 = Q^2, \qquad p \cdot q = - \half Q^2~.
}[symkin]
Here~$Q$ is a Lorentz-scalar with dimensions of energy; all dimensionless structure functions only depend on~$Q^2 \over M^2$. Note that~\symkin\ fixes the magnitude of the momenta~$p$ and~$q$, as well as the angle between them, but their directions are otherwise arbitrary. 

The analysis of appendix~\ref{ssjajaJBthreept} shows that imposing~\symkin, as well as~\jbconwimom\ and~\momspward\ allows two independent parity-even Lorentz structures. The first structure is annihilated by~$p^\mu$, i.e.~it is conserved; the second structure matches the right-hand side of the Ward identity~\momspward\ and is therefore determined by the structure function~${\mathsf J}\left({p^2 \over M^2}\right)$ in~\jbjbmomsp,
\eqna{
& \langle j^A_\mu(p)  j^A_\nu(q)  J^B_{\rho\sigma}(-p-q)\rangle \; \supset \;  {\hat \kappa_A \over 2 \pi Q^2} \; {\mathsf J}\bigg({Q^2 \over M^2}\bigg)  \bigg( \delta_{\mu\rho} \left(p_\nu+q_\nu\right)\left(p_\sigma-q_\sigma\right)\cr
&  \qquad - \delta_{\mu\sigma} \left(p_\nu + q_\nu\right)\left(p_\rho - q_\rho\right) + \delta_{\nu\rho} \left(p_\mu + q_\mu\right) \left(q_\sigma - p_\sigma\right) - \delta_{\nu\sigma} \left(p_\mu + q_\mu\right)\left(q_\rho - p_\rho\right) \bigg)~.
}[jajajbwardidstr]
As long as~$J_B^{(2)}$ is not redundant and~$\hat \kappa_A \neq 0$, the non-analytic structure in~\jajajbwardidstr\ contributes to the three-point function on the left-hand side at separated points in position space. For instance, if~${\mathsf J}\left({p^2 \over M^2}\right) = \mathsf J$ is a constant, the right-hand side of~\jajajbwardidstr\ is proportional to a pole~$\sim {\hat \kappa_A \mathsf J \over Q^2}$, which can only arise from separated points in position space. 

The Ward identities for the abelian 2-group symmetry~$U(1)_A^{(0)} \times_{\hat \kappa_A} U(1)_B^{(1)}$ discussed above were derived using the 2-group OPE in~\djjope, which in turn followed from the non-conservation equation~$d * j_A^{(1)} \sim \hat \kappa_A \, F_A^{(2)} \wedge * J_B^{(2)}$ in~\modconseq. In order to generalize these results to nonabelian and Poincar\'e 2-groups (see section~\ref{gsfbgf}), we need the corresponding non-conservation equations. As in the abelian case, they can be derived from the 2-group shifts of~$B^{(2)}$ in~\eqref{nonabbint} and~\eqref{ptwgpint}. Here we will use a shortcut: in section~\ref{ssecBC}, we constructed examples with abelian 2-group symmetry~$U(1)_A^{(0)} \times_{\hat \kappa_A} U(1)_B^{(1)}$, where~$U(1)_B^{(1)}$ was the magnetic 1-form symmetry with 2-form current~$J_B^{(2)} \sim * f_c^{(2)}$ in~\magcurr. The non-conservation equation in~\hybnoncons\ then agrees with the general formula~\modconseq\ if we also use the relation~$\hat \kappa_A = - \half \kappa_{A^2 C}$ from~\coeffmatch. We can immediately repeat this argument for nonabelian and Poincar\'e 2-groups by using the construction in section~\ref{ssecBCgen}. Starting with the non-conservation equations~\nonconssungravii\ and making the identifications in~\magcurr, \nonabgravkhat, we thus find the following equations for the nonabelian flavor current~$j_A^{(1)}$ and the antisymmetric part of the stress tensor,
\eqn{
d * j_A^{(1)}  = {\hat \kappa_A \over 8 \pi} \, dA^{(1)} \wedge * J_B^{(2)}~, \qquad T_{[ab]} = -{\hat \kappa_\SP \over 16 \pi} * \left( d \omega_{ab}^{(1)} \wedge *J_B^{(2)}\right)~.
}[nonconsgentgp]
Just as in the abelian case, this leads to operator-valued contact terms proportional to~$J_B^{(2)}$ in the OPE of~$d * j_A^{(1)}$ with another~$U(1)_A^{(0)}$ current, or in the OPE of~$T_{[ab]}$ with another stress tensor. These contact terms give rise to Ward identities that schematically read~$\langle (d*j_A^{(1)}) \,j_A^{(1)} J_B^{(2)}\rangle \, \sim \, \hat \kappa_A \, \langle J_B^{(2)} J_B^{(2)}\rangle$ and~$\langle T_{[ab]} T_{cd} J_B^{(2)}\rangle \, \sim \,\hat \kappa_{\SP} \, \langle J_B^{(2)} J_B^{(2)}\rangle$. As before, this implies that the corresponding 2-group symmetries (including the structure constants~$\hat \kappa_A, \hat \kappa_\SP$) are encoded in the characteristic three point functions~$\langle j_A^{(1)} j_A^{(1)} J_B^{(2)}\rangle$ and~$\langle T_{ab} T_{cd} J_B^{(2)}\rangle$ at separated points, unless~$J_B^{(2)}$ is a redundant operator. 

\subsec{Primary Currents and Unbroken 2-Group Symmetry in CFT}[sssprimcur]

We proceed to analyze the characteristic three-point functions introduced above in CFTs. For now, we assume that all currents are (non-redundant) conformal primaries. As we will see, this is equivalent to the assumption that the~$U(1)_B^{(1)}$ subgroup of the 2-group is spontaneously broken, while all other symmetries are preserved. In the abelian case, this would amount to the breaking pattern~$U(1)_A^{(0)} \times_{\hat \kappa_A} U(1)_B^{(1)} \; \rightarrow \; U(1)_A^{(0)}$. We will argue that this breaking pattern, and its analogues for other 2-groups, is inconsistent with the 2-group Ward identities, by showing that the characteristic three-point functions vanish at separated points. This establishes the claims around~\eqref{ssbpat} that the~$U(1)_B^{(1)}$ subgroup of a 2-group can only be spontaneously broken of the same is true of the entire 2-group symmetry. This scenario will be discussed in section~\ref{sssnonprimcurr} below. 

The~$U(1)_A^{(0)}$ current~$j_A^{(1)}$ and the~$U(1)_B^{(1)}$ current~$J_B^{(2)}$ are conformal primaries that satisfy the conservation equations 
\eqn{
d * j_A^{(1)} = 0~, \qquad d*J_B^{(2)} = 0~.
}[jajbcons]
They must therefore reside in short multiplets of the conformal group, since the conservation equations~\jajbcons\ constitute null descendants.  Along with unitarity, this determines the conformal scaling dimensions of the currents (see for instance~\cite{Mack:1975je,Minwalla:1997ka,Cordova:2016emh} and references therein),
\eqn{
\Delta \big({j^{(1)}_A}\big) = 3~, \qquad \Delta \big(J^{(2)}_B\big) = 2~.
}[jajbdim]
Note that the corresponding charges, which are obtained by integrating~$* j_A^{(1)}$ over 3-cycles and~$*J_B^{(2)}$ over 2-cycles, are dimensionless. 

We will rely on a special feature of four-dimensional CFTs:\footnote{~Some other physical consequences of this feature were discussed in~\cite{Argyres:1995xn, Cordova:2016xhm}.} a two-form current~$J_B^{(2)}$, with scaling dimension~$\Delta \big(J^{(2)}_B\big) = 2$, is not only conserved, but also necessarily closed. In fact, the structure of possible conformal null states implies that all three statements are equivalent (see~\cite{Mack:1975je,Minwalla:1997ka,Cordova:2016emh}),
\eqn{
\Delta \big(J^{(2)}_B\big) = 2 \quad \Longleftrightarrow \quad d* J_B^{(2)} = 0 \quad \Longleftrightarrow \quad d J_B^{(2)} = 0~. 
}[jbdimccc]
This implies that~$J_B^{(2)}$ is proportional to the field strength~$f^{(2)}$ of a free Maxwell field, or its dual~$J_B^{(2)} \sim * f^{(2)}$\,.\foot{~This determines whether~$J_B^{(2)}$ is the electric (e) or the magnetic (m) 2-form current of free Maxwell theory, which has~$U(1)^{(1)}_\text{e} \times U(1)^{(1)}_\text{m}$ global symmetry~\cite{Gaiotto:2014kfa} (see also appendix~\ref{appHooftFreeMax}). We can split the field strength into its self-dual and anti-self-dual parts, $f^{(2)} = f^{(2)+} + f^{(2)-}$, both of which are separately closed and conserved. Following the notation in table 26 of~\cite{Cordova:2016emh}, the operators~$f^{(2)\pm}$ are the conformal primaries of the multiplets~$[2;0]_2$ and $[0;2]_2$.} Here we choose the latter option, to match with~\magcurr. The operator equations for~$J_B^{(2)}$ in~\jbdimccc\ are the free Maxwell equations for~$f^{(2)}$. Therefore the action of~$J_B^{(2)} \sim * f^{(2)}$ on the vacuum creates a one-photon state. It follows that the~$U(1)_B^{(1)}$ symmetry is spontaneously broken, and the photon is the corresponding NG particle (see~\cite{Gaiotto:2014kfa} and references therein, as well as section~\ref{convgsbf} and appendix~\ref{appHooftFreeMax}). By contrast, the fact that~$j_A^{(1)}$ is a conformal primary means that~$U(1)_A^{(0)}$ is unbroken (see section~\ref{sssnonprimcurr} below). The presence of a free Maxwell field implies that the algebra of local CFT operators contains a closed subsector generated by the field strength~$f^{(2)}$. It follows that the theory has an unbroken, unitary~$\Z_2$ charge conjugation symmetry~$\mathsf C$, which only acts on the Maxwell subsector via~$f^{(2)} \rightarrow - f^{(2)}$, i.e.~$f^{(2)}$ is~$\mathsf C$-odd. All local operators from other sectors are not acted on by charge conjugation and are therefore~$\mathsf C$-even. 

Given our assumption that the currents~$j_A^{(1)}$, $J_B^{(2)}$ are conformal primaries, with scaling dimensions~\jajbdim, it is straightforward to impose the constraints of conformal symmetry on the characteristic three-point function, and to show that it must vanish at separated points,
\eqn{
\langle j_\mu^A(x) j_\nu^A(y) J^B_{\rho\sigma}(z)\rangle \sim  \, \langle j_\mu^A(x) j_\nu^A(y) \t f_{\rho\sigma}(z)\rangle = 0~.
}[jjfpmthpt]
Here~$\t f_{\rho\sigma} = \half \ep_{\rho\sigma\alpha\beta} f^{\alpha\beta}$ is the Hodge dual~$*f^{(2)}$ with its Lorentz indices written out. This result also holds if~$j_A^{(1)}$ is a nonabelian flavor current,  or if we replace one of the~$U(1)_A^{(0)}$ currents by a different abelian flavor current. There is a simple argument for~\jjfpmthpt\ based on charge conjugation: let~$j^{(1)}_\pm$ be the projections of~$j_A^{(1)}$ onto its~$\mathsf C$-even~($+$) and~$\mathsf C$-odd~($-$) parts, which also have scaling dimension~$\Delta\big(j_\pm^{(1)}\big) = 3$. The only~$\mathsf C$-odd operators contain an odd number of Maxwell field strength operators~$f^{(2)}$. Since~$\Delta\big(f^{(2)}\big) = 2$, this implies that the~$\mathsf C$-odd part of the current is necessarily a product~$j^{(1)}_- \sim \CO f^{(2)}$, where~$\CO$ is a conformal primary of dimension~$\Delta\big(\CO\big) = 1$ that belongs to the non-Maxwell sector of the CFT. (Since the sectors are decoupled, the product is non-singular.) Conformal unitarity bounds (see~\cite{Mack:1975je,Minwalla:1997ka,Cordova:2016emh}) imply that the only such operator~$\CO$ is a free scalar field, but this is not compatible with the fact that~$j_-^{(1)}$ and~$f^{(2)}$ transform in different Lorentz representations. Therefore~$j_-^{(1)} = 0$, and hence~$j_A^{(1)} = j^{(1)}_+$ is~$\mathsf C$-even. This implies that the characteristic three-point function in~\jjfpmthpt\ violates charge conjugation and must therefore vanish. 

An analogous result holds for the characteristic three-point function for Poincar\'e 2-group symmetry~$\SP \times_{\hat \kappa} U(1)_B^{(1)}$. If we assume that the stress tensor~$T_{\mu\nu}$ and the 2-form current~$J_B^{(2)}$ are conformal primaries, it can again be shown that the constraints of conformal symmetry force this three-point function to vanish at separated points, 
\eqn{
\langle T_{\mu\nu}(x) T_{\rho\sigma}(y) J^B_{\rho\sigma}(z)\rangle \sim  \, \langle T_{\mu\nu}(x) T_{\rho\sigma}(y) \t f_{\rho\sigma}(z) \rangle = 0~. 
}[ttfthptzer]
Alternatively, this result follows from charge conjugation: since~$\mathsf C$ commutes with the hamiltonian, it follows that~$T_{\mu\nu}$ must be~$\mathsf C$-even. The fact that~$f^{(2)}$ is~$\mathsf C$-odd then establishes~\ttfthptzer. 

As explained in section~\ref{secCharThrPt}, the fact that the characteristic three-point functions~\jjfpmthpt\ and~\ttfthptzer\ for conformal primary currents vanish at separated points implies one of the following two scenarios:

\begin{itemize}
\item[1.)] If~$J_B^{(2)}$ is not a redundant operator, its two-point function is non-vanishing at separated points and the Ward identities in section~\ref{secCharThrPt} imply that the 2-group structure constants~$\hat \kappa_A, \hat \kappa_\SP$ vanish and the 2-groups decompose into conventional product symmetries. In this scenario~$U(1)_B^{(1)}$ is spontaneously broken, because~$J_B^{(2)} \sim * f^{(2)}$ is a free Maxwell field.
\item[2.)] The current~$J_B^{(2)}$ is a redundant operator, which vanishes inside correlation functions at separated points. In particular, $U(1)_B^{(1)}$ is not spontaneously broken. This scenario is compatible with 2-group symmetry. 

\end{itemize}

\subsec{Non-Primary Currents and Spontaneous 2-Group Breaking}[sssnonprimcurr]

In this subsection we consider the abelian 2-group symmetry~$U(1)_A^{(0)} \times_{\hat \kappa_A} U(1)_B^{(1)}$ and show that~$U(1)_B^{(1)}$ can be spontaneously broken, as long as~$U(1)_A^{(0)}$, and hence the entire 2-group, are spontaneously broken as well. The results of setion~\ref{sssprimcur} show that this scenario cannot occur if all currents are conformal primaries, so we begin by relaxing this assumption. 

In a CFT, any local operator can be expressed as a linear combination of conformal primaries and descendants. For the currents~$j_A^{(1)}$ or~$J_B^{(2)}$, this gives rise to the following Hodge-like decompositions,
\eqn{
j_A^{(1)} = j_\text{C.P.}^{(1)} + d \chi + * d U^{(2)}~, \qquad J^{(2)}_B = J^{(2)}_\text{C.P.}   + d X^{(1)} + * dY^{(1)}~.
}[currdec]
The operators with subscript~C.P.~are conformal primaries, while the operators~$\chi$, $U^{(2)}$, $X^{(1)}$, $Y^{(1)}$ may themselves be linear combinations of primaries and descendants. The conservation of~$j^{(1)}_A$ separately requires~$d * j^{(1)}_\text{C.P.} = 0$ and~$d * d \chi =0$.\foot{~This is because the only null state condition for the conformal primary~$j^{(1)}_\text{C.P.}$ that is allowed by conformal representation theory and involves one derivative is the conservation equation~$d* j^{(1)}_\text{C.P.} = 0$.} This implies that~$\chi$ is a free scalar field of scaling dimension~$\Delta(\chi) = 1$. If the term~$d \chi \subset j_A^{(1)}$ is present in~\currdec, then the~$U(1)_A^{(0)}$ flavor symmetry is spontaneously broken, and~$\chi$ is the corresponding NG boson. The term~$* d U^{(2)} \subset j_A^{(1)}$ is an improvement term, which is automatically conserved and does not contribute to the~$U(1)_A^{(0)}$ charge.\foot{~\label{ft:nobdych} Conformal unitarity bounds require any operator contributing to~$U^{(2)}$ that is not annihilated by~$d$ to have scaling dimension~$>2$. Therefore~$*dU^{(2)}$ has a higher scaling dimension than~$j^{(1)}_\text{C.P.}$. It therefore decays more rapidly at long distances and cannot contribute to the~$U(1)_A^{(0)}$ charge.} 

We can repeat this discussion for~$J_B^{(2)}$. Its conservation separately requires~$d * J^{(2)}_\text{C.P.} = 0$, so that~$J^{(2)}_\text{C.P.} \sim * f^{(2)}$ is a free Maxwell field,\foot{~The only allowed conformal null state conditions for the primary~$J^{(2)}_\text{C.P.}$ that involve one derivative are the free Maxwell equations~$d * f^{(2)} = df^{(2)} = 0$ (see also the discussion around~\jbdimccc).} as well as~$d * d X^{(1)} = 0$. The latter condition is the free wave equation for~$X^{(1)}$. Thus~$X^{(1)}$ is a free field, which creates an on-shell, free, massless particle. The only possibility is~$X^{(1)} = d\phi$, where~$\phi$ is a free scalar field satisfying~$d * d \phi = 0$,\foot{~The massless free-field representations allowed by conformal representation theory were analyzed in~\cite{Siegel:1988gd,Minwalla:1997ka}. Of these, only spin-0 scalars, spin-$\half$ fermions, and spin-1 Maxwell fields are allowed by the Weinberg-Witten theorem~\cite{Weinberg:1980kq} (see also the recent discussion in~\cite{Cordova:2017dhq}).} but such an~$X^{(1)}$ does not contribute to~$J_B^{(2)}$ in~\currdec. The term~$* dY^{(1)} \subset J_B^{(2)}$ is an improvement term, which is automatically conserved and does not contribute to the~$U(1)_B^{(1)}$ charge.\foot{~\label{ft:nobdybch} Conformal unitarity bounds imply that~$*dY^{(1)}$ has a higher scaling dimension than~$J^{(2)}_\text{C.P.}$. It therefore decays more rapidly at long distances and does not contribute to the~$U(1)_B^{(1)}$ charge (see also footnote~\ref{ft:nobdych}).} 

Note that both~$j_A^{(1)}$ and~$J_B^{(2)}$ may contain improvement terms. In a given CFT, it is possible (and often convenient) to redefine the currents so that they are free of such terms. However, this may no longer be possible if we consider RG flows between different CFTs, because improvement terms can be generated along the flow. In a typical scenario, the UV currents are defined to be conformal primaries, without the improvement or NG terms in~\currdec.\foot{~The absence of NG bosons in the UV is expected because spontaneously broken symmetries are typically restored at high energies.} However, in the IR these currents can flow to non-primary operators, which may contain improvement terms. Moreover, $j_A^{(1)}$ mixes with the NG current~$d\chi$ if~$U(1)_A^{(0)}$ is spontaneously broken.

We now assume that~$U(1)_B^{(1)}$ is spontaneously broken, so that the conformal primary Maxwell term~$J_\text{C.P.}^{(2)} \sim * f^{(2)} \; \subset \; J_B^{(2)}$ is present in~\currdec. Our goal is to show that this assumption, together with 2-group symmetry, necessarily implies the presence of a NG term~$d\chi \subset j_A^{(1)}$ in~\currdec, so that~$U(1)_A^{(0)}$ is also spontaneously broken. We will establish this by demanding that the non-primary currents in~\currdec\ satisfy the 2-group Ward identity~\jjJwardid, which we repeat here for convenience,
\eqn{
{\d \over \d x_\mu} \big\langle j_\mu^A(x) j_\nu^A(y) J_{\rho\sigma}^B (z)\big\rangle = {\hat \kappa_A \over 2\pi} \, \d^\lambda \delta^{(4)}(x-y) \, \big\langle J^B_{\nu\lambda}(y) J^B_{\rho\sigma} (z) \big \rangle~.
}[twogpwardbis]
As in section~\ref{sssprimcur}, we will also use the charge conjugation symmetry~$\mathsf C$ of the free Maxwell sector, under which the field strength~$f^{(2)}$ is odd.

From the argument after~\jjfpmthpt\ in section~\ref{sssprimcur}, we know that the conformal primary contribution~$j_\text{C.P.}^{(1)} \subset j_A^{(1)}$ to the~$U(1)_A^{(0)}$ current is~$\mathsf C$-even. The same is true for~$\chi$, since it is a free field, decoupled from the Maxwell sector. We can decompose the improvement term~$U^{(2)}$ into its~$\mathsf C$-even part~$U^{(2)}_+$, and its~$\mathsf C$-odd part~$U^{(2)}_- \sim f^{(2)} \CO$, where~$\CO$ is a~$\mathsf C$-even operator. In principle, $\CO$ could contain various Lorentz representations, as well as arbitrary even powers of~$f^{(2)}$. Below we will see that~$\CO$ must have overlap with the NG boson~$\chi$, and ultimately only the term~$\chi \subset \CO$ will be important. Note that the~${\d \over \d x_\mu}$ derivative in~\twogpwardbis\ annihilates the improvement term in~$j_\mu^A$ at separated and coincident points. Together with~$\mathsf C$-invariance, this implies that only the terms~$\langle (j^{(1)}_\text{C.P.}(x) + d \chi(x)) \; (* dU^{(2)}_-) (y)  \; (*f^{(2)})(z)\rangle$ in the characteristic three-point function can contribute to the right-hand side of the Ward identity~\twogpwardbis. Wick-contracting~$f^{(2)}(y) \subset U^{(2)}_-(y)$ with~$f^{(2)}(z)$, we find a factorized Ward identity. Schematically, 
\eqna{
{\d \over \d x_\mu} \, \big\langle \big(j_\mu^\text{C.P.}(x) + \d_\mu \chi(x)\big) \;  \d^\lambda \CO(y) \big\rangle \; & \big\langle \t f_{\nu\lambda}(y) \; \t f_{\rho\sigma}(z) \big\rangle + \cdots \sim \cr
& \sim \hat \kappa_A \, \d^\lambda \delta^{(4)}(x-y) \, \big\langle \t f_{\nu\lambda}(y) \t f_{\rho\sigma}(z)\big\rangle + \cdots~.
}[factwardid]
Here the ellipses on both sides involve improvement terms~$* dY^{(1)} \subset J_B^{(2)}$. Since~$*dY^{(1)}$ has a higher scaling dimension than~$f^{(2)}$ (see footnote~\ref{ft:nobdybch}), the~$f^{(2)}$-dependent terms in~\factwardid\ are the leading long-distance effects. The conformal primary current~$j_\mu^\text{C.P.}$ only has a non-vanishing two-point function with itself, and such a two-point function cannot give rise to the~$\delta$-function on the right-hand side of~\factwardid. The only remaining possibility is that the free NG boson~$\chi$ has non-zero overlap with the operator~$\CO$, so that~$\d_\mu \langle \d_\mu \chi(x) \d^\lambda \CO(y) \rangle \sim \d^\lambda \d^2 \langle \chi(x) \chi(y) \rangle \sim \d^\lambda \delta^{(4)}(x-y)$. This shows that the Ward identity~\factwardid\ can only be satisfied in the presence of a NG boson term~$d\chi \subset j_A^{(1)}$, so that~$U(1)_A^{(0)}$ must be spontaneously broken. Note that, in addition to the NG boson~$\chi$, spontaneous~$U(1)_A^{(0)} \times_{\hat \kappa_A} U(1)_B^{(1)}$ breaking requires a~$\mathsf C$-odd improvement term~$* dU^{(2)}_- \subset j_A^{(1)}$, where $U^{(2)}_- \sim \chi \, f^{(2)}$. In section~\ref{ssGM}, we explore the simplest model that explicitly realizes this scenario.

\subsec{Constraints of 2-Group Symmetry on RG Flows}[sssRGcons] 

In sections~\sssprimcur\ and~\sssnonprimcurr, we have seen that the realization of 2-group symmetry in CFTs is highly constrained. For instance, we saw that the~$U(1)_B^{(1)}$ subgroup of an abelian 2-group~$U(1)_A^{(0)} \times_{\hat \kappa_A} U(1)_B^{(1)}$ can only be spontaneously broken if the same is true for~$U(1)_A^{(0)}$, and hence the entire 2-group. This is a manifestation of the general theme articulated in section~\ref{rgphtwgp}, according to which~$U(1)_A^{(0)}$ is not a good subgroup of the full 2-group. (This statement should be understood at the level of current algebra, see footnote~\ref{ft:localsgp}.) In this subsection, we consider another manifestation of the same theme, which involves the decoupling or emergence of the 2-group currents in the deep IR or the deep UV of RG flows with 2-group symmetry. 

Consider an RG flow with unbroken abelian 2-group symmetry~$U(1)_A^{(0)} \times_{\hat \kappa_A} \times U(1)_B^{(1)}$. As discussed in sections~\sssprimcur\ and~\sssnonprimcurr, this implies that~$J_B^{(2)}$ flows to zero in the deep IR, i.e.~it decouples from the low-energy theory and becomes a redundant operator. (More precisely, $J_B^{(2)}$ flows to a pure improvement term, which is a descendant that decays rapidly at long distances and does not contribute to the~$U(1)_B^{(1)}$ charge.) Let us assume that~$J_B^{(2)}$ decouples at an energy scale~$\sim E^\text{IR}(J^{(2)}_B)$. The flavor current~$j_A^{(1)}$ may persist all the way to the IR, or it may also decouple at another energy scale~$\sim E^\text{IR}(j^{(1)}_A)$. We will now argue that 2-group symmetry requires~$J_B^{(2)}$ to decouple first, 
\eqn{
E^\text{IR}(J^{(2)}_B) \; \gtrsim \; E^\text{IR}(j^{(1)}_A)~.
}[IRdecineq]
The reason this inequality is not sharp is that the decoupling scales~$E^\text{IR}(J^{(2)}_B)$ and~$E^\text{IR}(j^{(1)}_A)$ are themselves not sharply defined.

We can argue for~\IRdecineq\ using the non-conservation equation for~$j_A^{(1)}$ in~\modconseq. At energies below~$E^\text{IR}(j^{(1)}_A)$ the current~$j_A^{(1)}$ flows to zero and decouples. The same must therefore be true of~$d *j_A^{(1)}\sim \hat \kappa_A \, F_A^{(2)} \wedge *J_B^{(2)}$, and hence the operator~$J_B^{(2)}$ on the right-hand side. However, $J_B^{(2)}$ can only decouple at energies below~$E^\text{IR}(J^{(2)}_B)$, which establishes the inequality~\IRdecineq. Equivalently, we can examine the 2-group Ward identity~\jjJwardid\ that relates~$\langle j_A^{(1)} j_A^{(1)} J_B^{(2)}\rangle \sim \hat \kappa_A \langle J_B^{(2)} J_B^{(2)}\rangle$. The characteristic three-point function on the left-hand side decays exponentially at energies below~$E^\text{IR}(j^{(1)}_A)$, while the two-point function on the right-hand side decays exponentially at energies below~$E^\text{IR}(J^{(2)}_B)$. This again implies~\IRdecineq.  

So far we have mostly focused on the possible IR behavior of RG flows with 2-group symmetry. We will now examine such flows at high energies. There are two fundamentally different scenarios for the UV behavior of RG flows with 2-group symmetry:
\begin{itemize}
\item[1.)] If the 2-group symmetry is exact, it must persist up to arbitrarily high energies. In UV-complete theories, with CFT fixed points at short distances, we expect that~$J_B^{(2)}$ is redundant in the UV CFT. This follows from the results of section~\sssprimcur, because~$j_A^{(1)}$ and~$J_B^{(2)}$ should be conformal primaries at the UV fixed point.\foot{~As discussed around~\currdec, the only obstructions to the currents being conformal primaries are improvement terms and the mixing of~$j_A^{(1)}$ with a~$U(1)_A^{(0)}$ NG boson. In the UV, we can always redefine the currents so that they are free of improvement terms. Moreover, any spontaneously broken symmetry is typically restored at high energies, so that we do not expect NG bosons at the UV fixed point.} However, if~$J_B^{(2)}$ is redundant in the UV, it remains so along the entire RG flow.\foot{~An example is a purely topological theory with 2-group symmetry, such as the deformed~$\Z_{|q_C|}$ gauge theory discussed around~\bfdynint, as well as in section~\ref{ssTQFT}. Note that the redundant currents of such a TQFT can mix with the non-redundant currents of a CFT with conventional global symmetries.} 

Alternatively, the theory may not admit a UV completion with a CFT fixed point, in which case~$J^{(2)}_B$ can be a non-trivial operator. All abelian gauge theory examples constructed in section~\ref{SecBasics} (with~$J_B^{(2)} \sim * f^{(2)}$) fall into this category. Any attempt to UV-complete these models in QFT, e.g.~by embedding them into asymptotically-free nonabelian gauge theories, is incompatible with 2-group symmetry.  

\item[2.)] The 2-group symmetry may be emergent. In this case it is an accidental symmetry of the low-energy theory that is explicitly broken at short distances. This scenario is compatible with conventional UV completions; an example is discussed in section~\ref{SubsecQEDexamples}. 

\end{itemize}

In the second scenario, we would like to argue in favor of an approximate inequality between the energy scale $\sim E^\text{UV}(J^{(2)}_B)$ at which the 2-form current emerges, and the energy scale $\sim E^\text{UV}(j^{(1)}_A)$ at which the 1-form current emerges,
\eqn{
E^\text{UV}(J^{(2)}_B) \; \gtrsim \; E^\text{UV}(j^{(1)}_A)~.
}[uvemergineq]
This inequality states that a non-trivial 2-group symmetry (with~$\hat \kappa_A \neq 0$) can only emerge if~$J_B^{(2)}$ emerges at higher energies than~$j_A^{(1)}$. It is similar to~\IRdecineq, which constrains the possible decoupling of the currents in the IR. Both inequalities intuitively follow from the general principle reviewed at the beginning of this subsection, which states that~$U(1)_B^{(1)}$ is a good subgroup of the full~$U(1)_A^{(0)} \times_{\hat \kappa_A} U(1)_B^{(1)}$ 2-group symmetry, while this is not the case for~$U(1)_A^{(0)}$. However, the argument we present in favor of~\uvemergineq, which involves the background fields~$A^{(1)}$ and~$B^{(2)}$ that couple to the emergent currents, is not as straightforward (and therefore perhaps not as robust) as the argument for~\IRdecineq. 

The argument for~\uvemergineq\ is based on the observation (explained in section~\ref{SecGlobal}) that a 2-group shift~$B^{(2)} \rightarrow B^{(2)} + {\hat \kappa_A \over 2 \pi} \, \lambda_A^{(0)} \, F_A^{(2)}$ (see~\Btwogp) under~$U(1)_A^{(0)}$ background gauge transformations is only consistent if~$B^{(2)}$ is a 2-form background gauge field, which is also subject to~$U(1)_B^{(1)}$ background gauge transformations. Here we make the additional assumption that the emergent~$U(1)_A^{(0)}$ flavor symmetry (like all abelian symmetries in this paper) is compact. In the presence of a 2-group shift for~$B^{(2)}$, the ambiguity~$\lambda_A^{(0)} \sim \lambda_A^{(0)} + 2 \pi n$ (with~$n \in \Z$) of the~$U(1)_A^{(0)}$ gauge parameter leads to an ambiguity~$B^{(2)} \sim B^{(2)} +  \hat \kappa_A n F_A^{(2)}$. As explained in section~\ref{SecGlobal}, this unphysical ambiguity must be absorbed by~$U(1)_B^{(1)}$ background gauge transformations, which are only available if~$B^{(2)}$ is a background gauge field that couples to a conserved 2-form current. If the inequality~\eqref{uvemergineq} is violated, then~$j_A^{(1)}$ emerges as a conserved current when~$J_B^{(2)}$ is still a non-conserved 2-form operator, so that its source~$B^{(2)}$ is not subject to 1-form background gauge transformations. Then~$A^{(1)}$ is a standard~$U(1)_A^{(0)}$ background gauge field, but we cannot assign a 2-group shift to~$B^{(2)}$.

It is straightforward to extend the arguments above to nonabelian and Poincar\'e 2-groups. We would like to make a few comments about the Poincar\'e case. As in the abelian case, the Poincar\'e group~$\SP$ does not behave like a good subgroup of the full Poincar\'e 2-group~$\SP \times_{\hat \kappa_\SP} U(1)_B^{(1)}$. This leads to an analogue of the inequality~\uvemergineq, with  the stress tensor~$T_{\mu\nu}$ replacing the~$U(1)_A^{(0)}$ current~$j_A^{(1)}$. Explicitly, the inequality in the Poincar\'e case states that the scale~$\sim E^\text{UV}(T_{\mu\nu})$ at which the stress tensor emerges (if it does so at all) must be bounded from above by the emergence scale~$\sim E^\text{UV}(J^{(2)}_B)$ of the 2-form current,
\eqn{
E^\text{UV}(J^{(2)}_B) \; \gtrsim \; E^\text{UV}(T_{\mu\nu})~.
}[stemerge]
If we assume that the theory is Poincar\'e invariant and local at all energy scales, there should be a conserved stress tensor along the entire RG flow. In this case~$T_{\mu\nu}$ is not emergent, and~\stemerge\ implies that the same is true for~$J_B^{(2)}$. Therefore the entire Poincar\'e 2-group symmetry~$\SP \times_{\hat \kappa_\SP} U(1)_B^{(1)}$ is exact along the entire RG flow. It then follows from point 1.) above~\uvemergineq\ that such theories, with exact Poincar\'e 2-group symmetry and non-redundant~$J_B^{(2)}$, do not have UV completions as continuum QFTs, with standard CFT fixed points in the UV. A simple example of such a theory is explored in section~\ref{SubsecFermats}.

\newsec{Green-Schwarz Contact Terms and 2-Group 't Hooft Anomalies}[SecGSAnom]

In this section we present the details that underly our summary of 2-group 't Hooft anomalies in section~\ref{gscttwgpano}. As was pointed out there, GS contact terms and counterterms, which are discussed in section~\ref{secGScts}, play a crucial role in our analysis. In section~\ref{secModHooft} we review the approach of~\cite{Frishman:1980dq, Coleman:1982yg} to the~$\kappa_{A^3}$ 't Hooft anomaly for an ordinary~$U(1)_A^{(0)}$ flavor symmetry, which is based on an analysis of the~$\langle j_A^{(1)} j_A^{(1)} j_A^{(1)} \rangle$ three-point function in momentum space. In section~\ref{sssHoftAbTwGp} we reanalyze the~$\kappa_{A^3}$ 't Hooft anomaly in the presence of~$U(1)_A^{(0)} \times_{\hat \kappa_A} U(1)_B^{(1)}$ 2-group symmetry -- both from the point of view of the 2-group background gauge fields, and using the 2-group Ward identity satisfied by the~$\langle j_A^{(1)} j_A^{(1)} j_A^{(1)} \rangle$ correlator. More general 2-group anomalies are briefly discussed in section~\ref{ssecGenAnom}.

\subsec{Green-Schwarz Contact Terms and Counterterms}[secGScts]

In preparation for our discussion in section~\ref{sssHoftAbTwGp} below, we take a small detour and examine the~$\langle J_B^{(2)} j_A^{(1)}\rangle$ two-point function. This will lead to observables that we refer to as GS contact terms. They are four-dimensional analogues of the three-dimensional Chern-Simons contact terms analyzed in~\cite{Closset:2012vg,Closset:2012vp}. Due to the many similarities, we will keep the discussion brief and refer to~\cite{Closset:2012vg,Closset:2012vp} for additional details and background. The discussion in this subsection does not require 2-group symmetry and may be of independent interest.

In appendix~\ref{ssjbjatwopt}, it is shown that conservation of~$J^B_{\mu\nu}$ and~$j^A_\rho$ requires their two-point function in momentum space to take the following form, 
\eqn{
\langle J^B_{\mu\nu}(p) j^A_\rho(-p)\rangle = -{1 \over 2 \pi} {\mathsf K} \left({p^2 \over M^2}\right) \ep_{\mu\nu\rho\lambda} p^\lambda~.
}[jbjatwoptmain]
Here~${\mathsf K}\left({p^2\over M^2}\right)$ is a real, dimensionless structure function and~$M$ is some mass scale. In a CFT (or in a TQFT) this function must be a constant, $\mathsf K \left({p^2\over M^2}\right)= {\mathsf K}$, in which case~\jbjatwoptmain\ is linear in the momentum. It therefore gives rise to a contact term in position space,
\eqn{
\langle J^B_{\mu\nu}(x) j^A_\rho(0)\rangle = {i {\mathsf K} \over 2 \pi} \, \ep_{\mu\nu\rho\lambda} \d^\lambda \delta^{(4)}(x)~.
}[gscontact]
For reasons that will become apparent below, we refer to~\gscontact\ as a GS contact term. The fact that the~$\langle J_B^{(2)} j_A^{(1)}\rangle$ two-point function vanishes at separated points is required by the conformal Ward identities, because the two currents reside in different representations of the conformal group (they have different Lorentz quantum numbers and scaling dimensions). If~${\mathsf K} \neq 0$, these Ward identities are violated at coincident points. As we will explain momentarily, global issues may prevent us from setting~${\mathsf K}$ to zero using valid local counterterms, even though it is a pure contact term. This constitutes a kind of global conformal anomaly, which is similar to the superconformal anomaly for three-dimensional~$\CN=2$ theories analyzed in~\cite{Closset:2012vg,Closset:2012vp} (see~\cite{Chang:2017cdx} for a recent generalization to five dimensions). 

The GS contact term~\gscontact\ is closely related to the following GS counterterm, which is constructed out of the background gauge fields~$B^{(2)}$ and~$A^{(1)}$, 
\eqn{
S_\text{GS} = {i n \over 2 \pi} \int B^{(2)} \wedge F^{(2)}_A~, \qquad n \in \Z~.
}[gsctdef]
The quantization condition on~$n$ comes from the requirement that the GS counterterm~\gsctdef\ should be invariant (modulo~$2 \pi i \Z$) under large~$U(1)_B^{(1)}$ background gauge transformations~$B^{(2)} \; \rightarrow B^{(2)} + d\Lambda^{(1)}$, for which the gauge parameter~$\Lambda_B^{(1)}$ has non-trivial integer fluxes~${1 \over 2 \pi} \int_{\Sigma_2} d \Lambda^{(1)}_B \in \Z$. (Below, we will comment on the possibility of allowing non-integer~$n$.) Taking a functional derivative of~\gsctdef\ with respect to the background gauge fields (and using~\eqref{sajcoup}, \eqref{bjbcoup}), we find that adding the GS counterterm~\gsctdef\ to the action shifts the GS contact term~$\mathsf K$ in~\gscontact\ by the integer~$n$,
\eqn{
\mathsf K \; \rightarrow \; \mathsf K + n~, \qquad n \in \Z~.
}[gsconttshift]
Consequently, the integer part of~$\mathsf K$ is scheme dependent. By contrast, its fractional part, $\mathsf K~(\mod~1)$ is an intrinsic observable, which does not depend on the choice of regularization scheme. If this observable vanishes, then~$\mathsf K$ can be set to zero using a properly quantized GS counterterm~\gsctdef.

We would like to offer another perspective on the observable~$\mathsf K~(\mod~1)$. In the discussion around~\gsctdef\ and~\gsconttshift, we insisted on invariance under large~$U(1)_B^{(1)}$ gauge transformations. If we are willing to relax this requirement, we can add a GS counterterm~\gsctdef\ with a potentially non-integer coefficient~$n = - \mathsf K$ to set the GS contact term~$\mathsf K$ in~\gscontact. Under a~$U(1)_B^{(1)}$ gauge transformation, the partition function now picks up an anomalous phase,
\eqn{
Z[A^{(1)}, B^{(2)} + d \Lambda_B^{(1)}] = Z[A^{(1)}, B^{(2)}] e^{2 \pi i \mathsf K N}~, \quad N = {1 \over (2 \pi)^2} \int {d \Lambda^{(1)}_B} \wedge {F^{(2)}_A } \in \Z~.
}[anomlgbphz]
On suitable spacetime manifolds~$\CM_4$, the integer~$N$ can be made to take any value by appropriately choosing the~$U(1)_A^{(0)}$ background flux and the~$U(1)_B^{(1)}$ gauge parameter.\foot{~For instance, we can take~$\CM_4 = S^2 \times S^2$ with~$N$ units of~$A^{(1)}$ flux through one of the two-spheres, and one unit of~$\Lambda^{(1)}_B$ flux through the other.} Small, topologically trivial~$U(1)_B^{(1)}$ background gauge transformations have~$N = 0$, while large ones can have non-zero~$N$. The anomalous phase in~\anomlgbphz\ can be used to extract the fractional part~$\mathsf K~(\mod~1)$, which therefore remains observable, but it is not sensitive to the scheme-dependent integer part of~$\mathsf K$. This shows that the anomaly discussed below~\gscontact\ can be understood as a clash between conformal symmetry and invariance under large~$U(1)_B^{(1)}$ gauge transformations. The observable~$\mathsf K~(\mod~1)$ is the associated anomaly coefficient; whenever it is non-zero, the anomaly is present.

In light of the above discussion, it is tempting to think of the entire GS contact term~$\mathsf K$ as a (typically improperly quantized) GS term in the effective action for background fields,
\eqn{
W[A^{(1)}, B^{(2)}] \; \supset \;  {i {\mathsf K} \over 2 \pi} \int B^{(2)} \wedge F^{(2)}_A~.
}[gsctdefo]
Indeed, varying this term with respect to~$B^{(2)}$ and~$A^{(1)}$ correctly reproduces~\gscontact, and comparing with~\gsctdef\ gives~\gsconttshift. However, as we will illustrate below using a simple example, \gsctdefo\ cannot be taken at face value for arbitrary background field configurations and requires additional qualification. Nevertheless, it is occasionally a useful mnemonic for the GS contact term~\gscontact.

A simple example of a CFT for which the observable~${\mathsf K}~(\mod~1)$ is nonzero is topological~$\Z_p$ gauge theory. We use its presentation as a BF theory~\cite{Maldacena:2001ss,Banks:2010zn,Kapustin:2014gua}, which involves a dynamical~$U(1)^{(1)}_b$ gauge field~$b^{(2)}$, and a dynamical~$U(1)_c^{(0)}$ gauge field~$c^{(1)}$. We also couple these dynamical fields to the background fields~$B^{(2)}$ and~$A^{(1)}$. This leads to the following quadratic action,
\eqna{
S_\text{BF}[A^{(1)}, B^{(2)},\,& b^{(2)}, c^{(1)}] = {i p \over 2\pi}  \int b^{(2)}  \wedge f_c^{(2)}   + {i \over 2 \pi} \int  B^{(2)} \wedge f_c^{(2)}  \cr
& + {i q \over 2\pi} \int b^{(2)} \wedge F_A^{(2)} + {i n \over 2 \pi} \int B^{(2)} \wedge F_A^{(2)}~, \qquad p \in \Z_{\geq 1}~,~~q, n \in \Z~.
}[dymbf]
As was reviewed around~\bfdynint\ (see also section~\ref{ssTQFT}), this is the low-energy effective action of a~$U(1)_c^{(0)}$ gauge theory with an elementary scalar Higgs field of~$U(1)_c^{(0)}$ gauge charge~$p$ that also carries charge~$q$ under a~$U(1)_A^{(0)}$ flavor symmetry. The background fields~$A^{(1)}$ and~$B^{(2)}$ couple to the 1-form flavor current and the magnetic 2-form current, 
\eqn{
j_A^{(1)} = {i q \over 2 \pi} * d b^{(2)}~, \qquad J_B^{(2)} = {i \over 2 \pi} * f_c^{(2)}~.
}[bftopcurr]
In our definition of the theory, we have also included a bare GS counterterm~\gsctdef\ for the background fields. Note that all BF terms for dynamical and background fields in~\dymbf\ are properly quantized.

We would now like to compute~$\mathsf K$ in the theory~\dymbf. Using the expressions for the currents in~\bftopcurr, it can be computed using Feynman diagrams. Instead, we will compute it by attempting to integrate out the dynamical fields in the presence of the background fields, since this will allow us to clarify a few subtle points. The equations of motion set
\eqn{
p \, f_c^{(2)} + q \, F_A^{(2)} = 0~, \qquad p \, db^{(2)} + dB^{(2)} = 0~.
}[bfeom]
These equations can be used to express the currents~\bftopcurr\ in terms of background fields. This shows that the currents are redundant operators, as expected in a topological theory. As a result, none of the extended operators of the TQFT are charged under either~$U(1)_A^{(0)}$ or~$U(1)_B^{(1)}$. If we naively substitute~\bfeom\ back into~\dymbf, we obtain an effective action of the form~\gsctdefo\ for the background fields,
\eqn{
W[A^{(1)},B^{(2)}] =  {i {\mathsf K} \over 2 \pi} \int B^{(2)} \wedge F_A^{(2)}~, \qquad {\mathsf K} = -{q \over p} + n~.
}[gsfrakw]
Several comments are in order (for simplicity, we assume that~$p,q$ are relatively prime):
\begin{itemize}
\item The effective action~\gsfrakw\ takes the same form as a GS counterterm~\gsctdef, but if~$p \neq 1$ the coefficient~${\mathsf K}$ is fractional. This coefficient determines the GS contact term~\gscontact\ in the two-point function of~$J_B^{(2)}$ and~$j_A^{(1)}$. The freedom to change its integer part using a properly quantized GS counterterm is parametrized by~$n$, while its fractional part~$-{q \over p}~(\mod~1)$ is an intrinsic, scheme-independent property of the theory.

\item If~$p = 1$, the topological~$\Z_p$ gauge theory becomes invertible and describes a fully gapped phase with short-range entanglement. In this case~\gsfrakw\ reduces to a properly quantized GS counterterm~\gsctdef, as is expected on general grounds. 

\item The effective action~\gsfrakw\ is not invariant under large gauge transformations of~$B^{(2)}$, even though the original action~\dymbf\ was invariant under such transformations. The resolution of this apparent paradox is that~\gsfrakw\ is imprecise: it is not in general permissible to solve the equations of motion in~\bfeom\ as we did to obtain~\gsfrakw, because both~$c^{(1)}$ and~$A^{(1)}$ have integer fluxes. This manipulation is only valid when the flux of~$A^{(1)}$ is divisible by~$p$, in which case the fractional coefficient~$-{q \over p}$ in~\gsfrakw\ is harmless. In all other~$A^{(1)}$ flux sectors the equations of motion do not admit a solution, and hence the functional integral vanishes. With this caveat, the effective action is fully invariant under~$U(1)^{(1)}_B$ gauge transformations. See section~3.2 of~\cite{Closset:2012vp} for a closely related discussion.

\end{itemize}

Along RG flows, the structure function~${\mathsf K}\left({p^2 \over M^2}\right)$ in~\jbjatwoptmain\ interpolates between a GS contact term~\gscontact\ in the UV, ${\mathsf K}_\text{UV} = \lim_{p^2 \rightarrow \infty} {\mathsf K}\left({p^2 \over M^2}\right)$, and a different one in the IR, ${\mathsf K}_\text{IR} = \lim_{p^2 \rightarrow 0}{\mathsf K}\left({p^2 \over M^2}\right)$. The interpolating structure function~${\mathsf K}\left({p^2 \over M^2}\right)$ is scheme-independent modulo overall shifts by an integer, which are brought about by adding a properly quantized GS counterterm~\gsctdef. (Typically we imagine adjusting such counterterms in the UV.) Note that the difference~${\mathsf K}_\text{UV} - {\mathsf K}_\text{IR}$ is not affected by such shifts. It is therefore scheme-independent and can be extracted from the~$\langle J^B_{\mu\nu}(x) j^A_\rho(0)\rangle$ correlator at separated points.~(See~p.7 of~\cite{Closset:2012vp} for a closely related discussion.) By contrast, only the fractional parts of~${\mathsf K}_\text{UV}$ and~${\mathsf K}_\text{IR}$ are scheme independent.

\subsec{More on Conventional 't Hooft Anomalies}[secModHooft]

Consider a~$U(1)_A^{(0)}$ flavor symmetry with conserved current~$j_A^{(1)}$ and background gauge field~$A^{(1)}$. As explained in section~\ref{SecDescabelian}, such a flavor symmetry can have a reducible cubic~'t~Hooft anomaly characterized by the following 6-form anomaly polynomial
\eqn{
\CI^{(6)} = {1 \over (2 \pi)^3} \, {\kappa_{A^3} \over 3!} \, F^{(2)}_A \wedge F^{(2)}_A \wedge F^{(2)}_A~.
}[aaatHanom]
Here~$\kappa_{A^3}$ is the 't Hooft anomaly coefficient. Under a~$U(1)_A^{(0)}$ background gauge transformation, $A^{(1)} \; \rightarrow \; A^{(1)} + d \lambda_A^{(0)}$, the anomaly polynomial~\aaatHanom\ gives rise to the following anomalous c-number shift of the effective action (see for instance~\anomalies),
\eqn{
W[A^{(1)} + d \lambda^{(0)}_A] = W[A^{(1)}] + \CA_A~, \qquad \CA_A = {i \kappa_{A^3}  \over 24 \pi^2} \int \lambda_A^{(0)} \, F^{(2)}_A \wedge F^{(2)}_A~.
}[twogpacubanom]
Recall from section~\ref{ssecGenHooft} that~$\kappa_{A^3}$ is not affected by local counterterms, i.e.~it is an intrinsic, scheme-independent observable. Moreover, the anomalous variation~$\CA_A$ in~\twogpacubanom\ must be reproduced in any description of the theory. This implies 't Hooft anomaly matching, which states that~$\kappa_{A^3}$ is constant along RG flows and must match when it is computed using the UV or IR degrees of freedom,\footnote{~The argument for anomaly matching can be sharpend using inflow from a five-dimensional bulk (see also footnote~\ref{zeroinflow}), which plays the role of the spectator fermions in 't Hooft's original argument~\cite{tHooft:1979rat}. We can couple the four-dimensional theory on~$\CM_4$ to a non-dynamical theory on~$\CM_5$, with boundary~$\d \CM_5 = \CM_4$. The bulk action only involves the extension of the background field~$A^{(1)}$, $S_5 = - {i \kappa_{A^3} \over 24 \pi^2} \int_{\CM_5} \, A^{(1)} \wedge F_A^{(2)} \wedge F_A^{(2)}$. It is invariant under five-dimensional gauge transformations of~$A^{(1)}$ with support in the bulk, but shifts by~$-{i \kappa_{A^3} \over 24 \pi^2} \int_{\CM_4} \, \lambda_A^{(0)} \, F_A^{(2)} \wedge F_A^{(2)}$ if the gauge parameter~$\lambda_A^{(0)}$ has support on the boundary~$\CM_4$. This cancels the four-dimensional anomaly~\eqref{twogpacubanom}, so that the combined bulk-boundary system is anomaly free. Since this property is preserved under RG flow (e.g.~because we could imagine weakly gauging~$A^{(1)}$), and the bulk (being non-dynamical) always supplies the same 't Hooft anomaly, we conclude that the anomaly of the boundary theory is also unchanged along the entire RG flow.}
\eqn{
\kappa_{A^3}^\text{UV} = \kappa_{A^3}^\text{IR} = \kappa_{A^3}~.
}[hooftmatch]

We will now examine the consequences of the anomaly term~$\CA_A$ in~\twogpacubanom\ for correlation functions of the current~$j_A^{(1)}$. As usual, it leads to the a non-conservation equation for~$j_A^{(1)}$,
\eqn{
d * j_A^{(1)} = -{i \kappa_{A^3} \over 24 \pi^2} \, F_A^{(2)} \wedge F^{(2)}_A \qquad \Longleftrightarrow \qquad \d^\mu j^A_\mu = -{i \kappa_{A^3} \over 24 \pi^2} \, \ep^{\mu\nu\alpha\beta} \, \d_\mu A_\nu \, \d_\alpha A_\beta~.
}[aaanoncons]
Taking variational derivatives of~\aaanoncons\ with respect to~$-A^\nu(y)$, $-A^\rho(z)$ inserts~$j_\nu^A(y)$, $j_\rho^A(z)$ on the left-hand side, but leads to a c-number contact term on the right-hand side, 
\eqn{
{\d \over \d x_\mu} \, \langle j^A_\mu(x) j^A_\nu(y) j^A_\rho(z)\rangle = {i \kappa_{A^3} \over 12 \pi^2} \, \ep_{\nu\rho \alpha\beta} \, \d^\alpha \delta^{(4)}(x-y) \d^\beta \delta^{(4)}(x-z)~.
}[aaacontt]
In momentum space (see footnote~\ref{momcorrdef} or appendix~\ref{AppMomSpace}), this equation takes the form 
\eqn{
p_1^\mu \, \big\langle j_\mu^A(p_1) j_\nu^A(p_2) j^A_\rho(p_3)\big\rangle = - {\kappa_{A^3} \over 12 \pi^2} \, \ep_{\nu\rho\alpha\beta} \, p_2^\alpha p_3^\beta~, \qquad p_1 + p_2 + p_3 = 0~.
}[momsphooft]
Even though the right-hand side is a polynomial in the momenta, its presence leads to non-analytic  structures in the three-point function on the left-hand side. These structures contribute to the position-space correlator at separated points. This crucial feature of 't Hooft anomalies was thoroughly studied by the authors of~\cite{Frishman:1980dq, Coleman:1982yg}, and we now briefly recall some of their conclusions. 

Following~\cite{Frishman:1980dq, Coleman:1982yg} (see also the discussion around~\symkin\ above), we can simplify the analysis of~$\big\langle j_\mu^A(p_1) j_\nu^A(p_2) j^A_\rho(p_3)\big\rangle$ by specializing the momenta to configurations that satisfy
\eqn{
p_1^2 = p_2^2 = p_3^2 = Q^2~, \qquad p_1 + p_2 + p_3 = 0~. 
}[mombosesimp]
For such configurations, the parity-odd part of the~$j_A^{(1)}$ three-point function is controlled by a single dimensionless structure function~$\mathsf A\left({Q^2 \over M^2}\right)$ (see appendix~\ref{jajajaapp}),
\eqn{
\big\langle j_\mu^A(p_1) j_\nu^A(p_2) j^A_\rho(p_3)\big\rangle \; \supset \; {1 \over Q^2} \, {\mathsf A}\left({Q^2 \over M^2}\right) \left(\ep_{\mu\nu\alpha\beta} \, p_1^\alpha p_2^\beta \, p_{3\rho} + \ep_{\nu\rho\alpha\beta} \, p_2^\alpha p_3^\beta \, p_{1\mu} + \ep_{\rho\mu\alpha\beta} \, p_3^\alpha p_1^\beta \, p_{2\nu} \right)~,
}[poddjjjstr]
where~$M$ is some mass scale. Substituting into~\momsphooft\ shows that the structure function~$\mathsf A\left({Q^2 \over M^2}\right)$ reduces to a constant that is completely determined by the anomaly coefficient~$\kappa_{A^3}$, 
\eqn{
\mathsf A\left({Q^2 \over M^2}\right) = - {\kappa_{A^3} \over 12 \pi^2}~.
}[hooftpole]
If we insert this result back into~\poddjjjstr, we see that the entire parity-odd part of the three-point function is fixed by the anomaly to be a pole~$\sim{\kappa_{A^3} \over Q^2}$. This pole can be tracked along the entire RG flow, from the UV~(corresponding to~$Q^2 \rightarrow \infty$) to the IR~(corresponding to~$Q^2 \rightarrow 0$). The fact that the residue of this pole is always given by~$\kappa_{A^3}$ is another argument for the 't Hooft anomaly matching relation~\hooftmatch. Beyond that, we also learn that~$\kappa_{A^3}$ can be computed using only the massless, local degrees of freedom that are present in the theory, since only they can give rise to such a pole. Massive or topological degrees of freedom cannot contribute.\foot{~In fact, anomaly matching implies that even degrees of freedom that are massless, but obtain a mass under deformations that preserve~$U(1)_A^{(0)}$, do not contribute to~$\kappa_{A^3}$.}

The discussion above highlights the fact that the 't Hooft anomaly coefficient~$\kappa_{A^3}$ controls two different, a priori unrelated, quantities:
\begin{itemize}
\item[1.)] By definition, $\kappa_{A^3}$ determines the anomalous variation~$\CA_A$ of the effective action in~\twogpacubanom. The value of~$\kappa_{A^3}$ is not affected by local counterterms, and hence it is an intrinsic, scheme-independent observable. The anomalous variation~$\CA_A$, and hence~$\kappa_{A^3}$, is inert under RG flows, which leads to the 't Hooft anomaly matching condition~\hooftmatch.

\item[2.)] Via the anomalous Ward identity~\momsphooft, $\kappa_{A^3}$ also fixes the structure function~$\mathsf A\left({Q^2 \over M^2}\right)$ to the constant in~\hooftpole, which leads to a pole with residue~$\sim \kappa_{A^3}$ in the~$j_A^{(1)}$ three-point function~\poddjjjstr. This is consistent with anomaly matching~\hooftmatch\ and implies the stronger staement that only massless, local degrees of freedom contribute to~$\kappa_{A^3}$.  

\end{itemize}
As we will see below, both statements above are modified in the presence of 2-group symmetry, and the link between them is broken.

\subsec{'t Hooft Anomalies for Abelian 2-Group Symmetries}[sssHoftAbTwGp]

We will now repeat the analysis of the previous subsection for a theory with abelian 2-group symmetry~$U(1)_A^{(0)} \times_{\hat \kappa_A} U(1)_B^{(1)}$. Recall from section~\ref{ssecGenHooft} that 't Hooft anomalies are local c-number shifts of the effective action~$W[\CB]$ for background fields~$\CB$ under background gauge transformations that cannot be removed using local counterterms. Now the relevant background fields are the 2-group background fields~$A^{(1)}$ and~$B^{(2)}$, whose gauge transformations are given by~\eqref{atrans} and~\eqref{Btwogp},
\eqn{
A^{(1)} \quad \longrightarrow \quad A^{(1)} + d \lambda_A^{(0)}~, \qquad B^{(2)} \quad \longrightarrow \quad B^{(2)} + d \Lambda_B^{(1)} + {\hat \kappa_A \over 2 \pi} \, \lambda_A^{(0)} \, F^{(2)}_A~.
}[abtwgpbis]
Note that the analysis of candidate anomalies around~\aaatHanom\ and~\twogpacubanom\ remains valid in the presence of 2-group symmetry. This is because only the transformation rule for~$B^{(2)}$ is modified when~$\hat \kappa_A \neq 0$. As was pointed out in section~\ref{SecHooft} (see in particular the discussion at the end of section~\ref{ssecAnomPol}), the only possible anomaly 6-form polynomial that can be constructed using~$A^{(1)}$ and~$B^{(2)}$ is~\aaatHanom, which does not involve~$B^{(2)}$. However, the candidate anomalies that can be absorbed using local counterterm must be reanalyzed in light of~\abtwgpbis. 

The counterterm that will play a crucial role in our analysis of 2-group 't Hooft anomalies is the GS counterterm~\gsctdef\ introduced in section~\ref{secGScts} above, which we repeat here,
\eqn{
S_\text{GS} = {i n \over 2 \pi} \int B^{(2)} \wedge F^{(2)}_A~. 
}[gsctbis]
Under the 2-group background gauge transformation~\abtwgpbis, this term shifts by
\eqn{
S_\text{GS} \; \rightarrow \; S_\text{GS} + {i n \hat \kappa_A \over 4 \pi^2} \int \lambda_A^{(0)} \, F_A^{(2)} \wedge F_A^{(2)}~.
}[sgsctanom]
These formulas take exactly the same form as in the conventional GS mechanism, except that they lead to an 't Hooft anomaly involving background fields, rather than a gauge anomaly for dynamical fields.  Comparing~\sgsctanom\ to~\twogpacubanom, we conclude that adding the counterterm~\gsctbis\ shifts the anomaly coefficient~$\kappa_{A^3}$ as follows,
\eqn{
\kappa_{A^3} \; \rightarrow \; \kappa_{A^3} + 6 n \hat \kappa_A~.
}[kaaanshift]
This shows that~$\kappa_{A^3}$ is no longer scheme independent.

 In order to proceed, we must decide which values for~$n$ to allow in~\kaaanshift. Here we closely follow the discussion of GS counterterms in section~\ref{secGScts}. As was the case there, we can take two points of view:
\begin{itemize}
\item[1.)] If we insist on invariance under large~$U(1)_B^{(1)}$ gauge transformations, the counterterm coefficient~$n$ in~\gsctbis\ must be quantized, $n \in \Z$. It then follows from~\kaaanshift\ that the 't Hooft anomaly coefficient~$\kappa_{A^3}$ is only scheme-independent~$\mod~6 \hat \kappa_A$. Alternatively, only the fractional part~${\kappa_{A^3} \over 6 \hat \kappa_A}~(\mod~1)$ is intrinsic, while the integer part is scheme dependent and can be set to any value using a properly quantized GS counterterm. 

Once a particular scheme has been chosen, $\kappa_{A^3}$ is a well-defined number and the arguments for anomaly matching leading to~\hooftmatch\ apply; both the intrinsic and the scheme-dependent part of~$\kappa_{A^3}$ must match between the UV and the IR. It is convenient (but not essential) to choose a scheme where the integer part of~$\kappa_{A^3} \over 6 \hat \kappa_A$ vanishes. The anomalous shift~$\CA_A$ of the effective action under~$U(1)_A^{(0)}$ gauge transformations is then determined by the intrinsic observable~${\kappa_{A^3} \over 6 \hat \kappa_A}~(\mod~1)$. 

\item[2.)] If we give up on invariance under large~$U(1)_B^{(1)}$ gauge transformations, then~$n$ can take any value and we can add a GS counterterm with potentially fractional coefficient $n = -{\kappa_{A^3} \over 6 \hat \kappa_A}$ to set~$\kappa_{A^3} = 0$. In this case the theory is invariant under~$U(1)_A^{(0)}$ background gauge transformations but, exactly as in~\anomlgbphz, the partition function picks up an anomalous phase under large~$U(1)_B^{(1)}$ gauge transformations that is sensitive to the observable~${\kappa_{A^3} \over 6 \hat \kappa_A}~(\mod~1)$. Since this anomalous phase is also subject to matching, we again conclude that~${\kappa_{A^3} \over 6 \hat \kappa_A}~(\mod~1)$ must match between the UV and the IR.

\end{itemize}
Let us summarize our conclusions so far:
\begin{itemize}
\item Rather than being characterized by an arbitrary integer~$\kappa_{A^3} \in \Z$, the 2-group anomaly is only intrinsically meaningful~$\mod~6 \hat \kappa_A$. Equivalently, it is characterized by the fractional part~${\kappa_{A^3} \over 6 \hat \kappa_A}~(\mod~1)$, which must match along RG flows. 

\item Unlike the conventional~$\kappa_{A^3}$ anomaly, which only involves~$U(1)_A^{(0)}$ gauge transformations, the 2-group anomaly arises from a clash between~$U(1)_A^{(0)}$ and (large)~$U(1)_B^{(1)}$ gauge transformations: if~${\kappa_{A^3} \over 6 \hat \kappa_A}~(\mod~1) \neq 0$ we can preserve one or the other, but not both. As we will see below, as well as in section~\ref{ssTQFT}, this sensitivity to global issues makes it possible for non-trivial TQFTs to contribute to the 2-group 't Hooft anomaly.  
\end{itemize}

This general picture can be made more explicit in theories with 2-group symmetry that arise via the construction described in section~\ref{SecBasics}, from parent theories with a~$U(1)_A^{(0)} \times U(1)_C^{(0)}$ flavor symmetry and suitable mixed 't Hooft anomalies, by gauging~$U(1)_C^{(0)}$. Recall that the parent theory must have~$\kappa_{AC^2} = \kappa_{C^3} = 0$ (to ensure that the gauging is not obstructed, and that~$U(1)_A^{(0)}$ is not broken by an ABJ anomaly), but both~$\kappa_{A^2 C}$ and~$\kappa_{A^3}$ may be non-zero. In the parent theory, we are free to choose any basis to describe the two flavor symmetries. This freedom is restricted after gauging, but we can still shift the~$U(1)_A^{(0)}$ flavor charges by the~$U(1)_c^{(0)}$ gauge charges,
\eqn{
U(1)_A^{(0)} \; \rightarrow \; U(1)_A^{(0)} - n U(1)_c^{(0)}~.
}[aglcgaugshift]
If we use~$\kappa_{AC^2} = \kappa_{C^3} = 0$, this leads to the following changes in the 't Hooft anomalies,\foot{~TD would like to thank E.~Witten for a useful conversation suggesting the shifts in~\eqref{aglcgaugshift} and~\eqref{anomshift}.}
\eqn{
\kappa_{A^2 C} \; \rightarrow \; \kappa_{A^2 C}~, \qquad \kappa_{A^3} \; \rightarrow \; \kappa_{A^3} - 3 n \kappa_{A^2 C}~.
}[anomshift]
The fact that~$\kappa_{A^2 C}$ is unaffected ensures that the 2-group~constant~$\hat \kappa_A = - \half \kappa_{A^2 C}$ that emerges after gauging (see~\coeffmatch) is unambiguous. Moreover, the shift of~$\kappa_{A^3}$ in~\anomshift\ exactly coincides with~\kaaanshift, which was the result of adding the GS counterterm~\gsctbis. To see how this counterterm arises in the present context, note that the redefinition of~$U(1)_A^{(0)}$ in~\aglcgaugshift\ modifies the couplings of gauge fields to currents as follows,
\eqn{
A^{(1)} \wedge * j_A^{(1)} + c^{(1)} \wedge * j_C^{(1)} \quad \longrightarrow \quad A^{(1)} \wedge * \left( j_A^{(1)} - n j_C^{(1)}\right)+ c^{(1)} \wedge * j_C^{(1)}~.
}[cAredef]
In addition, both sides include the BF coupling~${i \over 2 \pi} \, B^{(2)} \wedge f^{(2)}_c$. The two-sides of~\cAredef\ can be made to agree by redefining~$c^{(1)} \; \rightarrow c^{(1)} + n A^{(1)}$, but due to the BF coupling this also generates a GS counterterm~\gsctbis. When~$n$ is an integer, this counterterm is properly quantized, and so are the shifted~$U(1)_A^{(0)}$ charges in~\aglcgaugshift\ and the redefined~$c^{(1)}$ gauge field. As in point~$2.)$ between~\kaaanshift\ and~\aglcgaugshift\ above, we can also contemplate fractional~$n$, but this requires a more careful treatment of global issues. 

We will now reanalyze the~$\langle j_A^{(1)} j_A^{(1)} j_A^{(1)}\rangle$ correlation function in the presence of 2-group symmetry. In light of the preceding discussion, we will insist on invariance under large~$U(1)_B^{(1)}$ gauge transformations, so that all GS counterterms~\gsctbis\ are properly quantized. This means that~$\kappa_{A^3}$ is scheme-independent~$\mod~6 \hat \kappa_A$, but once we fix a particular scheme~$\kappa_{A^3}$ is a well-defined number. With these caveats, the anomalous shift of the effective action is still given by~\twogpacubanom\ and leads to a contact term~\aaacontt\ in the Ward identity satisfied by the~$j_A^{(1)}$ three-point function. However, even in the absence of anomalies, this Ward identity is modified by the 2-group symmetry. Using the 2-group OPE in~\djjope, which reads~$\d^\mu j^A_\mu(x) j^A_\nu(y) = { \hat \kappa_A \over 2 \pi} \, \d^\lambda \delta^{(4)}(x-y) \, J^B_{\nu\lambda}(y)$, we find that the right-hand side of the Ward identity also contains terms that involve the~$\langle J_B^{(2)} j_A^{(1)}\rangle$ correlator analyzed in~section~\ref{secGScts},
\eqna{
{\d \over \d x_\mu} \, \langle j_\mu^A(x) j_\nu^A(y) j_\rho^A(z)\rangle =~ & {\hat \kappa_A \over 2 \pi} \, \Big( \d^\lambda \delta^{(4)} (x-y) \, \langle J_{\nu\lambda}^B(y)j_\rho^A(z)\rangle + \d^\lambda \delta^{(4)}(x-z) \, \langle j^A_\nu(y) J^B_{\rho\lambda}(z)\rangle\Big) \cr
& + {i \kappa_{A^3} \over 12 \pi^2} \, \ep_{\nu\rho \alpha\beta} \, \d^\alpha \delta^{(4)}(x-y) \d^\beta \delta^{(4)}(x-z)~.
}[fulltwgpward]
Fourier-transforming to momentum space (with~$p_1 + p_2 + p_3 = 0$), we find
\eqna{
p_1^\mu \, \big\langle j_\mu^A(p_1) j_\nu^A(p_2) j^A_\rho(p_3)\big\rangle =~&{\hat \kappa_A \over 2 \pi} \, \Big(p_1^\lambda \, \big \langle J_{\nu\lambda}^B(-p_3)  j^A_\rho(p_3)\big\rangle + p_1^\lambda \,  \big \langle J^B_{\rho\lambda} (-p_2) j_\nu^A(p_2)\big\rangle\Big) \cr
& - {\kappa_{A^3} \over 12 \pi^2} \, \ep_{\nu\rho\alpha\beta} \, p_2^\alpha p_3^\beta~. 
}[momtwgpwi]
Recall from~\jbjatwoptmain\ that the two-point functions on the right-hand side are parity odd (i.e.~they contain an explicit Levi-Civita~$\ep$-symbol) and determined by the structure function~$\mathsf K\left({p^2 \over M^2}\right)$. The entire right-hand side of~\momtwgpwi\ is therefore parity odd, so that the parity-even part of the~$j_A^{(1)}$ three-point function is annihilated by~$p_1^\mu$. To match the right-hand side, we must examine the parity-odd part of the three-point function. If we restrict the kinematics as in~\mombosesimp, it is entirely determined by the structure function~$\mathsf A\left({Q^2 \over M^2}\right)$ in~\poddjjjstr. Substituting into~\momtwgpwi, we find that
\eqn{
\mathsf A\left({Q^2 \over M^2}\right) = - {1 \over 12 \pi^2} \left(\kappa_{A^3} - 6 \hat \kappa_A \, {\mathsf K} \left({Q^2 \over M^2}\right)\right)~.
}[asftwgp]
Let us comment on some implications of this formula:
\begin{itemize}
\item The structure function~$\mathsf A\left({Q^2 \over M^2}\right)$ arises from the~$j_A^{(1)}$ three-point function at separated points, which is scheme-independent. If we add a GS counterterm~\gsctbis\ to the action, the anomaly coefficient~$\kappa_{A^3}$ shifts as in~\kaaanshift, $\kappa_{A^3} \; \rightarrow \kappa_{A^3} + 6 n \hat \kappa_A$, while the structure function~$\mathsf K\left({Q^2 \over M^2}\right)$ shifts as in~\gsconttshift, $\mathsf K\left({Q^2 \over M^2}\right) \; \rightarrow \; \mathsf K\left({Q^2 \over M^2}\right) + n$.  These contributions cancel in~\asftwgp, so that~$\mathsf A\left({Q^2 \over M^2}\right)$ remains unchanged, as had to be the case.

\item The structure function~$\mathsf A\left({Q^2 \over M^2}\right)$ can be used to define an effective, scale-dependent quantity~$\kappa_{A^3}^\text{eff.}\left({Q^2 \over M^2}\right)$, which only receives contributions from massless, local degrees of freedom (this definition should be compared to~\hooftpole),
\eqn{
\mathsf A\left({Q^2 \over M^2}\right) = - {1 \over 12 \pi^2} \, \kappa^\text{eff.}_{A^3}\left({Q^2 \over M^2}\right)~.
}[kappaeffdefn]
Without 2-group symmetry~$\hat \kappa_A = 0$ and~\kappaeffdefn\ reduces to~\hooftpole, $\kappa^\text{eff.}_{A^3}\left({Q^2 \over M^2}\right) =  \kappa_{A^3}$, where~$\kappa_{A^3}$ is the  conventional 't Hooft anomaly coefficient. If~$\hat \kappa_A \neq 0$, it follows from~\asftwgp\ that~$\kappa^\text{eff.}_{A^3}\left({Q^2 \over M^2}\right) \sim \mathsf A\left({Q^2 \over M^2}\right)$ evolves along the RG flow in a way that is correlated with the evolution of~$\mathsf K\left({Q^2 \over M^2}\right)$. In general the UV values~$\kappa^\text{eff.\,UV}_{A^3}$, $\mathsf K^\text{UV}$ (corresponding to~$Q^2 \rightarrow \infty$) and the IR values~$\kappa^\text{eff.\,IR}_{A^3}$, $\mathsf K^\text{IR}$ (corresponding to~$Q^2 \rightarrow 0$) do not match, i.e.~$\kappa^\text{eff.\,UV}_{A^3} \neq \kappa^\text{eff.\,IR}_{A^3}$, and similarly for~$\mathsf K^\text{UV, IR}$. By contrast, $\kappa_{A^3}^\text{UV} = \kappa^\text{IR}_{A^3} = \kappa_{A^3}$ satisfies the 't Hooft anomaly matching relation~\hooftmatch\ (see also point 1.) after~\kaaanshift). Substituting these UV and IR quantities into~\asftwgp, we find the following relations
\eqn{
\kappa^\text{eff.~UV, IR}_{A^3} =  \kappa_{A^3} - 6 \hat \kappa_A \, \mathsf K^\text{UV, IR}~.
}[uvircomp]

\item Recall from section~\secGScts\ that the GS contact terms~$\mathsf K^\text{UV, IR}$ can receive contributions from massive or topological degrees of freedom. The same must therefore be true of~$\kappa_{A^3}$, to ensure that the two contributions cancel in~\uvircomp, since~$\kappa_{A^3}^\text{eff.}\left(Q^2 \over M^2\right)$ only receives contributions from massless, local degrees of freedom. For example, a GS counterterm~\gsctbis, which can arise by integrating out massive states, contributes to both~$\kappa_{A^3}$ and~$\mathsf K$, but not to~$\mathsf A$ or~$\kappa^\text{eff.}_{A^3}$. Examples of non-trivial TQFTs with this property appeared in section~\ref{introex} and will be further discussed in section~\ref{ssTQFT}.

\item It follows from~\uvircomp\ that the effective UV-IR anomaly mismatch satisfies
\eqn{
\kappa_{A^3}^\text{eff.~UV} - \kappa_{A^3}^\text{eff.~IR} = - 6 \hat \kappa_A \left(\mathsf K^\text{UV} - \mathsf K^\text{IR}\right)~.
}[uviramism]
The differences on both sides are scheme independent and can be extracted from the $\langle j_A^{(1)} j_A^{(1)} j_A^{(1)}\rangle$ or~$\langle J_B^{(2)} j_A^{(1)}\rangle$ correlation functions at separated points. Equivalently, they can be computed by integrating out massive sates along the RG flow. Of course the actual 2-group 't Hooft anomaly~$\kappa_{A^3}$ satisfies UV-IR matching and drops out of the differences in~\uviramism, 
\end{itemize}

\subsec{Generalization to Nonabelian and Poincar\'e 2-Groups}[ssecGenAnom]

Here we briefly comment on 't Hooft anomalies in theories with nonabelian or Poincar\'e 2-group symmetries, where the 2-group shift of~$B^{(2)}$ takes the form in~\eqref{nonabbint} and~\eqref{ptwgpint}
\eqn{
B^{(2)} \quad \longrightarrow \quad B^{(2)} + {\hat \kappa_A \over 4\pi} \, \tr\left(\lambda_A^{(0)} \, dA^{(1)}\right) + {\hat \kappa_\SP \over 16 \pi} \, \tr \left(\theta^{(0)} \, d \omega^{(1)}\right)~.
}[nonabgravgsiii]
If the theory also has another, abelian flavor symmetry~$U(1)_Y^{(0)}$ (which may or may not participate in a 2-group), we can consider a GS counterterm~\gsctbis\ for~$B^{(2)}$ and the~$U(1)_Y^{(0)}$ field strength~$F_Y^{(2)}$,
\eqn{
S_\text{GS} = {i n \over 2 \pi} \int B^{(2)} \wedge F_Y^{(2)}~, \qquad n \in \Z~.
}[gsyct]
Substituting~\nonabgravgsiii\ into~\gsyct\ and comparing with~\nonabanom, \gravanom\ shows that the GS counterterm shifts the mixed 't Hooft anomalies~$\kappa_{A^2Y}$ (involving~$U(1)_Y^{(0)}$ and nonabelian background fields) and~$\kappa_{Y\SP^2}$ (involving~$U(1)_Y^{(0)}$ and background gravity fields) as follows,
\eqn{
\kappa_{A^2Y} \; \rightarrow \; \kappa_{A^2Y} - \hat \kappa_A n~, \qquad \kappa_{Y\SP^2} \; \rightarrow \; \kappa_{Y\SP^2} + 6 \hat \kappa_\SP n~.
}[nonabgravnshift]
As discussed around~\kaaanshift\ in section~\ref{sssHoftAbTwGp} above, this implies that some parts of these 't Hooft anomalies are scheme dependent. Similarly, as in~\fulltwgpward\ and~\momtwgpwi, the 2-group Ward identities relate~$\langle j_A^{(1)} j_A^{(1)} j_Y^{(1)}\rangle \sim \hat \kappa_A \langle J_B^{(2)} j_Y^{(1)}\rangle + \kappa_{A^2 Y}$ and~$\langle T_{\mu\nu} T_{\rho\sigma} j_Y^{(1)}\rangle \sim \hat \kappa_\SP \langle J_B^{(2)} j_Y^{(1)}\rangle + \kappa_{Y\SP^2}$, which can be used to generalize the discussion around~\asftwgp. 

Note that 2-group symmetries only affect the properties of reducible 't Hooft anomalies, such as~$\kappa_{A^2 Y}$ or~$\kappa_{Y\SP^2}$ in~\nonabgravnshift, or the reducible abelian~$\kappa_{Y^3}$ anomaly discussed in section~\ref{sssHoftAbTwGp} above. Irreducible 't Hooft anomalies, such as a~$\kappa_{A^3}$ anomaly for an~$SU(N)_A^{(0)}$ flavor symmetry, are not affected by 2-group symmetry. In particular, the anomaly coefficient~$\kappa_{A^3}$ is scheme-independent and fixes a certain structure function (proportional to the totally symmetric~$d^{abc}$ symbol associated with the cubic Casimir of~$SU(N)_A^{(0)}$) in the three-point function of the~$SU(N)_A^{(0)}$ current~$j_A^{(1)}$. This is expected from the analogy between 2-group symmetry and the conventional GS mechanism, since the latter can only be used to cancel reducible gauge anomalies.

\newsec{Examples}[SecExamples]

\subsec{Overview}[ssecExOverview]

In this section we illustrate our general observations about theories with 2-group symmetries using a variety of simple, explicit examples. All of our examples are weakly-coupled Lagrangian theories with scalars, fermions, and gauge fields. Many of them are renormalizable abelian gauge theories (albeit with Landau poles in the UV), while others (such as the~$\C\P^N$ models in section~\ref{ssCPN}) are non-renormalizable effective theories.  Given a model with 2-group symmetry, we can deform it and flow to new models with 2-group symmetry by tracking the RG flow. The deformations we consider in this section are mass terms, scalar potentials, and Yukawa couplings. To streamline the discussion below, we introduce the notation~$V_H(\rho)$ for a Higgs potential that leads to a vev~$\langle \rho \rangle = v$ around which the~$\rho$-fluctuations have mass~$M_H$, so that
\eqn{
V_H(\rho) = V_H(v) + \half M_H^2 \left(\rho-v\right)^2 + \cdots~.
}[higsspot]
Another natural deformation consists of gauging the 2-group background fields. We refer to section~\ref{SsecGauging} for a general discussion. Some of the models considered in this section have dynamical string excitations that are charged under the~$U(1)_B^{(1)}$ subgroup of the various 2-group symmetries. These strings will be discussed further in section~\ref{SecDefects}, where it is shown that 2-group symmetry fixes certain 't Hooft anomalies on their world sheets. 

\begin{figure}[h]
\centering
\includegraphics[trim=0 0cm 0cm 0,clip,height=12cm]{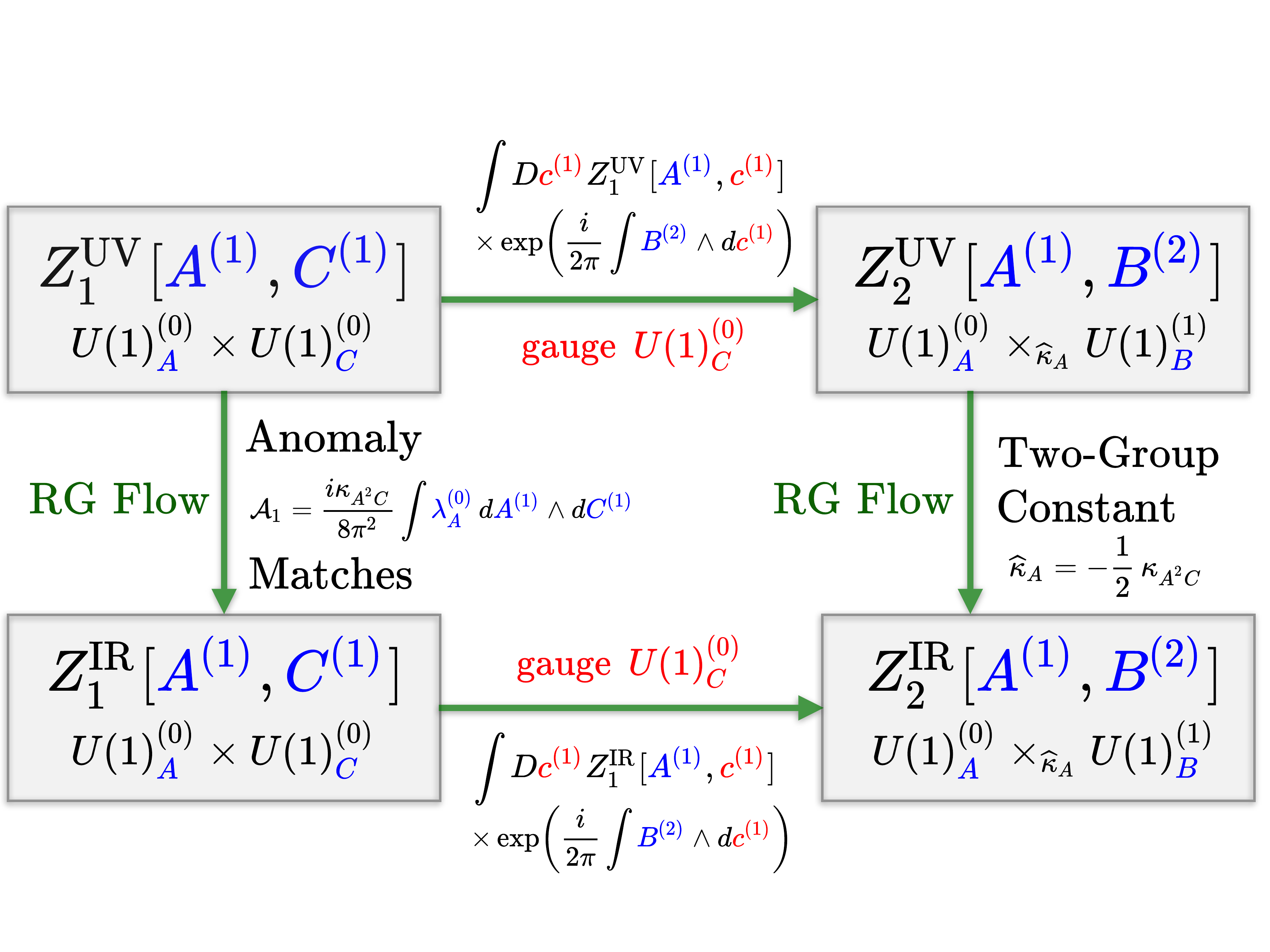}  

 \caption{Horizontal arrows represent the gauging of~$U(1)_C^{(0)}$ in a parent theory~$T_1$ with~$U(1)_A^{(0)} \times U(1)_C^{(0)}$ flavor symmetry and a mixed~$\kappa_{A^2 C}$ 't Hooft anomaly to produce a theory~$T_2$ with~$U(1)_A^{(0)} \times_{\hat \kappa_A} U(1)_B^{(1)}$ 2-group symmetry.  Vertical arrows represent the RG flows interpolating between~$T_{1}^\text{UV} \rightarrow T_{1}^\text{IR}$ and~$T_{2}^\text{UV} \rightarrow T_{2}^\text{IR}$. The diagram is commutative.} 
 \label{fig:gaugeRG}
\end{figure}

All our examples of theories with 2-group symmetry can be viewed as arising from the gauging construction explained in section~\ref{SecBasics}. (This is unavoidable, because this procedure has an inverse, which is explained in section~\ref{SsecGauging}.) Here we briefly recall how this works for the simplest abelian 2-group~$U(1)_A^{(0)} \times_{\hat \kappa_A} U(1)_B^{(1)}$ (see section~\ref{ssecBC}). The construction is illustrated by the right-pointing arrow in the top half of figure~\ref{fig:gaugeRG}. The starting point is a parent theory~$T_1$ with~$U(1)_A^{(0)} \times U(1)_C^{(0)}$ flavor symmetry and vanishing 't Hooft anomalies~$\kappa_{AC^2} = \kappa_{C^3} = 0$, but a nonzero~$\kappa_{A^2 C}$ 't Hooft anomaly. We can then gauge~$U(1)_C^{(0)} \rightarrow U(1)_c^{(0)}$ without spoiling the conservation of the~$U(1)_A^{(0)}$ current~$j_A^{(1)}$ via an ABJ anomaly. As explained in section~\ref{ssecBC}, we must ensure that the dynamical~$U(1)_c^{(0)}$ gauge fields have suitable kinetic terms by adjusting the counterterms for the~$U(1)_C^{(0)}$ background fields before gauging (this is not explicitly indicated in figure~\ref{fig:gaugeRG}). 

The resulting theory~$T_2$ has a new global symmetry~$U(1)_B^{(1)}$ associated with the magnetic 2-form current~$J_B^{(2)} = {i \over 2 \pi} \, f^{(2)}_c$. Here~$f^{(2)}_c$ is the field strength of the dynamical~$U(1)_c^{(0)}$ gauge field, which is closed because of the Bianchi identity, $d f^{(2)}_c = 0$. (See the discussion around~\eqref{magcurrint} and~\magcurr.) The mixed~$\kappa_{A^2 C}$ 't Hooft anomaly in~$T_1$ implies that the global symmetry of~$T_2$ is not a standard product of~$U(1)_A^{(0)}$ and~$U(1)_B^{(1)}$. Rather, these symmetries are fused into a non-trivial abelian 2-group~$U(1)_A^{(0)} \times_{\hat \kappa_A} U(1)_B^{(1)}$. As in~\coeffmatch, the 2-group structure constant is given by~$\hat \kappa_A = - \half \kappa_{A^2 C}$. As was mentioned in section~\ref{rgphtwgp}, and explained in section~\ref{SecCurrents}, $U(1)_B^{(1)}$ is a good subgroup of the 2-group, e.g.~we can spontaneously break~$U(1)_A^{(0)} \times_{\hat \kappa_A} U(1)_B^{(1)} \; \rightarrow U(1)_B^{(1)}$. This is not the case for~$U(1)_A^{(0)}$, and as we will review below, the spontaneous breaking pattern~$U(1)_A^{(0)} \times_{\hat \kappa_A} U(1)_B^{(1)} \; \rightarrow U(1)_A^{(0)}$ cannot occur. 

We would like to understand how the theory~$T_2$ evolves under RG flow, which is represented by the downward arrow in the right half of figure~\ref{fig:gaugeRG}. A useful complementary description of this RG flow comes from the fact that the diagram in figure~\ref{fig:gaugeRG} is commutative: we can either gauge~$U(1)_C^{(0)}$ in~$T_1$ and then follow the RG flow in~$T_2$, or we can first flow to low energies in~$T_1$ and gauge~$U(1)_C^{(0)}$ in the resulting IR effective theory to obtain a low-energy description of~$T_2$. One reason this perspective is useful is that~$T_1$ has a~$\kappa_{A^2 C}$ mixed 't Hooft anomaly for its~$U(1)_A^{(0)} \times U(1)_C^{(0)}$ flavor symmetry, which is subject to conventional 't Hooft anomaly matching. In~$T_2$, one consequence of this is that the 2-group structure constant~$\hat \kappa_A = - \half \kappa_{A^2 C}$ is inert under RG flow.\foot{~In fact, it is quantized, $\hat \kappa_A \in \Z$, see section~\ref{SecGlobal}.}  

We will use the simple models described below to exhibit various general features of the 2-group symmetric theory~$T_2$ and its RG flows:  

\begin{itemize}

\item[1.)] The possible realizations of~$U(1)_A^{(0)} \times_{\hat \kappa_A} U(1)_B^{(1)}$ in the IR of theory~$T_2$ were summarized in section~\ref{rgphtwgp}, and analyzed in sections~\ref{sssprimcur}, \ref{sssnonprimcurr}, and~\ref{sssRGcons}. Here we present the possibilities uncovered there from the perspective of 't Hooft anomaly matching for~$\kappa_{A^2 C}$ in the parent theory~$T_1$ with~$U(1)_A^{(0)} \times U(1)_C^{(0)}$ flavor symmetry, i.e.~we follow the arrows in the part of figure~\ref{fig:gaugeRG} that lies below the NW-SE diagonal.

\begin{itemize}
 
\item[1a.)] If~$U(1)_A^{(0)} \times U(1)_C^{(0)}$ is unbroken in~$T_1$, there are no NG bosons for either symmetry. 't Hooft anomaly matching for~$\kappa_{A^2 C}$ implies that both symmetries act non-trivially on some massless, local degrees of freedom in the deep IR. 

\smallskip

In~$T_2$, the dynamical~$U(1)_c^{(0)}$ gauge symmetry is IR-free: it is not higgsed, and the scale-dependent gauge coupling~$e^2 \left({p^2 \over M^2}\right)$ logarithmically runs to zero in the deep IR, because the massless degrees of freedom charged under~$U(1)_c^{(0)}$ ensure that the~$\beta$-function is strictly positive. Here~$M$ is some energy scale, which may be the UV cutoff. This implies that the two-point function of~$J_B^{(2)} \sim * f^{(2)}_c$ decays more rapidly at long distances than in free Maxwell theory (where~$e$ is constant), so that the operator~$J_B^{(2)} \sim * f^{(2)}_c$ effectively flows to zero (i.e.~it becomes redundant) in the deep IR. Schematically (and in particular, omitting all tensor structures),
\eqn{
\langle J_B^{(2)}(p) J_B^{(2)}(-p)\rangle  \sim  \langle f^{(2)}_c(p) f^{(2)}_c(-p)\rangle  \sim  e^2 \left({p^2 \over M^2}\right) \;\rightarrow \; 0 \quad \text{as} \quad p^2 \; \rightarrow \; 0~. 
}[irfreeeq]
Since there is no free Maxwell field at low energies that could act as a NG boson for~$U(1)_B^{(1)}$, it follows that the entire 2-group~$U(1)_A^{(0)} \times_{\hat \kappa_A} U(1)_B^{(1)}$, is unbroken. 

\smallskip

\item[1b.)] If~$U(1)_C^{(0)}$ is spontaneously broken in~$T_1$, there is an associated NG boson~$\chi$ that shifts as~$\chi \rightarrow \chi + \lambda_C^{(0)}$. Any mismatch in the~$\kappa_{A^2 C}$ 't Hooft anomaly can therefore be accounted for by a term~$\sim \int \chi \, F_A^{(2)} \wedge F_A^{(2)}$ in the low-energy effective action. Then~$U(1)_A^{(0)}$ can either act on the massless degrees of freedom (in which case it may be broken or unbroken), or it can only act on massive degrees of freedom and decouple in the deep IR. 

\smallskip

In~$T_2$, the~$U(1)_c^{(0)}$ gauge symmetry is higgsed, so that the photon acquires a mass~$m_\gamma$. Therefore~$J_B^{(2)} \sim * f_c^{(2)}$ decouples exponentially at long distances $\gtrsim m_\gamma^{-1}$ and is therefore unbroken. If~$U(1)_A^{(0)}$ does not act on massless, local degrees of freedom in the IR, the entire 2-group is unbroken. Note that this scenario is compatible with the theory having a gap and no massless, local IR degrees of freedom whatsoever. However, there may be non-local, topological degrees of freedom (see point 2.) below).  

\smallskip

If~$U(1)_A^{(0)}$ acts on massless, local degrees of freedom in the deep IR, it may or may not be spontaneously broken. In the former case the 2-group~$U(1)_A^{(0)} \times_{\hat \kappa_A} U(1)_B^{(1)}$ is unbroken, and in the latter case it is spontaneously broken to its~$U(1)_B^{(1)}$ subgroup. 

\smallskip

\item[1c.)] If~$U(1)_C^{(0)}$ is unbroken in~$T_1$, but $U(1)_A^{(0)}$ is spontaneously broken, there is a~$U(1)_A^{(0)}$ NG boson~$\chi$, which shifts as~$\chi \rightarrow \chi + \lambda_A^{(0)}$. In this case any mismatch in the~$\kappa_{A^2 C}$ 't Hooft anomaly can be accounted for by a term~$\sim \int \chi \, F_A^{(2)} \wedge F_C^{(2)}$ in the IR effective action. Additional massless degrees of freedom charged under~$U(1)_A^{(0)}$ or~$U(1)_C^{(0)}$ (with the exception of a NG boson for~$U(1)_C^{(0)}$, which is assumed to be unbroken) may be present in the IR, but this is not necessary.

\smallskip

In the absence of such additional degrees of freedom, the low-energy effective theory of~$T_2$ consists of a free~$U(1)_c^{(0)}$ Maxwell field, which spontaneously breaks~$U(1)_B^{(1)}$ (the free photon is the corresponding NG boson), and the NG boson~$\chi$ for the spontaneously broken~$U(1)_A^{(0)}$ symmetry. Therefore the entire 2-group~$U(1)_A^{(0)} \times_{\hat \kappa_A} U(1)_B^{(1)}$ is spontaneously broken in this Goldstone-Maxwell model. In this model, 2-group symmetry is realized via a particular improvement term~$\sim \hat \kappa_A * (f^{(2)}_c \wedge d \chi) \; \subset \; j_A^{(1)}$. This is the mechanism described around~\eqref{factwardid}. Even though it is a free theory, the Goldstone-Maxwell model has a number of interesting features, some of which are discussed in section~\ref{ssGM} below. In particular, we will see that the 2-group symmetry of the model is actually embedded inside a larger 3-group symmetry. 

\smallskip

The symmetry-breaking pattern is unmodified if there are additional massless degrees of freedom that are only charged under~$U(1)_A^{(0)}$. By contrast, if there are massless degrees of freedom charged under~$U(1)_c^{(0)}$, then the gauge coupling is IR-free and~$J_B^{(2)} \sim * f^{(2)}_c$ decouples at long distances (see the discussion around~\irfreeeq\ above). In this case the 2-group is spontaneously broken to its~$U(1)_B^{(1)}$ subgroup. 

\end{itemize}

\item[2.)] 2-group symmetry is compatible with all local degrees of freedom being massive, and the theory having a gap (see point 1b.) above). The IR theory can be a TQFT with 2-group symmetry and short- or long-range entanglement. (In the former case, the theory effectively has no non-trivial dynamical degrees of freedom and can be formulated using only background fields.) This remains true in the presence of certain reducible 2-group 't Hooft anomalies, which can be matched by the topological theory. RG flows to gapped theories are described in sections~\ref{sssQEDab} and~\ref{SubsecFermats}. A more detailed discussion of topological sectors with 2-group symmetry is in section~\ref{ssTQFT}.

\item[3.)] Some RG flows with 2-group symmetry naively appear to violate 't Hooft anomaly matching. More precisely, the effective 't Hooft anomaly~$\kappa_{A^3}^\text{eff.}$ defined in~\eqref{kappaeffdefn}, which is only sensitive to the massless, local degrees of freedom, may change along the RG flow. The actual 't Hooft anomaly~$\kappa_{A^3}$ must match, but in theories with 2-group symmetry it may receive contributions from gapped or topological degrees of freedom (see point~2.) above). As explained in section~\ref{sssHoftAbTwGp}, this discrepancy is due to the presence of GS contact terms (see section~\ref{secGScts}) in the Ward identity~\fulltwgpward. 

\end{itemize} 

We will also consider examples with 2-group symmetries that fuse~$U(1)_B^{(1)}$ with multiple abelian or nonabelian flavor symmetries, as well as with Poincar\'e symmetry. (See section~\ref{ssecBCgen} for a general discussion of how such examples arise from parent theories with suitable mixed 't Hooft anomalies.) As discussed at the end of section~\ref{sssRGcons}, exact 2-group symmetry is an obstruction to conventional UV completion in continuum QFT (see point 1.) below~\IRdecineq). As discussed below~\eqref{stemerge}, this is particularly dramatic in examples with Poincar\'e 2-group symmetry (see section~\ref{SubsecFermats}). However, if the 2-group symmetry is an emergent, accidental symmetry of the low-energy theory, which is explicitly broken at high energies, conventional UV completions are possible (see in particular point~2.) below~\IRdecineq). We will give explicit examples in section~\ref{sssQEDemerge}.

\subsec{QED-Like Models}[SubsecQEDexamples]

In this subsection we consider theories~$T_2$ that are mild generalizations of conventional QED. These are~$U(1)_c^{(0)}$ gauge theories with~$N_f$ flavors of massless Dirac fermions of~$U(1)_c^{(0)}$ charge~$q \in \Z$. In the absence of additional fields or interactions, this model has flavor symmetry~$G^{(0)} = SU(N_f)^{(0)}_L \times SU(N_f)^{(0)}_R$. We first focus on abelian subgroups~$U(1)_A^{(0)} \subset G^{(0)}$ and show that some of them belong to an abelian 2-group~$U(1)_A^{(0)} \times_{\hat \kappa_A} U(1)_B^{(1)}$. We then use the full flavor symmetry~$G^{(0)}$ to give examples of nonabelian 2-groups. Finally, we show that some of these QED-like examples have conventional UV completions as asymptotically-free nonabelian gauge theories, in which the 2-group symmetry is explicitly broken at high energies, but emerges as an accidental symmetry in the IR. 

\subsubsec{Examples of Abelian 2-Group Symmetries}[sssQEDab]

The simplest QED-like example with 2-group symmetry has~$N_f = 2$ flavors of Dirac fermions with~$U(1)_c^{(0)}$ gauge charge~$q$.\foot{~Massless QED with~$N_f = 1$ has no flavor symmetries, because~$U(1)_\text{axial}^{(0)}$ suffers from an ABJ anomaly.} We view the theory as arising from a parent theory~$T_1$ (see figure~\ref{fig:gaugeRG}) with~$U(1)_A^{(0)} \times U(1)_C^{(0)}$ flavor symmetry. The two Dirac fermions can be decomposed into four Weyl fermions~$\psi_\alpha^1, \psi_\alpha^2, \psi^3_\alpha,  \psi^4_\alpha$, whose~$U(1)_A^{(0)} \times U(1)_C^{(0)}$ flavor charges are summarized in table~\ref{tab:QEDabelian}. We also add a complex scalar field~$\phi$ with charges~$q_A, q_C$. 

\smallskip
\renewcommand{\arraystretch}{1.6}
\renewcommand\tabcolsep{6pt}
\begin{table}[H]
  \centering
  \begin{tabular}{ |c|c|c|c|c|c| }
\hline
{\bf Field} &  {$\psi^1_\alpha$} &  {$\psi^2_\alpha$} & $ \psi^3_\alpha$ & $ \psi^4_\alpha$ & {$\phi$} \\
\hline
\hline
$ U(1)_A^{(0)} $ &  $q_1$ & $q_2$ & $q_3$ & $q_4$ & $q_A$ \\
\hline
$U(1)_C^{(0)}$ &  $q$&  $q$ & $-q$ & $-q$ & $q_C$ \\
\hline
\end{tabular}
  \caption{Charge assignments in the parent theory~$T_1$. All charges are integers and~$q \neq 0$.}
  \label{tab:QEDabelian}
\end{table}

The anomaly coefficients are readily computed from the fermion charges in table~\ref{tab:QEDabelian},
\eqna{
& \kappa_{C^3} = \kappa_{C\SP^2} =  0~,\cr
& \kappa_{AC^2} = q^2 \left(q_1 + q_2 + q_3 + q_4\right)~,\cr
& \kappa_{A^2 C} = q \left(q_1^2 + q_2^2 - q_3^2 -q_4^2\right)~,\cr
& \kappa_{A^3} = q_1^3 + q_2^3 + q_3^3 + q_4^3~,\cr
& \kappa_{A\SP^2} = q_1 + q_2 + q_3 + q_4~.
}[abqedanomcs]
Since~$\kappa_{C^3}  = 0$, we are free to gauge~$U(1)_C^{(0)}$. However, we must also impose~$\kappa_{AC^2} = 0$, so that~$U(1)_A^{(0)}$ does not suffer from an ABJ anomaly after gauging. Since~$q \neq 0$, this requires
\eqn{
q_1 + q_2 + q_3 + q_4 = 0~.
}[noabjsimpqed]
This homogenous constraint admits three linearly independent solutions, which can be parametrized by~$(q_1, q_2, q_3) \in \Z^3$. Comparing with~\abqedanomcs, we see that this constraint also forces the mixed~$U(1)_A^{(0)}$-Poincar\'e 't Hooft anomaly to vanish, $\kappa_{A\SP^2} = 0$, while the remaining anomaly coefficients simplify as follows,
\eqna{
& \kappa_{A^2 C}  = -2 q\left(q_1 + q_3\right)\left(q_2 + q_3\right)~,\cr
& \kappa_{A^3} = -3 \left(q_1 + q_2\right)\left(q_1 + q_3\right) \left(q_2 + q_3\right)~.
}[simpqedanom]
Note that~$\kappa_{A^2 C} \in 2 \Z$ is even (and also that~$\kappa_{A^3} \in 6 \Z$, because~$\kappa_{A \SP^2} =0$) in accord with the discussion after~\coeffmatch\ and in appendix~\ref{AppQuantAnom}. 

We now follow the procedure explained in section~\ref{ssecBC} (see also the review below figure~\ref{fig:gaugeRG}) and gauge~$U(1)_C^{(0)} \rightarrow U(1)_c^{(0)}$ in the parent theory~$T_1$. The resulting theory~$T_2$ has abelian 2-group symmetry~$U(1)_A^{(0)} \times_{\hat \kappa_A} U(1)_B^{(1)}$, where~$U(1)_B^{(1)}$ is the magnetic~$1$-form symmetry with current~$J_B^{(2)} \sim *f^{(2)}_c$ that emerges as a result of the gauging. The 2-group structure constant is given by (see~\coeffmatch), 
\eqn{
\hat \kappa_A = - \half \kappa_{A^2 C} = q\left(q_1 + q_3\right)\left(q_2 + q_3\right)~.
}[simpabtsgpkap]
As explained in section~\ref{sssHoftAbTwGp}, between~\kaaanshift\ and~\aglcgaugshift, the~$\kappa_{A^3}$ 't Hooft anomaly of the parent theory~$T_1$ is truncated in~$T_2$, because only~$\kappa_{A^3}~(\mod~6 \hat \kappa_A)$ is scheme independent. The anomaly is therefore encoded by the fractional part~${\kappa_{A^3} \over 6 \hat \kappa_A}~(\mod~1)$. Using~\simpqedanom\ and~\simpabtsgpkap,
\eqn{
{\kappa_{A^3} \over 6 \hat \kappa_A} = - {q_1 + q_2 \over 2 q}~.
}[sqedtwgpan]
The integer part of this quantity is scheme dependent and can be adjusted using a properly quantized GS counterterm~\gsctbis (see also section~\ref{secGScts}). Equivalently, this follows from the discussion around~\anomshift: if we redefine the flavor symmetry~$U(1)_A^{(0)} \rightarrow U(1)_A^{(0)} - n U(1)_c^{(0)}$ (with~$n \in \Z$) by admixing an integer multiple of the gauge symmetry, then~$q_{1,2} \rightarrow q_{1,2} - n q$, so that the quantity in~\sqedtwgpan\ shifts by the integer~$n$. However, the fractional part of~\sqedtwgpan\ constitutes a scheme-independent, intrinsic 2-group 't Hooft anomaly,\foot{~As was explained between~\kaaanshift\ and~\aglcgaugshift, this anomaly should be thought of as mixed anomaly between~$U(1)_A^{(0)}$ and large~$U(1)_B^{(1)}$ gauge transformations. Demanding that~$n \in \Z$ preserves the latter.} which must match along RG flows. We will explicitly demonstrate this in various examples below. 

As long as~$\phi$ has a sufficiently positive mass term, so that it does not acquire a vev, the~$U(1)_c^{(0)}$ charged massless fermions in table~\ref{tab:QEDabelian} lead to a positive beta function and render the theory~$T_2$ IR free. Therefore~$J_B^{(2)} \sim * f^{(2)}_c$ decouples in the deep IR, and~$U(1)_A^{(0)} \times_{\hat \kappa_A} U(1)_B^{(0)}$ is unbroken. The parent theory $T_1$ consists of free fermions and a decoupled scalar~$\phi$ with zero vev, so that~$U(1)_A^{(0)} \times U(1)_C^{(0)}$ is unbroken. This is scenario 1a.) in section~\ref{ssecExOverview}.

If we add a Higgs potential~$V_H(\phi)$, as in~\higsspot, then~$\phi$ acquires a vev~$\langle \phi \rangle  = v$. As long as~$q_C \neq 0$, this spontaneously breaks~$U(1)_C^{(0)}$ in~$T_1$, and higgses the~$U(1)_c^{(0)}$ gauge symmetry to~$\Z_{|q_C|}$ in~$T_2$, so that~$U(1)_A^{(0)} \times_{\hat \kappa_A} U(1)_B^{(1)}$ is unbroken. This is scenario~1b.) in section~\ref{ssecExOverview}. If~$q_C = 0$, then~$U(1)_C^{(0)}$ is unbroken in~$T_1$, while~$U(1)_A^{(0)}$ is spontaneously broken. In~$T_2$, this means that the~$U(1)_c^{(0)}$ gauge symmetry is unbroken (and IR free), but the 2-group symmetry~$U(1)_A^{(0)} \times_{\hat \kappa_A} U(1)_B^{(1)}$ is spontaneously broken to its~$U(1)_B^{(1)}$ subgroup. This is scenario 1c.) in section~\ref{ssecExOverview}. 

Things get more interesting if we supplement the Higgs potential for~$\phi$ with Yukawa couplings of~$\phi$ to the fermions. We will consider two different deformations of this kind. If we add a Yukawa interaction of the form
\eqn{
 \SL_\text{Yukawa} = \lambda_{13} \, \phi \,  \psi^1 \psi^3 + \lambda_{24} \, \b \phi \,  \psi^2 \psi^4 + (\text{c.c.})~,
}[yukawa1] 
we must take the charges of~$\phi$ to be~$q_A = -(q_1 + q_3) = q_2 + q_4$ (which is compatible with~\noabjsimpqed) and~$q_C = 0$ in order to preserve~$U(1)_A^{(0)} \times U(1)_C^{(0)}$ in~$T_1$. If both Yukawa couplings~$\lambda_{13}, \lambda_{24}$ are non-zero, then the vev~$\langle \phi \rangle = v$ gives mass to all fermions. It also spontaneously breaks~$U(1)_A^{(0)}$, while~$U(1)_C^{(0)}$ remains unbroken. The IR of~$T_1$ therefore consists only of a~$U(1)_A^{(0)}$ NG boson that matches the~$\kappa_{A^2 C}$ 't Hooft anomaly via a term induced by integrating out the massive fermions. This is scenario 1c.) in section~\ref{ssecExOverview}. The theory~$T_2$ obtained after gauging is the Goldstone-Maxwell model, which describes spontaneous breaking of the full~$U(1)_A^{(0)} \times_{\hat \kappa_A} U(1)_B^{(1)}$ 2-group symmetry (see also section~\ref{ssGM}). If one of the Yukawa couplings in~\eqref{yukawa1} vanishes, only some of the fermions acquire a mass. 

We can also consider a different set of Yukawa couplings,
\eqn{
\SL'_\text{Yukawa} = \lambda_{12} \, \phi  \, \psi^1 \psi^2 + \lambda_{34} \, \b \phi \,  \psi^3 \psi^4 + (\text{c.c.})~,
}[yukawa2]
Now the charges of~$\phi$ are~$q_A = -(q_1 + q_2) = q_3 + q_4$ (see~\noabjsimpqed) and~$q_C = -2q$, which differ from those needed in~\eqref{yukawa1}. (We can therefore not combine the deformations~\eqref{yukawa1} and~\eqref{yukawa2} without introducing a second Higgs field.) Since~$q_C \neq 0$, the vev~$\langle \phi \rangle = v$ spontaneously breaks~$U(1)_A^{(0)} \times U(1)_C^{(0)}$ to a diagonal subgroup in theory~$T_1$. If the Yukawa coupligs~$\lambda_{12}, \lambda_{34}$ are both non-vanishing, then all fermions acquire masses.  This is scenario~1b.) in section~\ref{ssecExOverview}. In theory~$T_2$, the~$U(1)_c^{(0)}$ gauge symmetry is higgsed, so that the photon, along with the Higgs field~$\phi$ and all fermions, acquires a mass. Therefore the theory is gapped. Since~$\phi$ has charge~$q_C = -2 q \in 2 \Z$, it leaves a discrete subgroup~$\Z_{2|q|} \subset U(1)_c^{(0)}$ of the gauge symmetry unbroken. The low-energy theory is therefore a non-trivial topological~$\Z_{2 |q|}$ gauge theory, with unbroken~$U(1)_A^{(0)} \times_{\hat \kappa_A} U(1)_B^{(1)}$ 2-group symmetry (see section~\ref{ssTQFT} for additional details). 

The RG flow induced by the Yukawa couplings~\eqref{yukawa2} explicitly shows that a theory with 2-group symmetry can flow to a gapped phase. However, this flow is also naively inconsistent with conventional 't Hooft anomaly matching for~$\kappa_{A^3}$: in the UV, the fermions contribute the (generally nonzero) value for~$\kappa_{A^3}$ in~\simpqedanom, but the gapped IR theory superficially does not contribute. As explained at the end of section~\ref{sssHoftAbTwGp} (see also point 3.) in section~\ref{ssecExOverview}), this mismatch only affects the effective quantity~$\kappa_{A^3}^\text{eff.}$ defined in~\eqref{kappaeffdefn}. Recall that~$\kappa_{A^3}^\text{eff.}$ only receives contributions from massless, local degrees of freedom, just as the conventional~$\kappa_{A^3}$ 't Hooft anomaly in theories without 2-group symmetry. However, in the presence of 2-group symmetry, $\kappa_{A^3}^\text{eff.}$ generally does not match between the UV and the IR. By contrast, the genuine 2-group 't Hooft anomaly does satisfy anomaly matching, but can receive contributions from the topological~$\Z_{2 |q|}$ gauge theory in the IR. As we will explain in section~\ref{ssTQFT}, this topological theory contributes~${\kappa_{A^3} \over 6 \hat \kappa_A}  = -{q_A \over q_C} = -{q_1 + q_2 \over 2 q}$, which precisely matches the 2-group 't Hooft anomaly in~\sqedtwgpan. There we will also check that the difference between the effective quantity~$\kappa_{A^3}^\text{eff.}$ and the genuine 't Hooft anomaly~$\kappa_{A^3}$ is correctly accounted for by a non-trivial GS contact term~$\mathsf K$, as in~\eqref{uvircomp}. 

As before, we can keep some of the fermions massless by setting one of the Yukawa couplings in~\eqref{yukawa2} to zero. In this case the low-energy topological theory must account for the effective anomaly mismatch between the UV and IR fermion spectrum (see~section~\ref{ssTQFT}).

\subsubsec{Examples of Nonabelian 2-Group Symmetries}[sssQEDnonab]

We now generalize the model considered in the previous section from~$N_f = 2$ flavors of Dirac fermions to arbitrary~$N_f$. The parent theory~$T_1$ consists of~$2 N_f$ free, massless Weyl fermions, which we separate into two groups: $\psi^i_\alpha$ and~$\t \psi^{\t i}_\alpha$ with~$i, \t i = 1, \ldots, N_f$. We focus on a~$SU(N_f)^{(0)}_L \times SU(N_f)^{(0)}_R\times U(1)_C^{(0)}$ subgroup of the full~$U(2 N_f)$ flavor symmetry that exists in~$T_1$.\foot{~The commutant of~$U(1)_C^{(0)} \subset U(2N_f)$ that remains after gauging is~$SU(N_f)^{(0)}_L \times SU(N_f)^{(0)}_R \times U(1)^{(0)}_\text{axial}$. However, $U(1)^{(0)}_\text{axial}$ suffers from an ABJ anomaly after gauging.} The fermions~$\psi_\alpha^i$ carry charge~$q$ under~$U(1)_C^{(0)}$ and transform in the fundamental representation of~$SU(N_f)_L^{(0)}$, but are neutral under~$SU(N_f)_R^{(0)}$. Similarly, the fermions~$\t \psi^{\t i}_\alpha$ have~$U(1)_C^{(0)}$ charge~$-q$ and are neutral under~$SU(N_f)_L^{(0)}$, but transform in the fundamental representation of~$SU(N_f)_R^{(0)}$. These charge assignments are summarized in table~\ref{tab:QEDexample}. 

\smallskip
\renewcommand{\arraystretch}{1.6}
\renewcommand\tabcolsep{6pt}
\begin{table}[H]
  \centering
  \begin{tabular}{ |c|c|c|c| }
\hline
{\bf Field} &   $SU(N_f)_L^{(0)} $& $SU(N_f)_R^{(0)}$ &  {\bf $U(1)_C^{(0)}$} \\
\hline
\hline
$\psi^i_\alpha $ &  $\square$ & $\1$& $ q$ \\
\hline
$\t \psi^{\t i}_\alpha $ &  $\1$ & $\square$ &  $-q$ \\
\hline
\end{tabular}
  \caption{Transformation properties of fields in QED-like model with~$N_f$ massless flavors. Here~$\square$ denotes the fundamental representation of~$SU(N_{f})^{(0)}_{L, R}$.}
  \label{tab:QEDexample}
\end{table}

In the conventions of section~\ref{ssecNonabHooft}, the fermions in table~\ref{tab:QEDexample} give rise to the following anomaly 6-form polynomial (see~\nonabsixfm),
\eqna{
\CI^{(6)} = {1 \over (2 \pi)^3} \, \bigg(& -{i \kappa_{L^3} \over 3!} \, \tr\Big(F_L^{(2)} \wedge F_L^{(2)} \wedge  F_L^{(2)} \Big) -{\kappa_{L^2 C} \over 2!} \, \tr\Big(F_L^{(2)} \wedge F_L^{(2)}\Big) \wedge F_C^{(2)}   \cr
& -{i \kappa_{R^3} \over 3!} \, \tr\Big(F_R^{(2)} \wedge F_R^{(2)} \wedge  F_R^{(2)}\Big) - {\kappa_{R^2 C} \over 2!} \tr\Big(F_R^{(2)} \wedge F_R^{(2)}\Big) \wedge F_C^{(2)}\bigg)~.
}[lranompol]
Here~$F_L^{(2)}$,~$F^{(2)}_R$ and~$F^{(2)}_C$ are the field strength 2-forms for the backgrond~$SU(N_f)_L^{(0)}$, $SU(N_f)^{(0)}_R$, and~$U(1)_C^{(0)}$ flavor symmetries. Note that the~$\kappa_{C^3}$ anomaly vanishes. Comparing with~\sunfermionanom, we see that
\eqn{
\kappa_{L^3}  = \kappa_{R^3} = 1~, \qquad \kappa_{L^2 C} = -\kappa_{R^2 C} = q~.
}[lranomks]
The mixed anomalies~$\kappa_{L^2C}, \kappa_{R^2 C}$ are present for all~$N_f$, while the irreducible anomalies~$\kappa_{L^3}, \kappa_{R^3}$ are only present if~$SU(N_f)^{(0)}_{L,R}$ has a cubic Casimir, which happens for~$N_f \geq 3$. The diagonal subgroup~$SU(N_f)_\text{diag}^{(0)} \subset SU(N_f)^{(0)}_L \times SU(N_f)^{(0)}_R$ is  free of 't Hooft anomalies. This is reflected by the vanishing of~$\CI^{(6)}$ in~\lranompol\ upon setting~$F^{(2)}_L = - F^{(2)}_R$. 

If we follow the discussion in section~\ref{ssecBCgen} and gauge~$U(1)_C^{(0)}$, we obtain a theory~$T_2$ with nonabelian 2-group symmetry. For instance, $SU(N_f)_L^{(0)}$ participates in a nonabelian 2-group~$SU(N_f)^{(0)}_L \times_{\hat \kappa_L} U(1)_B^{(1)}$ with~$\hat \kappa_L = \kappa_{L^2 C} = q$ (see~\nonabgravkhat\ and~\lranomks). Similarly, $SU(N_f)_R^{(0)}$ is part of a nonabelian 2-group~$SU(N_f)_R^{(0)} \times_{\hat \kappa_R} U(1)_B^{(1)}$, with~$\hat \kappa_R = \kappa_{R^2 C} = -q$. We can summarize this by saying that the full global symmetry is the following 2-group,\foot{~The same 2-group symmetry (with~$q = 1$) arises in QCD with~$SU(N_c)$ gauge group and~$N_f$ massless quark flavors, if we gauge its~$U(1)^{(0)}_\text{baryon}$ flavor symmetry.}
\eqn{
\left(SU(N_f)^{(0)}_L \times SU(N_f)_R^{(0)}\right) \times_{\hat \kappa_L \, = \, q\,, \,\hat \kappa_R \,=\, -q} U(1)_B^{(1)}~.
}[nonabtwogpqedex]
Note that~$SU(N_f)_\text{diag.}^{(0)} \times U(1)_B^{(1)}$ is a good subgroup of~\nonabtwogpqedex. The abelian 2-group~$U(1)_A^{(0)} \times_{\hat \kappa_A} U(1)_B^{(1)}$ analyzed in section~\ref{sssQEDab} above is a subgroup of~\nonabtwogpqedex\ for~$N_f = 2$.\foot{~This is true up to an overall rescaling of the~$U(1)_A^{(0)}$ charges, as well as mixing of~$U(1)_A^{(0)}$ with~$U(1)_C^{(0)}$.} 

As explained in section~\ref{ssecGenAnom}, the irreducible 't Hooft anomalies~$\kappa_{L^3}$, $\kappa_{R^3}$ in~\lranomks\ that are present for~$N_f \geq 3$ are not modified in the presence of the 2-group symmetry~\nonabtwogpqedex. Correspondingly, they enjoy conventional 't Hooft anomaly matching, which requires massless, local degrees of freedom in the IR. In particular, this means that it is impossible to gap the model while preserving the full global symmetry. (See section~\ref{ssCPN} for some deformations of the model that are consistent with this claim.) The same conclusion holds for~$N_f = 2$, because of 't Hooft anomaly matching for the flavor-symmetry analog of the global anomaly described in~\cite{Witten:1982fp}, which is associated with~$\pi_4(SU(2)_{L,R}^{(0)})  = \Z_2$. Note that the deformation~\eqref{yukawa2}, which leads to a gapped theory, preserves the abelian 2-group~$U(1)_A^{(0)} \times_{\hat \kappa_A} U(1)_B^{(1)}$, but explicitly breaks~$SU(2)^{(0)}_{L,R}$.

\subsubsec{UV-Complete Models with Emergent 2-Group Symmetry}[sssQEDemerge]

As discussed at the end of sections~\ref{sssRGcons} and~\ref{ssecExOverview}, theories with exact 2-group symmetry cannot be UV completed in continuum QFT. However, this does not preclude the possibility of standard UV-complete theories with emergent 2-group symmetry at low energies, as long as the symmetry is explicitly violated at sufficiently high energies. We will briefly illustrate this using simple UV completions for some of the QED-like models that were shown to possess 2-group symmetry in subsections~\ref{sssQEDab} and~\ref{sssQEDnonab} above.

\smallskip
\renewcommand{\arraystretch}{1.6}
\renewcommand\tabcolsep{6pt}
\begin{table}[H]
  \centering
  \begin{tabular}{ |c|c|c| }
\hline
{\bf Field} &  $SU(N_{f})^{(0)}$ & $SU(2)_c^{(0)}$ \\
\hline
\hline
$\Psi^i_\alpha $ &  $\square $ & $\bf 2$ \\
\hline
$\Phi$  & $\1$  &  $\bf 3$ \\
\hline
\end{tabular}
  \caption{Field Content of the Georgi-Glashow Model}
  \label{tab:SU(2)completion}
\end{table}
\noindent

We consider a variant of the Georgi-Glashow model~\cite{Georgi:1972cj}. In this model, a low-energy~$U(1)_c^{(0)}$ gauge field arises from higgsing an~$SU(2)_c^{(0)}$ gauge theory using a real scalar field~$\Phi$ in the adjoint (i.e~triplet) representation~$\bf 3$ of~$SU(2)_c^{(0)}$. (We omit the~$SU(2)_c^{(0)}$ indices.) We also add~$N_f$ Weyl fermions~$\Psi^i_\alpha$ in the doublet representation~$\bf 2$ of~$SU(2)_c^{(0)}$. Here~$i = 1, \ldots,  N_f$ is a fundamental index for the~$SU(N_f)^{(0)}$ flavor symmetry of the model. As was shown in~\cite{Witten:1982fp}, the~$SU(2)_c^{(0)}$ gauge symmetry suffers from a gauge anomaly (associated with large~$SU(2)_c^{(0)}$ gauge transformations), unless the number~$N_f$ of fermion doublets is even, $N_f \in 2 \Z$. If we demand that the theory is also asymptotically free (and hence UV complete), $N_f$ must be an even integer in the range~$0 \leq N_f \leq 20$. 

In the low-energy~$U(1)_c^{(0)}$ gauge theory, every~$SU(2)_c^{(0)}$ fermion doublet gives rise to a Dirac fermion of~$U(1)_c^{(0)}$ charge~$1$. This is nothing but ordinary QED with~$N_f$ massless flavors, which coincides with the model in section~\ref{sssQEDnonab} if we set~$q = 1$ there. In the IR, the microscopic~$SU(N_f)^{(0)}$ flavor symmetry is enhanced to~$SU(N_f)_L^{(0)} \times SU(N_f)^{(0)}_R$ (with the microscopic symmetry embedded as the diagonal subgroup). Furthermore, the low-energy theory has an emergent~$U(1)_B^{(1)}$ symmetry, with magnetic current~$J_B^{(2)} \sim * f^{(2)}_{U(1)_c^{(0)}}$. Together, these symmetries form the nonabelian 2-group~\nonabtwogpqedex. At high energies, the emergent (i.e.~non-diagonal) part of the~$SU(N_f)^{(0)}_L \times SU(N_f)^{(0)}_R$ flavor symmetry, as well as the emergent~$U(1)_B^{(1)}$ symmetry, are both explicitly violated. For instance, at high energies, the~$U(1)_c^{(0)}$ field strength~$f^{(2)}_{U(1)_c^{(0)}}  \sim \tr\big(\Phi \, f^{(2)}_{SU(2)_c^{(0)}}\big)$ does not satisfy the Bianchi identity, so that~$J_B^{(2)}$ is explicitly broken. The energies scales at which the symmetries emerge are subject to the inequality~\uvemergineq.

\subsec{Chiral Fermat Model with Poincar\'e 2-Group Symmetry}[SubsecFermats]

In this subsection, $T_2$ will be a chiral~$U(1)_c^{(0)}$ gauge theory with Poincar\'e 2-group symmetry~${\mathscr P} \times_{\hat \kappa_{\mathscr P}} U(1)_B^{(1)}$, as well as abelian 2-group symmetries. For reasons explained below, we will refer to it as the Fermat model. Unlike (non-) abelian 2-group symmetry, which can emerge as an accidental symmetry in the IR of theories with conventional UV completions (see section~\ref{sssQEDemerge}), Poincar\'e 2-group symmetry obstructs such UV completions (see the discussion around~\eqref{stemerge}). Therefore, the Fermat model analyzed below does not have a UV completion in continuum QFT, with a conformal fixed point at short distances.

\smallskip
\renewcommand{\arraystretch}{1.6}
\renewcommand\tabcolsep{6pt}
\begin{table}[H]
  \centering
  \begin{tabular}{ |c|c|c|c|c|c| }
\hline
{\bf Field} &  {$\psi^1_\alpha$} &  {$\psi^2_\alpha$} & $ \psi^3_\alpha$ & $ \psi^4_\alpha$ & {$\phi$} \\
\hline
\hline
$ U(1)_A^{(0)} $ &  $q_1^A = x-4 y$ & $q_2^A = x$ & $q_3^A = -x$ & $q_4^A = y$ & $q_A$ \\
\hline
$U(1)_C^{(0)}$ &  $q^C_1 = 3$&  $q^C_2 = 4$ & $q^C_3 = 5$ & $q^C_4 = -6$ & $q_C$ \\
\hline
\hline
$ U(1)_X^{(0)} $ &  $q^X_1 = 1$ & $q^X_2 = 1$ & $q^X_3 = -1$ & $q^X_4 = 0$ & $-$ \\
\hline
$ U(1)_Y^{(0)} $ &  $q^Y_1 = -4$ & $q^Y_2 = 0$ & $q^Y_3 = 0$ & $q^Y_4 = 1$ & $-$ \\
\hline
\end{tabular}
  \caption{Charges in the parent theory~$T_1$ of the Fermat Model. Here~$x,y \in \Z$ are integers.}
  \label{tab:fermatsimp}
\end{table}

As before, we first consider a parent theory~$T_1$ (see figure~\ref{fig:gaugeRG}) with~$U(1)_A^{(0)} \times U(1)_C^{(0)}$ flavor symmetry and four Weyl fermions~$\psi_\alpha^1, \psi_\alpha^2, \psi^3_\alpha,  \psi^4_\alpha$, whose~$U(1)_A^{(0)} \times U(1)_C^{(0)}$ charge assignments are summarized in table~\ref{tab:fermatsimp}. We also include a complex scalar Higgs field~$\phi$ charged under~$U(1)_A^{(0)} \times U(1)_C^{(0)}$. We first consider the~$U(1)_C^{(0)}$ charges; the~$U(1)_A^{(0)}$ symmetry (including its relation to the~$U(1)_{X, Y}^{(0)}$ symmetries in table~\ref{tab:fermatsimp}) is discussed after~\eqref{fermyuk}. 

Since we would like to gauge~$U(1)_C^{(0)}$ to obtain a theory~$T_2$ with 2-group symmetry, we must ensure that the~$\kappa_{C^3}$ anomaly vanishes,
\eqn{
\kappa_{C^3} = \big(q_1^C\big)^3 + \big(q_2^C\big)^3 + \big(q_3^C\big)^3 + \big(q_4^C\big)^3 = 0~.
}[zerogauge]
The QED-like models considered in section~\ref{SubsecQEDexamples} are based on solutions of the form~$q^C_{2,4}  = - q^C_{1,3}$. Here we will focus on chiral solutions that assign different charges to the four Weyl fermions. Perhaps the simplest such solution is the one displayed in table~\ref{tab:fermatsimp}. Since the cubic equation~\zerogauge\ over the integers is an example of a Fermat equation,\foot{~This fact also makes it impossible to find a model with~$\kappa_{C^3} = 0$ based on three Weyl fermions with non-vanishing charges~$q^C_{1,2,3} \in \Z$, since this would require a non-trivial solution of~$(q^C_1)^3 + (q^C_2)^2 + (q^C_3)^3 = 0$ over the integers. Fermat's last theorem states that such solutions do not exist.} we refer to the resulting model as the Fermat model.\footnote{~Variants of the Fermat model can be constructed using other solutions of~\zerogauge, such as~\cite{wiki:xxx},
\eqna{
& q_1^C=1-(a-3b)(a^2+3b^2)~, \qquad q_2^C=(a^2+3b^2)^2-(a+3b)~,\cr
& q_3^C=(a+3b)(a^2+3b^2)-1~, \qquad q_4^C=(a-3b)-(a^2+3b^2)^2~.
}[genfermat]
Here~$a,b \in \Z$. Note that~$\kappa_C = 6 b (a^2+3 b^2-1) \in 6 \Z$, as expected on general grounds (see appendix~\ref{AppQuantAnom}). For certain choices of~$a,b$, the charges in~\genfermat\ have a common integer divisor, which can be scaled out. For instance, $a = b = 1$ gives~$q^C_{1,2,3,4} = (9, 12, 15, -18)$, which can be divided by~$3$ to obtain the charges in table~\ref{tab:fermatsimp}.} Note that there is a mixed 't Hooft anomaly between~$U(1)_C^{(0)}$ and Poincar\'e symmetry~$\mathscr P$,
\eqn{
\kappa_{C {\mathscr P}^2} = q_1^C + q_2^C + q_3^C + q_4^C  = 6~.
}[kgravfermat]
The fact that~$\kappa_{C {\mathscr P}^2} \in 6\Z$ is consistent with the general discussion below~\nonabgravkhat\ and in appendix~\ref{AppQuantAnom}. This implies that the theory~$T_2$ obtained by gauging~$U(1)_C^{(0)}$ has Poincar\'e 2-group symmetry~${\mathscr P} \times_{\hat \kappa_{\mathscr P}} U(1)_B^{(1)}$, with 2-group coefficient given by~\nonabgravkhat,
\eqn{
\hat \kappa_{\mathscr P} = - { \kappa_{C {\mathscr P}^2} \over 6} = -1~.
}[simpfermkhatg]
Note that~$\hat \kappa_{\mathscr P} \in \Z$ is required by global considerations (see section~\ref{SecGlobal}). 

As in the discussion around~\eqref{yukawa1} and~\eqref{yukawa2}, we can deform the model using the Higgs field~$\phi$. If we set~$q_C = 1$, we can add the following Yukawa couplings, 
\eqn{
\SL_\text{Yukawa} = \lambda_1 \, \b \phi^6 \big(\psi^1\big)^2 + \lambda_2 \, \b \phi^8 \big(\psi^2\big)^2   + \lambda_3 \, \b \phi^{10} \big(\psi^3\big)^2  + \lambda_4 \, \phi^{12} \big(\psi^4\big)^2  + (\text{c.c.})~.
}[fermyuk]
If we also add a potential~$V_H(\phi)$ (see~\higsspot), so that~$\phi$ acquires a vev~$\langle \phi \rangle = v$, then~$U(1)_c^{(0)}$ is higgsed and we can give mass to any fermion by dialing the Yukawa couplings~$\lambda_{1,2,3,4}$ in~\eqref{fermyuk}. If all of these couplings are present, the model is gapped. 

We will now analyze the flavor symmetries of the Fermat model. For this purpose, we consider the model without the Yukawa couplings~\fermyuk, and we also temporarily omit the Higgs field~$\phi$. Then any flavor symmetry~$U(1)_A^{(0)}$ only acts on the fermions~$\psi_\alpha^{1,2,3,4}$. In order to avoid an ABJ anomaly after gauging~$U(1)_C^{(0)}$, we must impose
\eqn{
\kappa_{A C^2} = 9 q_1^A + 16 q_2^A + 25 q^A_3 + 36 q^A_4 = 0~.
}[fermnoabj]
Different solutions of this homogenous constraint over the integers correspond to independent flavor symmetries. There are three linearly independent solutions of this kind. One of them is given by~$U(1)_C^{(0)}$ and corresponds to the charge vector~$(3,4,5,-6)$. We denote the other two linearly independent flavor symmetries by~$U(1)_X^{(0)}$ and~$U(1)_Y^{(0)}$, and assign them the charge vectors~$(1,1,-1,0)$ and~$(-4,0,0,1)$ (see table~\ref{tab:fermatsimp}). We use~$U(1)_A^{(0)}$ to denote a general integer linear combination~$x \, U(1)_X^{(0)} + y \, U(1)_Y^{(0)}$, parametrized by~$x, y \in \Z$.\foot{~For simplicity, we do not consider the most general mixing of~$U(1)_A^{(0)}$ with~$U(1)_C^{(0)}$, as we did around~\noabjsimpqed.} Below, we will again include the Higgs field~$\phi$, and assign it~$U(1)_A^{(0)} \times U(1)_C^{(0)}$ charges~$q_A, q_C$. 

Using the fermion charges in table~\ref{tab:fermatsimp}, we compute the remaining anomaly coefficients,
\eqna{
& \kappa_{A^{2} C}= 6 \left(2 x^2 - 4  xy + 7  y^2\right)~,\cr
& \kappa_{A^{3}}= \left(x - 3 y\right) \left(x^2 - 9   xy + 21   y^2\right)~, \cr
& \kappa_{A {\mathscr P}^2} = x- 3 y~.
}[anocoeffs3456]
Note that~$\kappa_{A^2 C}$ never vanishes for any~$x, y \in \Z$, because it is quadratic form with negative discriminant. However, both~$\kappa_{A^3}$ and~$\kappa_{A {\mathscr P}^2}$ vanish when~$x = 3 y$ (and nowhere else over the integers, because the quadratic factor of~$\kappa_{A^3}$ also has negative discriminant). Upon gauging~$U(1)_C^{(0)}$, the~$\kappa_{A^2 C}$ anomaly leads to an abelian 2-group symmetry $U(1)_A^{(0)} \times_{\hat \kappa_A} U(1)_B^{(1)}$, with 2-group structure constant~$\hat \kappa_A = - \half \kappa_{A^2 C}$ (see~\coeffmatch). Together with the Poincar\'e 2-group discussed around~\simpfermkhatg, the symmetry is
\eqn{
\left(U(1)_A^{(0)} \times {\mathscr P}\right) \times_{\hat \kappa_A \, , \,\hat \kappa_{\mathscr P}} U(1)_B^{(1)}~, \qquad \hat \kappa_A = - 3 \left(2x^2 - 4 xy + 7 y^2\right)~, \qquad \hat \kappa_{\mathscr P} = -1~.
}[totaltwgp]

We would now like to discuss the interplay of the~$\kappa_{A^3}$ and~$\kappa_{A {\mathscr P}^2}$ 't Hooft anomalies in~\eqref{anocoeffs3456} with the 2-group symmetry in~\totaltwgp. As was explained around~\kaaanshift\ and~\nonabgravnshift, a properly quantized GS counterterm~\gsctbis (see also section~\secGScts),
\eqn{
S_\text{GS} = {i n \over 2 \pi} \int B^{(2)} \wedge F^{(2)}_A~, \qquad n \in \Z~,
}[gscttris]
leads to the following joint shifts of the anomaly coefficients,
\eqn{
\kappa_{A^3} \; \rightarrow \; \kappa_{A^3} + 6 n \hat \kappa_A~, \qquad \kappa_{A {\mathscr P}^2} \; \rightarrow \; \kappa_{A {\mathscr P}^2}  + 6 n \hat \kappa_{\mathscr P}~, \qquad n \in \Z~.
}[aaaappshift]
The anomalies are only scheme-independent modulo these shifts. Once we chose a particular scheme,\foot{~For instance, since~$\hat \kappa_{\mathscr P} = -1$ (see~\totaltwgp), we can reduce~$\kappa_{A {\mathscr P}^2}~\mod~6$. Then any~$\kappa_{A^3} \in \Z$ is meaningful.} they must match between the UV and the IR. As discussed after~\eqref{yukawa2} (see also section~\ref{ssTQFT} below), the IR degrees of freedom that match the anomaly could either be massless and local, or topological. By contrast, the corresponding effective anomalies~$\kappa_{A^3}^\text{eff.}, \kappa^\text{eff.}_{A{\mathscr P}^2}$, which only detect massless, local degrees of freedom, need not match. This will be apparent below, where we exhibit deformations that gap the theory, but preserve the symmetry~\totaltwgp. 

The contribution of a gapped, topological sector to the anomalies is entirely due to a GS contact term~$\mathsf K_\text{TQFT}$ (see sections~\sssHoftAbTwGp\ and~\ref{ssTQFT}). Here we will be slightly imprecise and think of the GS contact term~$\mathsf K_\text{TQFT}$ as a GS counterterm~\gscttris\ with (possibly non-integer) coefficient. (A more careful treatment that distinguishes these two notions can be found in section~\secGScts.) Comparing with~\aaaappshift, we see that the contributions of a topological sector to~$\kappa_{A^3}$ and~$\kappa_{A {\mathscr P}^2}$ are not independent, 
\eqn{
\kappa^\text{TQFT}_{A^3} = 6 \, {\mathsf K}_\text{TQFT} \, \hat \kappa_A~, \qquad \kappa^\text{TQFT}_{A {\mathscr P}^2 } = 6 \, {\mathsf K}_\text{TQFT} \, \hat \kappa_{\mathscr P}~.
}[tqftanom]
It is therefore not possible to gap the Fermat model, unless the anomaly coefficients~$\kappa_{A^3}$,  $\kappa_{A {\mathscr P}^2}$ in~\eqref{anocoeffs3456} take the form~\tqftanom\ for some choice of~$\mathsf K_\text{TQFT}$. If this is not the case, it is necessary to break some of the symmetry in order to gap the model. For instance, the Yukawa couplings in~\fermyuk\ lead to a gapped theory, but they only preserve the Poincar\'e 2-group~${\mathscr P} \times_{\hat \kappa_{\mathscr P}} U(1)_B^{(1)}$ and explicitly break~$U(1)_A^{(0)}$.

The choices of~$x, y \in \Z$ in table~\ref{tab:fermatsimp} for which the 't Hooft anomalies~$\kappa_{A^3}$ and~$\kappa_{A {\mathscr P}^2}$ in~\eqref{anocoeffs3456} can in principle be saturated by a TQFT are determined by imposing~\tqftanom\ as a necessary condition. Using the explicit expressions in~\eqref{anocoeffs3456} and~\totaltwgp, this leads to the following possibilities, 
\eqn{
x = 0~,~y \in \Z \qquad \text{or} \qquad x = 3 y~,~y \in \Z~, \qquad \text{or} \qquad x = 3 p~,~y=5 p~,~p \in \Z~.
}[gapopt]
In each case, we will exhibit a deformation that gaps the model, but preserves the full 2-group symmetry in~\totaltwgp. The deformation consists of a Higgs potential~$V_H(\phi)$ for~$\phi$, which leads to a vev~$\langle \phi \rangle = v$. As long as the~$U(1)_c^{(0)}$ charge~$q_C$ of~$\phi$ is non-zero, the gauge symmetry is higgsed: both~$\phi$ and the~$U(1)_c^{(0)}$ photon acquire a mass. We also add Yukawa couplings that give mass to all fermions. These couplings turn out to preserve the full symmetry in~\totaltwgp\ if we judiciously dial the~$U(1)_A^{(0)}$ charge~$q_A$ of the Higgs field~$\phi$. As was already discussed around~\eqref{gsfrakw} and~\eqref{yukawa2}, and will be explained more fully in section~\ref{ssTQFT} below, the topological~$\Z_{|q_C|}$ gauge theory that remains in the deep IR after higgsing gives rise to a non-trivial GS contact term,
\eqn{
\mathsf K_\text{TQFT} = -{q_A \over q_C}~.
}[ktqftqaqc]
In every case we will explicitly check that this value indeed matches with the one in~\tqftanom. The~$\Z_{|q_C|}$ gauge theory in the IR has non-trivial dynamical degrees freedom if and only if~$|q_C| \geq 2$. In this case it can give rise to a fractional GS contact term in~\ktqftqaqc. If~$|q_C| = 1$, the TQFT does not contain any non-trivial dynamics and can be formulated using only the background fields. (In other words, it is invertible and the entanglement is short range.) The contact term~\ktqftqaqc\ must therefore be an integer.  

We now consider the three classes of possibilities for~$x, y \in \Z$ listed in~\gapopt:
\begin{itemize}
\item If~$x = 0$ and~$y \in \Z$, then the~$U(1)_A^{(0)}$ charges of the fermions in table~\ref{tab:fermatsimp}, the anomaly coefficients in~\eqref{anocoeffs3456}, and the 2-group structure constant~$\hat \kappa_A$ in~\totaltwgp\ reduce to
\eqn{
(q^A_{1},q^A_{2} ,q^A_{3} ,q^A_{4} ) = y \cdot  (-4, 0,0,1)~, \quad \kappa_{A^3} = -63 y^3~, \quad \kappa_{A {\mathscr P}^2} = - 3y~, \quad \hat \kappa_A = - 21 y^2~.
}[liszero]
We can therefore add the following Yukawa couplings, if we choose~$q_A$, $q_C$ accordingly, 
\eqn{
\SL_\text{Yukawa}^{(x = 0)} \sim \phi^4 \, \psi^1 \psi^3 + \b \phi \, \psi^2 \psi^4 + (\text{c.c.})~, \qquad q_A = y \in \Z~, \qquad q_C = -2~.
}[yuklzero]
From~\ktqftqaqc, we find~$\mathsf K_\text{TQFT} = - {q_A \over q_C} = \half \, y$, which agrees with~\tqftanom\ once we use~\liszero. The deep IR is described by a dynamical~$\Z_2$ gauge theory, which gives rise to a fractional GS contact term~$\mathsf K_\text{TQFT}$ when~$y$ is odd. 

\item If~$x = 3y$, for any~$y \in \Z$, the~$U(1)_A^{(0)}$ charges of the fermions in table~\ref{tab:fermatsimp}, the anomaly coefficients in~\eqref{anocoeffs3456}, and the 2-group structure constant~$\hat \kappa_A$ in~\totaltwgp\ are given by
\eqn{
(q^A_{1},q^A_{2} ,q^A_{3} ,q^A_{4} ) = y \cdot  (-1, 3,-3,1)~, \qquad \kappa_{A^3} =  \kappa_{A {\mathscr P}^2} = 0~, \qquad \hat \kappa_A = - 39 y^2~.
}[listhreem]  
The Yukawa couplings and~$q_A, q_C$ charges can then be chosen as follows,
\eqn{
\SL^{(x = 3y)}_\text{Yukawa} \sim \phi \, \psi^1 \psi^4 + \b \phi^3 \, \psi^2 \psi^3 + (\text{c.c.})~, \qquad q_A = 0~, \qquad q_C = 3~.
}[yuklthreem]
Note that~$\mathsf K_\text{TQFT} = -{q_A \over q_C} = 0$, which is consistent with the fact that the 't Hooft anomalies in~\listhreem\ vanish. The IR is described by a dynamical~$\Z_3$ gauge theory. 

\item If~$x = 3 p$ and~$y = 5p$, with~$p \in \Z$, the~$U(1)_A^{(0)}$ charges of the fermions in table~\ref{tab:fermatsimp}, the anomaly coefficients in~\eqref{anocoeffs3456}, and the 2-group structure constant~$\hat \kappa_A$ in~\totaltwgp\ take the form
\eqn{
(q^A_{1},q^A_{2} ,q^A_{3} ,q^A_{4} ) = p \cdot  (-17, 3,-3,5)~, \;\; \kappa_{A^3} =  -4788 p^3~, \;\; \kappa_{A {\mathscr P}^2} = -12 p~, \;\; \hat \kappa_A = - 399 p^2~.
}[lmpex] 
This leads to the following choice of Yukawa couplings and~$q_A, q_C$ charge assignements,
\eqn{
\SL^{(x = 3p \, , \, y = 5p)}_\text{Yukawa} \sim \phi^7 \, \psi_1 \psi_2 + \b \phi \,  \psi_3 \psi_4 + (\text{c.c.})~, \qquad q_A = 2 p~, \qquad q_C = -1~.
}[yuk35p]
In this example~$\mathsf K_\text{TQFT} = - {q_A \over q_C} = 2 p$, which agrees with~\tqftanom\ if we substitute the values in~\lmpex. Since~$|q_C| = 1$, the IR is described by an invertible TQFT for the background fields, and the GS contact term~$\mathsf K_\text{TQFT}$ is an integer. 

\end{itemize}

We would like to conclude our discussion of the Fermat model by showing that, in the absence of the Higgs field~$\phi$ or any Yukawa couplings, it has an even larger 2-group symmetry than~\totaltwgp. This will also illustrate the higher-rank abelian 2-groups described around~\eqref{multabbshift} and in section~\ref{ssecBCgen}. The Fermat model has an abelian flavor symmetry~$U(1)_X^{(0)} \times U(1)_Y^{(0)}$ of rank two (see table~\ref{tab:fermatsimp}). So far we have focused on its subgroup~$U(1)_A^{(0)} = x \, U(1)_X^{(0)} + y \, U(1)_Y^{(0)}$, but now we consider the interplay of~$U(1)_X^{(0)}$ and~$U(1)_Y^{(0)}$. The 't Hooft anomalies of the form~$\kappa_{I J C} = \sum_{i =1}^4 q_i^I q_i^J q^C_i$, with~$I, J \in \{X, Y\}$, are given by
\eqn{
\kappa_{X X C} = 12~, \qquad \kappa_{XYC} = -12~, \qquad \kappa_{Y Y C} = 42~.
}[kappaxyc]
Gauging~$U(1)_C^{(0)}$ therefore gives rise to a theory~$T_2$ whose global symmetry is a 2-group that involves a rank-2 abelian flavor symmetry, as well as Poincar\'e symmetry, 
\eqn{
\left(U(1)_X^{(0)} \times U(1)_Y^{(0)} \times {\mathscr P} \right) \times_{\hat \kappa_{IJ}\, , \, \hat \kappa_{\mathscr P} } U(1)_B^{(1)}~.
}[mattwogp]
The Poincar\'e 2-group coefficient~$\hat \kappa_{\mathscr P}$ is as in~\totaltwgp, but the abelian 2-group coefficients~$\hat \kappa_{IJ}$ now give rise to a non-diagonal, symmetric matrix, 
\eqn{
\left(\,\hat \kappa_{IJ} \right) = - \half \, \bigg(\,\begin{matrix}
 \kappa_{XXC} & \kappa_{XYC} \\[-14pt] \kappa_{XYC} & \kappa_{YYC}\\[-20pt]
 \phantom{.} & 
\end{matrix}\,\bigg) = \bigg(\,\begin{matrix}
-6 & 6  \\[-14pt] 6 & -21 \\[-23pt]
 \phantom{.} & 
\end{matrix}\,\bigg)~.
}[khatmat]

\subsec{$\C \P^N$ Models}[ssCPN]

Every~$U(1)_c^{(0)}$ gauge theory~$T_2$ without magnetic charges has a~$U(1)_B^{(1)}$ symmetry with current~$J_B^{(2)} \sim * f^{(2)}_c$. All such models can be deformed to a~$\C\P^1$ sigma model (coupled to other degrees of freedom), by adding two complex scalar fields~$\phi_{1,2}$ of~$U(1)_c^{(0)}$ charge~$+1$, and a Higgs potential~$V_H\left(\rho\right)$ that induces a vev~$\langle \rho\rangle = v$, with~$\rho^2 = |\phi_1|^2 + |\phi_2|^2$. In the parent theory~$T_1$, where~$U(1)_C^{(0)}$ is a global symmetry, the Higgs sector has an~$SO(4)^{(0)}$ flavor symmetry, which is spontaneously broken to~$SO(3)^{(0)}$ by~$\langle \rho \rangle  = v$. This leads to a sigma model with target space~$S^3$. The~$U(1)_C^{(0)} \subset SO(4)^{(0)}$  flavor symmetry rotates the Hopf fiber of the~$S^3$. After we gauge it, the NG boson that parametrizes the Hopf fiber is eaten by the~$U(1)_c^{(0)}$ photon, leaving only the~$\C\P^1$ base of the Hopf fibration in the IR. At low energies, the magnetic two-form current~$J_B^{(2)} \sim * f_c^{(0)}$ of the~$U(1)_c^{(0)}$ gauge theory flows to~$*\Omega^{(2)}$, where~$\Omega^{(2)}$ is the pullback to spacetime of the K\"ahler 2-form of the~$\C\P^1$ model. In particular, $\Omega^{(2)}$ is closed. Therefore the presence of a continuous~1-form global symmetry does not require abelian gauge fields (see also~\cite{Seiberg:2010qd, Banks:2010zn, Gaiotto:2014kfa}).

The deformation described above can applied to every~$U(1)_c^{(0)}$ gauge theory with 2-group global symmetry analyzed in sections~\ref{SubsecQEDexamples} and~\ref{SubsecFermats} above. This produces examples of~$\C\P^1$ models coupled to fermions, with (non-) abelian or Poincar\'e 2-group symmetries.

A variant of the preceding discussion leads to~$\C\P^N$ models with 2-group symmetry: we can deform conventional QED with~$N_f$ massless flavors, which was shown to possess nonabelian 2-group symmetry~\nonabtwogpqedex\ in section~\ref{sssQEDnonab}, by adding a Higgs field~$\Phi_i$ that is charged under both~$U(1)_c^{(0)}$ and~$SU(N_f)_L^{(0)}$. If~$\Phi_i$ acquires a vev, the gauge symmetry is higgsed, and~$SU(N_f)^{(0)}_L$ is spontaneously broken to~$S\left(U(1)_L^{(0)} \times U(N_f-1)^{(0)}_L\right)$, leading to a~$\C\P^{N_f-1}$ target space for the associated NG bosons. The resulting model (which also contains massless fermions) has the same 2-group symmetry as the original QED theory. As before, the magnetic~$U(1)_B^{(1)}$ current~$J_B^{(2)} \sim * f_c^{(0)}$ of QED  flows to~$*\Omega^{(2)}$ at low energies, with~$\Omega^{(2)}$ now the pullback to spacetime of the~$\C\P^{N_f-1}$ K\"ahler 2-form.

\subsec{Theories with Topological Sectors}[ssTQFT]

In previous subsections, we have discussed a variety of 2-group symmetric RG flows. In the IR, these flows either ended in gapped, topological theories, or in theories that contain a topological sector along with massless, local degrees of freedom. The TQFTs that appear in this context must also have 2-group symmetry. Moreover, as explained in section~\sssHoftAbTwGp, they can contribute to reducible 't Hooft anomalies via non-trivial GS contact terms. Here we will analyze these theories by thinking of them as arising from a parent theory~$T_1$ with~$U(1)_A^{(0)} \times U(1)_C^{(0)}$ global symmetry. 

Assume that~$T_1$ contains a complex scalar field~$\phi$ with~$U(1)_A^{(0)} \times U(1)_C^{(0)}$ charges~$q_A$ and~$q_C$. If~$\phi$ acquires a vev~$\langle \phi\rangle = v$, e.g.~via a Higgs potential~$V_H(\phi)$ (see~\higsspot), its radial mode typically becomes massive, but its phase~$\chi \sim \chi + 2 \pi$ remains light. At low energies, we can approximate~$\phi \approx v e^{i\chi}$, where~$\chi$ shifts as follows under~$U(1)_A^{(0)} \times U(1)_C^{(0)}$ transformations,
\eqn{
\chi \; \rightarrow \; \chi + q_A \lambda_A^{(0)} + q_C \lambda_C^{(0)}~.
}[chiacshift]
Thus~$\chi$ is the NG boson associated with the spontaneous breaking of~$U(1)_A^{(0)} \times U(1)_C^{(0)}$ to the~$U(1)_X^{(0)}$ subgroup under which~$\chi$ is invariant. 

At very low energies, $\chi$ is a nearly free scalar field described by the quadratic action
\eqn{
S^\chi_\text{kin.}[A^{(1)}, C^{(1)}, \chi] = v^2  \int \, \left(d\chi - q_A \, A^{(1)} - q_C\,  C^{(1)}\right) \wedge *\left(d\chi - q_A \, A^{(1)} - q_C\,  C^{(1)}\right)~.
}[chikinS]
There could also be another IR sector, which decouples from~$\chi$ at low energies (i.e.~the two sectors only interact through irrelevant operators), but is also charged under~$U(1)_A^{(0)} \times U(1)_C^{(0)}$,
\eqn{
S^\text{IR}[A^{(1)}, C^{(1)}] \; \supset \; \int \left(A^{(1)} \wedge * (j_{A}^\text{IR})^{(1)} + C^{(1)} \wedge * (j_{C}^\text{IR})^{(1)}\right)~.
}[starcurrs]
We will not need a detailed description of this sector, except the fact that it typically gives rise to all possible 't Hooft anomalies~$\kappa^\text{IR}_{C^3}$, $\kappa^\text{IR}_{AC^2}$, $\kappa^\text{IR}_{A^2 C}$, and~$\kappa_{A^3}^\text{IR}$. We choose the counterterms as in~\ctfix, so that the anomalies contributed by the IR sector under $U(1)_A^{(0)} \times U(1)_C^{(0)}$ background gauge transformations take the following form, 
\eqna{
& \CA^\text{IR}_{A} = {i \over 4 \pi^2} \int \lambda^{(0)}_A \, \bigg({\kappa^\text{IR}_{A^3} \over 3!} \, F^{(2)}_A \wedge F^{(2)}_A + {\kappa^\text{IR}_{A^2 C} \over 2!} \, F^{(2)}_A \wedge F^{(2)}_C + {\kappa^\text{IR}_{AC^2} \over 2!} \, F^{(2)}_C \wedge F^{(2)}_C \bigg)~,\cr
& \CA^\text{IR}_{C} = {i \kappa^\text{IR}_{C^3} \over 24 \pi^2} \int \lambda_C^{(0)} \, F_C^{(2)} \wedge F_C^{(2)}~.
}[staranom]
In general, the NG boson~$\chi$ also contributes to the anomalies, via suitable couplings to background fields, 
\eqna{
S^\chi_\text{ano.}  [A^{(1)},\,& C^{(1)}, \chi] = - {i \alpha \over 4 \pi^2} \int \left(d\chi- q_C \, C^{(1)}\right) \wedge A^{(1)} \wedge F^{(2)}_A \cr
& - {i \beta \over 4 \pi^2} \int \left(d\chi- q_C \, C^{(1)}\right) \wedge A^{(1)} \wedge F_C^{(2)} -{i \gamma \over 4 \pi^2} \int d\chi \wedge C^{(1)} \wedge F_C^{(2)}~.
}[chianomS]
These couplings give rise to the following~$U(1)_A^{(0)} \times U(1)_C^{(0)}$ anomalies,
\eqna{
& \CA^\chi_A = {i \over 4 \pi^2} \int \lambda_A^{(0)} \left( \alpha \, q_A \, F_A^{(2)} \wedge F_A^{(2)} + \left(\alpha q_C + \beta q_A\right) \, F_A^{(2)} \wedge F_C^{(2)} + \left(\beta q_C + \gamma q_A \right) \, F_C^{(2)} \wedge F_C^{(2)}\right)~,\cr
& \CA_C^\chi = {i \gamma q_C \over 4 \pi^2} \int \lambda_C^{(0)} \, F_C^{(2)} \wedge F^{(2)}_C~.
}[chianomA]
Adding the contributions in~\staranom\ and~\chianomS, the total 't Hooft anomalies are given by\foot{~Due to anomaly matching, these are both the UV and the IR values of the total 't Hooft anomalies.}
\eqna{
& \kappa_{A^3} = \kappa_{A^3}^\text{IR} + 6 \alpha q_A~, \cr
& \kappa_{A^2 C} = \kappa_{A^2 C}^\text{IR} + 2 \left(\alpha q_C + \beta q_A\right)~,\cr
& \kappa_{AC^2} = \kappa_{AC^2}^\text{IR} + 2 \left(\gamma q_A + \beta q_C\right)~,\cr
& \kappa_{C^3} = \kappa_{C^3}^\text{IR} + 6 \gamma q_C~.
}[uvirstaranom]
As on previous occasions, would like to eventually gauge~$U(1)_C^{(0)}$ without ruining~$U(1)_A^{(0)}$, and hence we assume that~$\kappa_{C^3} = \kappa_{AC^2} = 0$. Using~\uvirstaranom, we can then solve for~$\beta$ and~$\gamma$,
\eqn{
\beta = -{\kappa^\text{IR}_{AC^2} \over 2 q_C} + {\kappa^\text{IR}_{C^3} q_A \over 6 q_C^2}~, \qquad \gamma = -{\kappa^\text{IR}_{C^3} \over 6 q_C}~.
}[bgsolve]

It is instructive to dualize the NG boson~$\chi$ to a dynamical 2-form gauge field~$b^{(2)}$. As usual, the dual description can be derived by replacing~$d \chi$ in~\chikinS\ and~\chianomS\ with an unconstrained 1-form field~$u^{(1)}$. The fact that~$u^{(1)}$ should be closed on shell, with appropriately quantized periods~$\int_{\Sigma_1} u^{(1)} \in 2 \pi \Z$ around closed 1-cycles~$\Sigma_1$, is enforced by a dynamical~$U(1)_b^{(1)}$ gauge field~$b^{(2)}$, which acts as a Lagrange multiplier. We therefore consider the following action for the dynamical fields~$u^{(1)}$, $b^{(2)}$, as well as background fields, 
\eqna{
& \t S[A^{(1)}, C^{(1)}, u^{(1)}, b^{(2)}] = v^2 \int \left(u^{(1)} - q_A \, A^{(1)} - q_C \, C^{(1)}\right) \wedge *\left(u^{(1)} - q_A \, A^{(1)} - q_C \, C^{(1)}\right) \cr
& - {i \alpha \over 4 \pi^2} \int \left(u^{(1)} - q_C \, C^{(1)} \right) \wedge A^{(1)} \wedge F_A^{(2)} - {i \beta \over 4\pi^2}  \int \left(u^{(1)} - q_C \, C^{(1)} \right) \wedge A^{(1)} \wedge F_C^{(2)}  \cr
&  - {i \gamma \over 4 \pi^2} \int u^{(1)} \wedge C^{(1)} \wedge F_C^{(2)} + {i \over 2 \pi} \int b^{(2)} \wedge d u^{(1)}~.
}[dualitywb]
This action is invariant under~$U(1)_A^{(0)} \times U(1)_C^{(0)}$ background gauge transformations, up to the 't Hooft anomalies in~\chianomA, if we assign the following transformation rules to the dynamical fields,
\eqna{
& u^{(1)} \quad \longrightarrow \quad u^{(1)} + q_A \, d \lambda_A^{(0)} + q_C \, d \lambda_C^{(0)}~,\cr
& b^{(2)} \, \quad \longrightarrow  \quad b^{(2)} +  {\alpha \over 2 \pi} \, \lambda_A^{(0)} \, F_A^{(2)} + {\beta \over 2 \pi} \, \lambda_A^{(0)} \, F_C^{(2)} + {\gamma \over 2 \pi}  \, \lambda_C^{(0)} \, F_C^{(2)} ~.
}[ubtrans]
In addition, $b^{(2)} \rightarrow b^{(2)} + d \Lambda_b^{(1)}$ under~$U(1)_b^{(1)}$ gauge transformations. The transformation rule for~$u^{(1)}$ is expected, since it coincides with that of~$d \chi$ (see~\chiacshift). The GS shifts of~$b^{(2)}$ are required to cancel terms proportional to~$d u^{(1)}$, which does not vanish off shell. If we integrate out~$b^{(2)}$ in~\dualitywb, we can set~$u^{(1)} = d \chi$, with~$\chi \sim  \chi + 2 \pi$, which shows that~\dualitywb\ is physically equivalent to the sum of the NG actions~\chikinS\  and~\chianomS.

The dual description is obtained by instead integrating out the unconstrained 1-form~$u^{(1)}$ in~\dualitywb. It is convenient to define a modified 3-form field strength~$h^{(3)}$ for~$b^{(2)}$ that is invariant under the transformations in~\ubtrans (see~\Htwoabelian\ and section~\ref{SecGlobal}), 
\eqn{
h^{(3)} = db^{(2)} - {\alpha \over 2 \pi} \, A^{(1)} \wedge F_A^{(2)} - {\beta \over 2 \pi} \, A^{(1)} \wedge F_C^{(2)} - {\gamma \over 2\pi} \, C^{(1)} \wedge F_C^{(2)}~.
}[hdef]
The equation of motion for~$u^{(1)}$ that follow from the action~\dualitywb\ can then be written as
\eqn{
u^{(1)} - q_A  \,A^{(1)} - q_C \,C^{(1)} = -{i \over 4 \pi v^2} * h^{(3)}~.
}[weom]
Substituting back into~\dualitywb, we find
\eqna{
\t S[A^{(1)}, C^{(1)}, b^{(2)}] =~& {1 \over 16 \pi^2 v^2} \int * h^{(3)} \wedge h^{(3)} + {i \over 2 \pi} \int b^{(2)} \wedge \left(q_A \, F_A^{(2)} + q_C \, F_C^{(2)}\right) \cr
& -{i \gamma q_A \over 4 \pi^2} \int A^{(1)} \wedge C^{(1)} \wedge F_C^{(2)}~.
}[dualactac]
This is a dual description of the NG boson actions~\chikinS\ and~\chianomS\ in terms of a~$U(1)_b^{(1)}$ gauge field~$b^{(2)}$. The counterterm on the second line of~\dualactac, which is automatically supplied by the duality, ensures that the 't Hooft anomalies take the form~\chianomA. 

As in previous subsections, we can obtain a new, 2-group symmetric theory~$T_2$ by gauging~$U(1)_C^{(0)}$ in~$T_1$. We also include a suitable Maxwell kinetic term, with gauge coupling~$e$, for the~$U(1)_c^{(0)}$ field strength~$f_c^{(2)}$, as well as a coupling of the~$U(1)_B^{(1)}$ background field~$B^{(2)}$ to the magnetic current~$J_B^{(2)}\sim * f^{(2)}_c$. The~$U(1)_c^{(0)}$ photon acquires a mass~$m_\gamma \sim e v$ and can be integrated out. In the deep IR, we can then drop the Maxwell kinetic term, as well as the kinetic term~$\sim * h^{(3)} \wedge h^{(3)}$ in~\dualactac. The resulting low-energy theory takes the form of a BF theory, coupled to the IR sector described around~\starcurrs,
\eqna{
S_\text{BF}[A^{(1)}, B^{(2)}, b^{(2)}, c^{(1)}] = ~&{i \over 2 \pi} \int b^{(2)} \wedge \left(q_A \, F_A^{(2)} + q_C \, f_c^{(2)}\right) + {i \over 2 \pi} \int B^{(2)} \wedge f^{(2)}_c \cr
& - {i \gamma q_A \over 4 \pi^2} \int A^{(1)} \wedge c^{(1)} \wedge f_c^{(2)} + S^\text{IR}[A^{(1)}, c^{(1)}]~.
}[IRbf]
The terms in the first line of~\IRbf\ coincide with the BF description of~$\Z_{p}$ gauge theory coupled to 1- and 2-form background gauge fields in~\dymbf\ (see also~\cite{Maldacena:2001ss,Banks:2010zn,Kapustin:2014gua}), if we identify $p = q_C$ and~$q = q_A$. The transformations of~$A^{(1)}$ and~$c^{(1)}$ under background~$U(1)_A^{(0)}$ and dynamical~$U(1)_c^{(0)}$ gauge transformations are standard, while~$b^{(2)}$ undergoes the GS shifts in~\ubtrans, which now involve the dynamical field strength~$f_c^{(2)}$ and the background field strength~$F_A^{(2)}$. As usual, the~$\kappa_{A^2C}$ 't Hooft anomaly in~\uvirstaranom\ implies that the theory~\IRbf\ has~$U(1)_A^{(0)} \times_{\hat \kappa_A} U(1)_B^{(1)}$ 2-group symmetry, with~$\hat \kappa_A = - \half \kappa_{A^2 C}$. This leads to the 2-group shift~$B^{(2)} \rightarrow B^{(2)} + {\hat \kappa_A \over 2 \pi} \, \lambda_A^{(0)} \, F_A^{(0)}$ under~$U(1)_A^{(0)}$ background gauge transformations. 

We will now examine how the BF theory in~\IRbf\ realizes the~$\kappa_{A^3}$ 't Hooft anomaly of the parent theory T1. As on previous occasions, a GS contact term will play a crucial role. Integrating out~$b^{(2)}$ in~\IRbf\ enforces the constraint~$q_A \, F_A^{(2)} + q_C f_c^{(2)} = 0$, which encodes the unbroken subgroup~$U(1)_X^{(0)} \subset U(1)_A^{(0)} \times U(1)_c^{(0)}$ (see the discussion after~\chiacshift). Being imprecise about global issues (see section~\ref{secGScts} for details), we can therefore set~$c^{(1)} = -{q_A \over q_C} \, A^{(1)}$ and substitute back into~\IRbf,
\eqn{
S_\text{BF}[A^{(1)}, B^{(2)}] = {i \mathsf K_\text{TQFT} \over 2 \pi} \int B^{(2)} \wedge F_A^{(2)} + S^\text{IR}[A^{(1)}, -{q_A \over q_C} A^{(1)}]~, \qquad \mathsf K_\text{TQFT} = - {q_A \over q_C}~.
}[deepireff]
The~$\Z_{|q_C|}$ gauge theory therefore gives rise to a GS contact term~$-{q_A \over q_C}$, exactly as in~\gsfrakw (see also the discussions below~\bBshiftint\ and~\eqref{yukawa2}, as well as around~\ktqftqaqc). Using~\deepireff, it is straightforward to determine the anomaly under~$U(1)_A^{(0)}$ background gauge transformations. The GS contact term contributes an anomaly~$\CA_A^\text{GS}$ via the 2-group shift of~$B^{(2)}$,
\eqn{
\CA^\text{GS}_A = {i \mathsf K_\text{TQFT} \hat \kappa_A \over 4 \pi^2} \int \lambda_A^{(0)} \, F_A^{(2)} \wedge F_A^{(2)} = {i \kappa_{A^2 C} q_A \over 8 \pi^2 q_C} \int \lambda_A^{(0)} \, F_A^{(2)} \wedge F_A^{(2)}~.
}[gstqftanom]
In a theory that is gapped, without an additional IR sector, \gstqftanom\ is the only contribution of the low-energy theory to the 't Hooft anomaly. This observation played an important role in the discussions below~\eqref{yukawa2} and around~\ktqftqaqc.

If present, the IR sector contributes an anomaly~$\CA^\text{IR}_{A, \text{tot.}}$ which is the sum of two contributions: the anomaly~$\CA_A^\text{IR}$ in~\staranom, and the anomaly obtained from~$\CA_C^\text{IR}$ in~\staranom\ by substituting~$C^{(1)} = -{q_A \over q_C} \, A^{(1)}$ and~$\lambda_C^{(0)} = - {q_A \over q_C} \, \lambda_A^{(0)}$. The resulting total anomaly is given by
\eqna{
\CA^\text{IR}_{A, \text{tot.}} ~&= {i \over 8 \pi^2} \left({\kappa_{A^3}^\text{IR} \over 3} - {\kappa^\text{IR}_{A^2 C} q_A \over q_C} + {\kappa_{AC^2}^\text{IR} q_A^2 \over  q_C^2} - {\kappa^\text{IR}_{C^3} q_A^3 \over 3 q_C^3}\right) \int \lambda_A^{(0)} \, F_A^{(2)} \wedge F_A^{(2)} \cr
& = {i \over 8 \pi^2} \left({\kappa_{A^3} \over 3 } - {\kappa_{A^2 C} q_A \over q_C}\right) \int \lambda_A^{(0)} \, F_A^{(2)} \wedge F_A^{(2)}~.
}[totaliranom]
Here the second line was obtained from the first one using the formulas in~\uvirstaranom\ and~\bgsolve. If we add the anomalies~$\CA_A^\text{GS}$ and~$\CA^\text{IR}_{A, \text{tot.}}$ in~\gstqftanom\ and~\totaliranom, we find that the terms proportional to~$\kappa_{A^2 C}$ cancel, so that only the expected~$\kappa_{A^3}$ anomaly remains.

\subsec{The Goldstone-Maxwell Model of Spontaneous 2-Group Breaking}[ssGM]

In the previous subsection we considered parent theories~$T_1$ containing a NG boson with~$U(1)_A^{(0)} \times U(1)_C^{(0)}$ charges~$q_{A}$ and~$q_C$. We considered this theory, and its descendant~$T_2$ obtained by gauging~$U(1)_C^{(0)}$, under the assumption that~$q_C \neq 0$, which lead to topological~$\Z_{|q_C|}$ gauge theories with 2-group symmetry. Here we will analyze the case~$q_C = 0$, which is qualitatively very different. In the absence of additional charged degrees of freedom, the resulting model consists of a free~$U(1)_A^{(0)}$ NG boson~$\chi$ and a free~$U(1)_c^{(0)}$ Maxwell field, which plays the role of NG boson for the spontaneously broken~$U(1)_B^{(1)}$ symmetry. As discussed in point 1c.) in section~\ssecExOverview, this Goldstone-Maxwell (GM) model describes the low-energy dynamics of any theory in which the entire 2-group symmetry~$U(1)_A^{(0)} \times_{\hat \kappa_A} U(1)_B^{(1)}$ is spontaneously broken. The GM model is among the simplest examples of theories with 2-group symmetry. Nevertheless, it displays a number of subtle features. For instance, the~$U(1)_A^{(0)} \times_{\hat \kappa_A} U(1)_B^{(1)}$ 2-group symmetry of the model is embedded in an even larger 3-group symmetry, with an intricate anomaly structure. 

As before, we first discus the parent theory~$T_1$ with~$U(1)_A^{(0)} \times U(1)_C^{(0)}$ flavor symmetry and a mixed~$\kappa_{A^2 C}$ 't Hooft anomaly. We will assume that this is the only nonzero 't Hooft anomaly, so that~$\kappa_{C^3} = \kappa_{AC^2} =  \kappa_{A^3} = 0$. The vanishing of~$\kappa_{C^3}, \kappa_{AC^2}$ is needed in order to gauge~$U(1)_C^{(0)}$ without spoiling~$U(1)_A^{(0)}$. By contrast, the assumption that~$\kappa_{A^3} = 0$ is only for simplicity, and can be relaxed. Similarly, we will take the~$U(1)_A^{(0)}$ charge of~$\chi$ to be~$q_A = 1$, but it is straightforward to restore general~$q_A$. 

Under~$U(1)_A^{(0)} \times U(1)_C^{(0)}$ background gauge transformations, parametrized by~$\lambda_A^{(0)}$ and~$\lambda_C^{(0)}$, the NG boson~$\chi$ and the background gauge fields~$A^{(1)}$, $C^{(1)}$ shift as follows,
\eqn{
\chi \; \rightarrow \; \chi + \lambda_A^{(0)}~, \qquad A^{(1)} \; \rightarrow \; A^{(1)} + d \lambda_A^{(0)}~, \qquad C^{(1)} \; \rightarrow \; C^{(1)} + d \lambda_C^{(0)}~.
}[chiaone]
The quadratic action for~$\chi$ takes the same form as in~\chikinS\ and~\chianomS, except that we set~$q_A=1$, $q_C = \alpha = \gamma = 0$, ~$\beta = \half \kappa_{A^2C}$, and we integrate by parts in the anomalous term,
\eqn{
S_\chi[A^{(1)}, C^{(1)}, \chi] =  v^2 \int \left(d \chi - A^{(1)} \right) \wedge * \left(d \chi - A^{(1)}\right) + {i \kappa_{A^2 C} \over 8 \pi^2} \int \, \chi \, F_A^{(2)} \wedge F_C^{(2)}~.
}[chisimpac]
We now gauge~$U(1)_C^{(0)}$ and add a  Maxwell kinetic term (with gauge coupling~$e$) for the field strength~$f_c^{(2)}$, to obtain a theory~$T_2$ with 2-group symmetry --   the GM model,
\eqna{
S_\text{GM}[A^{(1)}, B^{(1)},\,& \chi, c^{(1)}] = v^2 \int \left(d \chi - A^{(1)} \right) \wedge * \left(d \chi - A^{(1)}\right) - {i \hat \kappa_A \over 4 \pi^2} \int \, \chi \, F_A^{(0)} \wedge f_c^{(2)} \cr
& + {1 \over 2 e^2} \int f^{(2)}_c \wedge * f^{(2)}_c + {i \over 2\pi} \int B^{(2)} \wedge f^{(2)}_c~, \qquad \hat \kappa_A = - \half \kappa_{A^2 C}~.
}[gmdef]
The GM model has~$U(1)_A^{(0)} \times_{\hat\kappa_A} U(1)_B^{(1)}$ abelian 2-group symmetry, with 2-group structure constant~$\hat \kappa_A = - \half \kappa_{A^2 C}$, and is therefore invariant under 2-group background gauge transformations (see for instance~\Btwogp\ or~\eqref{coeffmatch}), 
\eqn{
S_\text{GM}[A^{(1)} + d \lambda_A^{(0)}, B^{(2)} + d \Lambda^{(1)}_B + {\hat \kappa_A \over 2 \pi} \, \lambda_A^{(0)} \, F_A^{(2)}, \chi + \lambda_A^{(0)}, c^{(1)}] = S_\text{GM}[A^{(1)}, B^{(2)}, \chi, c^{(1)}]~.
}[twogpgm]
Note that there is no c-number 't Hooft anomaly under these transformations, because we assumed that~$\kappa_{A^3} = 0$. If the background fields~$A^{(1)}$ and~$B^{(2)}$ are set to zero, then~\gmdef\ reduces to a theory of two decoupled free fields: a NG boson~$\chi$, and a Maxwell field~$f^{(2)}_c$. However, in the presence of the background field~$A^{(1)}$, the dynamical fields~$\chi$ and~$f^{(2)}$ couple to each other. This coupling is responsible for the 2-group symmetry of the model. 

To see this explicitly, we examine the currents~$j_A^{(1)}$ and~$J_B^{(2)}$ in the GM model. We would like to verify that they satisfy the non-conservation equation~\modconseq, which we repeat here,
\eqn{
d * j_A  = {\hat \kappa_A \over 2 \pi} \, F_A^{(2)} \wedge * J_B^{(2)}~, 
}[twgpnconsbis]
and that their characteristic three-point function satisfies the 2-group Ward identity~\twgpwiint, \eqn{
{\d \over \d x_\mu} \langle j_\mu^A(x) j_\nu^A(y) J^B_{\rho\sigma}(z) \rangle = {\hat \kappa_A \over 2 \pi} \, \d^\lambda \delta^{(4)}(x-y) \langle J^B_{\nu\lambda}(y) J^B_{\rho\sigma}(z)\rangle~.
}[twgpwibis]
If we vary the action~\gmdef\ with respect to~$A^{(1)}$ and~$B^{(2)}$, we find that
\eqn{
j_A^{(1)} = - 2 v^2 \left(d \chi - A^{(1)}\right) -{i \hat \kappa_{A} \over 4 \pi^2} *\left(d \chi \wedge f_c^{(2)}\right)~, \qquad  J_B^{(2)} = {i \over 2 \pi} \, * f^{(2)}_c~.
}[japifmix]
The current~$j_A^{(1)}$ has several unusual features:
\begin{itemize}
\item Even though the theory is conformal in the absence of background fields, the current~$j_A^{(1)}$ is not a conformal primary operator, because the special conformal generators~$K_\mu$ annihilate~$f^{(2)}_c$, but not~$d \chi$.

\item Without background fields, the theory has a charge conjugation symmetry~$\mathsf C$, under which~$f^{(2)}_c$ is odd. The term in~$j_A^{(1)}$ that is bilinear in~$\chi$ and~$f^{(2)}_c$ makes it impossible to choose a~$\mathsf C$-transformation for~$\chi$ that renders the current~$\mathsf C$-even or -odd. It therefore violates the assumptions used to derive the vanishing of the characteristic three-point function in~\jjfpmthpt. 

\item The term in~$j_A^{(1)}$ that is bilinear in~$\chi$ and~$f^{(2)}_c$, which arises from the second term in the first line of~\gmdef, is automatically conserved: it is a pure improvement term. As we will explicitly see below, this term is responsible for the 2-group symmetry of the GM model, along the lines described around~\eqref{factwardid}. 
\end{itemize}
In order to compute the divergence of~$j_A^{(1)}$, we need the~$\chi$ equation of motion from~\gmdef,
\eqn{
2 v^2 \, d * \left(d\chi-A^{(1)}\right) = -{i \, \hat \kappa_A \over 4 \pi^2} \, F_A^{(2)} \wedge f^{(2)}_c = - { \hat \kappa_A \over 2 \pi} \, F_A^{(2)} \wedge * J^{(2)}_B~.
}[chieom]
Together with~\japifmix, this implies the 2-group non-conservation equation for~$j_A^{(1)}$ in~\twgpnconsbis. 

We would now like to explicitly verify the 2-group Ward identity~\twgpwibis. Using~\japifmix\ and restoring Lorentz indices, the~$U(1)_A^{(0)}$ current in the absence of background fields can be written as follows 
\eqn{
j_\mu^A = - 2 v^2 \, \d_\mu \chi - {\hat \kappa_A \over 2\pi} \, J^B_{\mu\alpha} \, \d^\alpha \chi~.
}[gmjalordef]
Applying Wick's theorem, the characteristic three-point function then takes the form
\eqna{
\langle j_\mu^A(x) j_\nu^A(y) J^B_{\rho\sigma}(z)\rangle =~& - {\hat \kappa_A \, v^2 \over  \pi} \,  \d_\nu \d^\alpha \langle \chi(x) \chi(y) \rangle \, \langle J^B_{\mu\alpha} (x) J^B_{\rho\sigma}(z)\rangle \cr
& - {\hat \kappa_A \, v^2 \over  \pi} \, \d_\mu \d^\beta \langle \chi(x) \chi(y)\rangle \, \langle J^B_{\nu\beta}(y) J^B_{\rho\sigma}(z)\rangle~.
}[gmchartpfn]
Since the~$\chi$ propagator satisfies~$\d^2\langle \chi(x) \chi(y)\rangle = - {1 \over 2 v^2} \delta^{(4)}(x-y)$ (see~\gmdef), and~$J_B^{(2)}$ is conserved, we conclude that~\gmchartpfn\ satisfies the Ward identity~\twgpwibis. 

As was already pointed out below~\twogpgm,  the GM model is free of 't Hooft anomalies under 2-group background gauge transformations of~$A^{(1)}$ and~$B^{(2)}$. However, as in free Maxwell theory (see appendix~\ref{appHooftFreeMax}), we can introduce another gauge field~$B^{(2)}_\text{e}$ that couples to the electric 2-form current. In order to make the notation more uniform, we will denote the magnetic 2-form gauge field~$B^{(2)}$, which has already appeared throughout this paper, by~$B^{(2)}_\text{m}$ for the remainder of this subsection. We will also denote the electric and magnetic 1-form symmetries of the theory by~$U(1)_\text{e}^{(1)}$ and~$U(1)_\text{m}^{(1)}$. Following the discussion for free Maxwell theory in appendix~\ref{appHooftFreeMax}, it is straightforward to introduce~$B^{(2)}_\text{e}$ into the GM action~\gmdef,\foot{~Compared to~\gmdef, we now write~$B^{(2)} = B^{(2)}_\text{m}$.}
\eqna{
S_\text{GM}&[A^{(1)}, B^{(2)}_\text{e}, B^{(2)}_\text{m}, \chi, c^{(1)}] = ~ v^2 \int \left(d \chi - A^{(1)} \right) \wedge * \left(d \chi - A^{(1)}\right) \cr
& + {i \over 2\pi} \int \left(B^{(2)}_\text{m} - {\hat \kappa_A \over 2 \pi} \, \chi \, F_A^{(2)}\right) \wedge f^{(2)}_c  + {1 \over 2 e^2} \int \left(f^{(2)}_c - B^{(2)}_\text{e}\right) \wedge * \left(f^{(2)}_c - B^{(2)}_\text{e}\right)~.
}[gmactbis]
The background and dynamical fields transform as follows under background gauge transformations for the 2-group~$U(1)_A^{(1)} \times_{\hat \kappa_A} U(1)^{(1)}_\text{m}$ and the electric 1-form symmetry~$U(1)^{(1)}_\text{e}$,
\begin{align}\label{symtransapp}
A^{(1)}  \qquad &\longrightarrow \qquad A^{(1)} + d \lambda_A^{(0)}~, \cr
B^{(2)}_\text{e}  \qquad &\longrightarrow \qquad B^{(2)}_\text{e} + d \Lambda_\text{e}^{(1)}~,\cr
B^{(2)}_\text{m} \qquad & \longrightarrow \qquad B_\text{m}^{(2)} + d\Lambda_\text{m}^{(1)} + {\hat \kappa_A \over 2 \pi} \, \lambda_A^{(0)} \, F^{(2)}_A~,\cr
\chi \qquad & \longrightarrow  \qquad \chi +\lambda_A^{(0)}~, \cr
c^{(1)} \qquad & \longrightarrow \qquad c^{(1)} + \Lambda_\text{e}^{(1)}~.
\end{align}
Under these transformations, the GM action~\gmactbis\ shifts by the following c-number,
\eqn{
S_\text{GM} \quad \longrightarrow \quad S_\text{GM} + {i \over 2 \pi} \int \Lambda^{(1)}_\text{e} \wedge d B^{(2)}_\text{m}~.
}[gmemshift]

Comparing with~\eqref{maxgtshift} in appendix~\ref{appHooftFreeMax}, we recognize~\gmemshift\ as the 't Hooft anomaly of free Maxwell theory. This anomaly is unavoidable, because it arises from the free~$\langle f^{(2)}_c f^{(2)}_c\rangle$ two-point function at separated points (see appendix~\ref{ssjbjbtwopt}). This constitutes a serious conundrum: the anomaly~\gmemshift\ arises via inflow from the five-dimensional action
\eqn{
S_5 = {i \over 2 \pi} \int_{\CM_5}  B^{(2)}_\text{e} \wedge d B^{(2)}_\text{m}~.
}[inflowbis]
As was mentioned around~\descenteq, experience suggests that all 't Hooft anomalies in local QFTs admit such a description in terms of anomaly inflow. However, the 2-group shift of~$B^{(2)}_\text{m}$ implies that the anomaly-inflow action~\inflowbis\ is not gauge invariant in the five-dimensional bulk. This problem cannot be fixed using the background fields in~\eqref{symtransapp}. 

The resolution is that we have not correctly identified the symmetry of the GM model:  as we will see shortly, the~$U(1)_A^{(0)} \times_{\hat \kappa_A} U(1)_\text{m}^{(1)}$ 2-group symmetry and the electric 1-form symmetry~$U(1)_\text{e}^{(1)}$ are in fact fused into an even larger 3-group symmetry, which is based on a~$U(1)_\Theta^{(2)}$ 2-form symmetry. The associated 3-form background gauge field~$\Theta^{(3)}$ couples to the tautologically conserved 3-form current~$* d \chi$ via
\eqn{
S_{\Theta\chi}[\Theta^{(3)}, \chi] = {i \over 2 \pi} \int \Theta^{(3)} \wedge d \chi~.
}[tchicoup]
We now postulate the following transformation rule to~$\Theta^{(3)}$,
\eqn{
\Theta^{(3)} \quad \longrightarrow \quad \Theta^{(3)} + d \Lambda_\Theta^{(2)} + {\hat \kappa_A \over 2 \pi} \, \Lambda_\text{e}^{(1)} \wedge F_A^{(2)}~.
}[ttrans]
Hence~$\Theta^{(3)}$ not only shifts under its own~$U(1)_\Theta^{(2)}$~$2$-form gauge transformation, parametrized by~$\Lambda_\Theta^{(2)}$, but also under the electric 1-form symmetry~$U(1)_\text{e}^{(1)}$, by an amount dictated by the~$U(1)_A^{(0)}$ field strength~$F_A^{(2)}$ and the 2-group structure constant~$\hat \kappa_A$. Therefore all gauge transformations are unified into a 3-group. Note that the~$U(1)_A^{(0)} \times_{\hat \kappa_A} U(1)_\text{m}^{(1)}$ 2-group is a good subgroup of this 3-group (this should be understood in the sense of footnote~\ref{ft:localsgp}), since the 3-group shift of~$\Theta^{(3)}$ in~\ttrans\ is only activated by~$U(1)_\text{e}^{(1)}$ background gauge transformations. 

The total action~$S_\text{tot.} = S_\text{GM} + S_{\Theta\chi}$, with~$S_\text{GM}$ in~\gmactbis\ and~$S_{\Theta\chi}$ in~\tchicoup, is now invariant under the gauge transformations in~\eqref{symtransapp} and~\ttrans, up to the following c-number 't Hooft anomalies,
\eqn{
S_\text{tot.} \quad \longrightarrow \quad S_\text{tot.} + {i \over 2 \pi} \int \Lambda_\text{e}^{(1)} \wedge d B^{(2)}_\text{m} + {i \over 2 \pi} \int \lambda_A^{(0)} \, d \Theta^{(3)} +{i \hat \kappa_A \over 4 \pi^2} \int \lambda_A^{(0)} \, d\Lambda^{(1)}_\text{e} \wedge F^{(2)}_A~.
}[threegpanom]
This includes the expected mixed electric-magnetic 1-form anomaly of free Maxwell theory, as well as the anomaly between the 1-form current~$d\chi$ and the three-form current~$* d \chi$ that arises from the structure of the free~$\langle \chi \chi\rangle$ two-point function. There is also an 't Hooft anomaly that mixes the Maxwell and the NG currents, which is proportional to the 2-group structure constant~$\hat \kappa_A$ and bilinear in the gauge parameters. The anomaly~\threegpanom\ arises via inflow from the following five-dimensional action, which is fully gauge invariant in the bulk,
\eqn{
S_5[A^{(1)}, \Theta^{(3)}, B^{(2)}_\text{e}, B^{(2)}_\text{m}] = {i \over 2 \pi} \int_{\CM_5} A^{(1)} \wedge d \Theta^{(3)} + {i \over 2\pi} \int_{\CM_5} B^{(2)}_\text{e} \wedge \left(dB^{(2)}_\text{m} -{\hat \kappa_A \over 2 \pi} \, A^{(1)} \wedge F_A^{(2)}\right)~.
}[threegpinfl]
The 3-group shift of~$\Theta^{(3)}$ in~\ttrans\ precisely cancels the bulk non-invariance of the second term under~$U(1)_\text{e}^{(1)}$ gauge transformations of~$B^{(2)}_\text{e}$. Therefore all 't Hooft anomalies arise from inflow, once the correct symmetry of the model has been identified.\foot{~A similar inflow puzzle was posed, and ultimately resolved, for discrete~$n$-group symmetries~\cite{Kapustin:2013uxa,Kapustin:2014zva,Thorngren:2015gtw}.} 

Finally, we should point out that if the GM model arises in the deep IR of a non-trivial RG flow with spontaneous 2-group breaking, the 3-group symmetry of the model is typically an emergent, accidental symmetry. This is due to the fact that the~$U(1)_\text{e}^{(1)}$ symmetry of free Maxwell theory is explicitly broken by electrically charged matter.

\newsec{Further Aspects of 2-Group Symmetries}[secfurthasp]

In this section we discuss global consistency conditions on 2-group background gauge fields, the gauging of these background fields, and the properties of strings and line defects in the presence of 2-group symmetry, all of which were briefly mentioned in previous sections. We also touch on two additional topics: the holographic dictionary for QFTs with 2-group symmetries, and the reduction of 2-group symmetries to lower dimensions.

\subsec{Global Properties of 2-Group Background Fields}[SecGlobal]

Consider an abelian~2-group~$U(1)_A^{(0)} \times_{\hat\kappa_A} U(1)_B^{(1)}$, with background gauge fields~$A^{(1)}, B^{(2)}$ that transform as in~\eqref{atrans} and~\Btwogp\ under background gauge transformations,
\eqn{
A^{(1)} \quad \longrightarrow \quad A^{(1)} + d \lambda_A^{(0)}~, \qquad B^{(2)} \quad \longrightarrow \quad B^{(2)} + d \Lambda_B^{(1)} + {\hat \kappa_A \over 2\pi} \, \lambda_A^{(0)} \, F_A^{(2)}~.
}[abtwgpbgftagain]
As explained in section~\ref{convgsbf}, the fact that~$U(1)_A^{(0)}$ and~$U(1)_B^{(1)}$ are compact implies that the gauge parameters~$\lambda_A^{(0)}$ and~$\Lambda_B^{(1)}$ have quantized, generally non-zero, periods around closed 1- and 2-cycles~$\Sigma_{1,2}$,  \eqn{
{1 \over 2 \pi} \int_{\Sigma_1} d \lambda_A^{(0)} \in  \Z~, \qquad {1 \over 2 \pi} \int_{\Sigma_2} d \Lambda^{(1)} \in  \Z~.
}[lLquant]
This means that the gauge parameters are ambiguous, e.g.~$\lambda_A^{(0)} \sim \lambda_A^{(0)} + 2 \pi$, but since they parametrize elements of the compact groups~$U(1)_A^{(0)}$ or~$U(1)_B^{(1)}$, these ambiguities should be invisible -- even at the level of the transformation rules in~\abtwgpbgftagain. This is the case for the transformation rule of~$A^{(1)}$, which only depends on~$d \lambda_A^{(0)}$. However, the 2-group shift of~$B^{(2)}$ implies that the ambiguity~$\lambda_A^{(0)} \sim \lambda_A^{(0)} + 2 \pi$ induces an ambiguity in $B^{(2)} \sim B^{(2)} + \hat \kappa_A \, F_A^{(2)}$. 

The resolution of this apparent paradox is that~$B^{(2)}$ also shifts under~$U(1)_B^{(1)}$ gauge transformations, $B^{(2)} \rightarrow B^{(2)} + d \Lambda_B^{(1)}$. Since the periods of~$F_A^{(2)}$ satisfy the same quantization condition~${1 \over 2 \pi} \int_{\Sigma_2} F_A^{(2)} \in \Z$ (see~\eqref{fsquant}) as those of~$d \Lambda_B^{(1)}$ in~\lLquant, it is possible to absorb the ambiguity~$B^{(2)} \sim B^{(2)} + \hat \kappa_A \, F_A^{(2)}$ if and only if the 2-group structure constant~$\hat \kappa_A$ is an integer, $\hat \kappa_A \in \Z$. This establishes the quantization condition for~$\hat \kappa_A$ that was mentioned throughout this paper.\foot{~As was discussed above~\abtwgpdef\ in section~\ref{gsfbgf}, the quantization of~$\hat \kappa_A$ also follows from the fact that it labels a group cohomology class~$\beta \in H^3(U(1)_A^{(0)}, U(1)_B^{(1)}) = \Z$.} It also shows that a 2-group shift of~$B^{(2)}$ is inconsistent under large~$U(1)_A^{(0)}$ background gauge transformations, unless~$B^{(2)}$ is separately invariant under~$U(1)_B^{(1)}$ background gauge transformations. This fact played an important role in establishing the inequality~\uvemergineq. 

It is straightforward to generalize the arguments above to nonabelian and Poincar\'e 2-groups, which give rise to the 2-group shifts of~$B^{(2)}$ in~\eqref{nonabbint} and~\eqref{ptwgpint}, 
\eqn{
B^{(2)} \quad \longrightarrow \quad B^{(2)} + {\hat \kappa_A \over 4 \pi} \tr \left(\lambda_A^{(0)} d A^{(1)}\right) + {\hat \kappa_\SP \over 16 \pi} \tr \left( \theta^{(0)} \, d \omega^{(1)}\right)~.
}[nonabpoincbagain]
Using a suitable~$U(1)^{(0)}$ subgroup of the Cartan torus and appealing to the abelian case described above leads to the quantization condition~$\hat \kappa_A \in \Z$. (The relative factor of~$2$ between~\abtwgpbgftagain\ and~\nonabpoincbagain\ is due to our conventions for nonabelian gauge fields, see section~\ref{ssecNonabHooft}). A similar, but slightly more involved, argument shows that consistency of~\nonabpoincbagain\ on non-trivial spacetime manifolds~$\CM_4$ requires the Poincar\'e 2-group structure constant to be quantized as well, $\hat \kappa_\SP \in \Z$. For instance, we can take~$\CM_4 = S^2 \times \R^2$ and cover~$S^2$ with two patches that overlap near the equator. The local frame rotation~$\theta^{(0)}$ that relates the two patches has monodromy around the equator of the~$S^2$. Using~\nonabpoincbagain, this leads to an ambiguity in~$B^{(2)}$ that can only be absorbed by a~$U(1)_B^{(1)}$ background gauge transformation of~$B^{(2)}$ if~$\hat \kappa_\SP \in \Z$. 

We will now show that the 2-group transformation rules in~\abtwgpbgftagain\ and~\nonabpoincbagain\ lead to restrictions on the topology of the spacetime manifold~$\CM_4$ and the allowed bundles for background~$1$-form gauge fields~$A^{(1)}$. For simplicity, we focus on a 2-group of the form~$\left(U(1)_A^{(0)} \times \SP\right) \times_{\hat \kappa_A, \hat \kappa_\SP} U(1)_B^{(1)}$, but it is straightforward to generalize the discussion. As was explained around~\eqref{Htwoabelian}, the 2-group shift of~$B^{(2)}$ implies that the conventional field strength~$d B^{(2)}$ is not invariant under~$U(1)_A^{(0)}$ background gauge transformations or local frame rotations, but we can instead consider a modified field strength~$H^{(3)}$ for~$B^{(2)}$ that also includes suitable Chern-Simons terms,
\eqn{
H^{(3)} =  dB^{(2)} - {\hat \kappa_A \over 2 \pi} \, \text{CS}^{(3)}(A) - \frac{\hat \kappa_{\SP}}{16\pi} \, \text{CS}^{(3)}(\omega)~.
}[modhdef]
Here~$\text{CS}^{(3)}(A) = A^{(1)} \wedge F_A^{(2)}$, while~$\text{CS}^{(3)}(\omega)$ is the gravitational Chern-Simons 3-form for the spin connection~$\omega^{(1)}$ defined in~\gravcs. Using the properties of the Chern-Simons terms under background gauge transformations (see in particular~\csvar), we find that~$H^{(3)}$ is gauge invariant. However, it is not closed; instead, it satisfies the following modified Bianchi identity (here we use~\dgravcsisrr), 
\eqn{
dH^{(3)}={\hat \kappa_A \over 2 \pi} \, F_A^{(2)} \wedge F_A^{(2)} + {\hat \kappa_\SP \over 16 \pi} \, \tr\left(R^{(2)} \wedge R^{(2)}\right)~.
}[modbianchi]

If we choose spacetime to be a compact riemannian manifold~$\CM_4$, we can integrate this equation over all of~$\CM_4$. Since the left-hand side of~\modbianchi\ is an exact 4-form, because~$H^{(3)}$ is well defined, it integrates to zero, so that
\eqn{
{\hat \kappa_A \over 2 \pi} \int_{\CM_4} F_A^{(2)} \wedge F^{(2)}_A + {\hat \kappa_\SP \over 16\pi} \int_{\CM_4} \tr\left(R^{(2)} \wedge R^{(2)}\right) = 0~.
}[topcons]
This topological constraint relates the Chern class $c_1(F^{(2)}_A) = {1 \over 2 \pi} \, F_A^{(2)}$ of the~$U(1)_A^{(0)}$ bundle to the signature of the spacetime manifold~$\CM_4$, which is proportional to~$\int_{\CM_4} \tr\big(R^{(2)} \wedge R^{(2)}\big)$. Constraints such as~\topcons\ are common in situations where a GS mechanism leads modifies the Bianchi identity as in~\modbianchi. This includes classic examples of string compactification~\cite{Candelas:1985en,Strominger:1986uh}.

It is instructive to examine the global restriction in~\topcons\ through the lens of section~\ref{SecBasics}, where theories with 2-group symmetry were constructed by gauging~$U(1)_C^{(0)}$ in parent theories with~$U(1)_A^{(0)} \times U(1)_C^{(0)}$ flavor symmetry and non-vanishing~$\kappa_{A^2 C}$ or~$\kappa_{C \SP^2}$ mixed 't Hooft anomalies. Naively, such anomalies appear to violate the~$U(1)_c^{(0)}$ gauge symmetry in the presence of the~$U(1)_A^{(0)}$ background field~$A^{(1)}$, or in the presence of a curved background metric. As we have seen this is not the case once we add the background gauge field~$B^{(2)}$, because its 2-group shifts effectively cancels the mixed anomalies via a GS mechanism. Equivalently, if we insist on~$U(1)_c^{(0)}$ gauge invariance, the presence of a mixed~$\kappa_{A^2 C}$ anomaly means that the background gauge field~$A^{(1)}$ is specified by more data than a conventional, geometric connection. Similarly, if we insist on~$U(1)_c^{(0)}$ gauge invariance, the presence of a mixed~$\kappa_{C \SP^2}$ anomaly implies that we must specify additional information, beyond a conventional riemannian metric, to place the theory on a curved 4-manifold. In both cases, the additional data is supplied by the background gauge field~$B^{(2)}$.   

The upshot is that there is no local obstruction to specifying any configuration for the background gauge field~$A^{(1)}$, or for the background metric. There is, however, a global topological constraint~\topcons\ on the allowed backgrounds, via the modified Bianchi identity~\modbianchi, which is an unavoidable consequence of 2-group symmetry. This constraint can be viewed as a global remnant of the original mixed~$\kappa_{A^2 C}$ and~$\kappa_{C \SP^2}$ anomalies. A similar phenomenon, which also involves the transmutation of a local anomaly into a global issue, arises in the context of 't Hooft anomalies for 2-group symmetries. As discussed in section~\sssHoftAbTwGp\ (especially between~\kaaanshift\ and~\aglcgaugshift), a reducible~$\kappa_{A^3}$ 't Hooft anomaly can superficially be removed by a GS counterterm~\gsctbis. However, the fractional part~${\kappa_{A^3} \over 6 \hat \kappa_A}~(\mod~1)$ persists as a genuine 't Hooft anomaly, due to a clash between conventional~$U(1)_A^{(0)}$ background gauge transformations, and topologically non-trivial~$U(1)_B^{(1)}$ background gauge transformations.

\begin{landscape}

\begin{figure}[h]
\vskip-40pt
\centering
\includegraphics[trim=0 0cm 0cm 0,clip,height=16.5cm]{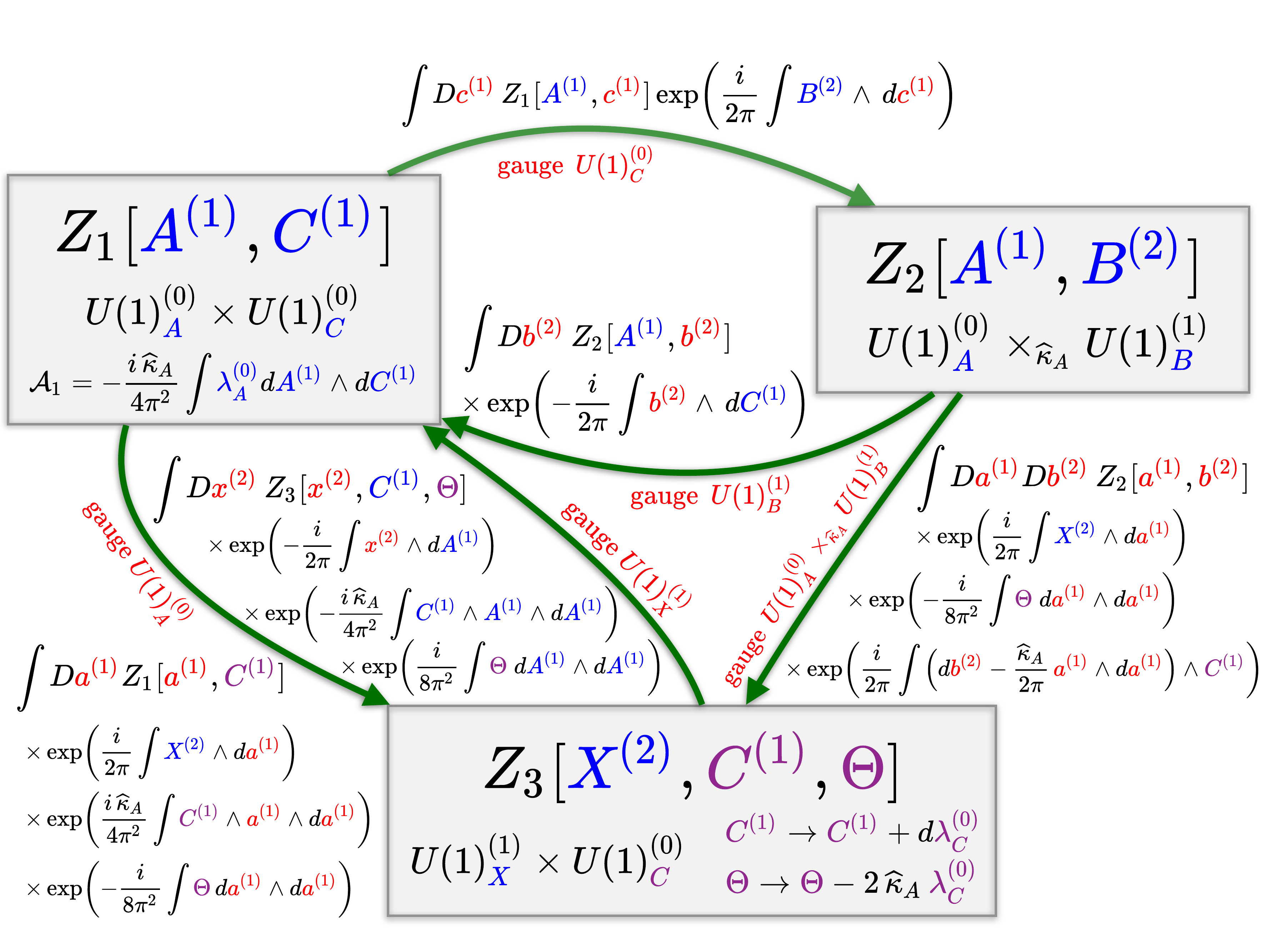}  

 \caption{Possible gaugings of a theory $T_2$ with~$U(1)_A^{(0)} \times_{\hat \kappa_A} U(1)_B^{(1)}$ 2-group symmetry and no 't Hooft anomalies. Theory $T_1$ is the parent theory with~$U(1)_A^{(0)} \times U(1)_C^{(0)}$ flavor symmetry and a mixed~$\kappa_{A^2 C} = - 2 \hat \kappa_A$ 't Hooft anomaly; it can be obtained from~$T_2$ by gauging~$U(1)_B^{(1)}$. Theory~$T_3$ results from~$T_2$ by gauging~$U(1)_A^{(0)} \times_{\hat \kappa_A} U(1)_B^{(1)}$, or from~$T_1$ by gauging~$U(1)_C^{(0)}$.} 
 \label{fig:gauging}
\end{figure}

\end{landscape}

\subsec{Gauging 2-Group Symmetries}[SsecGauging]

In this section we consider a theory with abelian 2-group symmetry~$U(1)_A^{(0)} \times_{\hat \kappa_A} U(1)_B^{(1)}$. (Generalizations to other 2-groups are straightforward.) In order to match the terminology introduced in section~\ref{ssecExOverview} (see in particular figure~\ref{fig:gaugeRG}) we will refer to this theory as theory 2, or simply~$T_2$. We use ~$T_1$ to refer to a parent theory with~$U(1)_A^{(0)} \times U(1)_C^{(0)}$ flavor symmetry and mixed~$\kappa_{A^2 C}$ 't Hooft anomaly, from which~$T_2$ arises by gauging~$U(1)_C^{(0)}$. As we will show below, such a parent theory~$T_1$ exists for every theory~$T_2$ with 2-group symmetry.  Later we will also introduce a third theory~$T_3$, which results form~$T_1$ by gauging~$U(1)_A^{(0)}$. Note that the possibility of simultaneously gauging both~$U(1)_A^{(0)}$ and~$U(1)_C^{(0)}$ in~$T_1$ is obstructed by its mixed~$\kappa_{A^2 C}$ 't Hooft anomaly. The three theories~$T_{1,2,3}$ are represented by grey boxes in figure~\ref{fig:gauging}. The purpose of this subsection is to supply a detailed explanation of this figure. 

Theory~$T_2$ has 2-group background gauge fields~$A^{(1)}$ and~$B^{(2)}$. We would like to understand what happens if we gauge the 2-group symmetry, or a subgroup thereof, by promoting the appropriate background gauge fields to dynamical gauge fields and doing the functional integral over their gauge orbits. It follows from the comments below~\Btwogp\ that it is not consistent to gauge~$U(1)_A^{(0)}$ without also gauging~$U(1)_B^{(1)}$, because the 2-group shift~$B^{(2)} \rightarrow B^{(2)} + {\hat \kappa_A \over 2 \pi} \, \lambda_A^{(0)} \, F_A^{(2)}$ mixes~$B^{(2)}$ with~$A^{(1)}$. This is consistent with the general principle we have encountered repeatedly, according to which~$U(1)_A^{(0)}$ is not a good subgroup of the full 2-group. (As always, this statement should be understood in the sense of current algebra, see footnote~\ref{ft:localsgp}.) However, $U(1)_B^{(1)}$ is a good subgroup, and it can be gauged by while keeping~$U(1)_A^{(0)}$ a global symmetry and~$A^{(1)}$ a background gauge field. The allowed possibilities are therefore to either gauge all of~$U(1)_A^{(0)} \times_{\hat \kappa_A} U(1)_B^{(1)}$, or to only gauge its~$U(1)_B^{(1)}$ subgroup,
\eqna{& U(1)_A^{(0)} \times_{\hat \kappa_A} U(1)_B^{(1)} \; \rightarrow \; U(1)_a^{(0)}\times _{\hat \kappa _A}U(1)_b^{(1)}~, \qquad A^{(1)} \; \rightarrow \; a^{(1)}~, \qquad B^{(2)} \; \rightarrow \; b^{(2)}~,\cr
& \hskip104pt \text{or} \cr
& U(1)_A^{(0)} \times_{\hat \kappa_A} U(1)_B^{(1)} \; \rightarrow \; U(1)_A^{(0)}\times _{\hat \kappa _A}U(1)_b^{(1)}~, \qquad B^{(2)} \; \rightarrow \; b^{(2)}~.
}[possgaug]
The fact that~$U(1)_A^{(0)}$ cannot by itself be gauged in~$T_2$ also follows from the parent theory~$T_1$. Since~$T_2$ is the result of gauging~$U(1)_C^{(0)}$ in~$T_1$, gauging~$U(1)_A^{(0)}$ in~$T_2$ amounts to simultaneously gauging~$U(1)_A^{(0)}$ and~$U(1)_C^{(0)}$ in~$T_1$. As was already mentioned above, this is not possible because~$T_1$ has a mixed~$\kappa_{A^2 C}$ 't Hooft anomaly. As we will see below, gauging the entire 2-group symmetry~$U(1)_A^{(0)} \times_{\hat \kappa_A} U(1)_B^{(1)}$ of~$T_2$ circumvents this problem by cancelling the mixed anomaly via a conventional GS mechanism for the dynamical gauge fields. 

Before considering the gaugings in~\possgaug, we recall, and expand on, some basic facts that have already been used in previous sections. (See~\cite{Gaiotto:2014kfa} and references therein for additional background. Until further notice, we assume that all possible 't Hooft anomalies are absent.) For the moment, we consider a simplification of the setup in figure~\ref{fig:gauging}, in which the parent theory only has a~$U(1)_C^{(0)}$ flavor symmetry. Gauging~$U(1)_C^{(0)}$ then leads to a new theory with a global~$U(1)_B^{(1)}$ symmetry, which arises from the magnetic 2-form current~$J_B^{(2)} = {i \over 2 \pi} * f_c^{(2)}$ (see~\eqref{magcurrint}). Here~$f_c^{(2)}$ is the~$U(1)_c^{(0)}$ Maxwell field strength. It is a useful fact that this procedure has an inverse: gauging the~$U(1)_B^{(1)}$ 1-form global symmetry of the~$U(1)_c^{(0)}$ gauge theory returns us to the original parent theory with~$U(1)_C^{(0)}$ flavor symmetry. In terms of the dynamical~$U(1)_b^{(1)}$ gauge field~$b^{(2)}$, the~$U(1)_C^{(0)}$ flavor current is given by~$j_C^{(1)} = {i \over 2 \pi} * d b^{(2)}$. 

The fact that gauging~$U(1)_C^{(0)}$ and~$U(1)_B^{(1)}$ are inverse operations can be made explicit by considering the partition functions of the two theories in the presence of their respective background gauge fields. As we have done throughout the paper, we can start with the partition function~$Z[C^{(1)}]$ of the parent theory in the presence of a~$U(1)_C^{(0)}$ background gauge field, and construct a new theory with partition function~$\t Z[B^{(2)}]$ that depends on a~$U(1)_B^{(1)}$ background gauge field~$B^{(2)}$ by gauging~$U(1)_C^{(0)}$. This involves coupling~$B^{(2)}$ to~$J_B^{(2)}$ and promoting~$C^{(1)} \rightarrow c^{(1)}$ to a dynamical~$U(1)_c^{(0)}$ gauge field, by doing a suitably gauge-fixed functional integral over its gauge orbits,
\eqn{\t Z[B^{(2)}] =\int Dc^{(1)} \, Z[c^{(1)}] \exp\left({i\over 2\pi}\int B^{(2)}\wedge dc^{(1)} \right)~.}[gaugesimp]
Here we use the notation~$\t Z$ because~\gaugesimp\ can be thought of as a functional Fourier transform.\foot{~Similar observations in three dimensions appear in~\cite{Kapustin:1999ha, Witten:2003ya}.} Any gauge-invariant terms for the dynamical gauge field~$c^{(1)}$, such as a Maxwell kinetic term, are included in~$Z[c^{(1)}]$. Prior to gauging, they correspond to gauge-invariant local counterterms for the~$U(1)_C^{(0)}$ background gauge field~$C^{(1)}$ in~$Z[C^{(1)}]$. Just as an ordinary Fourier transform, it is possible to invert~\gaugesimp\ by gauging~$U(1)_B^{(1)}$, i.e.~by promoting~$B^{(2)}$ to a dynamical gauge field~$b^{(2)}$ and doing an appropriately gauge-fixed functional integral over its gauge orbits,
\eqn{Z[C^{(1)}]=\int Db^{(2)} \, \t Z[b^{(2)}] \exp\left(-{i\over 2\pi}\int b^{(2)}\wedge dC^{(1)}\right)~.
}[btoA]
Note that the background gauge field~$C^{(1)}$ in~\btoA\ couples to the current~$j_C^{(1)} = {i \over 2 \pi} * d b^{(2)}$ described above. In order for~\btoA\ to be the inverse of~\gaugesimp\ (as will be established below), it is important that we do not include any additional terms for~$b^{(2)}$ in~\btoA, beyond what is supplied by~$\t Z[b^{(2)}]$. It may nevertheless be helpful (e.g.~for convergence or conceptual reasons) to include such additional terms. For instance, we can add a standard~$b^{(2)}$ kinetic term~$\sim {1 \over g^2} \, db^{(2)} \wedge * db^{(2)}$. Then~\btoA\ corresponds to the limit~$g \rightarrow \infty$.

To see that~\btoA\ is the inverse of~\gaugesimp, we substitute the expression for~$\t Z[B^{(2)}]$ in~\gaugesimp\ into the~$b^{(2)}$ functional integral on the right-hand side of~\btoA. This integral reduces to
\eqn{
\int D b^{(2)} \exp\left({i \over 2 \pi} \int b^{(2)} \wedge \left(d c^{(1)} - d C^{(1)} \right)\right) = \delta_\text{g.i.}\left(c^{(1)} - C^{(1)} \right)~.
}[gidf]
Here~$\delta_\text{g.i.}(c^{(1)} - C^{(1)})$ is a gauge-invariant~$\delta$-functional, which sets~$c^{(1)} = C^{(1)}$, up to (background) gauge transformations. It is normalized so that a suitably gauge-fixed functional integral over gauge orbits of~$c^{(1)}$ gives~$ \int Dc^{(1)} \delta_\text{g.i.}(c^{(1)} - C^{(1)}) = 1$ (see for instance the recent discussion in~\cite{Seiberg:2016gmd}). Using this fact, the remaining~$c^{(1)}$ functional integral on the right-hand side of~\btoA\ collapses to~$\int D c^{(1)} \delta_\text{g.i}(c^{(1)} - C^{(1)}) Z[c^{(1)}] = Z[C^{(1)}]$. A similar line of reasoning shows that substituting the expression for~$Z[C^{(1)}]$ in~\btoA\ into the~$c^{(1)}$ functional integral on the right-hand side of~\gaugesimp\ correctly reproduces~$\t Z[B^{(2)}]$. 

We will now repeat the preceding discussion in the presence of an additional~$U(1)_A^{(0)}$ flavor symmetry. If~$U(1)_A^{(0)}$ has no mixed 't Hooft anomalies with the~$U(1)_C^{(0)}$ flavor symmetry of the parent theory, it simply comes along for the ride. Instead, we will consider a parent theory~$T_1$ which has a non-zero mixed~$\kappa_{A^2 C}$ 't Hooft anomaly, while all other 't Hooft anomalies vanish. This theory resides in the top-left corner of figure~\ref{fig:gauging}. As usual, gauging~$U(1)_C^{(0)}$ int~$T_1$ then leads to a theory~$T_2$ (in the top-right corner of figure~\ref{fig:gauging}) with~$U(1)_A^{(0)} \times_{\hat \kappa_A} U(1)_B^{(1)}$ 2-group symmetry, with 2-group structure constant~$\hat \kappa_A = - \half\kappa_{A^2 C}$ (see for instance~\btwgpbiscoef). The partition function~$Z_2[A^{(1)}, B^{(2)}]$ of theory~$T_2$ in the presence of the 2-group background gauge fields~$A^{(1)}, B^{(2)}$ is obtained from the partition function~$Z_1[A^{(1)}, C^{(1)}]$ of theory~$T_2$ by gauging~$U(1)_C^{(1)}$, which is represented by the top, right-pointing arrow in figure~\ref{fig:gauging}, 
\eqn{Z_2[A^{(1)}, B^{(2)}]=\int Dc^{(1)}\, Z_1[A^{(1)}, c^{(1)}] \exp \left(\frac{i}{2\pi} \int B^{(2)}\wedge dc ^{(1)}\right)~.
}[Ztwoi]
This expression is analogous to~\gaugesimp. As was the case there, Maxwell kinetic terms (or any other gauge-invariant couplings) for~$c^{(1)}$ are included in~$Z_1[A^{(1)}, c^{(1)}]$.  

We saw around~\btoA\ and~\gidf\ that gauging~$U(1)_B^{(1)}$ in the absence of 2-group symmetry restores the parent theory with~$U(1)_C^{(0)}$ flavor symmetry. We will now repeat this discussion in theory~$T_2$, which has~$U(1)_A^{(0)} \times_{\hat  \kappa_A} U(1)_B^{(1)}$ 2-group symmetry. Our assumption that~$T_1$ only has a~$\kappa_{A^2 C}$ 't Hooft anomaly is equivalent to the assumption that~$T_2$ is free of 2-group 't Hooft anomalies (see section~\ref{sssHoftAbTwGp}, as well as below). We can therefore gauge~$U(1)_B^{(1)}$ and repeat (essentially verbatim) the discussion around~\btoA\ and~\gidf: we invert~\Ztwoi\ and reconstruct the partition function~$Z_1[A^{(1)}, C^{(1)}]$ of the parent theory~$T_1$ from the partition function~$Z_2[A^{(1)}, B^{(2)}]$ of~$T_2$ by gauging~$U(1)_B^{(1)}$ (see figure~\ref{fig:gauging}), 
\eqn{Z_1[A^{(1)}, C^{(1)}]=\int Db ^{(2)}~Z_2[A^{(1)}, b^{(2)}] \exp \left(-\frac{i}{2\pi} \int b^{(2)}\wedge dC^{(1)}\right)~.
}[Zonei]
This expression is manifestly invariant under~$U(1)_C^{(0)}$ background gauge transformations. Under a~$U(1)_A^{(0)}$ background gauge transformation, we have~$A^{(1)} \rightarrow A^{(1)} + d \lambda_A^{(0)}$. If we accompany this by a change of variables~$b^{(2)} \rightarrow b^{(2)} + {\hat \kappa_A \over 2 \pi} \, \lambda_A^{(0)} \, F_A^{(2)}$ in the functional integral, we can use the 2-group invariance of~$Z_2$ to conclude that only the exponential phase factor on the right-hand side of~\Zonei\ contributes, via a c-number phase factor. This amounts to the following anomalous shift~$\CA_1$ of the effective action~$W_1 = - \log Z_1$ under~$U(1)_A^{(0)}$,
\eqn{
\CA_1 = -{ i \hat \kappa_A \over 4 \pi^2} \int \lambda_A^{(0)} \, F_A^{(2)} \wedge F_C^{(2)} = {i \kappa_{A^2 C} \over 8 \pi^2}  \int \lambda_A^{(0)} \, F_A^{(2)} \wedge F_C^{(2)}~,
}[aoneanom]
which correctly reproduces the~$\kappa_{A^2 C}$ 't Hooft anomaly of~$T_1$, with the choice of counterterms~\ctfix\ used throughout this paper. 

The preceding discussion shows that any theory~$T_2$ with~$U(1)_A^{(0)} \times_{\hat \kappa_A} U(1)_B^{(1)}$ 2-group symmetry arises from a parent theory~$T_1$ with~$U(1)_A^{(0)} \times U(1)_C^{(0)}$ flavor symmetry and a mixed~$\kappa_{A^2 C} = - 2 \hat \kappa_A$ 't Hooft anomaly. The parent theory can be found explicitly by gauging the~$U(1)_B^{(1)}$  subgroup of the 2-group, as in~\Zonei. If~$T_2$ has a presentation as a~$U(1)_C^{(0)}$ gauging of~$T_1$, as is the case for the~$U(1)_c^{(0)}$ gauge theories considered in sections~\ref{SubsecQEDexamples} and~\ref{SubsecFermats}, gauging~$U(1)_B^{(1)}$ simply turns the dynamical~$U(1)_c^{(0)}$ gauge field back into a~$U(1)_C^{(0)}$ background gauge field, as in~\gidf. A more interesting example, which does not have such a presentation, is the~$\mathbb{CP}^{1}$ model with 2-group symmetry discussed in section~\ref{ssCPN}. If we gauge its~$U(1)_B^{(1)}$ symmetry, we can dualize the dynamical~$b^{(2)}$ gauge field to a periodic scalar, which reconstructs the Hopf fiber of the parent~$S^{3}$ sigma model. 

Recall from~\btoA\ that the coupling of~$C^{(1)}$ to~$b^{(2)}$ in~\Zonei\ amounts to taking the~$U(1)_C^{(0)}$ current to be~$j_C^{(1)} = {i \over 2 \pi} * b^{(2)}$. Here~$j_C^{(1)}$ is conserved, but not invariant under~$U(1)_A^{(0)}$ background gauge transformations, due to the 2-group shift of~$b^{(2)}$. As was explained around~\aoneanom, this reflects the~$\kappa_{A^2 C}$ 't Hooft anomaly of~$T_1$, given the particular choice of counterterms in~\ctfix. Another natural choice is to couple~$C^{(1)}$ in~\Zonei\ to a different current~$\t j_C^{(1)}$, which is defined in terms of the gauge-invariant field strength~$h^{(3)}$ of~$b^{(2)}$, 
\eqn{
\t j^{(1)}_C = \frac{i}{2\pi} * h^{(3)}~, \qquad h^{(3)} = db^{(2)} - {\hat \kappa_A \over 2 \pi} \, A^{(1)} \wedge F_A^{(2)}~, \qquad d* \t j^{(1)}_C = - \frac{i \hat \kappa _A}{4\pi ^2} \, F_A^{(2)} \wedge F_A^{(2)}~.
}[Joptions]
Observe that~$\t j_C^{(1)}$ is invariant under~$U(1)_A^{(0)}$ background gauge transformations, but not conserved in a~$U(1)_A^{(0)}$ background gauge field. This is an alternative presentation of the~$\kappa_{A^2 C}$ 't Hooft anomaly of~$T_1$, which differs from the one used above by a counterterm~$S \; \supset \; -{i \hat \kappa_A \over 4 \pi^2} \int C ^{(1)} \wedge A^{(1)} \wedge F_A^{(2)}$. Adding this counterterm amounts to setting~$s = \half \kappa_{A^2 C}$ in~\anomalies\ and~\eqref{anomconseq}, which renders the theory invariant under~$U(1)_A^{(0)}$ background gauge transformations, at the expense of replacing the conserved current~$j_C^{(1)}$ by~$\t j_C^{(1)}$. This presentation of the anomaly is mandatory if we want to gauge~$U(1)_A^{(0)}$, as we will do momentarily. 
 
We would now like to gauge the entire 2-group symmetry~$U(1)_A^{(0)} \times_{\hat \kappa_A} U(1)_B^{(1)}$ of~$T_2$. This gauging can be obstructed by a 2-group 't Hooft anomaly, which arises if the parent theory has a~$\kappa_{A^3}$ 't Hooft anomaly. (Here we are using a presentation of the anomaly where~$U(1)_B^{(1)}$ is preserved, while~$U(1)_A^{(0)}$ is anomalous, see section~\ref{sssHoftAbTwGp} for details.) This obstruction is present unless~$\kappa_{A^3} \equiv 0~(\mod~6 \hat \kappa_A)$,  in which case the genuine 2-group anomaly vanishes, while its scheme-dependent remainder can be set to zero using a properly quantized GS counterterm~\gsctdef. The condition~$\kappa_{A^3} \equiv 0~(\mod~6 \hat \kappa_A)$ also has a natural interpretation in the parent theory~$T_1$: if it holds, we can set the~$\kappa_{A^3}$ 't Hooft anomaly to zero by redefining~$U(1)_A^{(0)} \; \rightarrow \; U(1)_A^{(0)} - n \, U(1)_C^{(0)}$, as in~\aglcgaugshift\ and~\eqref{anomshift}. We therefore continue to assume that~$T_2$ is free of 2-group 't Hooft anomalies, and that~$T_1$ has vanishing~$\kappa_{A^3}$ 't Hooft anomaly, as we had done previously. 

We now consider the theory that arises by gauging~$U(1)_A^{(0)} \times_{\hat \kappa_A} U(1)_B^{(1)}$ in~$T_2$. (As we will show below, it is the same theory~$T_3$ that can obtained from~$T_1$ by gauging~$U(1)_A^{(0)}$.) On general grounds we expect that gauging~$U(1)_A^{(0)} \rightarrow U(1)_a^{(1)}$ will result in a new~$U(1)_X^{(1)}$ 1-form symmetry (with conserved current~$J_X^{(2)} = {i \over 2 \pi} * da^{(1)}$ and background gauge field~$X^{(2)}$), while gauging~$U(1)_B^{(1)} \rightarrow U(1)_b^{(1)}$ should give rise to a 0-form symmetry~$U(1)_C^{(0)}$, with background gauge field~$C^{(1)}$. However, the expected~$U(1)_C^{(0)}$ flavor symmetry suffers from an ABJ anomaly. Here the discussion around~\Joptions\ is relevant: since~$U(1)_a^{(0)}$ is now a dynamical gauge symmetry, it is not acceptable to couple the~$U(1)_C^{(0)}$ background gauge field~$C^{(1)}$ to the current~$j_C^{(1)} \sim * d b^{(2)}$, as we did in~\Zonei, because it is not gauge invariant. Instead we must use the gauge-invariant current~$\t j_C^{(1)}$ in~\Joptions. However, the non-conservation equation in~\Joptions\ now constitutes an ABJ anomaly for~$\t j_C^{(1)}$. This can be described by promoting the~$\Theta$-parameter, which appears in the action~$S \; \supset \; {i \Theta \over 8 \pi^2} \int da^{(1)} \wedge da^{(1)}$, to a background field and declaring that~$\Theta$ shifts under~$U(1)_C^{(0)}$ background gauge transformations (see the discussion around~\athetashift). Therefore~$T_3$ is obtained from~$T_2$ by following the arrow in the bottom-right corner of figure~\ref{fig:gauging}, according to which
\eqna{& Z_3[X^{(2)}, C^{(1)}, \Theta]=\int Da^{(1)} Db^{(2)} \, Z_2[a^{(1)}, b^{(2)}]  \, \exp\left(\frac{i}{2\pi} \int X ^{(2)}\wedge da^{(1)}\right)  \cr 
& \exp\left(-\frac{i }{8\pi ^2} \int  \Theta \, da^{(1)}\wedge da^{(1)}\right) \, \exp\left(\frac{i}{2\pi} \int \Big(db^{(2)}-\frac{\hat \kappa_A}{2\pi} \, a^{(1)}\wedge da^{(1)}\Big)\wedge C^{(1)}\right)~.
}[threefromtwo]
This partition function is invariant under~$U(1)_X^{(1)}$ gauge transformations of~$X^{(2)}$, as well as simultaneous shifts~$C^{(1)} \; \rightarrow \; C^{(1)} + d \lambda_C^{(0)}$ and~$\Theta \; \rightarrow \; \Theta - 2 \hat \kappa_A \lambda_C^{(0)}$ under~$U(1)_C^{(0)}$ background gauge transformations, which reflect the fact that~$U(1)_C^{(0)}$ suffers from an ABJ anomaly. 

In order to see that the theory obtained by gauging the entire 2-group symmetry of~$T_2$ really is~$T_3$, we evaluate the functional integral in~\threefromtwo\ in two steps: we first do the~$b^{(2)}$ integral, which amounts to gauging only~$U(1)_B^{(1)}$. As discussed around~\Zonei, this returns us to the parent theory~$T_1$. However, as discussed around~\Joptions, the fact that~$C^{(1)}$ in~\threefromtwo\ couples to the gauge-invariant current~$\t j_C^{(1)}$ furnishes the counterterm that renders~$T_1$ invariant under~$U(1)_A^{(0)}$ background gauge transformations, while leading to the non-conservation equation~$d* \t j_C^{(1)} \sim \hat \kappa_A \, F_A^{(2)} \wedge F_A^{(2)}$. It is now possible to gauge~$U(1)_A^{(0)}$, which leads to theory~$T_3$, with an ABJ anomaly for the~$U(1)_C^{(0)}$ current~$\t j_C^{(1)}$, as was claimed above. The gauging of~$U(1)_A^{(0)}$ in~$T_1$ to obtain~$T_3$ is represented by the arrow in the bottom-left corner of figure~\ref{fig:gauging}. Finally, we can close the loop by pointing out that this gauging can be inverted by gauging the~$U(1)_X^{(1)}$ symmetry of~$T_3$.

\subsec{Holographic Interpretation of 2-Group Symmetries}[ssecHologr]

Since the early days of the AdS/CFT correspondence~\cite{Maldacena:1997re,Gubser:1998bc,Witten:1998qj}, it has been understood that global symmetries of the boundary theory correspond to gauge symmetries in the bulk.\foot{~See~\cite{Witten:2017hdv} for an intuitive discussion of this basic point. A reexamination from the perspective of recent advances in bulk reconstruction (some of which are reviewed in~\cite{Harlow:2018fse}) will appear in~\cite{HOtoApp}.} The boundary values of the dynamical bulk gauge fields serve as background gauge fields that can be turned on in the boundary theory. In particular, the form of the corresponding background gauge transformations is dictated by the bulk gauge transformations. This leads to the following basic facts about the AdS$_5$ duals for symmetries of four-dimensional QFTs:
\begin{itemize}
\item[a)] A conventional~$U(1)_A^{(0)}$ flavor symmetry is represented by an abelian gauge field~$a^{(1)}$, with gauge symmetry~$a^{(1)} \rightarrow a^{(1)} + d \lambda_a^{(0)}$, which propagates in the bulk. It boundary value~$a^{(1)} | = A^{(1)}$ is a non-dynamical background gauge field that couples to the~$U(1)_A^{(0)}$ current~$j_A^{(1)}$ of the boundary theory. Gauge transformations that do not vanish on the boundary are non-trivial background gauge transformations that act on~$A^{(1)}$, rather than gauge redundancies. A bulk Maxwell term for~$a^{(1)}$ gives rise to the conformal~$\langle j_A^{(1)} j_A^{(1)} \rangle$ two-point function on the boundary.
 
\item[b)] The holographic dual of a~$U(1)_B^{(1)}$ symmetry was recently discussed in~\cite{Hofman:2017vwr} (see also~\cite{HOtoApp}). It is given by a 2-form gauge field~$b^{(2)}$, with gauge redundancy~$b^{(2)} \rightarrow b^{(2)} + d\Lambda_b^{(1)}$, whose boundary value~$b^{(2)}| = B^{(2)}$ is a background gauge field for the~$U(1)_B^{(1)}$ current~$J_B^{(2)}$ of the boundary theory. As emphasized in~\cite{Hofman:2017vwr}, a kinetic term~$\sim \int_{\text{AdS}_5} d b^{(2)} \wedge * d b^{(2)}$ does not lead to solutions that respect the symmetries of AdS$_5$.\foot{~This is similar to the fact that a free Maxwell field in~AdS$_3$, without a Chern-Simons term, does not admit solutions that respect the symmetries of~AdS$_3$.} Instead, $b^{(2)}$ behaves logarithmically near the boundary. In\cite{Hofman:2017vwr}, this was interpreted as the bulk manifestation of a logarithmic RG flow in an IR-free theory with 2-form current~$J_B^{(2)}$ and double-trace coupling~$\sim J_B^{(2)} \wedge * J_B^{(2)}$, just as in abelian gauge theories with charged matter, where~$J_B^{(2)} \sim * f^{(2)}$ is the magnetic 2-form current.

\item[c)] As reviewed in section~\sssprimcur, the only four-dimensional CFT that admits a 2-form current is free Maxwell theory, which possesses a~$U(1)_\text{e}^{(1)} \times U(1)_\text{m}^{(1)}$ global symmetry. The corresponding electric and magnetic 2-form currents, which are proportional to~$f^{(2)}$ and~$* f^{(2)}$ on the boundary, are represented by two propagating bulk 2-form gauge fields~$b^{(2)}_{\text{e}, \text{m}}$. The dynamics is governed by a topological action in the bulk,
\eqn{
S_5 = {i \over 2 \pi} \int_{\text{AdS}_5} b_\text{e}^{(2)} \wedge d b_\text{m}^{(2)}~. 
}[bbtopbulk]
The free Maxwell fields on the boundary arise from this topological theory as singleton, or edge, modes (see for instance~\cite{Maldacena:2001ss} and references therein). Note that~\bbtopbulk\ is a gauged version of the five-dimensional anomaly-inflow action~\inflowbis\ that captures the~$U(1)_\text{e}^{(1)} \times U(1)_\text{m}^{(1)}$ 't Hooft anomalies of free Maxwell theory~(see also appenidx~\ref{appHooftFreeMax}). 

\end{itemize}

Here we would like to comment on the holographic dictionary for a boundary QFT with 2-group symmetry, starting with the abelian case~$U(1)_A^{(0)} \times_{\hat \kappa_A} U(1)_B^{(1)}$. The boundary currents~$j_A^{(1)}$ and~$J_B^{(2)}$ are represented by 1- and 2-form gauge fields~$a^{(1)}$ and~$b^{(2)}$ in the bulk. As was reviewed above, the bulk gauge transformations should take the same form as the corresponding background gauge transformations on the boundary. For an abelian 2-group, this implies that~$a^{(1)}$ and~$b^{(2)}$ are subject to the following bulk gauge redundancy,
\eqn{
a^{(1)} \; \rightarrow \; a^{(1)} + d \lambda_a^{(0)}~, \qquad b^{(2)}  \; \rightarrow \; b^{(2)} + d \Lambda_b^{(1)} + {\hat \kappa_A \over 2 \pi} \, \lambda_a^{(0)} \, f_a^{(0)}~.
}[bulkgstwgp]
Therefore the bulk gauge transformations of~$b^{(2)}$ include a conventional GS shift that involves~$a^{(1)}$. Analogously, the bulk dual of theory with Poincar\'e 2-group symmetry involves a GS shift of~$b^{(2)}$ by the dynamical bulk gravity fields. 

While the Maxwell kinetic term for~$a^{(1)}$ is invariant under the bulk gauge transformations~\bulkgstwgp, the kinetic term~$\sim db^{(2)} \wedge * db^{(2)}$ is not invariant. Following the discussion around~\Htwoabelian, as well as in section~\SecGlobal, we can define a modified, gauge-invariant 3-form field strength,
\eqn{
h^{(3)} = db^{(2)} - {\hat \kappa_A \over 2 \pi} \, a^{(1)} \wedge f_a^{(2)}~.
}[hmodbulkdef]
Using~$h^{(3)}$, we can construct a gauge-invariant kinetic term proportional to
\eqn{
 \int_{\text{AdS}_5} * h^{(3)} \wedge h^{(3)} = \int_{\text{AdS}_5} * db^{(2)} \wedge db^{(2)} -{\hat \kappa_A \over \pi} \int_{\text{AdS}_5} * db^{(2)} \wedge a^{(1)} \wedge f_a^{(2)} + \CO\Big((a^{(1)})^4\Big)~.  
}[modhbulkin]
The second term, which contains the 2-group structure constant~$\hat \kappa_A$ and is an inevitable consequence of the bulk gauge symmetry~\bulkgstwgp, leads to a three-point coupling between the 2-form gauge field~$b^{(2)}$ and two 1-form gauge fields~$a^{(1)}$. This coupling has the required form to generate the characteristic three-point function~$\langle j_A^{(1)} j_A^{(1)} J_B^{(2)}\rangle$ associated with the boundary 2-group symmetry (which was discussed at length in sections~\ref{rgphtwgp} and~\ref{SecCurrents}) via a Witten diagram. 

As in point b) at the beginning of this subsection, the kinetic term for~$b^{(2)}$ in~\modhbulkin\ implies that~$b^{(2)}$ behaves logarithmically near the boundary, in a way that is not compatible with the symmetries of~AdS$_\text{5}$. However, unlike in point c) above, it is not possible to cure this behavior by including a second 2-form gauge field~$b^{(2)}_\text{e}$ in the bulk, which couples to~$b^{(2)} = b^{(2)}_\text{m}$ as in~\eqref{bbtopbulk}. The reason is that this coupling is not invariant under the GS shift of~$b^{(2)}$ in~\bulkgstwgp. This is consistent with our results in section~\sssprimcur: 2-group symmetry and conformal invariance are not compatible, as long as the currents~$j_A^{(1)}$ and~$J_B^{(2)}$ are conformal primaries. 

However, we saw in section~\sssnonprimcurr\ that there are conformally invariant models with 2-group symmetry in which the currents are not primaries and the 2-group symmetry is spontaneously broken. The simplest example of this kind is the free Goldstone-Maxwell (GM) model discussed in section~\ssGM. As was explained there, the non-invariance of the five-dimensional anomaly-inflow action~$\sim \int_{\CM_5} B^{(2)}_\text{e} \wedge dB^{(2)}_\text{m}$ for the Maxwell field under 2-group shifts of~$B^{(2)}_\text{m}$ is cured by the fact that the model has a larger 3-group symmetry. This gives rise to additional background gauge fields and a modified anomaly-inflow action~\threegpinfl\ in five dimensions. By analogy with conventional Maxwell theory (see the discussion around~\eqref{bbtopbulk} and in~\cite{Maldacena:2001ss}), it is plausible that a gauged version of this modified anomaly-inflow action furnishes a bulk representation of the GM model in terms of singleton edge modes, but we have not checked this in detail.

\subsec{Dimensional Reduction of 2-Group Symmetries}[ssecDimRed]

When a~$U(1)_B^{(1)}$ global symmetry in four dimensions is reduced to three dimensions, it splits into a 1-form symmetry and a 0-form symmetry~\cite{Gaiotto:2014kfa}. Here we will briefly examine what happens when~$U(1)_B^{(1)}$ is part of an abelian 2-group~$U(1)_A^{(0)} \times_{\hat \kappa_A }U(1)_B^{(1)}$.\foot{~We would like to thank N.~Seiberg for asking a question that led to the comments in this subsection.} We phrase the discussion in terms of the four-dimensional 2-group background gauge fields~$A_\mu$ and~$B_{\mu\nu}$. If we reduce along the~$x^4$ direction (and drop all~$\d_4$ derivatives), $A_\mu$ splits into a three-dimensional 1-form gauge field~$A_i~(i = 1,2,3)$, with gauge symmetry~$A_i \rightarrow A_i +  \d_i \lambda_A$ and field strength~$F^A_{ij} = \d_i A_j - \d_j A_i$, as well as a gauge-invariant scalar~$A_4$. If~$B_{\mu\nu}$ is a conventional 2-form gauge field, it splits into conventional 2- and 1-form gauge fields~$B_{ij}$ and~$B_{i 4}$ in three dimensions. This reflects the splitting of the four-dimensional~$U(1)_B^{(1)}$ symmetry into separate 1-form and 0-form symmetries that was mentioned above. 

However, if~$B_{\mu\nu}$ undergoes a 2-group shift under~$U(1)_A^{(0)}$ gauge transformations, the three-dimensional fields~$B_{ij}$ and~$B_{i4}$ transform as follows,
\eqna{
& B_{ij} \quad \longrightarrow \quad B_{ij} + \d_i \Lambda^B_{j} - \d_j \Lambda^B_i + {\hat \kappa_A \over 2 \pi} \, \lambda_A  \, F^A_{ij}~, \cr
& B_{i4} \quad \longrightarrow \quad B_{i4} + \d_i \Lambda^B_4 + {\hat \kappa_A \over 2 \pi} \, \lambda_A \, \d_i A_4~.
}[btwgpred]
Here the 1-form gauge parameter~$\Lambda^B_{\mu}$ in four dimensions splits into a three-dimensional 1-form~$\Lambda_i^B$ and a scalar~$\Lambda^B_4$. The first line of~\btwgpred\ shows that the three-dimensional 2-form gauge field~$B_{ij}$ inherits the 2-group shift of~$B_{\mu\nu}$ from four dimensions. By contrast, the term proportional to~$\hat \kappa_A$ in the~$B_{i4}$ gauge transformation on the second line of~\btwgpred\ can be removed by redefining the three-dimensional fields and their gauge symmetries as follows,
\eqn{
\t B_i = B_{i4} + {\hat \kappa_A \over 2 \pi} \, A_4 A_i~, \qquad \lambda_{\t B} = \Lambda^B_4 + {\hat \kappa_A \over 2 \pi} \, \lambda_A \, A_4~.
}[fieldredef]
Then~$\t B_i \rightarrow \t B_i + \d_i \lambda_{\t B}$ transforms like a conventional 1-form gauge field in three dimensions. Note that the field redefinition in~\fieldredef\ has no Lorentz-invariant uplift to four dimensions. We conclude that the reduction of~$U(1)_A^{(0)} \times_{\hat \kappa_A} U(1)_B^{(1)}$ 2-group symmetry to three dimensions leads to the same 2-group symmetry, as well as another, conventional 0-form symmetry associated with the background gauge field~$\t B_i$ in~\fieldredef. 

It is also interesting to go from a four-dimensional theory with~$U(1)_A^{(0)} \times_{\hat \kappa_A} U(1)_B^{(1)}$ 2-group symmetry to a two-dimensional theory. One way to do this involves dimensional reduction, as above. Alternatively we can compactify the four-dimensional theory on~$\R^2 \times \Sigma_2$, where~$\Sigma_2$ is a compact Riemann surface. We can consider sectors of fixed~$U(1)_B^{(1)}$ charge~$q_B$, i.e.~with~$\int_{\Sigma_2} * J_B^{(2)} = q_B$, by adding a counterterm~$\sim q_B \int B^{(2)} \wedge \text{vol}(\Sigma_2)$ and integrating over the background gauge field~$B^{(2)}$, which acts as a chemical potential for the~$U(1)_B^{(1)}$ symmetry. Due to the 2-group shift of~$B^{(2)}$, the resulting two-dimensional theory on~$\R^2$ has a~$U(1)_A^{(0)}$ 't Hooft anomaly~$\sim q_B \hat \kappa_A \int_{{\mathbb R}^2} \lambda_A^{(0)} F_A^{(2)}$. Alternatively, as explained in section~\ref{SsecGauging}, integrating over~$B^{(2)}$ transforms the theory into its parent with~$U(1)_A^{(0)} \times U(1)_C^{(0)}$ global symmetry and a mixed~$\kappa_{A^2 C}$ 't Hooft anomaly. From this point of view, the counterterm fixes the flux of the~$U(1)_C^{(0)}$ background field strength through~$\Sigma_2$ to be~$q_B$. In this flux sector, the 't Hooft anomaly in two dimensions arises from the four-dimensional~$\kappa_{A^2 C}$ anomaly via
\eqn{
\kappa_{A^2 C} \int_{{\mathbb R}^2 \times \Sigma_2} \lambda_A^{(2)} \, F_A^{(2)} \wedge F_C^{(2)} \;  \sim  \; q_B \kappa_{A^2 C} \int_{{\mathbb R}^2}  \lambda_A^{(0)} \, F_A^{(2)}~.
}[tdfdfluxes]
A different relation between 2-group symmetries in four dimensions and 't Hooft anomalies in two dimensions appears in subsection~\ref{SecDefects} below.

\subsec{2-Group Symmetries, Strings, and Line Defects}[SecDefects]

As reviewed in section~\ref{convgsbf}, the basic objects that are charged under a~$U(1)_B^{(1)}$ symmetry are dynamical strings and line defects. Here we will examine what happens to these objects if~$U(1)_B^{(1)}$ is part of a 2-group, starting with dynamical strings and an abelian 2-group~$U(1)_A^{(0)} \times_{\hat \kappa_A} U(1)_B^{(1)}$. Consider a string that extends along its two-dimensional worldsheet~$\Sigma_2$, and let~$\Sigma'_2$ be the transverse two-dimensional space. The presence of the string is characterized by the 1-form string charge~$q_B = \int_{\Sigma'_2} * J_B^{(2)}$. We can now take the non-conservation equation~\eqref{modconseq} that characterizes the 2-group,
\eqn{
d * j_A^{(1)} = {\hat \kappa_A \over 2 \pi} \, F_A^{(2)} \wedge * J_B^{(2)}~,
}[twgpnonconsbisbis]
and integrate it over the transverse directions~$\Sigma'_2$ to obtain the following equation on the string world sheet~$\Sigma_2$,\foot{~\label{ft:2daanom} In two-dimensions, a complex Weyl fermion of chirality~$\sigma = \pm1$ and~$U(1)_A^{(0)}$ charge~$q$ yields an 't Hooft anomaly~$\kappa_{A^2}^\text{2d} = \sigma q^2$. The corresponding non-conservation equation takes the form~$d * j_A^{(0)} = -{\kappa^\text{2d}_{A^2} \over 4 \pi} \, F_A^{(2)}$.}
\eqn{
d* j_A^{(1)}\big|_{\Sigma_2} = -{ \kappa_{A^2}^\text{2d} \over 4\pi} \, F_A^{(2)}~, \qquad  \kappa^\text{2d}_{A^2} = - 2 \, \hat \kappa_A \, q_B~, \qquad q_B = \int_{\Sigma'_2} * J_B^{(2)}~.
}[wshooft]
The world-sheet theory therefore has a two-dimensional~$U(1)_A^{(0)}$ 't Hooft anomaly~$\kappa_{A^2}^\text{2d}$, which is determined by the 2-group structure constant~$\hat \kappa_A$ of the four-dimensional theory and the total string charge~$q_B$. Similarly, a nonabelian 2-group induces a world-sheet 't Hooft anomaly for the nonabelian flavor symmetry, while a Poincar\'e 2-group~$\SP \times_{\hat \kappa_\SP} U(1)_B^{(1)}$ leads to a non-zero gravitational anomaly on the string world sheet,\foot{~\label{ft:2dclcrdef} A complex fermion of chirality~$\sigma = \pm 1$ contributes~$c_L - c_R = \sigma$.}
\eqn{
c_L - c_R = - 6 \, \hat \kappa_\SP \, q_B~.
}[wsgravanom]

Many of the theories with 2-group symmetry discussed in section~\ref{SecExamples} have solitonic strings charged under~$U(1)_B^{(1)}$. For instance, the~$\C\P^N$ models in section~\ssCPN\ have skyrmion strings described by maps of the transverse plane~$\Sigma'_2$ into the~$\C\P^N$ target space that wrap the non-trivial 2-cycle~$\C\P^1 \subset \C\P^N$. The string charge~$q_B$ is the degree of this map. Similarly, in sections~\sssQEDab\ and~\SubsecFermats, we considered~$U(1)_c^{(0)}$ gauge theories with Higgs fields, which admit ANO strings. Here the string charge is given by the magnetic flux in the transverse plane, $q_B = {1 \over 2 \pi} \int_{\Sigma'_2} f_c^{(2)}$. The general formulas~\wshooft\ and~\wsgravanom\ apply to all of these examples. We will now show this more explicitly for examples with ANO strings, by examining the fermion zero modes on the string. Our discussion is similar to that in~\cite{Witten:1984eb}, except that we take~$U(1)_A^{(0)}$ to be a flavor symmetry, rather than a gauge symmetry. 

As in sections~\sssQEDab\ and~\SubsecFermats, we consider examples with a single Higgs field~$\phi$. For simplicity, we take its charges under the~$U(1)_c^{(0)}$ gauge and the~$U(1)_A^{(0)}$ flavor symmetry to be~$q_C = -1$ and~$q_A = 0$.\foot{~Once we assume that~$q_C = -1$, we can always redefine~$U(1)_A^{(0)}$ by an integer multiple of the~$U(1)_c^{(0)}$ gauge symmetry to set~$q_A  = 0$ (see the discussion around~\aglcgaugshift).} If we add a suitable potential~$V_H(\phi)$ (see~\higsspot), then~$\phi$ acquires a vev~$\langle \phi\rangle = v$, and the~$U(1)_c^{(0)}$ gauge symmetry is higgsed. Consider an ANO string of~$U(1)_B^{(1)}$ charge (i.e.~magnetic flux)~$q_B$ stretched along~$x^3$, and located at~$x^1 = x^2 = 0$ in the transverse plane, for which we introduce a complex coordinate~$z = x^1 + ix^2 = |z|e^{i\theta}$. The profile of the~$U(1)_c^{(0)}$ gauge field~$c^{(1)}$ and the Higgs field~$\phi$ in the~$z$-plane takes the following asymptotic form, which is valid far away from the string,
\begin{equation}\label{vortexi}
c_z = \frac{q_B}{2i z} +\cdots~,  \qquad \phi = v e^{i q_B \theta} +\cdots~, \qquad |z|\to \infty~.
\end{equation}
The phase of the Higgs field has monodromy~$2 \pi q_B$ as we traverse a large~$S^1$ in the~$z$-plane.

Now assume that the theory has chiral fermions, which we separate into~$\psi^i$ and~$\t \psi^i$ for notational purposes. We denote their~$U(1)_c^{(0)}$ charges by~$q_c^i, \, \t q_c^i \in \Z$, but take their~$U(1)_A^{(0)}$ charges~$q_A^i$ and~$\t q_A^i = - q_A^i$ to be equal and opposite. The Yukawa couplings then take the following schematic form, 
\eqn{
\SL_\text{Yukawa} = \sum_i \, \lambda_i \, \phi^{q_c^i + \t q_c^i} \, \psi^i \t \psi^i + (\text{c.c.})~. 
}[anoyuk]
Since~$\phi$ is neutral under~$U(1)_A^{(0)}$ and~$q_A^i + \t q_A^i = 0$, these Yukawa couplings preserve the~$U(1)_A^{(0)}$ flavor symmetry, as well as the gauge symmetry. The expression in~\anoyuk\ is only valid if~$q_c^i + \t q_c^i \geq 0$. In every term where~$q_c^i + \t q_c^i <0$, we must replace~$\phi^{q_c^i + \t q_c^i} \rightarrow (\b \phi)^{-(q_c^i + \t q_c^i)}$ to ensure that only positive powers of~$\phi$, $\b \phi$ appear. Once~$\phi$ acquires a vev, $\langle \phi \rangle = v$, all fermions in~\anoyuk\ are massive. 

In the presence of the string, the fermions~$\psi^i$ and~$\t \psi^i$ have normalizable zero modes~\cite{Jackiw:1981ee,Weinberg:1981eu}, which propagate on the string worldsheet but are localized in the transverse direction.\foot{~The Yukawa couplings~\anoyuk\ ensure that the zero modes of~$\psi^i$ and~$\t \psi^i$ decay exponentially rapidly, as~$\sim \exp\left(- v|\lambda_i|  |z|\right)$, at large  transverse distances,~$|z| \rightarrow \infty$, and are therefore normalizable.}  By examining the index of the world-sheet Dirac operator, as well as from other considerations (see for instance~\cite{Jackiw:1981ee,Weinberg:1981eu}), it can be shown that the number of zero modes, and their chirality, is determined by the string charge~$q_B$ and the gauge charges~$q_c^i, \t q_c^i$ of the bulk fermions. Explicitly, $\psi^i$ has~$| q_B q_c^i|$ zero modes of chirality~$\sigma_i = \text{sgn} ( q_B q_c^i)$ and~$\t \psi^i$ has~$|q_B \t q_c^i|$ zero modes of chirality~$\t \sigma_i = \text{sgn} (q_B \t q_c^i)$.\foot{~Here~$\sigma = +1$ indicates left-movers and~$\sigma = -1$ corresponds to right-movers.} Since the zero modes carry the same~$U(1)_A^{(0)}$ flavor charges~$q_A^i$ and~$\t q_A^i = -q_A^i$ as their parent fermions, we can directly evaluate the corresponding 't Hooft anomaly on the string world sheet (see footnote~\ref{ft:2daanom}),
\eqn{
\kappa_{A^2}^\text{2d} = \sum_i \, (q_A^i)^2 \left( |q_B q_c^i| \sigma_i +  |q_B \t q_c^i| \t \sigma_i\right) = q_B \sum \, (q_A^i)^2 \left(q_c^i + \t q_c^i\right) = q_B \kappa_{A^2 C} = - 2 \, \hat \kappa_A \, q_B~.
}[wsanomzm]
Here we have used the expression~$\hat \kappa_A = - \half \kappa_{A^2 C}$ for the 2-group structure constant (see~\coeffmatch), which leads to agreement between~\wsanomzm\ and the general formula~\wshooft\ for the world-sheet 't Hooft anomaly that was derived on the basis of 2-group symmetry. Similarly, the gravitational anomaly on the string world sheet can be computed as follows (see footnote~\ref{ft:2dclcrdef}), 
\eqn{
c_L - c_R = \sum_i \, \left( |q_B q_c^i| \sigma_i +  |q_B \t q_c^i| \t \sigma_i\right) = q_B \sum_i q_c^i = q_B \kappa_{C \SP^2} = - 6 \, \hat \kappa_\SP \, q_B~,
}[clcrws]
in agreement with~\wsgravanom. Here we have used the relation~$\hat \kappa_\SP = -{1 \over 6} \kappa_{C \SP^2}$ from~\nonabgravkhat. 

In the~$U(1)_c^{(0)}$ gauge theory examples discussed above, the line defects that carry magnetic 1-form charge~$U(1)_B^{(1)}$ are 't Hooft lines. In the Higgs phase, an 't Hooft line extended along the time direction can serve as an endpoint for ANO strings. Similarly, a spatially extended 't Hooft line at a fixed moment in time creates an ANO string. As discussed above, these strings have chiral zero modes and 't Hooft anomalies on their worldsheets. One might suspect that this leads to some unusual, or even pathological, features of 't Hooft lines in theories with 2-group symmetry, but we are not aware of any such pathologies. 

To make this more concrete, we consider 't Hooft lines in the Goldstone-Maxwell (GM) model (see section~\ssGM). In the absence of background fields, the model reduces to a free NG boson~$\chi$ and a free Maxwell field~$f_c^{(2)}$. An 't Hooft line~$H_n(L)$ of integer charge~$n \in \Z$ can then be written as an open surface operator (see appendix~\ref{appHooftFreeMax}), 
\eqn{
H_n(L) = \exp\left({ 2 \pi n \over e^2} \int_{\Sigma_2} * f_c^{(2)}\right)~.
}[la]
Here~$\Sigma_2$ is a 2-cycle with boundary~$\d \Sigma_2 = L$. It can be viewed as the worldsheet of an unobservable Dirac string that is needed to properly define the magnetic monopole singularity characterizing the 't Hooft defect. Note that~\la\ does not depend on the choice of~$\Sigma_2$, because the source-free Maxwell equations imply~$d * f_c^{(2)} = 0$. Since~$n \in \Z$, this remains true in the presence of Wilson lines with electric charge~$m \in \Z$ (see appendix~\ref{appHooftFreeMax}).

In the presence of the 2-group background fields~$A^{(1)}$ and~$B^{(2)}$, the action of the GM model takes the form~\gmdef, which leads to the following equation of motion for~$f_c^{(2)}$,
\eqn{
d \left({1 \over e^2} * f_c^{(2)} + {i \over 2 \pi} \Big(B^{(2)} - {\hat \kappa_A \over 2 \pi} \, \chi \, F_A^{(2)}\Big) \right) = 0~.
}[bggmeom]
We can therefore modify the definition of the 't Hooft line in~\la\ as follows,
\eqn{
H_n(L) = \exp\left({2 \pi n \over e^2} \int_{\Sigma_2} * f_c^{(2)} + i n \int_{\Sigma_2} \Big(B^{(2)} - {\hat \kappa_A \over 2 \pi} \, \chi \, F_A^{(2)}\Big) \right)~.
}[labis]
Note that this reduces to~\la\ if the background fields~$A^{(1)}$ and~$B^{(2)}$ are set to zero. It follows from~\bggmeom\ that~\labis\ does not depend on the choice of~$\Sigma_2$, and as before, this even remains true in the presence of integer-charge Wilson lines. The term~$\sim \int_{\Sigma_2} \chi \, F_A^{(2)}$ that involves the NG boson induces a~$U(1)_A^{(0)}$ 't Hooft anomaly on the Dirac-string world sheet~$\Sigma_2$, but this anomaly is cancelled by the 2-group shift of the surface counterterm~$\sim \int_{\Sigma_2} B^{(2)}$. Without this cancellation, it would be possible to detect the Dirac string, and~\labis\ would not define a genuine line operator. As in conventional Maxwell theory, the surface counterterm also ensures that~$H_n(L)$ transforms with the correct charge~$n$ under the magnetic~$U(1)_B^{(1)}$ symmetry (see appendix~\ref{appHooftFreeMax}).


\ack{\smallskip We are grateful to~F.~Benini, L.~Bhardwaj, S.~Gukov, E.~D'Hoker, D.~Harlow, P.-S.~Hsin, Z.~Komargodski, N.~Seiberg, K.~Ohmori, H.~Ooguri, Y.~Tachikawa, J.~Trnka, and E.~Witten for helpful discussions. C.C. is supported by the Marvin L.~Goldberger Membership at the Institute for Advanced Study, and DOE grant de-sc0009988. The work of TD is supported by the National Science Foundation under grant number PHY-1719924, and by the John Templeton Foundation under award number 52476. KI is supported by DOE grant DE-SC0009919 and the Dan Broida Chair.
}


\begin{appendices}

\newsec{Quantization of Some 't Hooft Anomaly Coefficients}[AppQuantAnom]

Consider two abelian flavor symmetries and Poincar\'e symmetry,\foot{~It is straightforward to extend the arguments in this appendix to more general flavor symmetries.}
\eqn{
U(1)_A^{(0)} \times U(1)_C^{(0)} \times \SP~.
}[gsgp]
The possible 't Hooft anomaly coefficients are~$\kappa_{A^3}$, $\kappa_{A^2 C}$, $\kappa_{AC^2}$, $\kappa_{C^3}$, $\kappa_{A \SP^2}$, and~$\kappa_{C \SP^2}$. In theories of free fermions~$\psi^i_\alpha$ with~$U(1)_A^{(0)} \times U(1)_C^{(0)}$ charges~$q_A^i$ and~$q_C^i$, they are given by
\eqna{
&\kappa_{A^3} = \sum_i (q^i_A)^3~, \qquad \kappa_{A^2 C} = \sum_i (q^i_A)^2 q_C^i~,\qquad  \kappa_{A C^2} = \sum_i q^i_A (q_C^i)^2~,\cr
& \kappa_{C^3} = \sum_i (q^i_C)^3~,\qquad  \kappa_{A \SP^2} = \sum_i q_A^i~, \hskip46pt \kappa_{C \SP^2} = \sum_i q_C^i~.
}[fermks]
Since the flavor symmetries are compact, all charges are integers, $q^i_A, q_C^i\in \Z$, and hence the same is true of the various~$\kappa$'s in~\fermks. 

The anomaly coefficients are not completely independent; they satisfy the constraints
\eqn{
\kappa_{A^3} \equiv \kappa_{A\SP^2} ~(\mod~6)~, \qquad \kappa_{C^3} \equiv \kappa_{C\SP^2} ~(\mod~6)~, \qquad \kappa_{A^2 C} \equiv \kappa_{AC^2} ~(\mod~2)~, 
}[kappacons]
To see this, reduce the formulas in~\fermks\ $\mod~2$ and~$\mod~3$. Since~$(q_A^i)^3$ is odd if and only if~$q_A^i$ is odd, we have~$(q_A^i)^3 \equiv q_A^i~(\mod~2)$. The same result is true~$\mod~3$ (this can be checked by examining the cases~$q_A^i \equiv 0,1,2~(\mod~3)$ in turn), and hence also~$\mod~6$. Summing over charges gives the first constraint in~\kappacons, and replacing~$A \rightarrow C$ gives the second one. Finally, note that~$(q_A^i)^2 q_C^i$ and~$q_A^i (q_C^i)^2$ are both even, and hence vanish~$(\mod~2)$, unless~$q_A^i$, $q_C^i$ are both odd, in which case they are both equal to~$1~(\mod~2)$.  Summing over charges establishes the third constraint in~\kappacons.  In this paper, we have often assumed that 
\eqn{
\kappa_{C^3} = \kappa_{AC^2} = 0~,
}[kvanish]
so that~$U(1)_C^{(0)}$ can be gauged without ruining~$U(1)_A^{(0)}$ through an ABJ anomaly. Together with the general constraints in~\kappacons, this assumption leads to stronger quantization conditions for the anomaly coefficients~$\kappa_{A^2 C}$ and~$\kappa_{C \SP^2}$,
\eqn{
\kappa_{A^2 C}  \in 2 \Z~, \qquad \kappa_{C \SP^2} \in 6 \Z~.
}[strongkquant]

We will now show that the anomaly coefficients are integers satisfying~\kappacons\ without appealing to free fermions. This can be argued in a variety of ways (see for instance~\cite{AlvarezGaume:1984dr} for additional details). Here we will do so from the perspective of the five-dimensional action~$S_5$ for the~$U(1)_A^{(0)}$ and~$U(1)_C^{(0)}$ background gauge fields~$A^{(1)}$,  $C^{(1)}$ and the spin connection~$\omega^{(1)}$ that gives rise to the anomalies via inflow. As explained in section~\ref{ssecAnomPol}, $S_5 = 2 \pi i \int \CI^{(5)}$ consists of various Chern-Simons terms that arise from the anomaly 6-form polynomial~$\CI^{(6)}$ via the descent equation~$\CI^{(6)} = d \CI^{(5)}$. This anomaly polynomial takes the form 
\eqn{
\CI^{(6)} = {\kappa_{A^3} \over 6} \,\mathsf X^{(6)}_{A A A} + {\kappa_{A^2 C} \over 2} \, \mathsf X^{(6)}_{AAC} + {\kappa_{A C^2} \over 2} \, \mathsf X^{(6)}_{ACC} + {\kappa_{C^3} \over 6} \, \mathsf X^{(6)}_{CCC} + \kappa_{A \SP^2} \, \mathsf Y^{(6)}_{A} + \kappa_{C \SP^2} \, \mathsf Y^{(6)}_C~.
}[isixxydef]
Here we have defined the following wedge products of Chern and Pontryagin densities,
\eqn{
\mathsf X^{(6)}_{IJK}  = {1\over (2 \pi)^3} \, F_I^{(2)} \wedge F_J^{(2)} \wedge F_K^{(2)}~, \qquad   \mathsf Y^{(6)}_I = {1 \over 384 \pi^3} \, F_I^{(2)} \wedge \tr\left(R^{(2)} \wedge R^{(2)}\right)~,
}[xyijkdef]
with~$I, J, K \in \{A, C\}$. Applying descent to~\isixxydef, and choosing the counterterms as in~\ctfix\ and~\gravdescff, leads to the following Chern-Simons terms in five dimensions,
\eqna{
& S_5[A^{(1)}, C^{(1)}, \omega^{(1)}] = 2 \pi i \int_{\CM_5} \CI^{(5)} = {i \kappa_{A^3} \over 24 \pi^2} \int_{\CM_5}  A^{(1)} \wedge F_A^{(2)} \wedge F_A^{(2)} \cr
& + {i \kappa_{A^2 C}\over 8 \pi^2} \int_{\CM_5}    A^{(1)} \wedge F_A^{(2)} \wedge F_C^{(2)}  + {i \kappa_{A C^2}\over 8 \pi^2} \int_{\CM_5}  A^{(1)} \wedge F_C^{(2)} \wedge F_C^{(2)} + {i \kappa_{C^3}\over 24 \pi^2} \int_{\CM_5}  C^{(1)} \wedge F_C^{(2)} \wedge F_C^{(2)}\cr
& + {i \kappa_{A \SP^2} \over 192 \pi^2} \int_{\CM_5} \text{CS}^{(3)}(\omega) \wedge F_A^{(2)} + {i \kappa_{C \SP^2} \over 192 \pi^2} \int_{\CM_5} \text{CS}^{(3)}(\omega) \wedge F_C^{(2)}~.  
}[infloactapp]
Here~$\text{CS}^{(3)}(\omega)$ is the gravitational Chern-Simons 3-form defined in~\gravcs. Demanding that the Chern-Simons terms in~\infloactapp\ are well defined on any oriented five-manifold~$\CM_5$ with a spin structure, and for arbitrary~$U(1)_A^{(0)}$ and~$U(1)_C^{(0)}$ connections, leads to quantization conditions for their coefficients.\foot{~If~$\CM_5$ is not spin, there are more stringent quantization conditions than those discussed below.} 

One way to see this involves extending~$\CM_5$ to a oriented, spin six-manifold~$\CM_6$, with boundary~$\d \CM_6 = \CM_5$. Similarly, the connections~$A^{(1)}, C^{(1)}$ are also extended over~$\CM_6$. We can then define~$S_5 = 2 \pi i \int_{\CM_6} \CI^{(6)}$, where~$\CI^{(6)} = d\CI^{(5)}$ is the 6-form anomaly polynomial. In general, different six-dimensional extensions can lead to to different answers for~$S_5$. In order to ensure that~$S_5$ only depends on five-dimensional data, we demand that all extensions give the same answer, up to integer multiples of~$2 \pi i$. By a standard argument, which involves gluing two different extensions along~$\CM_5$, this translates into the requirement that
\eqn{
\int_{\CM_6} \CI^{(6)}\in \Z~,
}[anompolquant]
for any closed, oriented six-manifold~$\CM_6$ with a spin structure. 

We must therefore determine the integrality properties of the 6-forms~$\mathsf X^{(6)}_{IJK}$ and~$\mathsf Y^{(6)}_I$ defined in~\xyijkdef\ that appear in the anomaly polynomial~\isixxydef. Since~$\CM_6$ is spin, these are constrained by the Atiyah-Singer index theorem. Let~$\slashed D$ be the spin-$\half$ Dirac operator on~$\CM_6$ that couples to the connections~$A^{(1)}$, $C^{(1)}$ with charges~$q_A, q_C \in \Z$. The Atiyah-Singer theorem states that the index of~$\slashed D$, which is necessarily  an integer, is given by 
\eqn{
\mathsf I(\slashed D) = \int_{\CM_6} \hat A \, \exp\left({1 \over 2 \pi} (q_A \, F_A^{(2)} + q_C \, F_C^{(2)}) \right)~. 
}[indexthm]
In the conventions of~\cite{AlvarezGaume:1984dr}, the~$\hat A$-genus is given by~$\hat A = 1 + {1 \over 192 \pi^2} \tr\left(R^{(2)} \wedge R^{(2)}\right) + \cdots$, so that~\indexthm\ has the following expansion in terms of~$\mathsf X^{(6)}_{IJK}$ and~$\mathsf Y^{(6)}_I$ (see~\xyijkdef),
\eqna{
\mathsf I(\slashed D) =~& {q_A^3 \over 6  } \int_{\CM_6} \mathsf X^{(6)}_{AAA} + {q_A^2 q_C \over 2 } \int_{\CM_6} X^{(6)}_{AAC}  + {q_A q_C^2 \over 2 } \int_{\CM_6} X^{(6)}_{ACC} + {q_C^3 \over 6 } \int_{\CM_6} X^{(6)}_{CCC}  \cr
& +  {q_A} \int_{\CM_6} \mathsf Y^{(6)}_A + {q_C } \int_{\CM_6} \mathsf Y^{(6)}_C~.
}[indexexp]
Since~$\mathsf I(\slashed D) \in\Z$, we obtain various quantization conditions by choosing different~$q_A$ and~$q_C$:
\begin{itemize}
\item If~$q_A = 1$ and~$q_C = 0$, or vice versa, we find that
\eqn{
 {1 \over 6 } \int_{\CM_6} \mathsf X^{(6)}_{III} +   \int_{\CM_6} \mathsf Y^{(6)}_{I} \in \Z~, \qquad  I \in \{A, C\}~.
}[aaagquant]

\item If we choose~$q_A = q_C = 1$ and subtract the integer combination in~\aaagquant\ for both~$I = A$ and~$I = C$, we find that 
\eqn{
 {1 \over 2 } \int_{\CM_6} \left( \mathsf X^{(6)}_{AA C} + \mathsf X^{(6)}_{ACC} \right) \in \Z~.
}[aacaccquant]
\end{itemize}

Independently of the index theorem, $\mathsf X^{(6)}_{IJK}$ has integer periods (even if~$\CM_6$ is not spin),
\eqn{
\int_{\CM_6} \mathsf X^{(6)}_{IJK} \in \Z~, \qquad I, J, K \in \{A,C\}~. 
}[intperiods]
This is because~$X^{(6)}_{IJK}$ was defined as a product of Chern classes, $c_1(F^{(2)}_{A,C}) = {1 \over 2 \pi} \, F_{A,C}^{(2)}$, in~\xyijkdef. Together with~\eqref{aaagquant}, the constraint~\intperiods\ implies that
\eqn{
6 \int_{\CM_6} Y^{(6)}_{I} \in \Z~, \qquad I \in \{A, C\}
}[sixyint]
By combining the quantization conditions~\eqref{aaagquant}, \eqref{aacaccquant}, \eqref{intperiods}, and~\sixyint, we find that the most general anomaly polynomial~$\CI^{(6)}$ that satisfies~\anompolquant\ is given by
\eqna{
\CI^{(6)} =~&   \sum_{I \in \{A, C\} } \left(\ell_I X^{(6)}_{III} + 6 m_I Y^{(6)}_I + n_I \, \Big({1 \over 6} \, X^{(6)}_{III} + Y_I^{(2)}\Big)\right) \cr
& + p \, \mathsf X^{(6)}_{AAC} + q \, \mathsf X^{(6)}_{ACC} + {r \over 2} \, \left( \mathsf X^{(6)}_{AA C} + \mathsf X^{(6)}_{ACC} \right)~, \qquad \ell_I, m_I, n_I, p, q, r \in \Z~.
}[intlincomb]
Comparing with~\isixxydef, we see that the anomaly coefficients can be expressed as
\eqna{
& \kappa_{A^3} = 6 \ell_A + n_A~, \qquad \kappa_{A^2C} = 2 p + r~, \hskip45pt \kappa_{AC^2} = 2 q + r~,\cr
& \kappa_{C^3} = 6 \ell_C + n_C~, \qquad \kappa_{A \SP^2} = 6 m_A + n_A~, \qquad \kappa_{C \SP^2} = 6 m_C + n_C~.
}[anomcoeff]
This implies the constraints in~\kappacons, and if we assume~\kvanish, also those in~\strongkquant.

\newsec{Select Current Correlation Functions in Momentum Space}[AppMomSpace]

In this appendix, we analyze several two- and three-point correlation functions of~$1$-form and~$2$-form currents that are needed in the main text (mostly in sections~\ref{SecCurrents} and~\ref{SecGSAnom}). Working in four-dimensional, euclidean momentum space, we present the decomposition of these correlators into Lorentz-invariant structure functions. We then use this decomposition to discuss some properties of interest, including possible 't Hooft anomalies.

As in footnote~\ref{momcorrdef}, our conventions are that the momentum-space two-point function~$\langle \CA(p) \CB(-p)\rangle$ of two local operators~$\CA(x), \CB(x)$ is given by
\eqn{
\langle \CA(p) \CB(-p)\rangle = \int d^4 x \, e^{-i  p \cdot x} \, \langle \CA(x) \CB(0)\rangle~,
}[momtwptdef]
while the momentum-space three-point function~$\langle \CA(p)\CB(q) \CC(-p-q)\rangle$ of three local operators~$\CA(x), \CB(y), \CC(z)$ is defined by
 \eqn{
\langle \CA(p)\CB(q) \CC(-p-q)\rangle = \int d^4x \, d^4 y \, e^{-i(p \cdot x + q \cdot y)} \, \langle \CA(x) \CB(y) \CC(0)\rangle~.
}[momthreeptdef]

\subsec{The~$\langle J_B^{(2)} J_{B'}^{(2)}\rangle$ Two-Point Function}[ssjbjbtwopt]

We first consider the two-point function~$\langle J^B_{\mu\nu}(p) J^{B'}_{\alpha\beta}(-p)\rangle$ of two distinct~$2$-form currents~$J^B_{\mu\nu} = J^B_{[\mu\nu]}$ and~$J^{B'}_{\alpha\beta} = J^{B'}_{[\alpha\beta]}$. In position space, their mass dimension is~$[J^B_{\mu\nu}] = [J^{B'}_{\alpha\beta}] =2$, and hence the momentum-space two-point function is dimensionless. Prior to imposing any conservation equations, the most general Lorentz structures that can appear are given by
\eqna{
\langle J^{B}_{\mu\nu}(p) J^{B'}_{\alpha\beta}(-p)\rangle =~& A(p^2) \ep_{\mu\nu\alpha\beta} + B(p^2) \left(p_\mu p_\alpha \delta_{\nu\beta} - p_\nu p_\alpha \delta_{\mu\beta} - p_\mu p_\beta \delta_{\nu\alpha} + p_\nu p_\beta \delta_{\mu\alpha} \right) \cr 
& + C(p^2) \left(\delta_{\mu\alpha} \delta_{\nu\beta} - \delta_{\mu\beta} \delta_{\nu\alpha}\right) + D(p^2) \left(\ep_{\mu\nu\alpha \rho} p^\rho p_\beta - \ep_{\mu\nu\beta\rho} p^\rho p_\alpha\right)~.
}[jjpstrfns]
Here~$A,B,C,D$ are Lorentz-invariant structure functions,\foot{~Note that another Lorentz structure proportional to~$\ep_{\alpha\beta\mu\rho} p^\rho p_\nu - \ep_{\alpha\beta\nu\rho} p^\rho p_\mu$ can be reduced to a linear combination of the~$A$ and~$D$ structures using the Schouten identity~$\ep_{[\mu\nu\alpha\beta} p_{\gamma]} = 0$. This is related to the discussion after~\eqref{bbpanompol} below.} whose mass dimensions are $[A] = [C] = 0$, $[B] = [D] = -2$.

In momentum space, the fact that~$\d^\mu J^B_{\mu\nu} = \d^\alpha J^{B'}_{\alpha\beta} = 0$ at separated points implies that
\eqn{
p^\mu \langle J^{B}_{\mu\nu}(p) J^{B'}_{\alpha\beta}(-p)\rangle~\sim~0~, \qquad p^\alpha \langle J^{B}_{\mu\nu}(p) J^{B'}_{\alpha\beta}(-p)\rangle~\sim~0~.
}[jjpconsmom]
Here the notation~$X \sim Y$ means that the expressions~$X$ and~$Y$ are equal, up to a polynomial expression in the momenta. Such polynomials correspond to~$\delta$-function contact terms in position space, or their derivatives. In the context of~\eqref{jjpconsmom}, they violate current conservation at coincident points and indicate a possible 't Hooft anomaly. 

If we apply~\eqref{jjpconsmom} to~\eqref{jjpstrfns}, we find that
\eqn{
A(p^2)~\sim~0~, \qquad C(p^2)~\sim~-p^2 B(p^2)~, \qquad p^2D(p^2)~\sim~0~.
}[jjpccsfimp]
By tuning local counterterms in the background fields~$B^{(2)}$ and~$B'^{(2)}$ that couple to the conserved currents~$J_B^{(2)}$ and~$J^{(2)}_{B'}$, we can adjust the contact terms in their two-point function.\foot{~Occasionally, some physical principle may restrict the space of allowed counterterms, and hence the freedom to adjust or remove certain contact terms in correlation functions. Some examples appear in sections~\ref{secGScts} and~\ref{sssHoftAbTwGp}.}   This amounts to shifting the structure functions in~\jjpstrfns\ by polynomials in~$p^2$. Using such shifts, we can set~$A(p^2) = 0$, $C(p^2) = - p^2 B(p^2)$, and~$D(p^2) = {i \kappa_{B B'} \over 4 \pi p^2}$, where~$\kappa_{B B'}$ is a dimensionless constant. (The normalization is for future convenience, see below.) Finally, using the fact that~$[B] = -2$, we can write~$B(p^2) = {1 \over p^2} {\mathsf J}\left({p^2 \over M^2}\right)$, where~${\mathsf J}$ is a dimensionless structure function and~$M$ is some mass scale. Substituting back into~\eqref{jjpstrfns}, we obtain
\eqna{
\langle J^B_{\mu\nu}(p) J^{B'}_{\alpha\beta}(-p)\rangle = {1 \over p^2} \,{\mathsf J}&\left({p^2 \over M^2}\right) \Big( p_\mu p_\alpha \delta_{\nu\beta} - p_\nu p_\alpha \delta_{\mu\beta} - p_\mu p_\beta \delta_{\nu\alpha} + p_\nu p_\beta \delta_{\mu\alpha} \cr
& - p^2 \delta_{\mu\alpha} \delta_{\nu\beta} + p^2 \delta_{\nu\alpha} \delta_{\mu\beta}\Big)  + {i \kappa_{B B'} \over 4 \pi p^2} \left(\ep_{\mu\nu\alpha\rho} p^\rho p_\beta - \ep_{\mu\nu\beta\rho} p^\rho p_\alpha\right)~.
}[jjpfinal]
Reflection positivity implies that the structure function~${\mathsf J}$ and the constant~$\kappa_{B B'}$ are real. 

The term proportional to~$\kappa_{B B'}$ in~\jjpfinal\ is annihilated by~$p^\mu$, but not by~$p^\alpha$, corresponding to a non-trivial polynomial on the right-hand side of the second equation in~\jjpconsmom,
\eqn{
p^\mu \langle J^B_{\mu\nu}(p) J^{B'}_{\alpha\beta}(-p)\rangle = 0~, \qquad p^\alpha \langle J^B_{\mu\nu}(p) J^{B'}_{\alpha\beta}(-p)\rangle = - {i \kappa_{B B'} \over 4 \pi} \, \ep_{\mu\nu\beta \rho} p^\rho~.
}[anomct]
This indicates a mixed 't Hooft anomaly between the two currents. At the level of the anomaly 6-form polynomial, it is captured by a term (see also the discussion in~\cite{Gaiotto:2014kfa}),\foot{~Applying descent leads to~$\CI^{(5)} = -{\kappa_{B B'}\over 4 \pi^2} B'^{(2)} \wedge dB^{(2)}$. From this it follows that the anomaly under~$U(1)_{B'}^{(1)}$ gauge transformations, parametrized by~$\Lambda_{B'}^\mu$, is given by~$\CA_{B'} = - {i \kappa_{B B'} \over 4 \pi} \int d^4 x \, \ep_{\mu\alpha\beta\gamma} \Lambda_{B'}^\mu \d^\alpha B^{\beta\gamma}$. This implies the non-conservation equation~$\d^\mu J^{B'}_{\mu\nu} = {i \kappa_{B B'} \over 4 \pi} \, \ep_{\nu\alpha\beta\gamma} \d^\alpha B^{\beta\gamma}$, and hence a contact term in~$\d_y^\alpha \langle J^B_{\mu\nu}(x) J^{B'}_{\alpha\beta}(y)\rangle = {i \kappa_{B B'} \over 4 \pi} \, \ep_{\mu\nu\beta\rho} \d^\rho \delta^{(4)}(x-y)$. In momentum space, this becomes the second equation in~\anomct.}
\eqn{
\CI^{(6)} \; \supset \; {\kappa_{B B'}  \over (2\pi)^2} \, d B^{(2)} \wedge dB'^{(2)}~. 
}[bbpanompol]
Given the choice of contact terms in~\jjpfinal, we find that~$J^B_{\mu\nu}$ is conserved at separated and coincident points, but conservation of~$J^{B'}_{\alpha\beta}$ is violated by contact terms. By adjusting the counterterm~$\int_{\CM_4} B^{(2)} \wedge B'^{(2)}$ we can redefine the contact terms so that~$J^B_{\mu\nu}$ is anomalous and~$J^{B'}_{\alpha\beta}$ is conserved. This is consistent with the general discussion of reducible anomalies after~\redanom. Finally, arguments analogous to those in appendix~\ref{AppQuantAnom} show that the anomaly coefficient~$\kappa_{B B'}$ is quantized, $\kappa_{B B'} \in \Z$, so that~${i \kappa_{B B'} \over 2 \pi} \int_{\CM_5} B'^{(2)} \wedge dB^{(2)}$ is invariant under large~$1$-form gauge transformations of~$B^{(2)}$, $B'^{(2)}$. The quantization condition implies that~$\kappa_{B B'}$ does not depend on continuously variable couplings, and that it is inert under RG flows. 

If the two currents~$J^B_{\mu\nu}$ and~$J^{B'}_{\alpha\beta}$ are identical (i.e~$B = B'$), there are additional constraints on~$\langle J^B_{\mu\nu}(p) J^{B'}_{\alpha\beta}(-p)\rangle$ from Bose symmetry, which exchanges
\eqn{
\mu\nu\, , \,p \quad \longleftrightarrow \quad \alpha\beta\, , \,-p~.
}[jjpbose]
In terms of the structure functions in~\jjpstrfns, such an exchange leaves~$B(p^2)$ and~$C(p^2)$ invariant, but shifts~$A(p^2) \rightarrow A(p^2)+p^2 D(p^2)$ and~$D(p^2) \rightarrow -D(p^2)$. Therefore, Bose symmetry sets~$D(p^2) = 0$, which in turn implies the vanishing of the mixed 't Hooft anomaly coefficient~$\kappa_{B B'} = 0$. The absence of an 't Hooft anomaly for a single~$2$-form current immediately follows from the anomaly polynomial~\bbpanompol, because~$dB^{(2)} \wedge dB'^{(2)}$ vanishes if~$B^{(2)} = B'^{(2)}$.

\subsec{The~$\langle J^{(2)}_B j^{(1)}_A\rangle$ Two-Point Function}[ssjbjatwopt]

We examine the two-point function~$\langle J^B_{\mu\nu}(p) j^A_\rho(-p)\rangle$ between a 2-form current~$J^B_{\mu\nu} = J^B_{[\mu\nu]}$ and a 1-form current~$j^A_\rho$. In position space, their mass dimensions are~$[J^B_{\mu\nu}] = 2$ and~$[j_\rho^A] = 3$. It follows that the momentum-space two-point function has mass dimension~$+1$. Before imposing the conservation laws, the decomposition into Lorentz structures takes the form
\eqn{
\langle J^B_{\mu\nu}(p) j^A_\rho(-p)\rangle = {\mathsf T} \left({p^2 \over M^2}\right) \left(\delta_{\mu\rho} p_\nu - \delta_{\nu\rho} p_\mu\right) -{1 \over 2 \pi} {\mathsf K} \left({p^2 \over M^2}\right) \ep_{\mu\nu\rho\lambda} p^\lambda~.
}[jbjalorst]
Here~${\mathsf T}, {\mathsf K} $ are dimensionless, Lorentz-invariant structure functions, and~$M$ is a mass scale. In accordance with section~\secGScts, the normalization of~${\mathsf K} $ is such that a properly quantized Green-Schwarz counterterm $S_\text{GS} = {i n \over 2 \pi} \int B^{(2)} \wedge F^{(2)}_A~(n \in \Z)$ shifts~${\mathsf K} \rightarrow {\mathsf K} + n$. As in the discussion around~\jjpconsmom, conservation of~$J^B_{\mu\nu}$ and~$j^A_\rho$ at separated points implies
\eqn{
p^\mu \langle J^B_{\mu\nu}(p) j^A_\rho(-p)\rangle~\sim~0~, \qquad p^\rho \langle J^B_{\mu\nu}(p) j^A_\rho(-p)\rangle~\sim~0~.
}[jbjacons]
Imposing these conditions on~\jbjalorst\ leads to~${\mathsf T}  \sim 0$, so that~${\mathsf T}$ is a polynomial in~$p^2$, which can be set to zero using local counterterms.

\subsec{The~$\langle j_A^{(1)} j_A^{(1)} J_B^{(2)}\rangle$ Three-Point Function}[ssjajaJBthreept]

Here we consider the three-point function~$\langle j^A_\mu(p) j^A_\nu(q) J^B_{\rho\sigma}(-p-q)\rangle$ of two identical 1-form currents~$j_\mu^A$ and~$j_\nu^A$, as well as a 2-form current~$J^B_{\mu\nu} = J^B_{[\mu\nu]}$. In position space, the currents have mass dimensions~$[j_A^{(1)}] = 3$ and~$[J_B^{(2)}] = 2$, and hence the momentum-space three-point function is dimensionless. Bose symmetry exchanges
\eqn{
\mu\, , \, p \quad \longleftrightarrow \quad \nu\, , \, q~.
}[bosejajaJB]
We would like to decompose the three-point function into Lorentz structures. Here we distinguish between parity-odd structures, which contain an~$\ep$-symbol, and parity-even structures, which do not. We will only consider the parity-even structures, since only these are needed in section~\ref{SecCurrents}. Moreover, we restrict the momenta to the Bose-symmetric locus
\eqn{
p^2 = q^2 = (p+q)^2 = Q^2~, \qquad p\cdot q = - \half Q^2~.
}[symkinii]
All structure functions only depend on~$Q^2$ and are therefore invariant under~\bosejajaJB. This simplifies the enumeration Lorentz structures that are compatible with Bose symmetry.

Before imposing conservation laws or Ward identities, the parity-even part of the three-point function can involve the following Lorentz structures, 
\eqna{
& \langle j^A_\mu(p) j^A_\nu(q) J^B_{\rho\sigma}(-p-q)\rangle = A(Q^2) \left(\delta_{\mu\rho} p_\nu p_\sigma - \delta_{\mu\sigma} p_\nu p_\rho + \delta_{\nu\rho} p_\mu p_\sigma - \delta_{\nu\sigma} p_\mu p_\rho\right) \cr
& + \left(A(Q^2) \rightarrow B(Q^2)~,~p \rightarrow q\right) + C(Q^2) \Big(\delta_{\mu\rho} p_\nu q_\sigma - \delta_{\mu\sigma} p_\nu q_\rho + \delta_{\nu\rho} q_\mu p_\sigma - \delta_{\nu\sigma} q_\mu p_\rho\Big) \cr
& + \left(C(Q^2) \rightarrow D(Q^2)~,~p \leftrightarrow q\right) + E(Q^2) \left(p_\mu p_\nu - q_\mu q_\nu\right)\left(p_\rho q_\sigma - p_\sigma q_\rho\right)~.
}[jajaJBstrfns]
Since the three-point function is dimensionless, the structure functions~$A, B, C, D, E$ have mass dimensions~$[A] = [B] = [C] = [D] = -2$ and~$[E] = -4$. We would now like to impose conservation of~$J^B_{\rho\sigma}$, as in~\jbconwimom, and the 2-group Ward identity~\momspward\ satisfied by~$j^A_\mu$. The former constraint leads to\foot{~Here, as in section~\ref{secCharThrPt}, we impose conservation equations and Ward identities at separated and coincident points. In momentum space, this means that these relations hold exactly, rather than up to polynomials in the momenta (see the discussion around~\jjpconsmom).}
\eqn{
A(Q^2) = -C(Q^2)~, \qquad B(Q^2) = -D(Q^2)~, \qquad A(Q^2) + B(Q^2) =  \half Q^2 E(Q^2)~,
}[consone]
while the latter one imposes the following relation,
\eqn{
A(Q^2) + 2 B(Q^2) = - {\hat \kappa_A \over 2 \pi Q^2} \, {\mathsf J} \left({Q^2 \over M^2}\right)~. 
}[constwo]
Here~${\mathsf J}$ is the structure function that controls the~$J_B^{(2)}$ two-point function, as in~\jjpfinal, which appears on the right-hand side of the Ward identity~\momspward. 

Note that~\consone\ and~\constwo\ are linear equations for the structure functions~$A, B, C, D, E$, while~${\mathsf J}$ can be viewed as an inhomogenous source term. Consequently, the general solution of these equations can be obtained by adding to the general solution of the homogenous system (with~${\mathsf J} = 0$) any particular solution of the inhomogenous equations:
\begin{itemize}
\item The general solution of the homogenous system, with~${\mathsf J} = 0$, can be parametrized by a single structure function, which we take to be~$E(Q^2)$. Then
\eqn{
A(Q^2) = -C(Q^2) = Q^2 E(Q^2)~, \qquad B(Q^2) = - D(Q^2) = -\half Q^2 E(Q^2)~,
}
which leads to the following Lorentz structure,
\eqna{  
 \langle j^A_\mu(p) j^A_\nu(q) & J^B_{\rho\sigma}(-p-q)\rangle \; \supset \;  E(Q^2)  \bigg( \left(p_\mu p_\nu - q_\mu q_\nu\right) \left(p_\rho q_\sigma - p_\sigma q_\rho\right) \cr
& + Q^2 \delta_{\mu\rho} \big(p_\nu + \half q_\nu\big)\left(p_\sigma -q_\sigma\right) - Q^2 \delta_{\mu\sigma} \big(p_\nu + \half q_\nu\big)\left(p_\rho - q_\rho\right) \cr
& + Q^2 \delta_{\nu\rho} \big(q_\mu + \half p_\mu\big) \left(q_\sigma - p_\sigma\right)  - Q^2 \delta_{\nu\sigma} \big(q_\mu + \half p_\mu\big) \left(q_\rho - p_\rho\right) \bigg)~.
}[parevcons]
Using~\symkinii, it is straightforward to verify that this structure is annihilated by both $p^\mu$ and~$(p+q)^\rho$.

\item We also need a particular solution to the inhomogenous system, where the source~${\mathsf J}$ is turned on. Since the structure in~\parevcons\ was parametrized by~$E(Q^2)$, it is convenient to chose a particular inhomogenous solution with~$E(Q^2) = 0$, 
\eqn{
A(Q^2) = -B(Q^2) = - C(Q^2) = D(Q^2)  = {\hat \kappa_A \over 2 \pi Q^2} \, {\mathsf J}\left({Q^2 \over M^2}\right)~, \qquad E(Q^2) = 0~. 
}
This gives rise to the Lorentz structure in~\jajajbwardidstr, 
\eqna{
& \langle j^A_\mu(p)  j^A_\nu(q)  J^B_{\rho\sigma}(-p-q)\rangle \; \supset \;  {\hat \kappa_A \over 2 \pi Q^2} \; {\mathsf J}\bigg({Q^2 \over M^2}\bigg)  \bigg( \delta_{\mu\rho} \left(p_\nu+q_\nu\right)\left(p_\sigma-q_\sigma\right)\cr
&  \qquad - \delta_{\mu\sigma} \left(p_\nu + q_\nu\right)\left(p_\rho - q_\rho\right) + \delta_{\nu\rho} \left(p_\mu + q_\mu\right) \left(q_\sigma - p_\sigma\right) - \delta_{\nu\sigma} \left(p_\mu + q_\mu\right)\left(q_\rho - p_\rho\right) \bigg)~.
}[jajajbwardidstrbis]
Again one can use~\symkinii\ to check that this structure is annihilated by~$(p+q)^\rho$ and satisfies the Ward identity~\momspward.  

\end{itemize}

\subsec{The~$\langle j_A^{(1)} j_A^{(1)} j_A^{(1)}\rangle$ Three-Point Function}[jajajaapp]

Here we consider the three-point function~$\langle j^A_\mu(p_1) j^A_\nu(p_2) j^A_\rho(p_3) \rangle$ of three identical 1-form currents~$j_A^{(1)}$. The momenta satisfy~$p_1 + p_2 + p_3$. In position space, these currents have mass dimension~$[j_A^{(1)}] = 3$, and hence the momentum-space three-point function has dimension~$+1$. Bose symmetry arbitrarily permutes the pairs 
\eqn{
\mu~,~p_1 \quad \longleftrightarrow \quad \nu~,~p_2 \quad \longleftrightarrow \quad \rho~,~p_3~.
}[bosejajaja]
We would like to decompose the three-point function into Lorentz structures. For our purposes, it suffices to focus on parity-odd structures, which contain an explicit~$\ep$-symbol. We can simplify the action~\bosejajaja\ of Bose symmetry by following~\cite{Coleman:1982yg} and specializing the momenta to configurations that satisfy (see also appendix~\ref{ssjajaJBthreept} above),
\eqn{
p_1^2 = p_2^2 = p_3^2 = Q^2~, \qquad p_1 + p_2 + p_3 = 0~,
}[bosemomapp]
where~$Q$ is a Lorentz-scalar quantity with dimensions of energy. A more general analysis, which is also valid away from these special momenta, was carried out in~\cite{Frishman:1980dq}. All structure functions only depend on~$Q^2$ and are therefore invariant under the Bose exchanges~\bosejajaja. There is in fact a unique parity-odd Lorentz structure that satisfies this requirement (see for instance section~2 of~\cite{Coleman:1982yg}),
\eqna{
\langle j_\mu^A(p_1) j_\nu^A(p_2) j_\rho^A(p_3)\rangle \; \supset \; {1 \over Q^2} \, {\mathsf A}\left({Q^2 \over M^2}\right) \bigg(\ep_{\mu\nu\alpha\beta} \, p_1^\alpha p_2^\beta  p_{3\rho} + \ep_{\nu\rho\alpha\beta} \, p_2^\alpha p_3^\beta p_{1\mu} + \ep_{\rho\mu\alpha\beta} \, p_3^\alpha p_1^\beta p_{2\nu}\bigg)~.
}[poddjajajasf]
Here~$\mathsf A$ is a dimensionless structure function, and~$M$ is some mass scale. Contracting both sides with~$p_1^\mu$ leads to
\eqn{
p_1^\mu \, \langle j_\mu^A(p_1) j_\nu^A(p_2) j_\rho^A(p_3)\rangle = {\mathsf A} \left({Q^2 \over M^2}\right) \, \ep_{\nu\rho\alpha\beta} \, p_2^\alpha p_3^\beta~.
}[pjajajaanom]
This formula is a basic ingredient in our analysis of 't Hooft anomalies in sections~\ref{secModHooft} and~\ref{sssHoftAbTwGp}.

\newsec{Aspects of Free Maxwell Theory}[appHooftFreeMax]

In this appendix we briefly recall some basic facts about Maxwell theory, i.e.~free~$U(1)_c^{(0)}$ gauge theory with field strength~$f^{(2)}_c = dc^{(1)}$ (see~\cite{Gaiotto:2014kfa} for additional details and references). The theory has two 1-form global symmetries, one electric (e) and one magnetic (m), 
\eqn{
U(1)_\text{e}^{(1)} \times U(1)_\text{m}^{(1)}~.
}[fmaxsym]
The corresponding currents are
\eqn{
J_\text{e}^{(2)} = -{1 \over e^2} \, f^{(2)}_c~, \qquad J^{(2)}_\text{m} = {i \over 2 \pi} * f^{(2)}_c~. 
}[fmaxcurr]
They are conserved if we use the source-free Maxwell equations, $d * f^{(2)}_c = d f^{(2)}_c = 0$.  

The background fields that couple to the currents in~\fmaxcurr\ are~$B^{(2)}_\text{e}$ and~$B^{(2)}_\text{m}$, with 1-form background gauge transformations parametrized by~$\Lambda^{(1)}_{\text{e}, \text{m}}$ (see also the discussion around~\eqref{bgtshift} and~\eqref{LamHflux}),
\eqn{
B^{(2)}_{\text{e},\text{m}} \; \rightarrow \; B^{(2)}_{\text{e},\text{m}} + d\Lambda^{(1)}_\text{\text{e},\text{m}}~.
}[ofgtem]
The electric description of the theory is based on the dynamical~$U(1)^{(0)}_c$ gauge field~$c^{(1)}$, which satisfies~$f^{(2)}_c = d c^{(1)}$. It shifts under~$U(1)_\text{e}^{(1)}$ background gauge transformations, but is neutral under~$U(1)_\text{m}^{(1)}$, and hence the same is true for~$f^{(2)}_c$, 
\eqn{
c^{(1)} \quad \longrightarrow \quad c^{(1)} + \Lambda_\text{e}^{(1)}~, \qquad f_c^{(2)} \quad \longrightarrow \quad f_c^{(2)} + d \Lambda_\text{e}^{(1)}~.
}[cflameshift]
We take the action of the theory coupled to background fields to be\foot{~For simplicity, we do not include a~$\theta$-term.}
\eqn{
S[B^{(2)}_\text{e}, B^{(2)}_\text{m}, c^{(1)}] = {1 \over 2 e^2} \int \left( f^{(2)}_c- B^{(2)}_\text{e}\right) \wedge *\left( f^{(2)}_c- B^{(2)}_\text{e}\right) + {i \over 2 \pi} \int B^{(2)}_\text{m} \wedge f^{(2)}_c~.
}[maxacbf]
This action includes the couplings~$S \supset \int \big( B^{(2)}_\text{e} \wedge * J_\text{e}^{(2)} + B^{(2)}_\text{m} \wedge * J_\text{m}^{(2)}\big)$ to the currents in~\fmaxcurr, as well as a seagull counterterm~$\sim \int B^{(2)}_\text{e} \wedge * B^{(2)}_\text{e}$, which ensures that the kinetic term is invariant under~\cflameshift. The second term is invariant under~$B^{(2)}_m$ gauge transformations, since~$f^{(2)}_c = d c^{(1)}$ is automatically closed. However, it leads to a c-number shift under~$U(1)_\text{e}^{(1)}$ background gauge transformations, so that
\eqn{
S\Big[B^{(2)}_\text{e} + d \Lambda_\text{e}^{(1)}, B^{(2)}_\text{m} + d \Lambda_\text{m}^{(1)}, c^{(1)} + \Lambda_\text{e}^{(1)}\Big] = S[B^{(2)}_\text{e}, B^{(2)}_\text{m}, c^{(1)}] + {i \over 2 \pi} \int \Lambda_\text{e}^{(1)} \wedge dB ^{(2)}_\text{m}~.
}[maxgtshift]

The shift in~\maxgtshift\ constitutes a mixed 't Hooft anomaly between~$U(1)^{(1)}_\text{e}$ and~$U(1)_\text{m}^{(1)}$, which cannot be removed using local counterterms. It can be viewed as arising (via inflow) from the following five-dimensional topological action for the background fields, 
\eqn{
S_5[B^{(2)}_\text{e}, B^{(2)}_\text{m}] = {i \over 2 \pi} \int_{\CM_5} B^{(2)}_e \wedge dB^{(2)}_\text{m}~,
}[bebminflow]
or equivalently, from a term~$\CI^{(6)} \supset {1 \over 4 \pi^2} \,dB^{(2)}_\text{e} \wedge dB^{(2)}_\text{m}$ in the  6-form anomaly polynomial. As is typical of mixed anomalies, we can change the presentation of the anomaly by adding local counterterms. If we integrate~\bebminflow\ by parts and add a four-dimensional counterterm~$\sim \int_{\CM_4} B^{(2)}_\text{e} \wedge B^{(2)}_\text{m}$ to cancel the resulting boundary contribution (which amounts to replacing~$f^{(2)}_c \rightarrow f^{(2)}_c - B^{(2)}_\text{e}$ in the last term of~\maxacbf), the five-dimensional action becomes~$\sim \int_{\CM_5} B^{(2)}_\text{m} \wedge d B^{(2)}_\text{e}$. This is invariant under~$U(1)_\text{e}^{(1)}$, but gives rise to an 't Hooft anomaly under~$U(1)_\text{m}^{(1)}$ background gauge transformations.  

It is instructive to examine electric-magnetic duality in the presence of the background fields~$B^{(2)}_{\text{e}, \text{m}}$. As usual, we dualize~$f^{(2)}_c$ by considering an extended theory that includes a Lagrange multiplier~$\t c^{(1)}$,
\eqn{
\t S[B^{(2)}_\text{e}, B^{(2)}_\text{m}, c^{(1)}, \t c^{(1)}] = S[B^{(2)}_\text{e}, B^{(2)}, c^{(1)}] - {i \over 2 \pi} \int d \t c^{(1)} \wedge f_c^{(2)}~. 
}[sextd]
The Lagrange multiplier~$\t c^{(1)}$ is also a 1-form gauge field, associated with its own~$U(1)^{(0)}_{\t c}$ gauge symmetry. Integrating over~$\t c^{(1)}$ enforces the Bianchi identity on~$f^{(2)}_c$, which is now an unconstrained two-form field. Moreover, summing over the fluxes of~$\t c^{(1)}$, which satisfy the usual quantization condition~${1 \over 2 \pi} \int_{\Sigma_2} d \t c^{(1)} \in \Z$, ensures that the fluxes of~$f^{(2)}_c$ satisfy the same quantization condition. (This requirement fixes the normalization of the coupling between~$\t c^{(1)}$ and~$f_c^{(2)}$ in~\sextd.) In order to maintain invariance under background gauge transformations (up to the 't Hooft anomaly in~\maxgtshift), we must assign the following shift to~$\t c^{(1)}$ under~$U(1)_\text{m}^{(1)}$ gauge transformations,
\eqn{
\t c^{(1)} \quad \longrightarrow \quad \t c^{(1)} + \Lambda_\text{m}^{(1)}~.
}[cmagshift]
We can now integrate out the unconstrained two-form~$f^{(2)}_c$ using its equation of motion,
\eqn{
* \left(f^{(2)}_c - B^{(2)}_\text{e}\right) = {i e^2 \over 2 \pi} \left( d \t c^{(1)} - B^{(2)}_\text{m}\right)~.
}[dualityeom]
Substituting back into~\sextd, we obtain a dual presentation of the theory in terms of the magnetic gauge field~$\t c^{(1)}$,
\eqn{
\t S[B^{(1)}_\text{e}, B^{(1)}_\text{m}, \t c^{(1)}] = {e^2 \over 8 \pi^2} \int \left(d\t c^{(1)} - B^{(2)}_\text{m}\right) \wedge * \left( d \t c^{(1)} - B^{(2)}_\text{m}\right) -{i \over 2 \pi} \int B^{(2)}_\text{e} \wedge \left(d\t c^{(1)} - B^{(2)}_\text{m}\right)~.
}[magaction]
We can therefore identify the magnetic coupling~$\t e^2 = {4 \pi^2 \over e^2}$ and the currents in~\fmaxcurr, which are given by~$J_\text{e}^{(2)} = - {i \over 2 \pi} * d\t c^{(1)} = - \t J_\text{m}^{(2)}$ and~$J_\text{m}^{(2)} = - {1 \over \t e^2} \, d \t c^{(1)} = \t J_\text{e}^{(2)}$.\foot{~The relative sign in the transformation of the currents is a standard property of electric-magnetic duality.} Note that the duality automatically generates a counterterm~$\sim \int B^{(2)}_\text{e} \wedge B^{(2)}_\text{m}$ in~\magaction, which ensures that the mixed~$U(1)_\text{e}^{(1)}$\,-\,$U(1)_\text{m}^{(1)}$ 't Hooft anomaly takes the same form as in~\maxgtshift. 

Wilson loops~$W_m(L)$ of charge~$m \in \Z$ and 't Hooft loops~$H_n(L)$ of charge~$n \in \Z$ are defined as holonomies of the electric and magnetic gauge fields~$c^{(1)}$ and~$\t c^{(1)}$ around a closed 1-cycle~$L$, 
\eqn{
W_m(L) = \exp\left(i m \int_L c^{(1)}\right)~, \qquad H_n(L) = \exp\left(i n \int_L \t c^{(1)}\right)~.
}[whloops]
Electric-magnetic duality exchanges~$c^{(1)} \leftrightarrow \t c^{(1)}$, and hence~$W_m(L) \leftrightarrow H_{m}(L)$. Using~\cflameshift\ and~\cmagshift, we see that the charges of~$W_m(L)$ and~$H_n(L)$ under~$U(1)^{(1)}_\text{e} \times U(1)_\text{m}^{(1)}$ are~$(m,0)$ and~$(0,n)$, respectively. It is often useful to express the loop operators in~\whloops\ as open surface operators, which are obtained by integrating~$f^{(2)}_c$ and~$* f^{(2)}_c$ over a 2-cycle~$\Sigma_2$ with boundary~$\d \Sigma_2 = L$.  In the electric description, where~$f^{(2)}_c = dc ^{(1)}$, this is straightforward for the Wilson loop, 
\eqn{
W_m(L) = \exp\left(i m \int_{\Sigma_2} f^{(2)}_c\right)~.
}[wilsurf]
Similarly, we can use the duality relation~\dualityeom\ to obtain the following presentation of the 't Hooft loop in terms of purely electric variables,
\eqn{
H_n(L) = \exp\left({2 \pi n \over e^2} \int_{\Sigma_2} * \left(f^{(2)}_c -B^{(2)}_\text{e}\right) + i n \int_{\Sigma_2} B^{(2)}_\text{m}\right)~.
}[hooftsurf]
Here~$\Sigma_2$ can be viewed as the worldsheet of an unobservable Dirac string used to define the 't Hooft loop. The surface counterterms~$\sim B^{(2)}_{\text{e}, \text{m}}$ ensure some important properties of~$H_n(L)$:
\begin{itemize}
\item It transforms correctly, with charges~$(0,n)$, under~$U(1)_\text{e}^{(1)} \times U(1)_\text{m}^{(1)}$.
\item It is invariant under small deformations of the bounding surface~$\Sigma_2$, because the 2-form integrand in the exponent of~\hooftsurf\ is closed. This follows from the equations of motion for the action~\maxacbf,
\eqn{
d \left( {1 \over e^2} * \left(f^{(2)}_c - B_\text{e}^{(2)}\right) + {i \over 2 \pi} \, B^{(2)}_\text{m}\right) = 0~.
}[bgfmeom]
Since this statement holds in the presence of the background fields~$B^{(2)}_{\text{e}, \text{m}}$, it also applies to deformations of~$\Sigma_2$ that cross insertions of the field strength~$f^{(2)}_c$ and its dual~$* f^{(2)}_c$, which are obtained by taking variational derivatives with respect to~$B^{(2)}_{\text{e}, \text{m}}$. This shows that the Dirac string cannot be detected using such insertions.

\item The independence of~$H_n(L)$ on the choice of~$\Sigma_2$ continues to hold in the presence of charged Wilson lines. To see this, consider another 2-cycle~$\Sigma'_2$ with~$\d \Sigma'_2 = L$. The union of~$\Sigma_2$ and the orientation-reversal of~$\Sigma'_2$ is a closed 2-cycle, which can be viewed as the boundary~$\d \Sigma_3$ of its interior~$\Sigma_3$. Using~$\Sigma'_2$ rather than~$\Sigma_2$ in~\hooftsurf\ changes the exponent by an integral over~$\d \Sigma_3$, which can be evaluated using Stokes' theorem and the equations of motion~\bgfmeom. In the presence of a Wilson line with charge~$m$, \bgfmeom\ acquires a term~$i m \delta^{(3)}_L$ on its right-hand side.\foot{~Here~$\delta^{(3)}_L$ is a 3-form current with support on the line, so that~$\int_L c^{(1)} = \int_{\CM_4} c^{(1)} \wedge \delta^{(3)}_L$.} Therefore, using~$\Sigma'_2$ rather than~$\Sigma_2$ multiplies~$H_n(L)$ by a phase~$e^{i \varphi}$, with~$\varphi = 2 \pi m_\text{tot.} n$. Here~$m_\text{tot.}$ is the net charge of all Wilson lines passing through the region~$\Sigma_3$. Since each Wilson line carries integer charge, it follows that~$m_\text{tot.} \in \Z$, so that~$e^{i \varphi} = 1$. The Dirac string thus remains unobservable in the presence of Wilson lines and the background fields~$B^{(2)}_{\text{e},\text{m}}$.

\end{itemize}

\end{appendices}


\bibliography{bibliography}

\providecommand{\href}[2]{#2}\begingroup\raggedright\begin{thebibliography}{10}

\bibitem{Gaiotto:2014kfa}
D.~Gaiotto, A.~Kapustin, N.~Seiberg, and B.~Willett, ``{Generalized Global
  Symmetries},'' \href{http://dx.doi.org/10.1007/JHEP02(2015)172}{{\em JHEP}
  {\bfseries 02} (2015) 172},
\href{http://arxiv.org/abs/1412.5148}{{\ttfamily arXiv:1412.5148 [hep-th]}}.

\bibitem{Closset:2012vg}
C.~Closset, T.~T. Dumitrescu, G.~Festuccia, Z.~Komargodski, and N.~Seiberg,
  ``{Contact Terms, Unitarity, and F-Maximization in Three-Dimensional
  Superconformal Theories},''
  \href{http://dx.doi.org/10.1007/JHEP10(2012)053}{{\em JHEP} {\bfseries 10}
  (2012) 053},
\href{http://arxiv.org/abs/1205.4142}{{\ttfamily arXiv:1205.4142 [hep-th]}}.

\bibitem{Closset:2012vp}
C.~Closset, T.~T. Dumitrescu, G.~Festuccia, Z.~Komargodski, and N.~Seiberg,
  ``{Comments on Chern-Simons Contact Terms in Three Dimensions},''
  \href{http://dx.doi.org/10.1007/JHEP09(2012)091}{{\em JHEP} {\bfseries 09}
  (2012) 091},
\href{http://arxiv.org/abs/1206.5218}{{\ttfamily arXiv:1206.5218 [hep-th]}}.

\bibitem{our6dtoapp}
C.~Cordova, T.~T. Dumitrescu, and K.~A. Intriligator ,~to appear.

\bibitem{Dumitrescu:2011iu}
T.~T. Dumitrescu and N.~Seiberg, ``{Supercurrents and Brane Currents in Diverse
  Dimensions},'' \href{http://dx.doi.org/10.1007/JHEP07(2011)095}{{\em JHEP}
  {\bfseries 07} (2011) 095},
\href{http://arxiv.org/abs/1106.0031}{{\ttfamily arXiv:1106.0031 [hep-th]}}.

\bibitem{Green:1984sg}
M.~B. Green and J.~H. Schwarz, ``{Anomaly Cancellation in Supersymmetric D=10
  Gauge Theory and Superstring Theory},''
\href{http://dx.doi.org/10.1016/0370-2693(84)91565-X}{{\em Phys. Lett.}
  {\bfseries 149B} (1984) 117--122}.

\bibitem{Green:1987mn}
M.~B. Green, J.~H. Schwarz, and E.~Witten, {\em {Superstring Theory Vol. 2:
  Loop Amplitudes, Anomalies and Phenomenology}}.
\newblock Cambridge University Press,
1987.
\newblock

\bibitem{Polchinski:1998rr}
J.~Polchinski, {\em {String Theory Vol. 2: Superstring Theory and Beyond}}.
\newblock Cambridge University Press,
1998.
\newblock

\bibitem{Candelas:1985en}
P.~Candelas, G.~T. Horowitz, A.~Strominger, and E.~Witten, ``{Vacuum
  Configurations for Superstrings},''
\href{http://dx.doi.org/10.1016/0550-3213(85)90602-9}{{\em Nucl. Phys.}
  {\bfseries B258} (1985) 46--74}.

\bibitem{Strominger:1986uh}
A.~Strominger, ``{Superstrings with Torsion},''
\href{http://dx.doi.org/10.1016/0550-3213(86)90286-5}{{\em Nucl. Phys.}
  {\bfseries B274} (1986) 253}.

\bibitem{Kapustin:2013uxa}
A.~Kapustin and R.~Thorngren, ``{Higher symmetry and gapped phases of gauge
  theories},''
\href{http://arxiv.org/abs/1309.4721}{{\ttfamily arXiv:1309.4721 [hep-th]}}.

\bibitem{baez2004higher}
J.~C. Baez and A.~D. Lauda, ``Higher-dimensional algebra v: 2-groups,'' {\em
  Version} {\bfseries 3} (2004) 423--491.

\bibitem{Baez:2004in}
J.~Baez and U.~Schreiber, ``{Higher gauge theory: 2-connections on
  2-bundles},''
\href{http://arxiv.org/abs/hep-th/0412325}{{\ttfamily arXiv:hep-th/0412325
  [hep-th]}}.

\bibitem{Schreiber:2008}
U.~Schreiber and K.~Waldorf, ``{Connections on non-Abelian Gerbes and their
  Holonomy},'' {\em Theory Appl. Categ.} {\bfseries 28} (2013) 476--540,
  \href{http://arxiv.org/abs/0808.1923}{{\ttfamily arXiv:0808.1923 [math.DG]}}.

\bibitem{Dijkgraaf:1989pz}
R.~Dijkgraaf and E.~Witten, ``{Topological Gauge Theories and Group
  Cohomology},''
\href{http://dx.doi.org/10.1007/BF02096988}{{\em Commun. Math. Phys.}
  {\bfseries 129} (1990) 393}.

\bibitem{Tachikawa:2017gyf}
Y.~Tachikawa, ``{On gauging finite subgroups},''
\href{http://arxiv.org/abs/1712.09542}{{\ttfamily arXiv:1712.09542 [hep-th]}}.

\bibitem{Sharpe:2015mja}
E.~Sharpe, ``{Notes on generalized global symmetries in QFT},''
  \href{http://dx.doi.org/10.1002/prop.201500048}{{\em Fortsch. Phys.}
  {\bfseries 63} (2015) 659--682},
\href{http://arxiv.org/abs/1508.04770}{{\ttfamily arXiv:1508.04770 [hep-th]}}.

\bibitem{tHooft:1979rat}
G.~'t~Hooft, ``{Naturalness, chiral symmetry, and spontaneous chiral symmetry
  breaking},''
\href{http://dx.doi.org/10.1007/978-1-4684-7571-5_9}{{\em NATO Sci. Ser. B}
  {\bfseries 59} (1980) 135--157}.

\bibitem{Mack:1975je}
G.~Mack, ``{All unitary ray representations of the conformal group SU(2,2) with
  positive energy},''
\href{http://dx.doi.org/10.1007/BF01613145}{{\em Commun. Math. Phys.}
  {\bfseries 55} (1977) 1}.

\bibitem{Minwalla:1997ka}
S.~Minwalla, ``{Restrictions imposed by superconformal invariance on quantum
  field theories},'' \href{http://dx.doi.org/10.4310/ATMP.1998.v2.n4.a4}{{\em
  Adv. Theor. Math. Phys.} {\bfseries 2} (1998) 783--851},
\href{http://arxiv.org/abs/hep-th/9712074}{{\ttfamily arXiv:hep-th/9712074
  [hep-th]}}.

\bibitem{Cordova:2016emh}
C.~Cordova, T.~T. Dumitrescu, and K.~Intriligator, ``{Multiplets of
  Superconformal Symmetry in Diverse Dimensions},''
\href{http://arxiv.org/abs/1612.00809}{{\ttfamily arXiv:1612.00809 [hep-th]}}.

\bibitem{Frishman:1980dq}
Y.~Frishman, A.~Schwimmer, T.~Banks, and S.~Yankielowicz, ``{The Axial Anomaly
  and the Bound State Spectrum in Confining Theories},''
\href{http://dx.doi.org/10.1016/0550-3213(81)90268-6}{{\em Nucl. Phys.}
  {\bfseries B177} (1981) 157--171}.

\bibitem{Coleman:1982yg}
S.~R. Coleman and B.~Grossman, ``{'t Hooft's Consistency Condition as a
  Consequence of Analyticity and Unitarity},''
\href{http://dx.doi.org/10.1016/0550-3213(82)90028-1}{{\em Nucl. Phys.}
  {\bfseries B203} (1982) 205--220}.

\bibitem{Maldacena:2001ss}
J.~M. Maldacena, G.~W. Moore, and N.~Seiberg, ``{D-brane charges in five-brane
  backgrounds},'' \href{http://dx.doi.org/10.1088/1126-6708/2001/10/005}{{\em
  JHEP} {\bfseries 10} (2001) 005},
\href{http://arxiv.org/abs/hep-th/0108152}{{\ttfamily arXiv:hep-th/0108152
  [hep-th]}}.

\bibitem{Banks:2010zn}
T.~Banks and N.~Seiberg, ``{Symmetries and Strings in Field Theory and
  Gravity},'' \href{http://dx.doi.org/10.1103/PhysRevD.83.084019}{{\em Phys.
  Rev.} {\bfseries D83} (2011) 084019},
\href{http://arxiv.org/abs/1011.5120}{{\ttfamily arXiv:1011.5120 [hep-th]}}.

\bibitem{Kapustin:2014gua}
A.~Kapustin and N.~Seiberg, ``{Coupling a QFT to a TQFT and Duality},''
  \href{http://dx.doi.org/10.1007/JHEP04(2014)001}{{\em JHEP} {\bfseries 04}
  (2014) 001},
\href{http://arxiv.org/abs/1401.0740}{{\ttfamily arXiv:1401.0740 [hep-th]}}.

\bibitem{Gukov:2013zka}
S.~Gukov and A.~Kapustin, ``{Topological Quantum Field Theory, Nonlocal
  Operators, and Gapped Phases of Gauge Theories},''
\href{http://arxiv.org/abs/1307.4793}{{\ttfamily arXiv:1307.4793 [hep-th]}}.

\bibitem{Kapustin:2013qsa}
A.~Kapustin and R.~Thorngren, ``{Topological Field Theory on a Lattice,
  Discrete Theta-Angles and Confinement},''
  \href{http://dx.doi.org/10.4310/ATMP.2014.v18.n5.a4}{{\em Adv. Theor. Math.
  Phys.} {\bfseries 18} no.~5, (2014) 1233--1247},
\href{http://arxiv.org/abs/1308.2926}{{\ttfamily arXiv:1308.2926 [hep-th]}}.

\bibitem{Kapustin:2014zva}
A.~Kapustin and R.~Thorngren, ``{Anomalies of discrete symmetries in various
  dimensions and group cohomology},''
\href{http://arxiv.org/abs/1404.3230}{{\ttfamily arXiv:1404.3230 [hep-th]}}.

\bibitem{Thorngren:2015gtw}
R.~Thorngren and C.~von Keyserlingk, ``{Higher SPT's and a generalization of
  anomaly in-flow},''
\href{http://arxiv.org/abs/1511.02929}{{\ttfamily arXiv:1511.02929
  [cond-mat.str-el]}}.

\bibitem{Bhardwaj:2016clt}
L.~Bhardwaj, D.~Gaiotto, and A.~Kapustin, ``{State sum constructions of
  spin-TFTs and string net constructions of fermionic phases of matter},''
  \href{http://dx.doi.org/10.1007/JHEP04(2017)096}{{\em JHEP} {\bfseries 04}
  (2017) 096},
\href{http://arxiv.org/abs/1605.01640}{{\ttfamily arXiv:1605.01640
  [cond-mat.str-el]}}.

\bibitem{BeniniToApp}
F.~Benini, C.~Cordova, and P.-S. Hsin, ``On 2-group global symmetries and their
  anomalies.'' ,~to appear.

\bibitem{etingof2009fusion}
P.~Etingof, D.~Nikshych, V.~Ostrik, and E.~Meir, ``{Fusion categories and
  homotopy theory},''
\href{http://arxiv.org/abs/0909.3140}{{\ttfamily arXiv:0909.3140 [math.QA]}}.

\bibitem{Barkeshli:2014cna}
M.~Barkeshli, P.~Bonderson, M.~Cheng, and Z.~Wang, ``{Symmetry, Defects, and
  Gauging of Topological Phases},''
\href{http://arxiv.org/abs/1410.4540}{{\ttfamily arXiv:1410.4540
  [cond-mat.str-el]}}.

\bibitem{Barkeshli:2017rzd}
M.~Barkeshli and M.~Cheng, ``{Time-reversal and spatial reflection symmetry
  localization anomalies in (2+1)D topological phases of matter},''
\href{http://arxiv.org/abs/1706.09464}{{\ttfamily arXiv:1706.09464
  [cond-mat.str-el]}}.

\bibitem{Seiberg:1997zk}
N.~Seiberg, ``{New theories in six-dimensions and matrix description of M
  theory on T**5 and T**5 / Z(2)},''
  \href{http://dx.doi.org/10.1016/S0370-2693(97)00805-8}{{\em Phys. Lett.}
  {\bfseries B408} (1997) 98--104},
\href{http://arxiv.org/abs/hep-th/9705221}{{\ttfamily arXiv:hep-th/9705221
  [hep-th]}}.

\bibitem{Witten:1995gx}
E.~Witten, ``{Small instantons in string theory},''
  \href{http://dx.doi.org/10.1016/0550-3213(95)00625-7}{{\em Nucl. Phys.}
  {\bfseries B460} (1996) 541--559},
\href{http://arxiv.org/abs/hep-th/9511030}{{\ttfamily arXiv:hep-th/9511030
  [hep-th]}}.

\bibitem{Schwarz:1995zw}
J.~H. Schwarz, ``{Anomaly - free supersymmetric models in six-dimensions},''
  \href{http://dx.doi.org/10.1016/0370-2693(95)01610-4}{{\em Phys. Lett.}
  {\bfseries B371} (1996) 223--230},
\href{http://arxiv.org/abs/hep-th/9512053}{{\ttfamily arXiv:hep-th/9512053
  [hep-th]}}.

\bibitem{Ohmori:2014kda}
K.~Ohmori, H.~Shimizu, Y.~Tachikawa, and K.~Yonekura, ``{Anomaly polynomial of
  general 6d SCFTs},'' \href{http://dx.doi.org/10.1093/ptep/ptu140}{{\em PTEP}
  {\bfseries 2014} no.~10, (2014) 103B07},
\href{http://arxiv.org/abs/1408.5572}{{\ttfamily arXiv:1408.5572 [hep-th]}}.

\bibitem{Cordova:2015vwa}
C.~Cordova, T.~T. Dumitrescu, and X.~Yin, ``{Higher Derivative Terms, Toroidal
  Compactification, and Weyl Anomalies in Six-Dimensional (2,0) Theories},''
\href{http://arxiv.org/abs/1505.03850}{{\ttfamily arXiv:1505.03850 [hep-th]}}.

\bibitem{Cordova:2015fha}
C.~Cordova, T.~T. Dumitrescu, and K.~Intriligator, ``{Anomalies,
  renormalization group flows, and the a-theorem in six-dimensional (1, 0)
  theories},'' \href{http://dx.doi.org/10.1007/JHEP10(2016)080}{{\em JHEP}
  {\bfseries 10} (2016) 080},
\href{http://arxiv.org/abs/1506.03807}{{\ttfamily arXiv:1506.03807 [hep-th]}}.

\bibitem{AlvarezGaume:1985ex}
L.~Alvarez-Gaume, ``{An Introduction to Anomalies},'' in {\em {International
  School of Mathematical Physics: 6th Course: Fundamental Problems of Gauge
  Field Theory Erice, Italy, July 1-14, 1985}}.
\newblock
1985.
\newblock

\bibitem{Weinberg:1996kr}
S.~Weinberg, {\em {The quantum theory of fields. Vol. 2: Modern applications}}.
\newblock Cambridge University Press,
2013.
\newblock

\bibitem{Harvey:2005it}
J.~A. Harvey, ``{TASI 2003 lectures on anomalies},''
\newblock 2005.
\newblock
\href{http://arxiv.org/abs/hep-th/0509097}{{\ttfamily arXiv:hep-th/0509097
  [hep-th]}}.
\newblock

\bibitem{Wess:1971yu}
J.~Wess and B.~Zumino, ``{Consequences of anomalous Ward identities},''
\href{http://dx.doi.org/10.1016/0370-2693(71)90582-X}{{\em Phys. Lett.}
  {\bfseries 37B} (1971) 95--97}.

\bibitem{Bardeen:1984pm}
W.~A. Bardeen and B.~Zumino, ``{Consistent and Covariant Anomalies in Gauge and
  Gravitational Theories},''
\href{http://dx.doi.org/10.1016/0550-3213(84)90322-5}{{\em Nucl. Phys.}
  {\bfseries B244} (1984) 421--453}.

\bibitem{Adler:1969gk}
S.~L. Adler, ``{Axial vector vertex in spinor electrodynamics},''
\href{http://dx.doi.org/10.1103/PhysRev.177.2426}{{\em Phys. Rev.} {\bfseries
  177} (1969) 2426--2438}.

\bibitem{Bell:1969ts}
J.~S. Bell and R.~Jackiw, ``{A PCAC puzzle: $\pi^0 \rightarrow \gamma \gamma$
  in the sigma model},''
\href{http://dx.doi.org/10.1007/BF02823296}{{\em Nuovo Cim.} {\bfseries A60}
  (1969) 47--61}.

\bibitem{Adler:1969er}
S.~L. Adler and W.~A. Bardeen, ``{Absence of higher order corrections in the
  anomalous axial vector divergence equation},''
\href{http://dx.doi.org/10.1103/PhysRev.182.1517}{{\em Phys. Rev.} {\bfseries
  182} (1969) 1517--1536}.

\bibitem{Coleman:1985zi}
S.~R. Coleman and B.~R. Hill, ``{No More Corrections to the Topological Mass
  Term in QED in Three-Dimensions},''
\href{http://dx.doi.org/10.1016/0370-2693(85)90883-4}{{\em Phys. Lett.}
  {\bfseries 159B} (1985) 184--188}.

\bibitem{AlvarezGaume:1984dr}
L.~Alvarez-Gaume and P.~H. Ginsparg, ``{The Structure of Gauge and
  Gravitational Anomalies},''
  \href{http://dx.doi.org/10.1016/0003-4916(85)90087-9}{{\em Annals Phys.}
  {\bfseries 161} (1985) 423}.
[Erratum: Annals Phys.171,233(1986)].

\bibitem{AlvarezGaume:1983ig}
L.~Alvarez-Gaume and E.~Witten, ``{Gravitational Anomalies},''
\href{http://dx.doi.org/10.1016/0550-3213(84)90066-X}{{\em Nucl. Phys.}
  {\bfseries B234} (1984) 269}.

\bibitem{Peccei:1977ur}
R.~D. Peccei and H.~R. Quinn, ``{Constraints Imposed by CP Conservation in the
  Presence of Instantons},''
\href{http://dx.doi.org/10.1103/PhysRevD.16.1791}{{\em Phys. Rev.} {\bfseries
  D16} (1977) 1791--1797}.

\bibitem{Argyres:1995xn}
P.~C. Argyres, M.~R. Plesser, N.~Seiberg, and E.~Witten, ``{New N=2
  superconformal field theories in four-dimensions},''
  \href{http://dx.doi.org/10.1016/0550-3213(95)00671-0}{{\em Nucl. Phys.}
  {\bfseries B461} (1996) 71--84},
\href{http://arxiv.org/abs/hep-th/9511154}{{\ttfamily arXiv:hep-th/9511154
  [hep-th]}}.

\bibitem{Cordova:2016xhm}
C.~Cordova, T.~T. Dumitrescu, and K.~Intriligator, ``{Deformations of
  Superconformal Theories},''
  \href{http://dx.doi.org/10.1007/JHEP11(2016)135}{{\em JHEP} {\bfseries 11}
  (2016) 135},
\href{http://arxiv.org/abs/1602.01217}{{\ttfamily arXiv:1602.01217 [hep-th]}}.

\bibitem{Siegel:1988gd}
W.~Siegel, ``{All Free Conformal Representations in All Dimensions},''
\href{http://dx.doi.org/10.1142/S0217751X89000819}{{\em Int. J. Mod. Phys.}
  {\bfseries A4} (1989) 2015}.

\bibitem{Weinberg:1980kq}
S.~Weinberg and E.~Witten, ``{Limits on Massless Particles},''
\href{http://dx.doi.org/10.1016/0370-2693(80)90212-9}{{\em Phys. Lett.}
  {\bfseries 96B} (1980) 59--62}.

\bibitem{Cordova:2017dhq}
C.~Cordova and K.~Diab, ``{Universal Bounds on Operator Dimensions from the
  Average Null Energy Condition},''
\href{http://arxiv.org/abs/1712.01089}{{\ttfamily arXiv:1712.01089 [hep-th]}}.

\bibitem{Chang:2017cdx}
C.-M. Chang, M.~Fluder, Y.-H. Lin, and Y.~Wang, ``{Spheres, Charges,
  Instantons, and Bootstrap: A Five-Dimensional Odyssey},''
\href{http://arxiv.org/abs/1710.08418}{{\ttfamily arXiv:1710.08418 [hep-th]}}.

\bibitem{Witten:1982fp}
E.~Witten, ``{An SU(2) Anomaly},''
\href{http://dx.doi.org/10.1016/0370-2693(82)90728-6}{{\em Phys. Lett.}
  {\bfseries 117B} (1982) 324--328}.

\bibitem{Georgi:1972cj}
H.~Georgi and S.~L. Glashow, ``{Unified weak and electromagnetic interactions
  without neutral currents},''
\href{http://dx.doi.org/10.1103/PhysRevLett.28.1494}{{\em Phys. Rev. Lett.}
  {\bfseries 28} (1972) 1494}.

\bibitem{wiki:xxx}
Wikipedia, ``Euler's sum of powers conjecture,'' 2017.
\newblock
  \url{https://en.wikipedia.org/w/index.php?title=Euler%27s_sum_of_powers_conjecture&oldid=812866572}.

\bibitem{Seiberg:2010qd}
N.~Seiberg, ``{Modifying the Sum Over Topological Sectors and Constraints on
  Supergravity},'' \href{http://dx.doi.org/10.1007/JHEP07(2010)070}{{\em JHEP}
  {\bfseries 07} (2010) 070},
\href{http://arxiv.org/abs/1005.0002}{{\ttfamily arXiv:1005.0002 [hep-th]}}.

\bibitem{Kapustin:1999ha}
A.~Kapustin and M.~J. Strassler, ``{On mirror symmetry in three-dimensional
  Abelian gauge theories},''
  \href{http://dx.doi.org/10.1088/1126-6708/1999/04/021}{{\em JHEP} {\bfseries
  04} (1999) 021},
\href{http://arxiv.org/abs/hep-th/9902033}{{\ttfamily arXiv:hep-th/9902033
  [hep-th]}}.

\bibitem{Witten:2003ya}
E.~Witten, ``{SL(2,Z) action on three-dimensional conformal field theories with
  Abelian symmetry},''
\href{http://arxiv.org/abs/hep-th/0307041}{{\ttfamily arXiv:hep-th/0307041
  [hep-th]}}.

\bibitem{Seiberg:2016gmd}
N.~Seiberg, T.~Senthil, C.~Wang, and E.~Witten, ``{A Duality Web in 2+1
  Dimensions and Condensed Matter Physics},''
  \href{http://dx.doi.org/10.1016/j.aop.2016.08.007}{{\em Annals Phys.}
  {\bfseries 374} (2016) 395--433},
\href{http://arxiv.org/abs/1606.01989}{{\ttfamily arXiv:1606.01989 [hep-th]}}.

\bibitem{Maldacena:1997re}
J.~M. Maldacena, ``{The Large N limit of superconformal field theories and
  supergravity},'' \href{http://dx.doi.org/10.1023/A:1026654312961}{{\em Int.
  J. Theor. Phys.} {\bfseries 38} (1999) 1113--1133},
  \href{http://arxiv.org/abs/hep-th/9711200}{{\ttfamily arXiv:hep-th/9711200
  [hep-th]}}.
[Adv. Theor. Math. Phys.2,231(1998)].

\bibitem{Gubser:1998bc}
S.~S. Gubser, I.~R. Klebanov, and A.~M. Polyakov, ``{Gauge theory correlators
  from noncritical string theory},''
  \href{http://dx.doi.org/10.1016/S0370-2693(98)00377-3}{{\em Phys. Lett.}
  {\bfseries B428} (1998) 105--114},
\href{http://arxiv.org/abs/hep-th/9802109}{{\ttfamily arXiv:hep-th/9802109
  [hep-th]}}.

\bibitem{Witten:1998qj}
E.~Witten, ``{Anti-de Sitter space and holography},''
  \href{http://dx.doi.org/10.4310/ATMP.1998.v2.n2.a2}{{\em Adv. Theor. Math.
  Phys.} {\bfseries 2} (1998) 253--291},
\href{http://arxiv.org/abs/hep-th/9802150}{{\ttfamily arXiv:hep-th/9802150
  [hep-th]}}.

\bibitem{Witten:2017hdv}
E.~Witten, ``{Symmetry and Emergence},''
\href{http://arxiv.org/abs/1710.01791}{{\ttfamily arXiv:1710.01791 [hep-th]}}.

\bibitem{Harlow:2018fse}
D.~Harlow, ``{TASI Lectures on the Emergence of the Bulk in AdS/CFT},''
\href{http://arxiv.org/abs/1802.01040}{{\ttfamily arXiv:1802.01040 [hep-th]}}.

\bibitem{HOtoApp}
D.~Harlow and H.~Ooguri ,~to appear.

\bibitem{Hofman:2017vwr}
D.~M. Hofman and N.~Iqbal, ``{Generalized global symmetries and holography},''
\href{http://arxiv.org/abs/1707.08577}{{\ttfamily arXiv:1707.08577 [hep-th]}}.

\bibitem{Witten:1984eb}
E.~Witten, ``{Superconducting Strings},''
\href{http://dx.doi.org/10.1016/0550-3213(85)90022-7}{{\em Nucl. Phys.}
  {\bfseries B249} (1985) 557--592}.

\bibitem{Jackiw:1981ee}
R.~Jackiw and P.~Rossi, ``{Zero Modes of the Vortex - Fermion System},''
\href{http://dx.doi.org/10.1016/0550-3213(81)90044-4}{{\em Nucl. Phys.}
  {\bfseries B190} (1981) 681--691}.

\bibitem{Weinberg:1981eu}
E.~J. Weinberg, ``{Index Calculations for the Fermion-Vortex System},''
\href{http://dx.doi.org/10.1103/PhysRevD.24.2669}{{\em Phys. Rev.} {\bfseries
  D24} (1981) 2669}.

\end{thebibliography}\endgroup

\end{document}